\documentclass[twoside,openright]{report}

\usepackage{HDR}
\usepackage{textcomp}			
\usepackage[nottoc,numbib]{tocbibind}
\usepackage{upgreek}				
\usepackage[utf8]{inputenc} 		
\usepackage[T1]{fontenc}    		
\usepackage[dvipsnames]{xcolor}	
\usepackage{hyperref}       		
\hypersetup{
    colorlinks = true,
    linkcolor = {NavyBlue},		
    anchorcolor = .,
    citecolor = {NavyBlue},		
    filecolor = .,				
    menucolor = .,
    runcolor = {cyan}, 
    urlcolor = {NavyBlue},		
    }
\usepackage{url}            
\usepackage{array,booktabs}       
\usepackage{amsfonts}       
\usepackage{nicefrac}       
\usepackage{microtype}      
\usepackage{fancyhdr}       
\usepackage{graphicx}       
\usepackage{setspace} 
\usepackage{soul} 
\usepackage{ragged2e} 
\usepackage[toc,page]{appendix}
\usepackage{longtable}
\usepackage{eso-pic}
\usepackage{tikz}
\usepackage[superscript]{cite}

\graphicspath{{./Figures/}}     

\DeclareUnicodeCharacter{2009}{0}

\begin{document}

\newpage
\pagenumbering{roman}
\thispagestyle{empty}

\begin{figure}[h!]
  \centering
  \includegraphics[width=1\columnwidth]{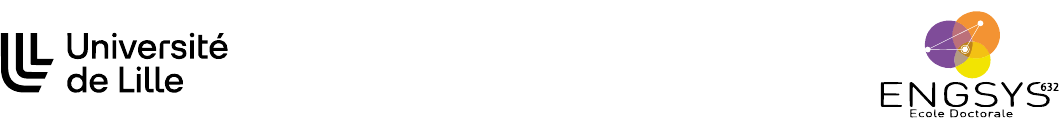}
\end{figure}

\centering 

\linespread{3}
\bf{\textsc{Universit\'{e} de Lille}} \\

\linespread{3}
\normalfont Doctoral School ENGSYS (632) \\
\linespread{1.5}
Institut d'\'{E}lectronique, de Micro\'{e}lectronique et de Nanotechnologie Research Unit \\

\linespread{3}
Habilitation to conduct research defended by \bf{\textsc{S\'{e}bastien Pecqueur}} \\

\linespread{3}
\normalfont Defended on the \bf{26.6.2025} \\
\linespread{1.5}
\normalfont Academic Field \bf{Engineering Sciences} \\

\setlength{\parskip}{3em}
\normalfont \huge Organic Electronic Classifiers for Sensing\\

\vspace*{\fill}
\normalsize \raggedright
Committee members --------------------------------------------------------------------------------------------- \\
\singlespacing
\begin{tabular}{ c m{3cm} m{9.7cm} }
Referees & Esma ISMAILOVA & Maitre de Conf\'{e}rences of the {E}cole des Mines de Saint-\'{E}tienne at the Centre of Microelectronics in Provence (CMP), Gardanne, France\\[10pt] 
 & Nicolas CLEMENT & Directeur de Recherche CNRS at the Laboratory for Analysis and Architecture of Systems (LAAS), Toulouse, France\\[10pt] 
 & Yannick COFFINIER & Directeur de Recherche CNRS at the Institute of Electronics, Microelectronics and Nanotechnology (IEMN), Lille, France\\[15pt] 
Examiners & Kamal LMIMOUNI & Professeur des Unversit\'{e}s of the Universit\'{e} de Lille at the Institute of Electronics, Microelectronics and Nanotechnology (IEMN), Lille, France\\[10pt] 
& Giorgio MATTANA & Maitre de Conf\'{e}rences of Paris Cit\'{e} University at the Interfaces, Treatments, Organization and Systems Dynamics (ITODYS) laboratory, Paris, France\\[15pt] 
Garantor & Fabien ALIBART & Charg\'{e} de Recherche CNRS at the Nanotechnologies and Nanosystems Laboratory (LN2) and the Universit\'{e} de Sherbrooke, Canada \\ 
\end{tabular}

\newpage 
\thispagestyle{empty}
\

\newpage
\thispagestyle{empty}

\begin{figure}[h!]
  \centering
  \includegraphics[width=1\columnwidth]{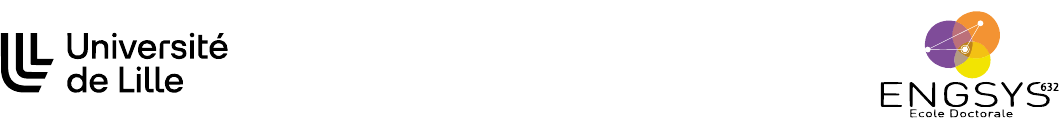}
\end{figure}

\centering 

\linespread{3}
\bf{\textsc{Universit\'{e} de Lille}} \\

\linespread{3}
\normalfont \'{E}cole doctorale ENGSYS (632) \\
\linespread{1.5}
Unité de Recherche Institut d'\'{E}lectronique, de Micro\'{e}lectronique et de Nanotechnologie \\

\linespread{3}
Habilitation \`{a} diriger les recherches pr\'{e}sent\'{e}e par \bf{\textsc{S\'{e}bastien Pecqueur}} \\

\linespread{3}
\normalfont Soutenue le \bf{26.6.2025} \\
\linespread{1.5}
\normalfont Discipline \bf{Sciences pour l’Ing\'{e}nieur} \\

\setlength{\parskip}{3em}
\normalfont \huge Classifieurs \'{E}lectroniques Organiques pour la D\'{e}tection\\

\vspace*{\fill}
\normalsize \raggedright
Composition du jury --------------------------------------------------------------------------------------- \\
\singlespacing
\begin{tabular}{ c m{3cm} m{9cm} }
Rapporteurs: & Esma ISMAILOVA & Maitre de Conf\'{e}rences de l'\'{E}cole des Mines de Saint-\'{E}tienne au Centre Micro\'{e}lectronique de Provence (CMP), Gardanne, France\\[10pt] 
 & Nicolas CLEMENT & Directeur de Recherche CNRS au Laboratoire d'Analyse et d'Architecture des Syst\`{e}mes (LAAS), Toulouse, France\\[10pt] 
 & Yannick COFFINIER & Directeur de Recherche CNRS \`{a} l'Institut d'\'{E}lectronique, Micro-\'{e}lectronique et Nanotechnologie (IEMN), Lille, France\\[15pt] 
Examinateurs: & Kamal LMIMOUNI & Professeur des Universit\'{e}s de l'Universit\'{e} de Lille \`{a} l'Institut d'\'{E}lectronique, Micro-\'{e}lectronique et Nanotechnologie (IEMN), Lille, France\\[10pt] 
& Giorgio MATTANA & Maitre de Conf\'{e}rences de l'Universit\'{e} Paris-Cit\'{e} au laboratoire  Interfaces, Traitements, Organisation et Dynamique des Syst\`{e}mes (ITODYS), Paris, France\\[15pt] 
Garant: & Fabien ALIBART & Charg\'{e} de Recherche CNRS au Laboratoire Nanotechnologies et Nanosyst\`{e}mes (LN2) et l'Universit\'{e} de Sherbrooke, Canada \\ 
\end{tabular}

\newpage 
\thispagestyle{empty}
\

\newpage
\thispagestyle{empty}

\topskip70pt

This habilitation to conduct research has been prepared at the following research unit. \\

\singlespacing
\singlespacing
\singlespacing
\singlespacing
\singlespacing
\singlespacing
\singlespacing
\singlespacing
\singlespacing
\singlespacing
\singlespacing
\singlespacing

\begin{figure}[h]
\centering
\begin{tabular}{cc}
\begin{minipage}{35em} \bf Institut d'\'{E}lectronique de Micro\'{e}lectronique et de Nanotechnologie  \\ \normalfont (IEMN) - UMR 8520 \\ Avenue Henri Poincar\'{e}, cit\'{e} scientifique \\ CS 60069, 59652 Villeneuve d’Ascq Cedex, FRANCE\end{minipage} & \includegraphics[scale=0.7]{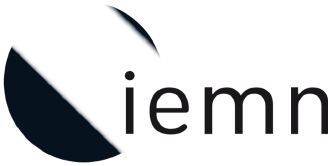}\\
\end{tabular}
\end{figure}

\singlespacing
\singlespacing
\singlespacing
\singlespacing
\singlespacing
\singlespacing
\singlespacing
\singlespacing
\singlespacing
\singlespacing
\singlespacing
\singlespacing
\singlespacing
\singlespacing
\singlespacing
\singlespacing
\singlespacing
\singlespacing
\singlespacing
\singlespacing
\singlespacing
\singlespacing
\singlespacing
\singlespacing

 \justifying \textit{I, S\'{e}bastien Pecqueur, am the sole author of this monograph. Neither texts nor graphical contents result of any generative artificial intelligence. All figures have been edited specifically for the purpose of this habilitation to not infringe copyrights with publishers where parts of the mentioned results are reported in peer-reviewed articles. All reported content has been generated from raw data produced by different scientists and funded by different public organizations contracted with the CNRS. Detailed information is specified in the appendices listed in the table of contents of this monograph. SP, 5.3.2025}

\newpage 
\thispagestyle{empty}
\

\newpage
\thispagestyle{empty}

\lhead{Abstract}
\rhead{\thepage}
\renewcommand{\headrulewidth}{0.4pt}
\cfoot{}

\setlength{\parskip}{1em}

\centering \textbf{\textsc{Organic Electronic Classifiers for Sensing}} \\

\singlespacing

\centering {\sc {Preamble}} \\

\normalfont \justifying This monograph describes nine years of research carried out at the Institute for Electronics, Micro-electronics and Nanotechnologies (IEMN), developed around defining a generic concept for detection, filling a void between metrological sensors and biological senses, sensing an environment's qualities along with their measurable properties in information generation technologies. The first chapter introduces fundamental notions of recognition for complex environments, such as for their chemistry, for which organic semiconductors can embed two new functionalities in consumer electronics. The second and third chapters mostly summarize contributions to the state-of-the-art literature on these matters: in the second chapter, on studying conducting polymers as both chemical detectors and conductimetric transducers, and the third chapter, on studying electropolymerization sensitivity to conceptualize evolutionary electronics. The fourth chapter presents several results on the conception of several "classifiers" exploiting both functionalities: in tasks aiming at integrating different sensitivities at a very small scale, at broadening sensing devices' receptive fields based on experience, and at physically engraving the experience data in a sensing hardware. Along with this monograph are also associated four appendices, summarizing different elements related to the context of this research.\


\hrulefill

\centering \textbf{\textsc{Classifieurs \'{E}lectroniques Organiques pour la D\'{e}tection}} \\

\singlespacing

\centering {\sc {Pr\'{e}ambule}} \\

\normalfont \justifying Ce monographe d\'{e}crit neuf ann\'{e}es de recherches men\'{e}es \`{a} l’Institut d'\'{E}lectronique de Micro\'{e}lectronique et de Nanotechnologie (IEMN), articul\'{e}es autour de la d\'{e}finition un concept g\'{e}n\'{e}rique de syst\`{e}mes de d\'{e}tection, comblant un vide entre capteurs m\'{e}trologiques et sens biologiques, qui int\`{e}grerait intrins\`{e}quement la nature qualitative d'un environnement \`{a} ses propri\'{e}t\'{e}s mesurables dans les technologies de g\'{e}n\'{e}ration de l'information. Le premier chapitre introduit des notions fondamentales autour de la reconnaissance d'environnements complexes, notamment dans leur chimie, pour laquelle les semi-conducteurs organiques permettent d'apporter deux nouvelles fonctionnalit\'{e}s dans l'\'{e}lectronique grand public. Les deuxi\`{e}me et troisi\`{e}me chapitres r\'{e}sument essentiellement des contributions apport\'{e}es \`{a} l'\'{e}tat de l'art sur ces sujets : dans le deuxi\`{e}me chapitre, autour de l'\'{e}tude des polym\`{e}res conducteurs en tant que d\'{e}tecteurs chimiques et transducteurs conductim\'{e}triques, et dans le troisi\`{e}me chapitre, autour de l'\'{e}tude de la sensibilit\'{e} de l'\'{e}lectropolym\'{e}risation pour concevoir de l'\'{e}lectronique \'{e}volutive. Le quatri\`{e}me chapitre expose plusieurs r\'{e}sultats sur la conception de diff\'{e}rents « classifieurs » utilisant ces deux fonctionnalit\'{e}s : dans des projets visant à int\'{e}grer des sensibilit\'{e}s vari\'{e}es \`{a} tr\`{e}s petite \'{e}chelle, agrandir le champs r\'{e}ceptif des syst\`{e}mes de d\'{e}tection en fonction de l'exp\'{e}rience, et graver physiquement l'information de cette exp\'{e}rience dans le mat\'{e}riel des syst\`{e}mes de d\'{e}tection. Quatre annexes sont \'{e}galement associ\'{e}es \`{a} ce monographe, rassemblant diff\'{e}rents \'{e}l\'{e}ments associ\'{e}s au contexte de cette recherche.\\

\bf Mots cl\'{e}s: \normalfont polym\`{e}res conducteurs, d\'{e}tection chimique, imp\'{e}dance spectro-\'{e}lectrochimique, classification \\

\hrulefill

\vspace*{\fill}

\bf Institut d'\'{E}lectronique de Micro\'{e}lectronique et de Nanotechnologie \normalfont (IEMN) - UMR 8520 \\
Avenue Henri Poincar\'{e}, cit\'{e} scientifique CS 60069, 59652 Villeneuve d’Ascq Cedex, FRANCE 

\newpage 
\thispagestyle{empty}
\

\newpage
\thispagestyle{empty}

\begin{figure}[h!]
  \centering
  \includegraphics[width=1\columnwidth]{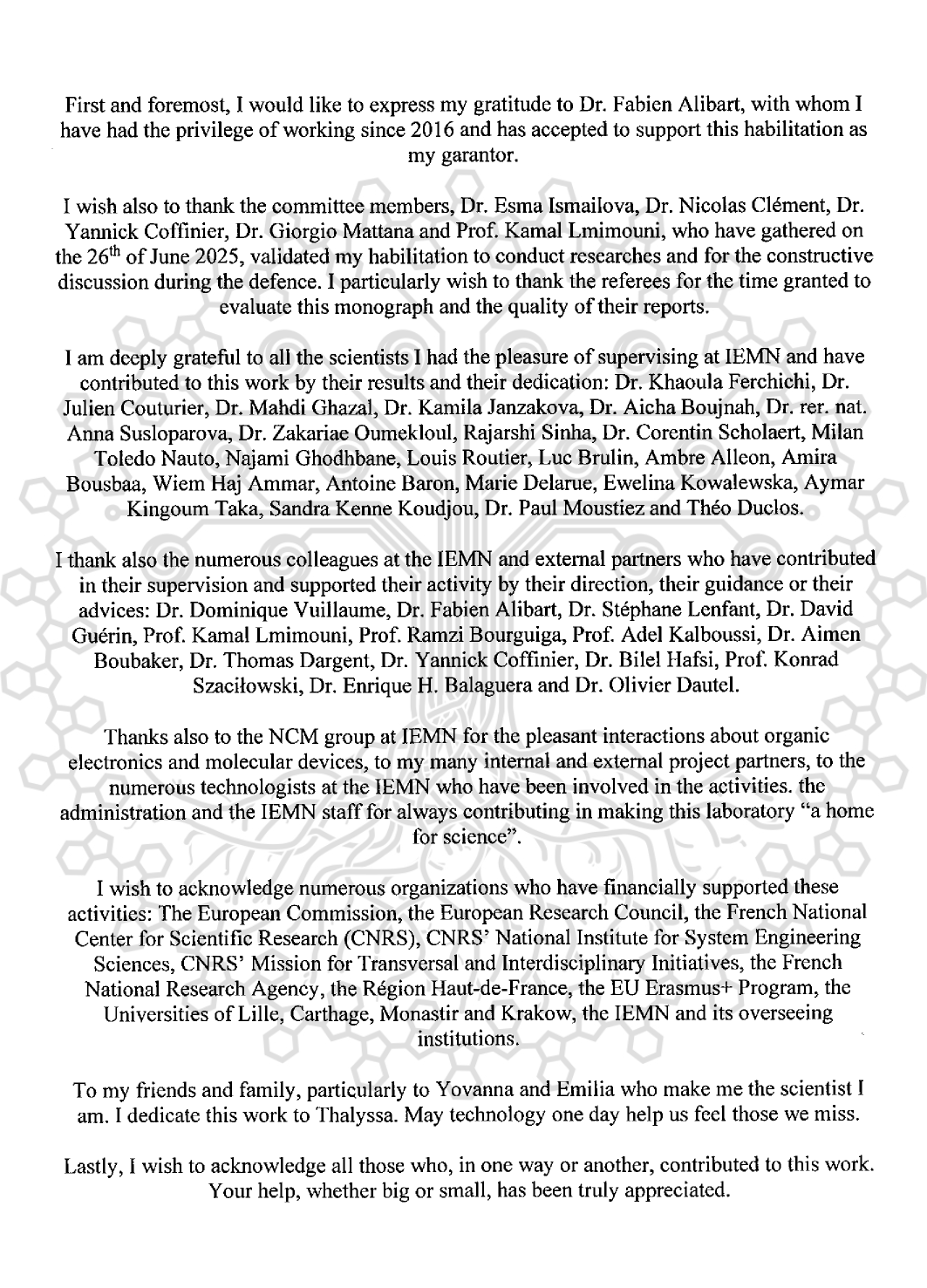}
\end{figure}

\
\newpage 
\thispagestyle{empty}
\

\lhead{Contents}
\rhead{\thepage}
\renewcommand{\headrulewidth}{0.4pt}
\cfoot{}

\tableofcontents

\vspace*{\fill}

\newpage 
\thispagestyle{empty}
\

\newpage
\pagenumbering{arabic}
\lhead{}
\rhead{}
\renewcommand{\headrulewidth}{0pt}
\cfoot{\thepage}

\AddToShipoutPictureBG{%
\def\Image{}%
  \ifnum\value{page}=1\relax
    \def\Image{\includegraphics[height=2cm]{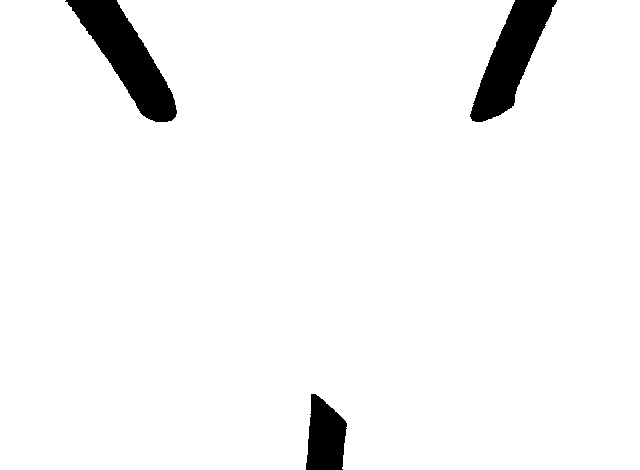}}
\else
  \ifnum\value{page}=3\relax
    \def\Image{\includegraphics[height=2cm]{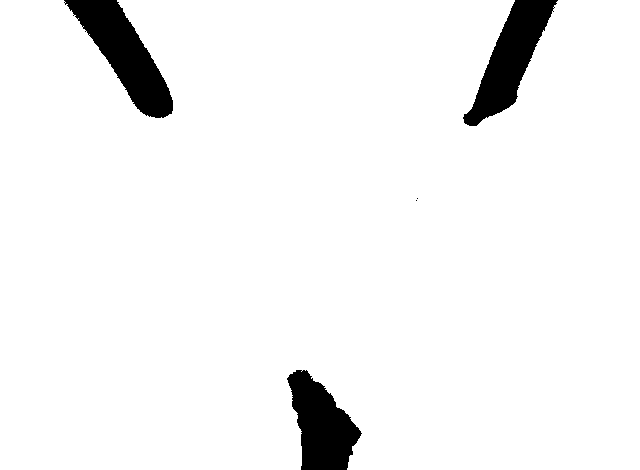}}
\else
  \ifnum\value{page}=5\relax
    \def\Image{\includegraphics[height=2cm]{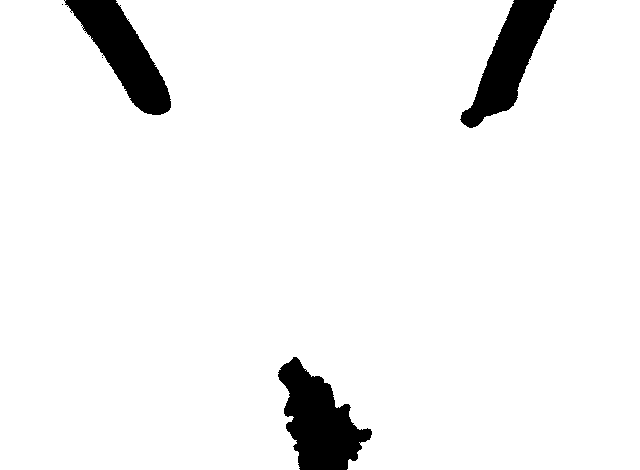}}
\else
  \ifnum\value{page}=7\relax
    \def\Image{\includegraphics[height=2cm]{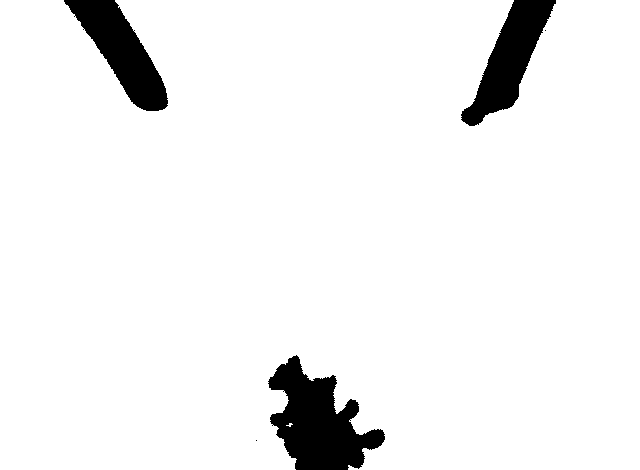}}
\else
  \ifnum\value{page}=9\relax
    \def\Image{\includegraphics[height=2cm]{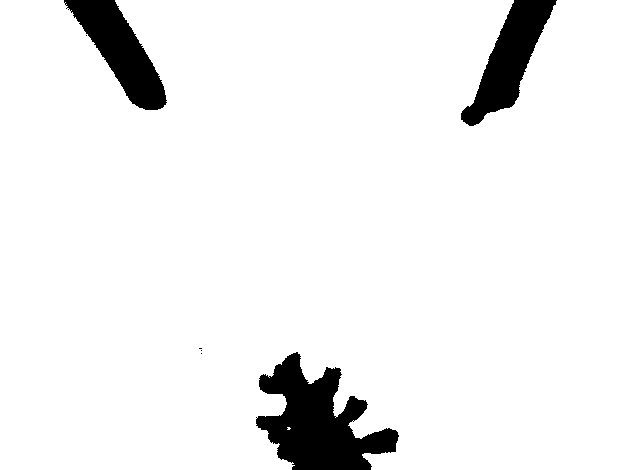}}
\else
  \ifnum\value{page}=11\relax
    \def\Image{\includegraphics[height=2cm]{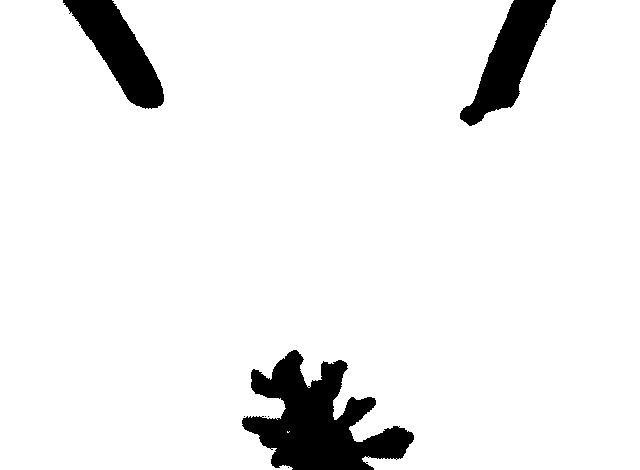}}
\else
  \ifnum\value{page}=13\relax
    \def\Image{\includegraphics[height=2cm]{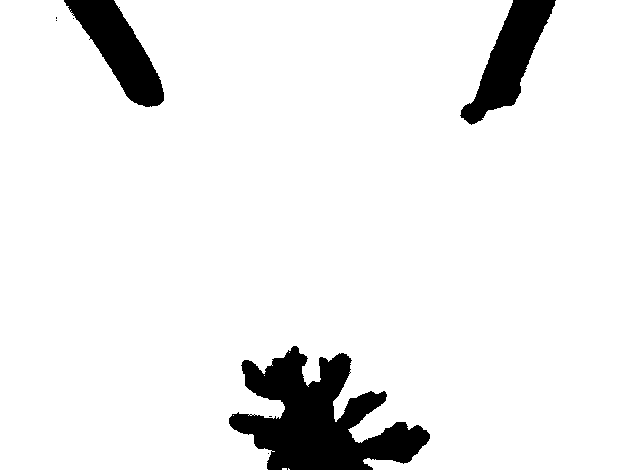}}
\else
  \ifnum\value{page}=15\relax
    \def\Image{\includegraphics[height=2cm]{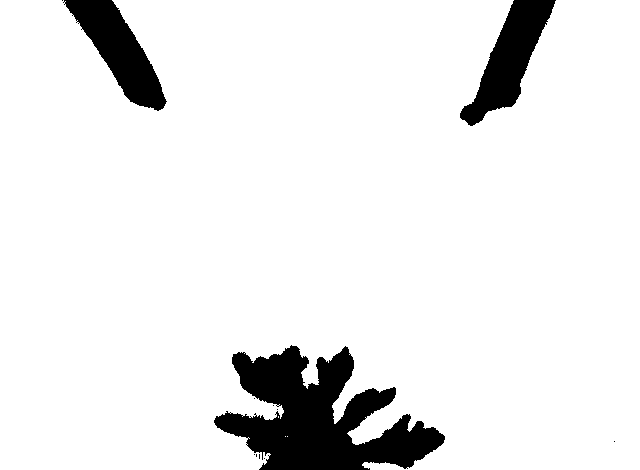}}
\else
  \ifnum\value{page}=17\relax
    \def\Image{\includegraphics[height=2cm]{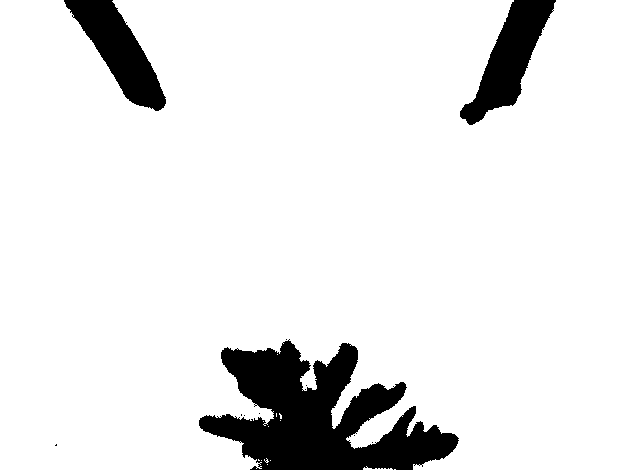}}
\else
  \ifnum\value{page}=19\relax
    \def\Image{\includegraphics[height=2cm]{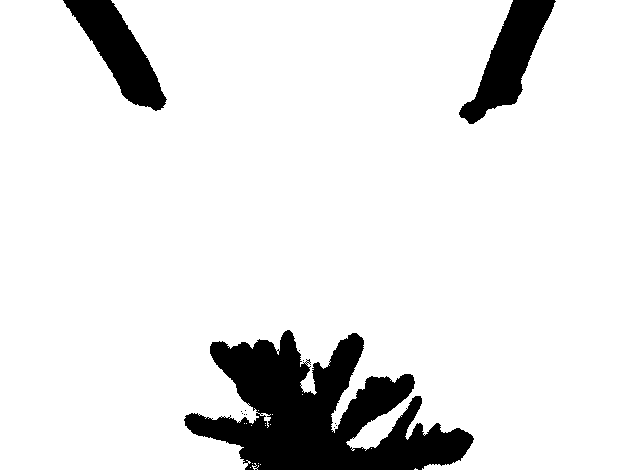}}
\else
  \ifnum\value{page}=21\relax
    \def\Image{\includegraphics[height=2cm]{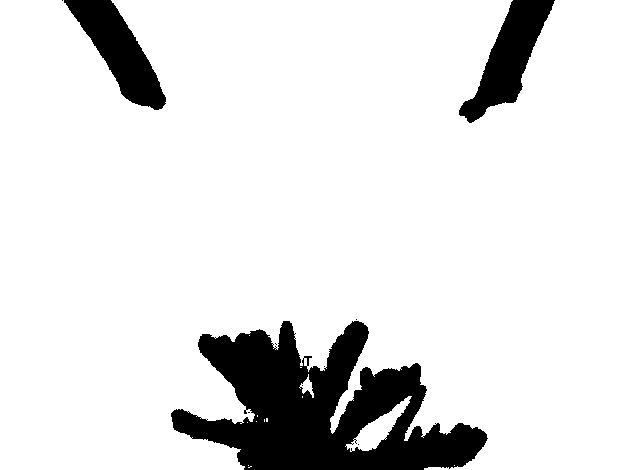}}
\else
  \ifnum\value{page}=23\relax
    \def\Image{\includegraphics[height=2cm]{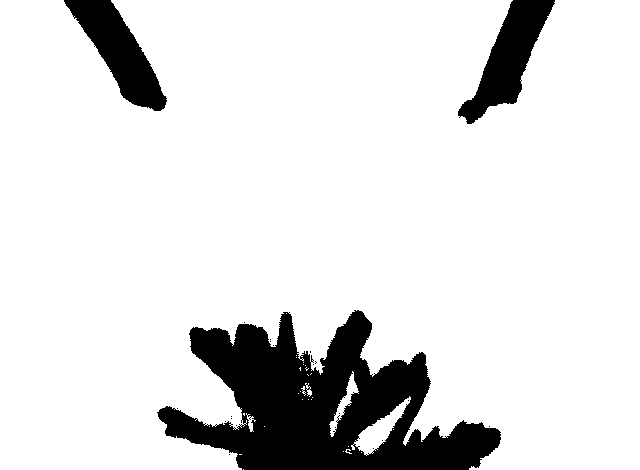}}
\else
  \ifnum\value{page}=25\relax
    \def\Image{\includegraphics[height=2cm]{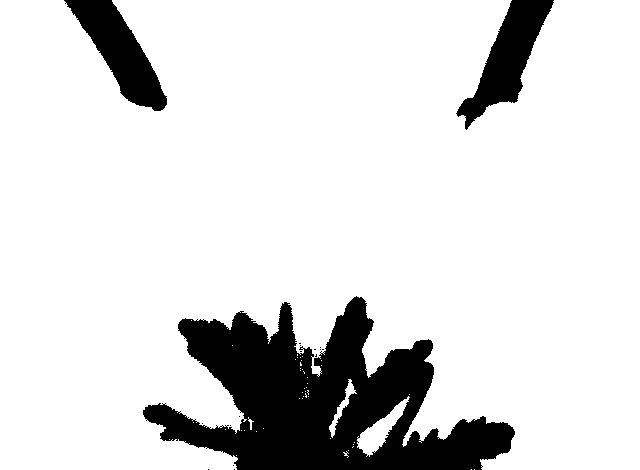}}
\else
  \ifnum\value{page}=27\relax
    \def\Image{\includegraphics[height=2cm]{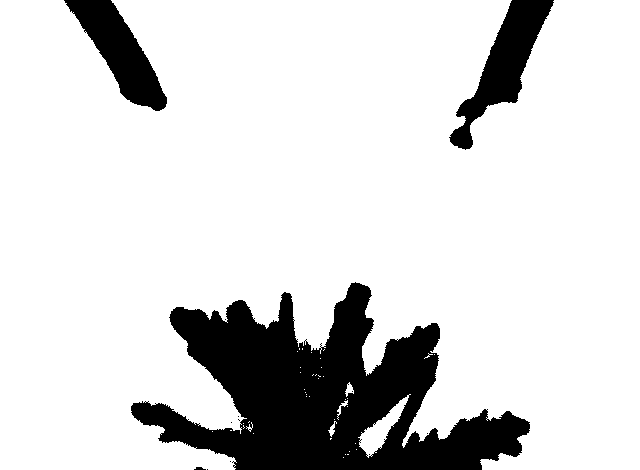}}
\else
  \ifnum\value{page}=29\relax
    \def\Image{\includegraphics[height=2cm]{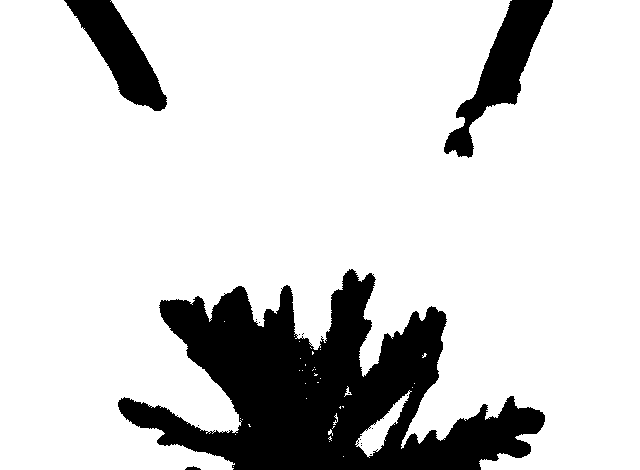}}
\else
  \ifnum\value{page}=31\relax
    \def\Image{\includegraphics[height=2cm]{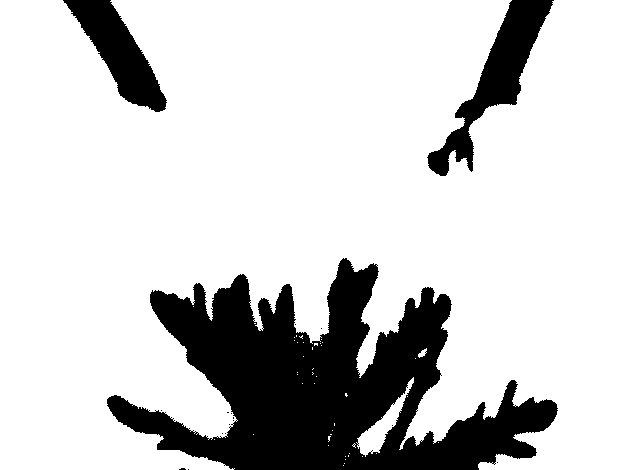}}
\else
  \ifnum\value{page}=33\relax
    \def\Image{\includegraphics[height=2cm]{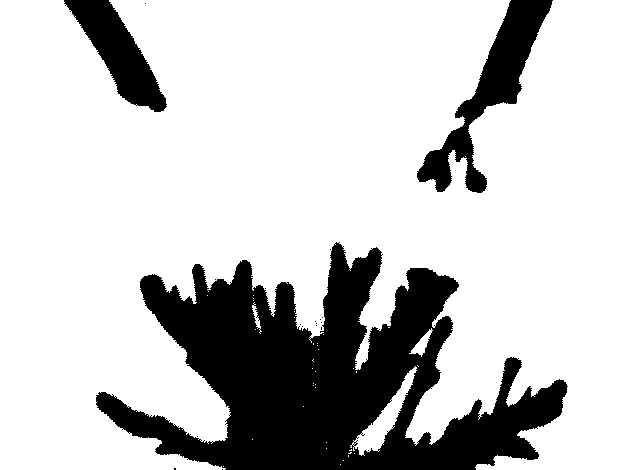}}
\else
  \ifnum\value{page}=35\relax
    \def\Image{\includegraphics[height=2cm]{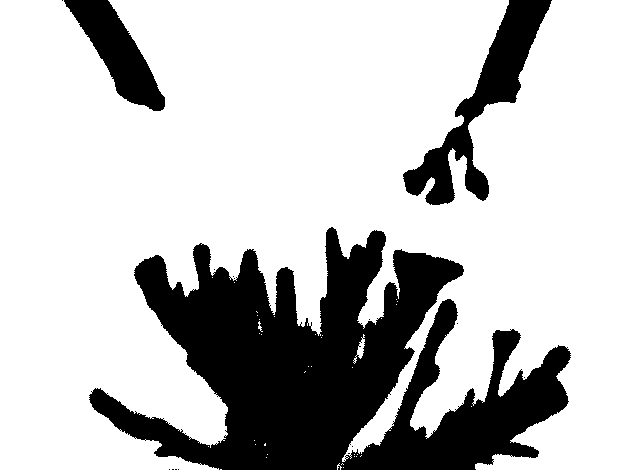}}
\else
  \ifnum\value{page}=37\relax
    \def\Image{\includegraphics[height=2cm]{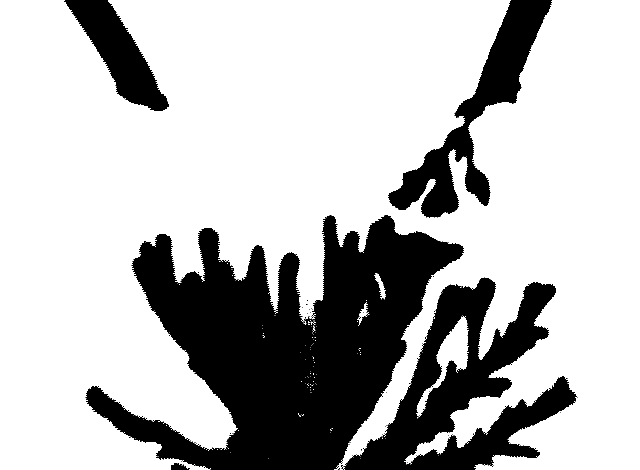}}
\else
  \ifnum\value{page}=39\relax
    \def\Image{\includegraphics[height=2cm]{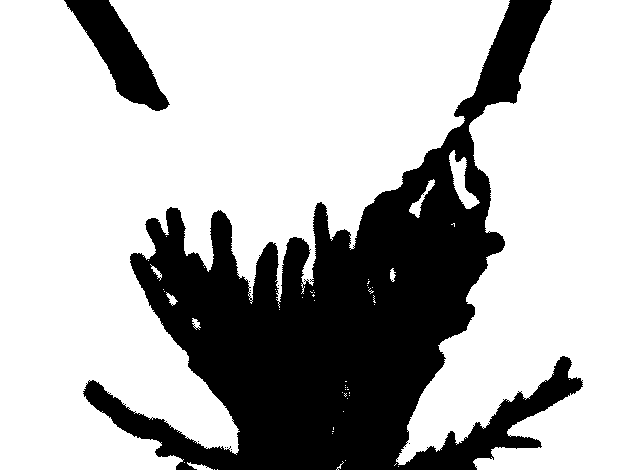}}
\else
  \ifnum\value{page}=41\relax
    \def\Image{\includegraphics[height=2cm]{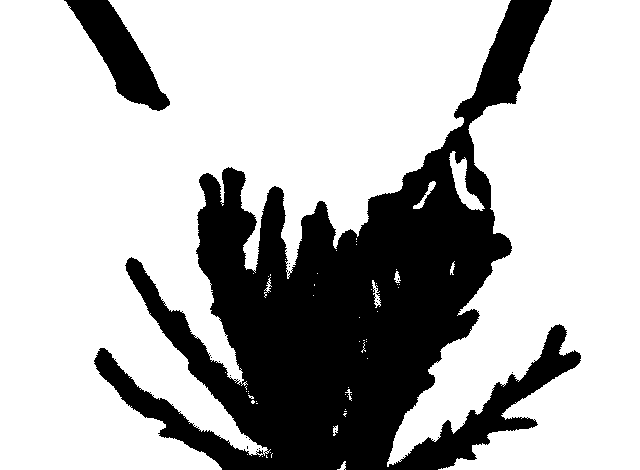}}
\else
  \ifnum\value{page}=43\relax
    \def\Image{\includegraphics[height=2cm]{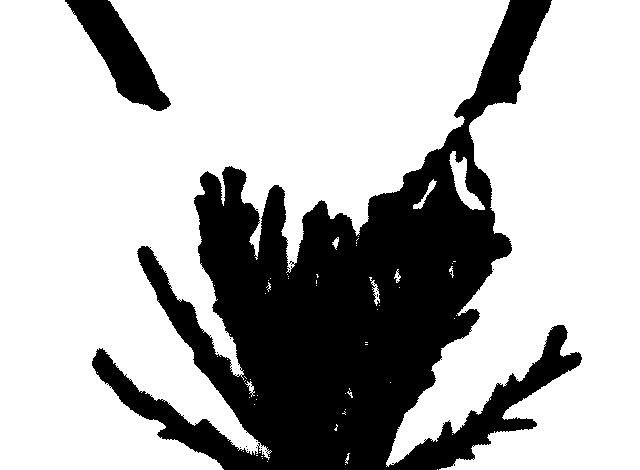}}
\else
  \ifnum\value{page}=45\relax
    \def\Image{\includegraphics[height=2cm]{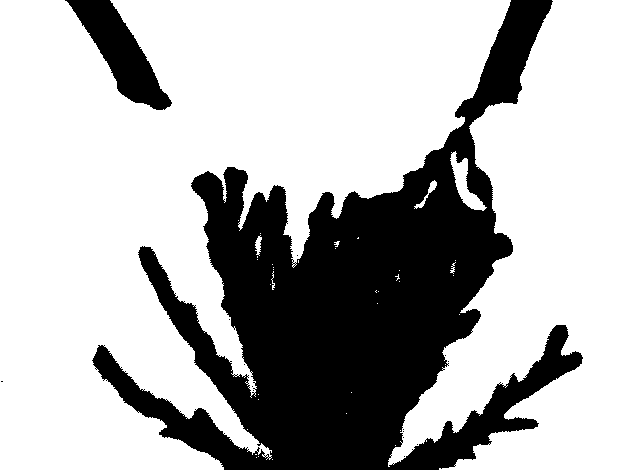}}
\else
  \ifnum\value{page}=47\relax
    \def\Image{\includegraphics[height=2cm]{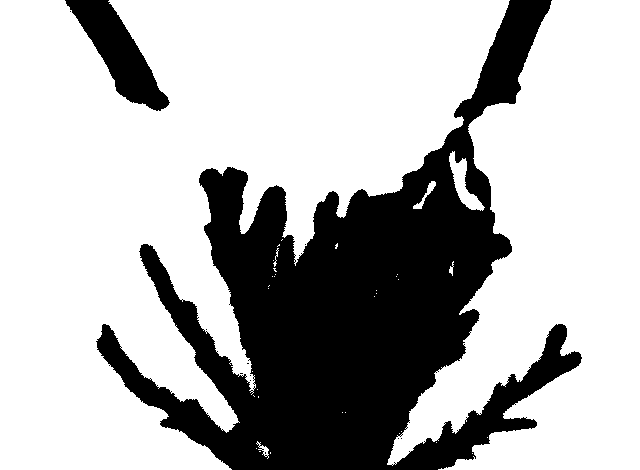}}
\else
  \ifnum\value{page}=49\relax
    \def\Image{\includegraphics[height=2cm]{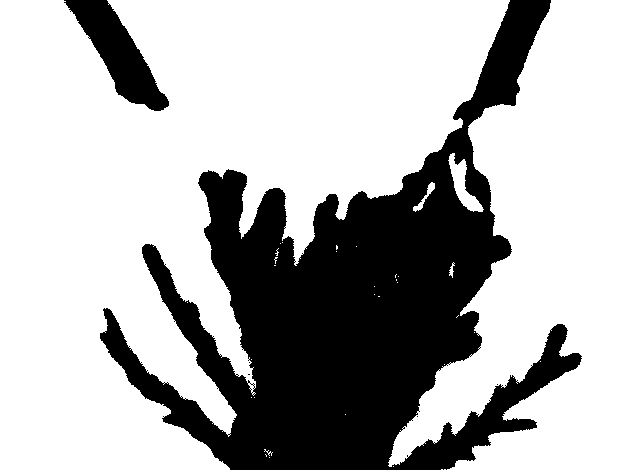}}
\else
  \ifnum\value{page}=51\relax
    \def\Image{\includegraphics[height=2cm]{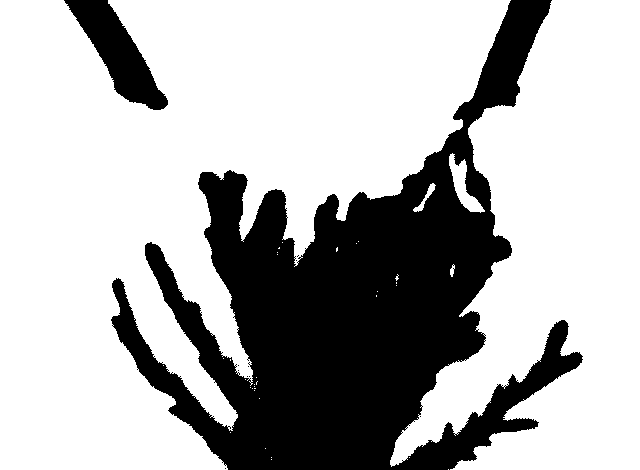}}
\else
  \ifnum\value{page}=53\relax
    \def\Image{\includegraphics[height=2cm]{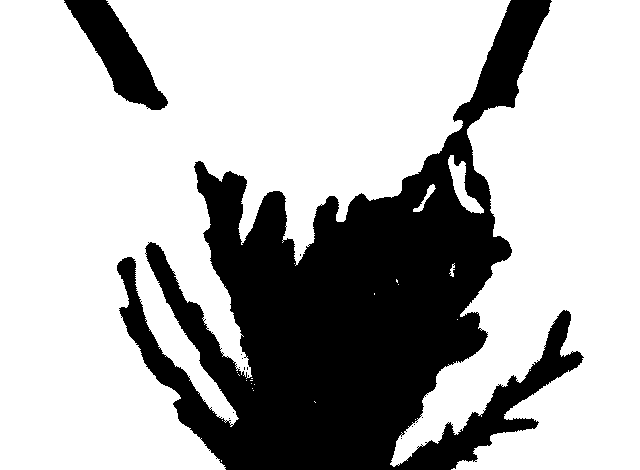}}
\else
  \ifnum\value{page}=55\relax
    \def\Image{\includegraphics[height=2cm]{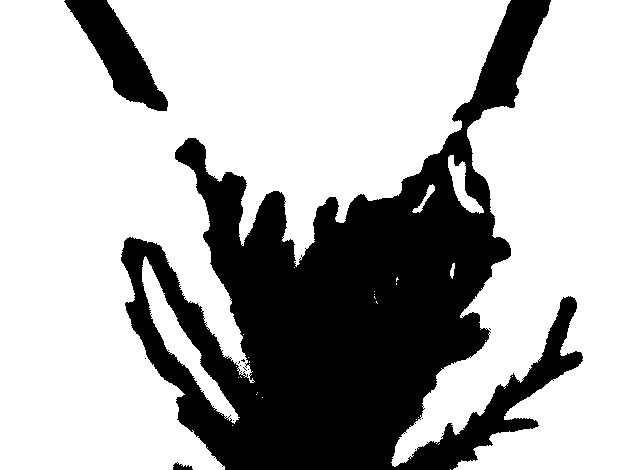}}
\else
  \ifnum\value{page}=57\relax
    \def\Image{\includegraphics[height=2cm]{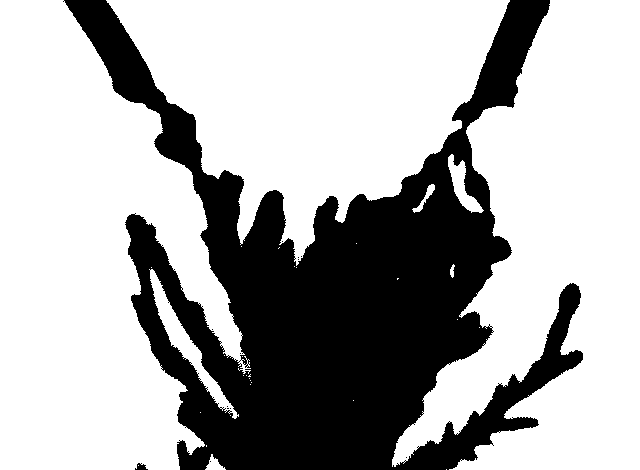}}
\else
  \ifnum\value{page}=59\relax
    \def\Image{\includegraphics[height=2cm]{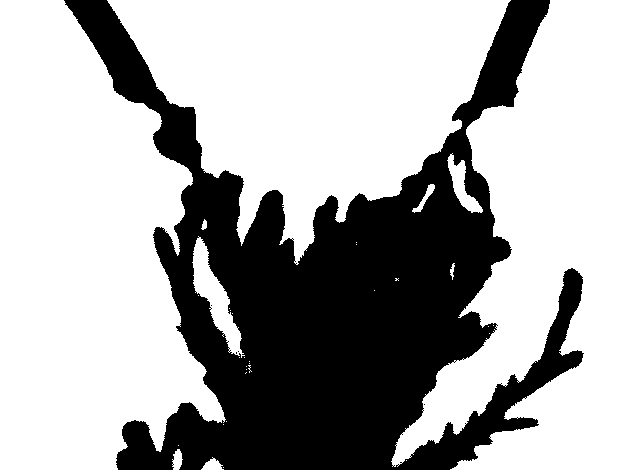}}
\else
  \ifnum\value{page}=61\relax
    \def\Image{\includegraphics[height=2cm]{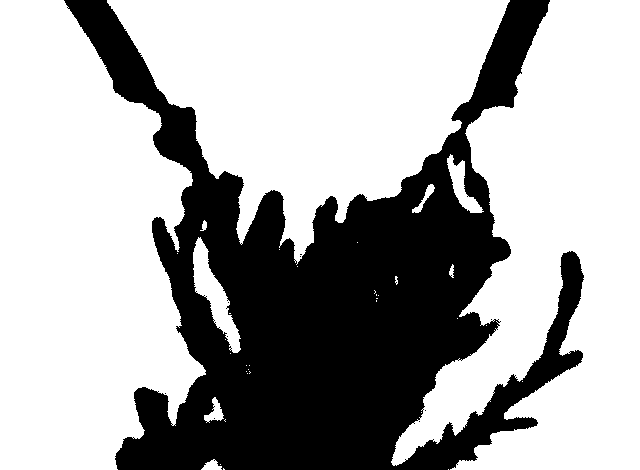}}
\else
  \ifnum\value{page}=63\relax
    \def\Image{\includegraphics[height=2cm]{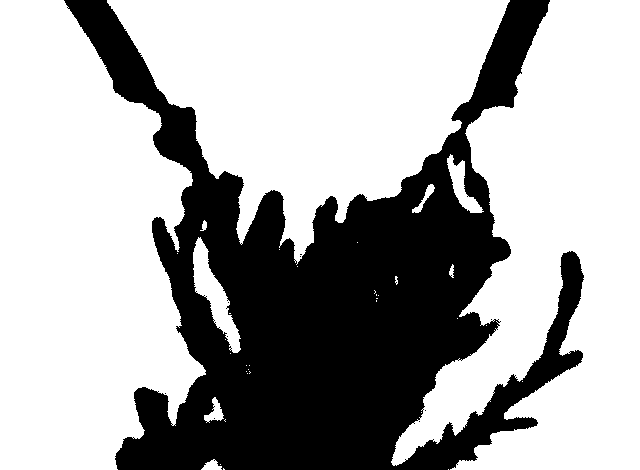}}
\else
  \ifnum\value{page}=65\relax
    \def\Image{\includegraphics[height=2cm]{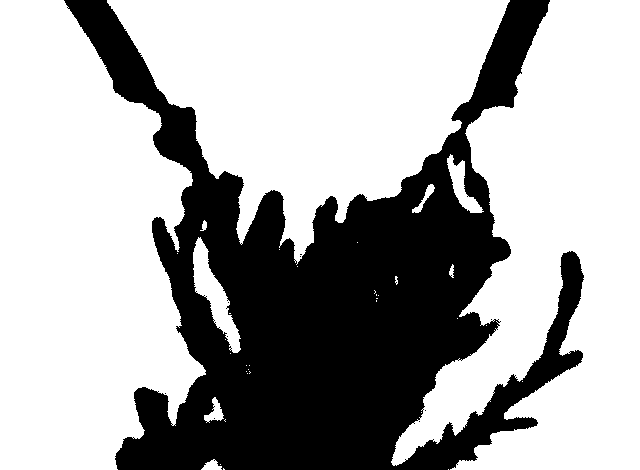}}
  \fi  \fi  \fi  \fi  \fi  \fi  \fi  \fi  \fi  \fi  \fi  \fi  \fi  \fi  \fi  \fi  \fi  \fi  \fi  \fi  \fi  \fi  \fi  \fi  \fi  \fi  \fi  \fi  \fi  \fi  \fi  \fi  \fi    
\begin{tikzpicture}[remember picture,overlay]
  \node[anchor=south east,inner sep=0pt] at 
    (current page.south east)   
    {\Image};
\end{tikzpicture}
}

\chapter{Context \& Motivations}
\label{Ch1}
\setcounter{page}{1}

\textit{From primitive senses to sophisticated sensors, sensing is a mean to understand the world. Now, many technologies measure what is imperceptible to us. With the Internet-of-Things (IoT), sensors are in every electronic devices, and with artificial intelligence (AI), all collected information can give us back complex responses: it is important to control this feedback in its quality and quantity. In customer electronics, chemistry remains difficult to sense, but so important to understand from our environment. Chemicals are created every day and molecules have so many impacts on us. It is possible to make devices with molecules too, organic semiconductors are already in some of our pockets. However, they are not used to sense chemistry yet. Now electronics has to reinvent, inspiring from nature to process information and manufacture for ensuring global stability of our environment, conducting polymers can reinvent sensing with "classifiers" to answer:}

\centering {\sc {--- Can sensors help us sensing beyond our senses? ---}}\\

\normalfont \justifying \textit{\\
This chapter introduces on the complexity of our senses (Sec.\ref{Ch1.1}), describes what conventional sensors tend to do (Sec.\ref{Ch1.2}) and opens to what organic semiconductors may have to offer (Sec.\ref{Ch1.3}).}\\\

\section{\textit{sentio ergo sum}: From Senses to Sensors, or \textit{vice versa}}
\label{Ch1.1}

Sensing makes us aware about ourselves among others in a physical environment. We understand more of it everyday as we generate, share and store more information now. \textbf{Yet, there are no tools to sense each thing that we feel and perceive from our environment in its whole complexity}. Natural beings have different sensitivities, so it is essential to preserve biodiversity as well as diversity of cults \& opinions. Our anthropic activity threatens both of these more and more, physically and culturally, as technologies evolve. So, it is essential to reconsider the way technologies influence us to help feeling better in our biotope, in particular how sensing technologies should generate information from our environment and from ourselves. \textbf{Our senses are no ideal sensors of truth}: we can be easily fooled by ourselves or by others. As \textbf{biochemical beings}, our relationship with \textbf{chemistry is very insidious}: environment's chemistry does not stop at our skin and can penetrate and threaten our health and our judgment. \textbf{Molecules are also ubiquitous information carriers}: we can quantify precisely a few of them, but, they are so many that counting them all to have a qualitative picture of an environment appears barely feasible.\\[3pt]
In the following, an overview is made on how we feel and perceive our environment, how it can help us to understand it, or not (Subs.\ref{Ch1.1.1}) and why chemistry is such a challenging environmental information to classify (Subs.\ref{Ch1.1.2}).

\subsection{Duality between Sensations \& Perceptions}
\label{Ch1.1.1}

\begin{figure}
  \centering
  \includegraphics[width=1\columnwidth]{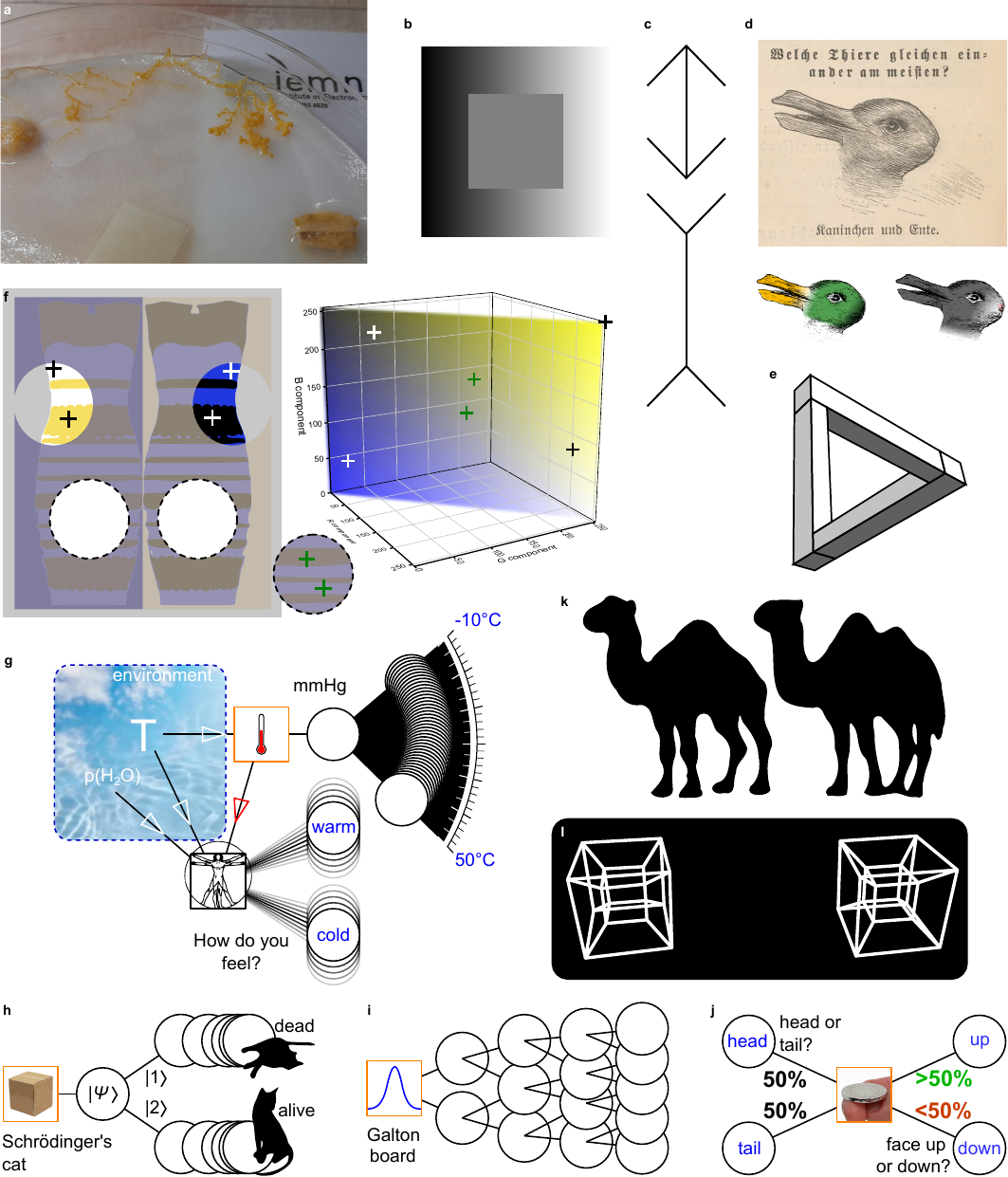}
  \caption{\textbf{Perceptions' Limits \& Cognitive Biases $\vert$ a,} \textit{physarum polycephalum} adapting to its environment by sensing gradients of nutriments.\cite{Tero2010,Adamatzky2010} \textbf{b-c,} Two perceptive optical illusions: isochromatic (\textbf{b}, described by Goethe, 1810).\cite{Goethe1810} and M\"uller-Lyer's isometric illusion (\textbf{c}, as depicted by M\"uller-Lyer, 1889).\cite{MuellerLyer1889} \textbf{d,} Reproduction of the duck-rabbit,\cite{1892} which perception is interchangeable and can be biased by data manipulation (colored versions below). \textbf{e,} Penrose triangle:\cite{Penrose1958} an example of paradox of an apparently 3D object which cannot be defined as "faced-up" or "faced-down". \textbf{f,} Illustration of "{\#}thedress" phenomenon,\cite{Brainard2015,LaferSousa2015,Winkler2015,FeitosaSantana2018} when perceptions of object's "true colors" depend on individuals' sensitivity to a background. \textbf{g,} Illustrative discrepancy between metrological sensors and perceptive individuals. \textbf{h,} Schr\"odinger's cat, which full state is imperceptible through measurements.\cite{Leggett2005} \textbf{i,} Galton's Quincunx as cascades of equiprobable two-state random generators, converging to non-equiprobable density states, projecting a binomial law to a normal distribution in space.\cite{Galton1886,Galton1889} \textbf{j,} Coin-toss which states' non-equiprobabilty is class-definition dependent (as reported in Barto\v{c} \textit{et al.} 2024).\cite{Bartos2024} \textbf{k,} Two-dimensional projections of a camel and a dromedary: an example of a projection-sensitive perception. \textbf{l,} Stereoscopic view of a tesseract, as a prosthetic projection on two-eye sight of a four-dimensional object.\cite{Noll1965}}
  \label{fig:fig0.1.1}
\end{figure}

Since 3.7 billion years of evolution,\cite{Rosing1999,Hassenkam2017} living organisms need to sense features of their environment to scavenge enough resources and survive long enough to reproduce (Fig.\ref{fig:fig0.1.1}.a). \textbf{Among them, we are sensitive beings in a physical world: rational but intuitive}. As an aim to understand better our environment, we construct physical concepts of it, based on (1) what \textbf{we sense}, (2) what \textbf{we perceive} and (3) \textbf{tools}.\\[3pt]
$\bullet $ \textbf{We all sense differently}. It is a physical aptitude allowing us to absorb an incoming information on our nervous system. Animals from different species have different senses,\cite{Griffin1944,Kalmijn1966,Baker1987,Marshall2011,Marshall2014} and genes on members of a same specie can induce the expression of different receptors for a same sensory function.\cite{Neitz2011,Trimmer2019,Dioszegi2019} Genetically identical twins can also have different experiences conditioning their physical abilities or disabilities to sense, but it is strictly related an aptitude of the nervous system, not what it has memorized.\\[3pt]
$\bullet $ \textbf{We all perceive differently}. Because our sensitivities are complex, numerous \textbf{cognitive biases and subjective constancies} exist, distorting perceptions from what it actually is: like illusions of colors (Fig.\ref{fig:fig0.1.1}.b) or sizes (Fig.\ref{fig:fig0.1.1}.c) for sight. Some patterns are also particularly ambiguous to categorize so even a single individual can classify a same stimuli in different ways (Fig.\ref{fig:fig0.1.1}.d), or never recognize it at all like in the case of a paradox (Fig.\ref{fig:fig0.1.1}.e). Perceptions are consequences of past and present incoming information, creating \textbf{projections} as \textbf{feelings}. Feelings are expressions of stresses on our endocrine system from a complex neurotransmitters activity: glands release different hormones, inducing a feedback on the nervous system, promoting its activity to converge to \textbf{different classes} of moods. Because we do not have the same endocrine organs, nor the same experience imprinted in our brain, perceptions are unique: they are strictly related to an aptitude of the endocrine system and a state of the brain memory conditioned from past experiences. \textbf{Classes are personal} and a few of them are named with culturally built semantic notions, so individuals can communicate on what they perceive. However, because semantic classes are not universal, one may strongly disagree on perceptions. "{\#}thedress" is a quite recent example of a color perception disagreement over a huge set of people distributed all over the world at the same time (Fig.\ref{fig:fig0.1.1}.f). This example highlights the \textbf{difference between the information classes we perceive and the physical significance of input information we sense}: sensations and perceptions are not the same.\\[3pt]
$\bullet $ \textbf{To agree on what we perceive, we build tools that sense}: sensors, analyzers and different information generators (Fig.\ref{fig:fig0.1.1}.g). \textbf{They project information to something perceptible}: sounds, images or moving dial pointers. Significant discoveries were made by hardware designs, to help us feeling what is far or small, sometimes centuries after theories.\cite{Collaboration2019,Schaffer1979} Not everything can be sensed without being perturbed (like qubit - Fig.\ref{fig:fig0.1.1}.h)\cite{Leggett2005}. This conflicts with sensing \textbf{information that is intrinsic to an environment} when users are considered as an environmental component. Things can be detected without necessarily be measurable. However, devices can detect statistical information (probability of existence: something that has no definite value at a given time) into perceptible measurements (Fig.\ref{fig:fig0.1.1}.i). Despite an inherent level of randomness, such information generators can be widely accepted for classifications supporting decision making (Fig.\ref{fig:fig0.1.1}.j): information uncertainty is not always a source of classification unacceptability. The quality of a classification \textbf{relies highly on sample preparation and instrumental setting}. Depending on the projection of environment samples on a set instrument, classes can be discriminated or not (Fig.\ref{fig:fig0.1.1}.k). Depending on the settings, classifiers can even project information on a higher informational space to observe imperceptible classes of environments, just by a \textbf{proper hardware physical adaptation} (Fig.\ref{fig:fig0.1.1}.l).\\[3pt]
If tools help sensing more information, perceptions of it shall be globally accepted. Now, technologies promote more information traffic, we are exposed to more complex information in our environment from many devices, sometimes too much to handle, with fuzzy or highly specialized semantic notions, misleading us in categorizing informational classes. Information can also be pruned to bias its acceptability. Originally, sensing technologies are not meant to consider how information will be exploited as long as they properly measure a physical carrier. But since sensing technologies are more involved in machine-learning supported decision-making and public acceptance will strongly depend on their performance for such applications, it is necessary to consider informational figure of merits to select new sensing technologies aiming pattern classification.\\[3pt]
\textbf{In the whole monograph, the strategy for sensing technology acceptance is envisaged by conditioning hardware to directly project information classes, ideally without software support, to make classifier to be invulnerable to data manipulation nor thieving.}

\subsection{Complexity of Molecular Fields in Chemical Environments}
\label{Ch1.1.2}

\begin{figure}
  \centering
  \includegraphics[width=1\columnwidth]{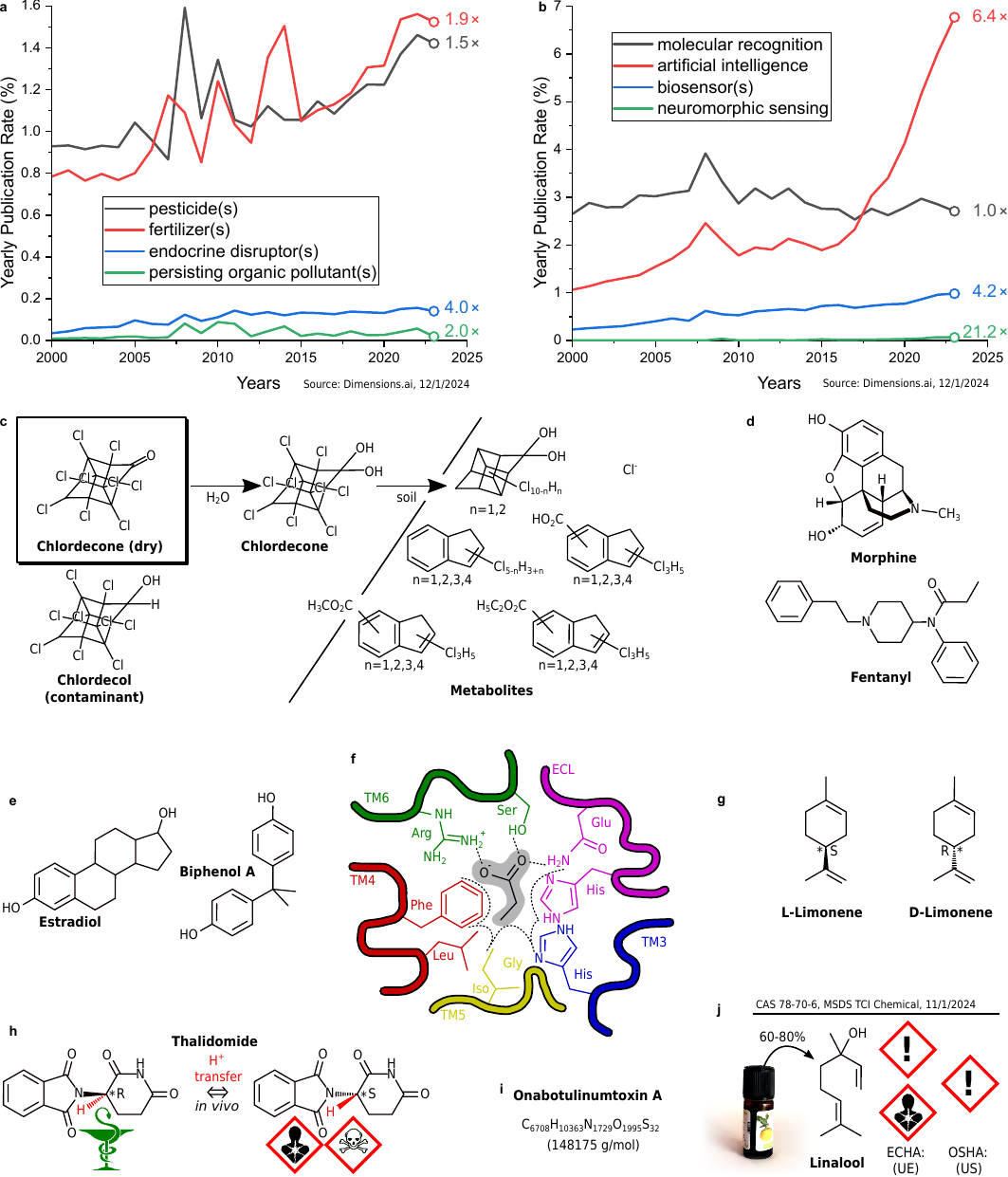}
  \caption{\textbf{Current Context and Peculiarities on Chemical Classification $\vert$ a,} Publication rates of four different classes of environmental pollution in the scientific literature of the XXI\textsuperscript{st} century. \textbf{b,} Publication rates of four different technologies applied for environment sensing in the scientific literature of the XXI\textsuperscript{st} century. \textbf{c,} An "eternal" persisting organic pollutant which decomposes \textit{in naturo} to metabolites which are structurally very different from their parent molecule \cite{Chevallier2019}. \textbf{d,} Comparison of two opioids which have very similar geopolitical impacts and neurotoxicological effects, but very different chemical structures \cite{Sun2021}. \textbf{e,} Comparison of an estrogen steroid and a synthetic xenohormone acting as an endocrine-disruptor, but with a very different molecular structure \cite{Rubin2011}. \textbf{f,} Complexation environment of propionate (as the deprotonated form of the odoriferous propionic acid molecule) on a human olfactory receptor \cite{Billesboelle2023}. \textbf{g,} Two enantiomers which are perceived with very distinctive smells by humans.\cite{Friedman1971} \textbf{h,} A sedative and antiemetic drug which racemizes \textit{in vivo} causing congenital malformations \cite{Tokunaga2018}. \textbf{i,} Molecular formula and molar mass of Botox\textsuperscript{\textregistered}
 illustrating the structural complexity of the most powerful neurotoxic protein ever reported \cite{Kanehisa2000}. \textbf{j,} A reprotoxic chemical in Europe but not in America.}
  \label{fig:fig0.1.2}
\end{figure}

\textbf{Molecules are simpler to detect than chemistry is to be recognized}. Like temperature, pressure or light, chemistry is an important environmental property to perceive: It \textbf{impacts health} and \textbf{influences decisions}. Chemistry is no state nor wave: it forms a huge set of ambivalent objects that are both \textbf{physical carriers} and \textbf{features of information}. Their uniqueness is defined by the \textbf{structure of molecules}. Molecules have a mass, their bonds vibrate and their chemical groups can react changing their enthalphy: \textbf{they can be detected with various analytical techniques}. Molecules may compose only a small feature of a pattern belonging to a complex class, which some are to be detected with technologies (Fig.\ref{fig:fig0.1.2}.a-b).\\[3pt]
Molecules are \textbf{transient} in environments: they may react, creating other molecules (Fig.\ref{fig:fig0.1.2}.c): detecting a particular structure may practically be challenging but not particularly relevant. Molecules are physically \textbf{invasive}: they penetrate skin, mucosa and interfere with our biochemistry.\cite{PruessUstuen2011,Yang2015} Their perception can be antagonist to their effects: hedonicity, umami and pleasence may insidiously condition addictions or alter rationality/consciousness.\cite{Nestler2004,Zhang2010,Bontempi2024} Also, they may be repulsive but lacks of them may be physiologically detrimental.\cite{Morini2024} \\[3pt]
\textbf{There is not always a trivial relationship between molecular structures and impacts defining their class}: Drugs with very distinct chemical features may activate the same receptors and induce similar demographic consequences (Fig.\ref{fig:fig0.1.2}.d). This difference raises important challenges for chemical pollution assessment and how using chemicals without predicting their impacts \textit{a priori} (Fig.\ref{fig:fig0.1.2}.e). The hazard of a structure can also be environment-dependent (Fig.\ref{fig:fig0.1.2}.h). Chemicals can be classified by their effects rather than their structure, but such categorizing remains people-dependent, case-specific and generates conflicts of usage (Fig.\ref{fig:fig0.1.2}.j).\\[3pt]
Chemicals are also substantially different to other physical information carriers, as they are \textbf{numerous}. Just the set of synthetically possible molecules with only nine elements of the periodic table and less than 17~atoms contains over ten millions of different structures. \cite{Ruddigkeit2012, Buehlmann2020} And yet, one of the most powerful neurotoxin ever reported is a protein, made of only four elements, but over 2$\times$10\textsuperscript{4} atoms, (Fig.\ref{fig:fig0.1.2}.h): Metrology on single molecules is highly important to target specific threats, however, \textbf{it seems unrealistic to invent at least one selective detectors for each molecule human or nature create, nor to verify ultimately their selectivity in practice}.\cite{Muir2006,Brown2008,Howard2010,Scheringer2012, Breivik2012,Schymanski2017,ReppasChrysovitsinos2018,Wang2020}\\

Despite various sizes and shapes, many molecules can be sensed through \textbf{our senses}. \\[3pt]
Chemo-perception is generically associated to various functions involving classification of molecules, by the expression of receptors that are sensitive to the environment chemistry of cells they are bonded to their membrane. \textbf{Receptors} directly interact with our exposome to inform us on diverse states of an environment, while deeper receptors governing our biochemistry and physiology may require the least interference with exogenous molecules to be fully functional.\\[3pt]
The sophistication of chemical-recognition relies on the \textbf{structural complementarity} between sensed molecules and macromolecular receptors made of \textbf{only 21 aminoacid bases} (Fig.\ref{fig:fig0.1.2}.f). This complementarity is subtitle enough to be \textbf{enantiospecific}, as chiral molecules can smell different (Fig.\ref{fig:fig0.1.2}.g).
Each receptor on a neuron has the \textbf{same transduction mechanism}: modulating the ability of a channel on a cell's membrane to transport electrochemical energy, triggered by a conformal change of the receptor host with a molecular guest. Unlike electronics, transduction is not processed via a single kind of information carrier (electrons): Projection is parallelized on an input node of a physical classifier with about a \textbf{hundred of different neuro-transmitters} defining the neural activity at the level of each cell and/or globally on multiple cells in a network of them.\\[3pt]
In the case of human olfaction, \textbf{339} different G-coupled proteins sensitize \textbf{6$\times$10\textsuperscript{6}} olfactory receptor neurons.\cite{Moran1982,Purves2003,Malnic2004,March2022} Smelling/tasting allows animals to rescue,\cite{Fenton1992,Greatbatch2015} find resources,\cite{Jackson2021,Zhong2023} assess peoples' anxiety,\cite{Wilson2022} or retrieve corpses.\cite{Martin2023}. Well trained dogs can discriminate sick patients' sweat from healthy ones', faster than nowadays analytical techniques.\cite{Williams1989,Essler2021} Before modern medical tools, ancient physicians used to rely on all their senses to diagnose patient health, smell and taste included.\cite{Harpas2021} \\[3pt]
Because of the high dimensionality of such perception field as human olfaction, the \textbf{representation of fundamental/primary scents in a vast informational space is still a challenge to map}.\cite{Sell2006,Haddad2008,Haddad2008a,Kumar2015,Keller2017,Zhou2018,Licon2019,Lee2023}

\section{When Sensing with Sensors makes no more Sense}
\label{Ch1.2}

"Sens" is an affix used in many English terms by several communities who project \textbf{various information} when referring to senses, sensors, to sense, sensitivity or sensing. Experimental research on "sensing" may seem senseless if limited to "feel or experience something without being able to explain exactly how" (\href{https://dictionary.cambridge.org/dictionary/english/sensing}{Cambridge dictionary}). It is however not always simple to \textbf{qualify undefinable information} as physical measurements (Subs.\ref{Ch1.1.1}), nor to \textbf{quantify innumerable information} with accepted metrics (Subs.\ref{Ch1.1.2}).\\[3pt]
To attempt normalizing devices to sense up to these limits, \textbf{classifiers} are presented here in this section as an extension of metrological sensors hardware (Subs.\ref{Ch1.2.1}) to devices aiming pattern classification (Subs.\ref{Ch1.2.2}), as a continuum with classifiers from software engineering.\\

\subsection{Chemical Sensors and Physical Analyzers to Measure Analytes}
\label{Ch1.2.1}

\begin{figure}
  \centering
  \includegraphics[width=1\columnwidth]{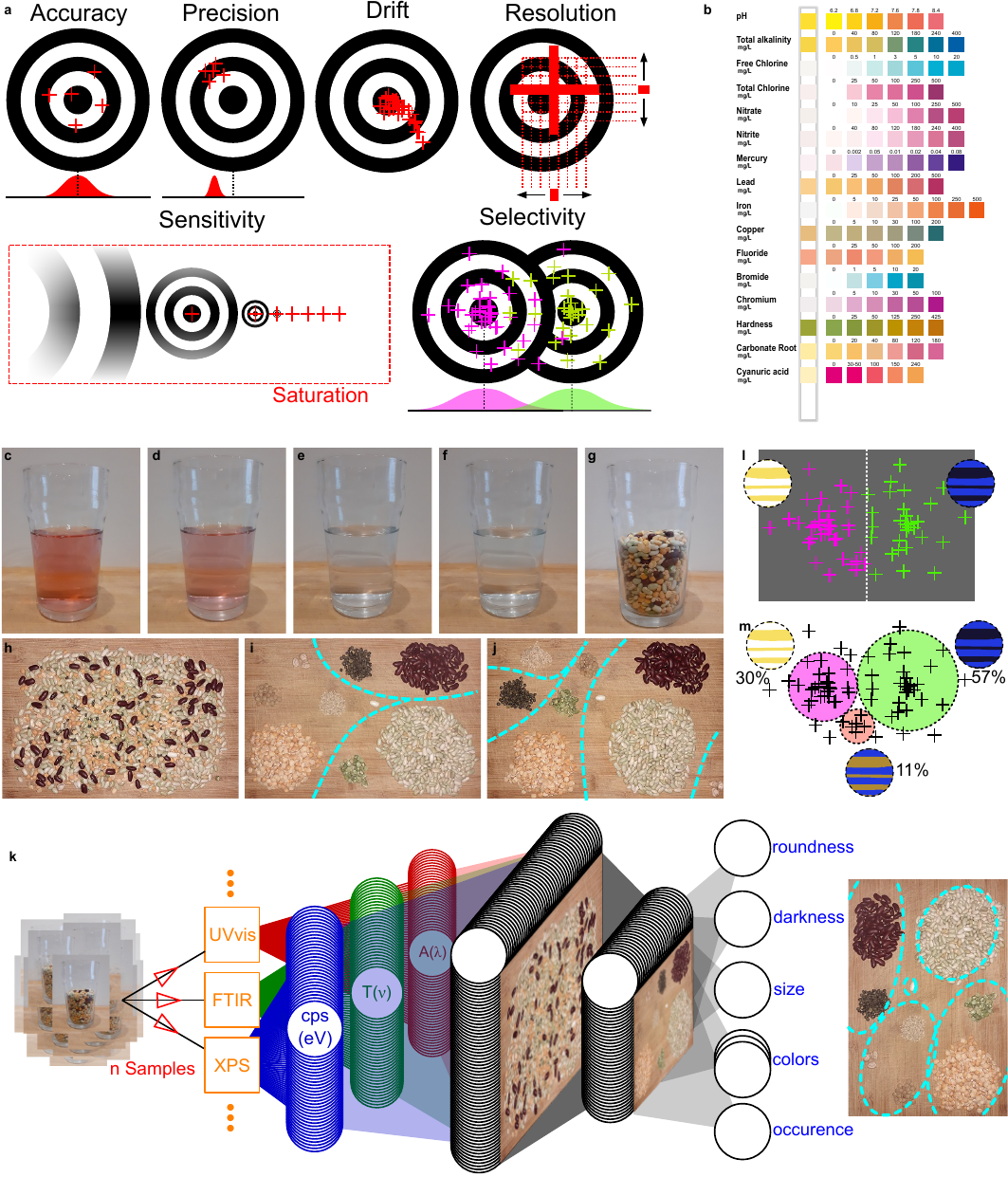}
  \caption{\textbf{Sensing Chemistry $\vert$ a,} Representation of a few figures of merit for sensors:\cite{IEEE2018} accuracy, precision, drift, resolution, saturation, sensitivity and selectivity. \textbf{b,} Colorimetric test strip scaling different analytes used for water quality assessment. \textbf{c-f,} Gradual dilutions of a colored analyte in water, for which metrology could be used with a spectrophotometer and Beer-Lambert law as a quantitative model. \textbf{g,} Constitutionally complex sample. \textbf{h,} Projection of a complex sample. \textbf{i-j,} Two other projections which allow better discrimination on different criteria. \textbf{k,} Representation of an information classification used in chemometrics,\cite{Brown1996} involving the contributions of various analyzers to identify informational features on samples composing a database. \textbf{l,} Representation for a binary discrimination of "{\#}thedress" if asking individuals composing a cohort if they perceive "{\#}thedress" as gold/white or as blue/black. \textbf{m,} Identification of information classes, where Lafer-Sousa \textit{et al.} evidenced statistically under-represented sets of populations experiencing different perceptions of "{\#}thedress".\cite{LaferSousa2015}}
  \label{fig:fig0.2.1}
\end{figure}

Chemistry is hard to sense in electronics because of the lack of chemical specificity in its materials. Interfacing molecules with functional materials in microelectronics is complex.\cite{Joo2008} Technologies are either \textbf{chemical sensors} or electrically-controlled \textbf{analyzers}. For sensors, (bio)molecular \textbf{receptors/probes/guests} detect chemical \textbf{analytes/targets/hosts} to measure their content in samples.\cite{Schneider2008,Liu2009,Pinalli2013,Busschaert2015} A probe immobilized on a \textbf{transducer} must have affinity with one single target to promote \textbf{selectivity}: a major figure of merit of sensors (Fig.\ref{fig:fig0.2.1}.a). Chemical \textbf{detectors} do not always require transducers and can be seen by eye if colorimetric (Fig.\ref{fig:fig0.2.1}.b).\cite{Lim2009,Li2016,Li2019,Maho2020,Gaggiotti2020} On sensors, suitable transducers to read them can be electrochemical,\cite{Grieshaber2008} optical,\cite{McDonagh2008} or acoustic.\cite{Grate2000} Based on the reactions' reversibility of targets with probes, detectors are either single-use or not.\cite{Kruppa2006} Their use is limited within targets' stability boundaries: Biochemical detectors are fragile, not long-lasting and require care for handling and storage.\cite{Matange2021} \textbf{They are hardly suitable for customer electronics}. Analyzers rely on physical mechanisms to exalt a specific physical feature on molecules. By varying a setting over a range, molecular fingerprints are scanned to compose normative databases. A non-exhaustive list of them includes spectrometers of mass, UV-visible light (UV-vis), Raman, Fourier-transform infrared (FTIR), X-ray photoemission (XPS) and chromatographs. They are laboratory instruments, highly automated nowadays, but require qualified users to operate them to avoid misuses and false-interpretations. Because of this, their high cost and the difficulty to miniaturize them, \textbf{they are also hard to implement in customer electronics}.\\[3pt]
Both chemical sensors and analyzers can serve qualitative and quantitative assessments. If samples are simple in their composition, targets may not screen each other: correlating their activity rates to their concentration allows using sensors for metrology (Fig.\ref{fig:fig0.2.1}.c-f). If samples are complex in their composition (Fig.\ref{fig:fig0.2.1}.g), quantifying each component requires stronger means to efficiently project sensed measurements on an informational space (Fig.\ref{fig:fig0.2.1}.h-j): chemical sensors need to "filter" the different constituents by accumulating responses of \textbf{many probes on an array},\cite{Potyrailo2016,Albert2000} and analyzers to enrich their response by \textbf{increasing the number of physical features} to sense (Fig.\ref{fig:fig0.2.1}.k).\cite{Hierlemann2008}\\[3pt]
Disregarding the quality of hardware, effective information projections are diverse (Fig.\ref{fig:fig0.2.1}.i-j). If one aims at projecting based on the knowledge of well-defined molecules as information classes, \textbf{one finds only what one seeks} (Fig.\ref{fig:fig0.2.1}.l). It is the case for chemical sensors where probes are pre-selected to sensitize arrays and ease sample discrimination with chosen chemical markers. But if information is collected from large sets of samples, its organization in a space can drive discriminating samples by their structure (Fig.\ref{fig:fig0.2.1}.m).\cite{Chapman2020} If classes are defined on an informational basis and not on a physical one (semantically, not based on molecular properties), classifiers discriminate complex samples identified as generic sets of structurally characteristic environments, that are vague in their molecular compositions: "pollution" for instance.\cite{Wang2020,Naidu2021} In some applications, quantitative and multi-parametric assessments are often means to few-score classifications, if no binary classifications of true positives and negatives: disease staging for instance.\cite{Gonnella1984}\\[3pt]
\textbf{In the Chap.\ref{Ch2} of this monograph, organic semiconductors are studied as non selective but specifically sensitive materials,\cite{Peveler2016} versatile in their structure to be both molecular detectors and electrical transducers for conductimetric sensing arrays (Subs.\ref{Ch2.2.1}) and in impedance analyzers (Subs.\ref{Ch2.2.2}).}

\subsection{Information Generators in the Internet-of-Things to Monitor our Activities}
\label{Ch1.2.2}

\begin{figure}
  \centering
  \includegraphics[width=1\columnwidth]{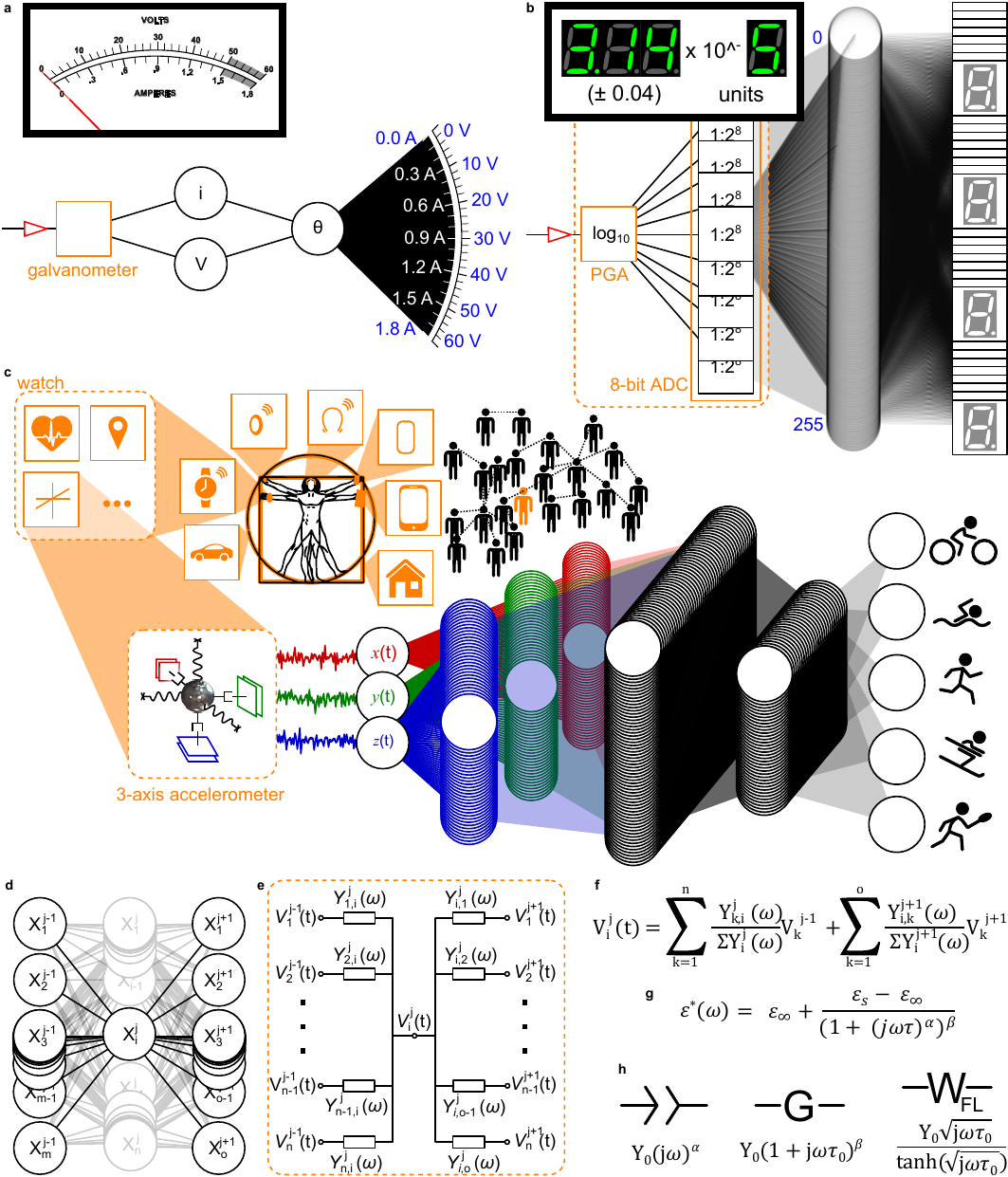}
  \caption{\textbf{Connected Sensors $\vert$ a,} Classifiers for an analog sensor, physically projecting an infinite number of classes as a dial pointer position in a galvanometer. \textbf{b,} Classifiers for a four-digit seven-segment display, to convert an analog voltage value to a finite number of classes. \textbf{c,} Classifiers of classifiers: depiction of a three-axis accelerometer as a classifier for motion activities, as information generator in an IoT device. \textbf{d,} Generic structure of a three-layer classifier as a nodal network. \textbf{e,} Analogy in analog electronics with a voltage divider. \textbf{f,} Expression of i\textsuperscript{th} node on j\textsuperscript{th} layer's voltage state of the physical classifier in Fig.\ref{fig:fig0.2.2}.e as an expression of directly connected nodes' voltage, pondered by the admittance of each interconnection. \textbf{g,} Generic expression of the Havriliak-Negami permittivity for degenerated Debye relaxations observed in diffusion-limited electrochemical circuits.\cite{Havriliak1966} \textbf{h,} Three well-referenced electrochemical circuit elements,\cite{Cole1941,Davidson1951,Wang1987,Franceschetti1991,Jacquelin1991,Boukamp2003} non-existent in conventional electronics, which parameters can be modified by electropolymerization.}
  \label{fig:fig0.2.2}
\end{figure}

Sensors usages have considerably changed over time. First, most of them are not analog (Fig.\ref{fig:fig0.2.2}.a), so any measurement defined in $\mathbb{R}$ with a physical unit is projected on a display as an information set in $\mathbb{Q}$: Any digital sensor is a classifier outputting a restricted number of possibilities (Fig.\ref{fig:fig0.2.2}.b). Close enough to any physical meaning, \textbf{classifiers' information is no exact measurement, only an acceptable projection}.\\[3pt]
Nowadays' machine-learning performances on IoT devices is such that software classifiers can \textbf{acceptably} project physically non-trivial information on screens of customer electronic hardware.\cite{Baldassarre2023} These are based on high-complexity \textbf{models} and a lot of information to feed models for \textbf{calibration}. Parts of information in these classifiers come from sensors. If cameras and microphones are still involved partly in measurements, physical sensors like accelerometers serve strictly as \textbf{data generators} (Fig.\ref{fig:fig0.2.2}.c): They map dynamical patterns of motion-based activities into classes of physical practices in many customer electronic items. Their underlying model does not rely on Newton’s Second Law of motion of a massive body in a three dimensional damped oscillator, but a highly parametric algorithm fitted to a \textbf{phenomenological model} projecting information that has a high probability to be accepted by an end-user.\cite{Yang2010,Farrahi2019} Stressing the point that many physical sensors do not practically need physical theories anymore to operate, it questions on how related physical properties are to information-based figures of merits of pattern classification tasks: \textbf{do good sensors in IoT make good sensing classifiers, or can poorly manufactured sensors perform as good, or even better?}\\[3pt]
Using sensors and other sensitive information generators (keyboards, mouses, touchscreens), performances of such classifiers depend also strongly on their \textbf{acceptance}. A classifier relies on computational means to exploit a model and a database to calibrate it. Because some models are extremely complex, they may require remote connections to servers: to compute information and display results on simple electronic objects, and to access sufficiently large amount of proprietary data \textbf{to train} (in a sense of continuously calibrating) \textbf{remote classifiers} recognizing user-specific classes. This raises two important questions regarding classifiers' acceptance: how to ensure \textbf{classifiers' non-vulnerability} to information thieving,\cite{Pathan2006} and how clarifying \textbf{information rights \& ownership} between multiple data generators enriching databases and multiple foreign technology providers?\cite{Janecek2018} If not sufficiently trained, classifiers can unacceptably be wrong, which further diminishes their use, increase lacks of training, further diminishing their performances...\\[3pt]
The reciprocal dependency between well-trained classifiers and their users' participating to their training has created an interdependent relationship, unusual with conventional sensors. For metrology, experimentalists interfere the least with measurements to not intrude the studied environments. In case classifiers deliver user-specific information, users are part of the environment they study on. An important consequence is the creation of \textbf{information feedback loops} which may bias information classification in data manipulation: as an illustrative example, click farms have appeared in the last decade to generate false information to bias public opinions on interpreting satisfaction metrics. Not only affecting simple information generators, it also concerns the acceptance of sensing technologies: two famous examples is Simon Weckert's "Google Map Hack", whose artwork demonstrated a limit of sensing in remote classifiers by faking a traffic jam,\cite{Weckert2020} and the "dieselgate" scandal,\cite{ECA2019} for which counterfeiting NO\textsubscript{x} sensors' readout created massive public nonacceptance in pollution control policies.\\[3pt]
Currently, remote classifiers' information traffic is mostly dominated by shared images/videos and number/text contents, increased by queries to generative AI chatbots. \textbf{Private information related to our exogenous and endogenous chemistry is no subject to remote classifiers limitations yet}. As remote chemical information classification  could impact building phenomenological models on environmental (bio)chemical threats spreading or living species extinctions, it is crucial that these limits do not affect sensing technologies' societal acceptances prior their introduction into the IoT.\\[3pt] 
As mean to protect information generators while diminishing their dependency to external computing resources, \textbf{in-sensor and near-sensor computing} considers integrating memory and processing resources on sensing nodes to lower classification costs and encoding generated data, making them more independent.\cite{Zhou2020}\\[3pt] 
\textbf{In the Chap.\ref{Ch3} of this monograph, in-sensor (Sec.\ref{Ch3.1}) and near-sensor (Sec.\ref{Ch3.2}) electropolymerization is presented as \textit{in materio} information processing mechanism to calibrate sensing arrays by modifying their electrochemical admittance in a network (Fig.\ref{fig:fig0.2.2}.d), and exploiting electrochemical relaxations of organic semiconductor materials (Fig.\ref{fig:fig0.2.2}.g-h) in an analog electrochemical circuit (Fig.\ref{fig:fig0.2.2}.e-f).}

\section{Another Angle for Organic Semiconductors}
\label{Ch1.3}

Since the discovery of organic semiconductors, electronics has drastically changed, our relationship to technologies has evolved, the impact we have on our environment has deepen. In a context where these materials have made it to the industry for at least a decade, and in the perspective of improving information classification while protecting those who generate information, two axis are presented to introduce these materials to next-generation electronic classifiers.\\[3pt]
In the following, the technological evolution of organic semiconductors is overviewied with emphasize on specificities that are relevant for environment classification (Subs.\ref{Ch1.3.1}) and conducting polymers are presented as materials sharing properties of silicon and biopolymers to change how electronics help up better interface our environment without compromising it (Subs.\ref{Ch1.3.2}).\\

\subsection{A Wide Field for Molecular Conductors of Electricity}
\label{Ch1.3.1}

\begin{figure}
  \centering
  \includegraphics[width=1\columnwidth]{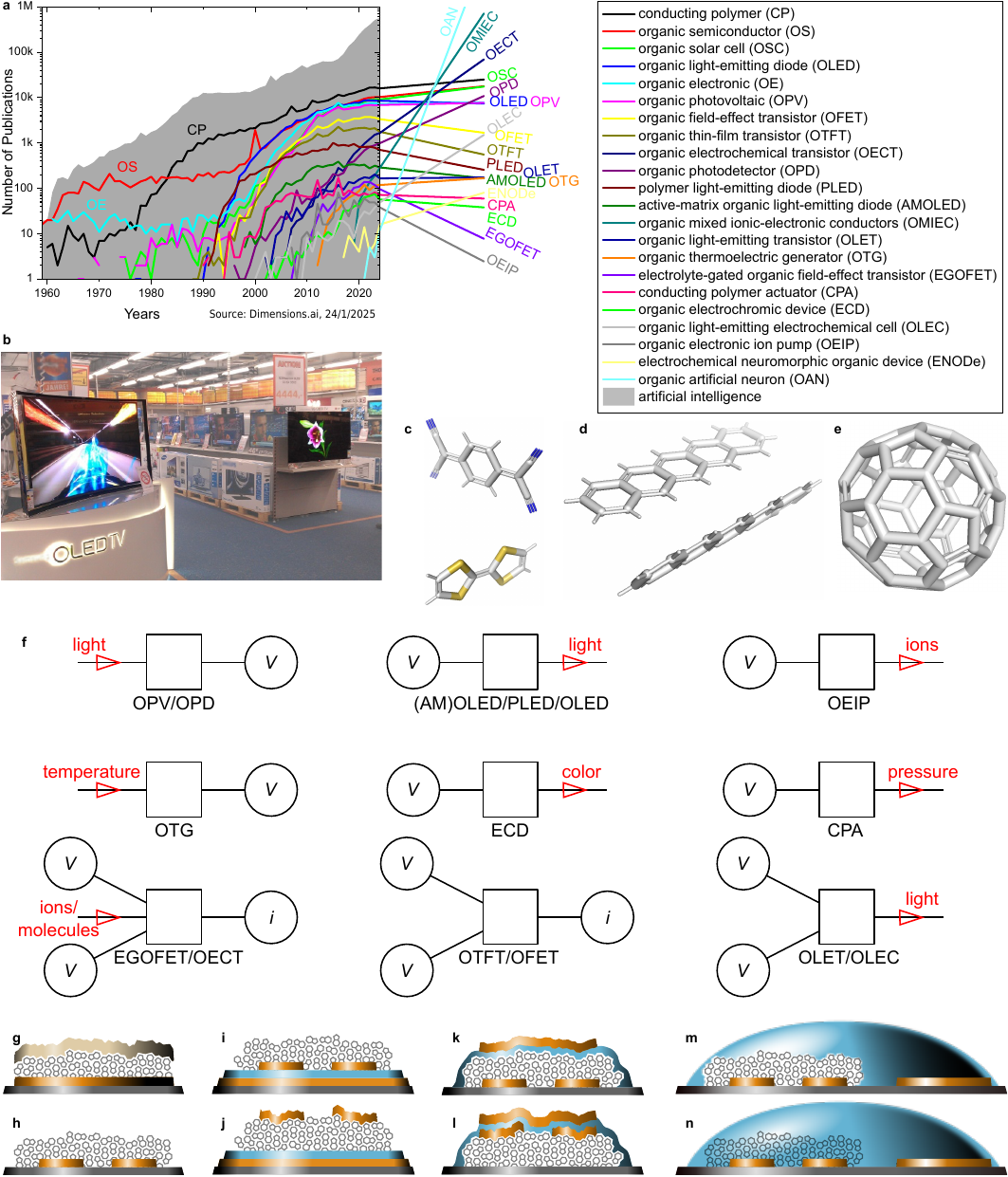}
  \caption{\textbf{65 Years of Organic Semiconductors $\vert$ a,} Number of scientific publications referring to a list of organic semiconductor technologies. The slopes beyond the graph are 20 years linear estimations from the past four-year publication rates. \textbf{b,} Photograph of two OLED TVs (presented at Pecqueur's PhD defense ten years ago)\cite{Pecqueur2014}. \textbf{c,} Co-crystal of TTF-TCNQ (structure from Kistenmacher \textit{et al.})\cite{Kistenmacher1974}. \textbf{d}, Two pentacene molecules in a crystal (structure from Schiefer \textit{et al.})\cite{Schiefer2008}. \textbf{e,} C\textsubscript{60} fullerene (structure from David \textit{et al.})\cite{David1991}. \textbf{f,} Structure of different organic semiconductor technologies as classifiers. \textbf{g-h,} Dipole structures: stacked (\textbf{g}) and in-plane (\textbf{h}). \textbf{i-l,} Dielectric transistor structures: bottom-gate/bottom-contact (\textbf{i}), bottom-gate/top-contact (\textbf{j}), top-gate/bottom-contact (\textbf{k}) and top-gate/top-contact (\textbf{l}). \textbf{m-n,} Electrolytic transistor structures: an EGOFET in a liquid electrolyte, with an in-plane gate electrode (\textbf{m}) and an OECT in a liquid electrolyte, with an in-plane gate electrode (\textbf{n}).}
  \label{fig:fig0.3.1}
\end{figure}

\textbf{Organic semiconductor} (OS) science has matured over at least 65 years of researches (Fig.\ref{fig:fig0.3.1}.a): Some compose broadly distributed devices as industrial products in consumer electronics, such as organic light-emitting diode (OLED) displays, now in many smartphones and TV for at least ten years (Fig.\ref{fig:fig0.3.1}.b). Technologies investigated since the late 80's, like organic solar cells (OSCs) for organic photovoltaics (OPV), are still important research topics, associating OS with other materials to respond to ever-growing societal energetic needs. Other fruits of OS research may have come too early in R{\&}D: such as organic thin-film transistors (OTFTs) and organic field-effect transistors (OFETs), envisioned to be an alternative to silicon (Si) in the early 2000's. Back then, the state of the art for the complementary metal-oxide semiconductor (CMOS) fabrication process was still at the 130-nm node: Developments by narrowing OS' density of states and \textbf{reducing disorder} over large wafers was more risky than going down Moore’s law. All these technologies have in common to be associated with at least one OS, if not many, between electrodes in a device.\\[3pt]
\textbf{Organic electronics'} (OE) premises started in early 60's with the assessment of tetracyanoquinodimethane's (TCNQ) electrical conductance as an adduct with an electron donor, among which,\cite{Melby1962} with tetrathiafulvalene (TTF) forming TTF-TCNQ (Fig.\ref{fig:fig0.3.1}.c) discovered more than ten years later: it crystallized an interest for charge-transfer complexes as supramolecular conductors of electricity.\cite{Coleman1973,Anderson1973,Kirtley2008}. Also from early 60's, Wassermann \textit{et al.} characterized unsaturated macromolecules' electrical conductance,\cite{Wasserman1960,French1961,Murphy1961,Wassermann1961} poly(thiophene) among them,\cite{Armour1967} 17+ years before polyacetylene was reported as a semiconductor by 2000 prized Nobel laureates in Chemistry, and their coworkers.\cite{Shirakawa1977} Since then, various materials were studied, among which, pentacene (Fig.\ref{fig:fig0.3.1}.d) which is probably the most referenced small molecule semiconductor, and C\textsubscript{60} (Fig.\ref{fig:fig0.3.1}.e) which is also an sp\textsuperscript{2}-carbon allotrope owing its exceptional electron mobility to intermolecular orbital hybridization,\cite{Acquah2017} and not only intramolecular $\uppi$-$\upsigma$-$\uppi$ conjugation like 2D graphene.\\[3pt] 
Thanks to uncountable chemical variations on such molecules, different physical properties can be optimized in regard to the needs of different  technologies (Fig.\ref{fig:fig0.3.1}.a). Among these technologies, one can discern the ones which are sensitive to the environment (light, temperature, ions or molecule), the ones generating information different than electrical (light, color, pressure or ions) and the ones processing only electrons (n-type) and/or holes (p-type) in an OS (Fig.\ref{fig:fig0.3.1}.f). Specific technologies includes devices of different architectures: diodes feature OS confined between two electrodes in a stack (Fig.\ref{fig:fig0.3.1}.g) or over two planar electrodes on a substrate (Fig.\ref{fig:fig0.3.1}.h). OTFTs/OFETs can have four different configurations, where bottom-gate architectures (Fig.\ref{fig:fig0.3.1}.i-j) do not bury the OS under a dielectric thin-film like top-gate ones (Fig.\ref{fig:fig0.3.1}.k-l). Electrolyte-gated organic field-effect transistors (EGOFETs) have an OS in contact with an electrolyte medium (Fig.\ref{fig:fig0.3.1}.m). In the specific case of an organic electrochemical transistor (OECT) featuring an OS as an organic mixed ionic–electronic conductor (OMIEC), the OS is even more intimately coupled with a liquid electrolyte solution (Fig.\ref{fig:fig0.3.1}.n).\\[3pt]
\textbf{To exploit OS for their chemical sensitivity, sensing elements in this monograph have a coplanar two-electrode architecture to sense gas-phase environments and the one of an OECT for liquid phases supporting an electrolyte. They involve also p-type conducting polymers (CP) for their stability in O\textsubscript{2} and H\textsubscript{2}O containing environments, and are used at room temperature and atmospheric pressure.}

\subsection{What if Electronics could Feel and Grow}
\label{Ch1.3.2}

\begin{figure}
  \centering
  \includegraphics[width=1\columnwidth]{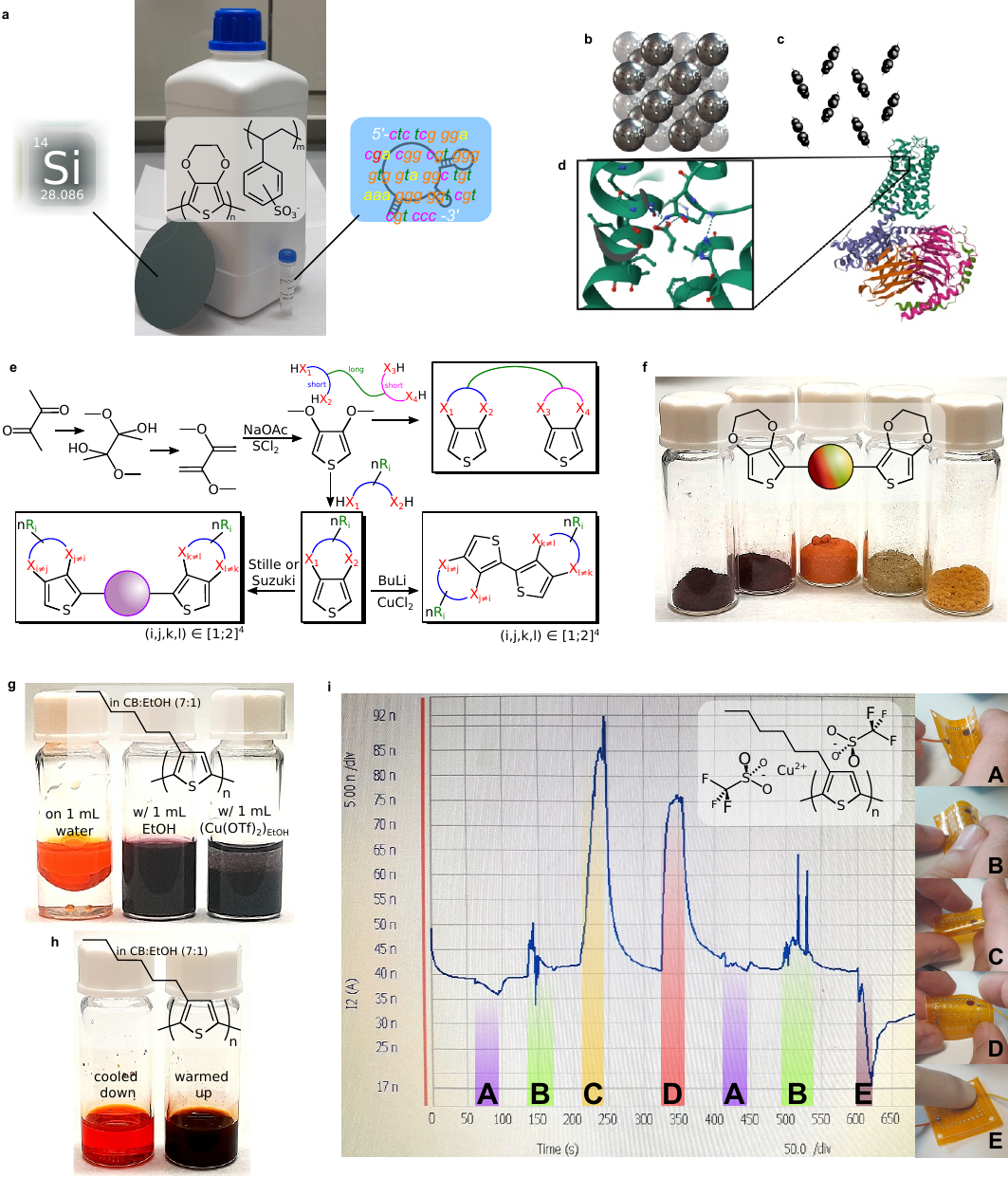}  \caption{\textbf{Specificity of Conducting Polymers $\vert$ a,} Photograph of a 3"-Si wafer (about 20~€ each), 1~L of CP formulation (400~€ the bottle) and 33.1~nmol of an aptamer (about 200~€ for 532~$\upmu$g). \textbf{b,} Si crystal (1~nm\textsuperscript{2}). \textbf{c,} Pentacene crystal (1 nm\textsuperscript{2} from Schiefer \textit{et al.})\cite{Schiefer2008}. \textbf{d,} Structure of the Human olfactory receptor OR51E2 bound to propionate.\cite{Billesboelle2022,Billesboelle2023} Scales between \textbf{b}, \textbf{c} and \textbf{d} are comparable. \textbf{e,} From the synthesis of 3,4-dimethoxythiophene,\cite{McDonald1979,vonKieseritzky2004} typical steps to prepare different electropolymerizable CP precursors. \textbf{f,} Color-dependency of CP monomers with their chemistry (here, five EDOTyl end-capped monomers as dried powders). \textbf{g,} CP sensitivity to its solvent and its doped state: A poly(3-hexylthiophene) (P3HT) solution which has no affinity with water (left), can suspend in ethanol (middle), and for which the presence of the p-dopant promote the polymer precipitation (right). \textbf{h,} Memory-effect of CP thermochromism. Left: a P3HT solution in ambient which has been warm-up until total dissolution. Right: the same solution stored at 5°C, keeping a state of CP aggregates suspension after three days in ambient. \textbf{i,} Electrosensitivity of a drop-casted CP with the mechanical stresses it experiences when coated on a flexible substrate.}
  \label{fig:fig0.3.2}
\end{figure}

Interestingly, 1977’s first paper on reporting the semiconducting behavior of polyacetylene is also about sensitizing it to different doping gases at ambient temperature.\cite{Shirakawa1977} CPs are different from Si as semiconductor and are also different from biopolymers as sensitive macromolecules (Fig.\ref{fig:fig0.3.2}.a-d). Si requires more energy than OSes to process due to the covalence of Si-Si bonds in a crystal, stronger than OS' intermolecular interactions. Si has a high melting point to reach above 1400°C and maintain a high thermal control for days to let an ingot crystallize.\cite{Dalaker2010} Saw slicing ingots to wafers generates about 40\%-loss of starting material.\cite{Li2021} Furthermore processes on wafers to oxidize gate dielectrics or to diffuse dopants require more energy to heat, cool down, maintain low vacuum, and generate more materials/chemicals/ultra-pure-water wastes. Stronger interactions in Si require more energy than OS to prepare it, but the activation energy to disturbed it is also higher: OS' lower glass transition temperature makes OS' charge carriers more sensitive to disorder in ambient, with transient and broad property dispersion in energy and space. If worth the costs, Si is an ideal substrate to control electronic materials' topology. As molecular materials, OS' sensitivity is less specific than DNA/RNA or proteins: Macromolecular structures of oligo-nucleotides or -peptides are dictated by an ultimate control of their monomeric sequences in a sufficiently stable environment. They compose biological cells whose organization is locally disordered, but globally robust.\cite{Kitano2004} Because of the high control in their sequences, these macromolecules are however very expensive to fabricate, in a very low-mass throughput process (sold for about 100~€ per 100~$\upmu$g). They also require specific handling to keep their folding,\cite{Dill2008,Maxwell2005} and cares under specific chemical conditions in the cold, for a short lifetime. \textbf{At a frontier between both realms, CP can be the right candidates to sensitize electronics classifiers as both molecular detectors \& electrical transducers}.\\[3pt] 
p-type CPs, and poly(styrene sulfonate) doped poly(3,4-ethylenedioxythiophene) (PEDOT:PSS) formulations in particular, can show durable performances in ambient and harsh environmental conditions.\cite{Duc2018,Tumova2023,Bian2023} CP's sensitivity to molecules and ions has often been reported.\cite{Migdalski1996,Zhu1996,Perepichka2002,Holliday2006,VillarroelMarquez2020,Lu2021} They are not intrinsically selective, but their design is versatile and can be tuned via many synthetic ways (Fig.\ref{fig:fig0.3.2}.e) to feature small chemical groups or to graph large biopolymer fragments.\cite{Dong2007,Berezhetska2015,Ghosh2018,Leclercq2019} Their field of sensitivity extends beyond environment's chemistry: First, their properties are also sensitive to the chemistry of their own monomers (Fig.\ref{fig:fig0.3.2}.f). CPs can be sensitive to the way they are process: they are chemically (Fig.\ref{fig:fig0.3.2}.g) and physically sensitive (Fig.\ref{fig:fig0.3.2}.h).\cite{McQuade2000} On a device, a CP's response is not binary and expresses analog patterns, gathering complex information on environmental stimuli it experiences (Fig.\ref{fig:fig0.3.2}.i). Depending on how environments imprint materials' properties, \textbf{CPs can enable multi-sensory functions to an electrical hardware on an electronic device}. \\[3pt]
Our ecosystem has evolved since the discovery of CPs: our relationship to electronics and information processing has greatly changed, affecting our environmental footprint. To increase information processing performances after Moore's law death, "beyond CMOS" strategies are pursued.\cite{IEEE_beyondCMOS_2023} Among them, neuromorphic electronics proposes to parallelize computing in hardware like artificial neural network (ANN) do in software. In this field, CPs were investigated for at least 20 years,\cite{Pecqueur2018c} first by exploiting charge delays in EGOFETs.\cite{Erokhin2005,Erokhin2007} By mimicking the way biological synapses operate, CPs exploit slow charge carrier mobilities as a resource and not a constraint for computation, unlike in serial communication. Other figures of merit are being revised, such as physical property heterogeneity are an advantage for OS classifiers,\cite{Pecqueur2018b} where dispersion enhances classification and is no technological bottleneck. A last figure of merit probably to deconstruct in the next years is the need to keep physical properties steady to program material only in their electron states. Natural classifiers are very heterogeneous structures, evolutionary by nature through their whole lifetime: organisms exploit topological plasticity as a resources, unlike electronics for which hardware is immotile with devices that should not break nor leak. CP \textbf{electropolymerizaton} could be an interesting way to manufacture electronics from information-carrying voltage inputs,\cite{Gerasimov2019} as a material/energy cost-effective way to both fabricate hardware and process information. As a bottom-up fabrication process, it consumes strictly materials needed for structuring, and requires only a low-voltage energy source. Like most recent developments on PEDOT:PSS additive manufacturing,\cite{Yuk2020,Jordan2021,Ryan2022,Martinelli2022,Liu2023,Zhou2023} it has a \textbf{great potential for installing electronic manufactures,\cite{EU2023} in territories with minimal resources and strong regulations on waste emissions\cite{EU2020}}.\\[3pt]
\textbf{The whole monograph presents results supporting the use of CPs to sensitize electronics (Chap.\ref{Ch2}) and to make it evolvable (Chap.\ref{Ch3}), before implementing both functions on practical classifiers (Chap.\ref{Ch4}).}

\chapter{Sensitive Semiconductors}
\label{Ch2}

\textit{This chapter gathers aspects of my research involving detection, transduction and information processing of one or many integrated sensing elements coated with conducting polymers.\\ The discussion is centered on four main notions: the environment to detect (Sec.\ref{Ch2.1}), the measurement to be transduced (Sec.\ref{Ch2.2}), the  model for a classifier (Sec.\ref{Ch2.3}) and its calibration technique (Sec.\ref{Ch2.4}).}\\

\section{Metrology: Counting Needles in Haystacks}
\label{Ch2.1}

Prior imagining technologies to supervise their design by themselves, it is important to wonder how we produce them by ourselves. In the particular case of sensing chemistry, this is no easy task: Detecting chemistry is both about \textbf{identifying} and \textbf{quantifying} molecules. If each detectable molecule is considered unique among others, counting them all is a challenge. Trying to reduce this large number of unique molecules into a smaller number of classes grouping chemicals belonging to a same family is complicated to apprehend: Shall only major components be counted? Shall decomposed products be counted? On what ground these classes are defined (size, polarity, chemical reactivity)? These multiple questions condition the implementation of an efficient experimental methodology to screen multiple environments and highlight a main difference with chemical sensing compared to sensors measuring more elementary physical properties like temperature or pressure: \textbf{there is more than one information carrier to count}. This is a concern to not neglect with higher technological readiness to consider the numerous impacts potential molecular \textbf{interferents} can have on a readout when sensing real-world environment samples. Therefore, a first concern lies on how presenting a reduced set of "good" representatives of relevant environments to assess a sensing element specificity, without ourselves biasing the analysis results, intentionally or not (Subs.\ref{Ch2.1.1}). A second concern lies on the potential biases related to the selection of relevant parameters to measure from a sensing element. Sensing hardware can be stimulated in many ways to exalt \textbf{a measurement} carrying an information from a perturbed environment. The quality of a measurement conditions the merits of a sensing element, despite the fact that many sorts of measurements could be made with a specific sensing element: proper selections must be made on this ground as well. In particular, on what basis shall a particular physical property of an equation model for an assumed transduction mechanism be considered as \textbf{a relevant information feature} for sensing, and what would be the benefit to consider a sensing element as \textbf{an information generator} rather than a metrological sensor for sensing chemistry (Subs.\ref{Ch2.1.2}).\\

\textit{This section is mostly associated to two studies published in} Pecqueur \textit{et al.} \textit{\href{https://doi.org/10.3390/s17030570}{Sensors}} \textbf{17}(\textit{3}), 570 (2017).\\

\subsection{Observed Environments: What, How Much and How?}
\label{Ch2.1.1}

\begin{figure}
  \centering
  \includegraphics[width=1\columnwidth]{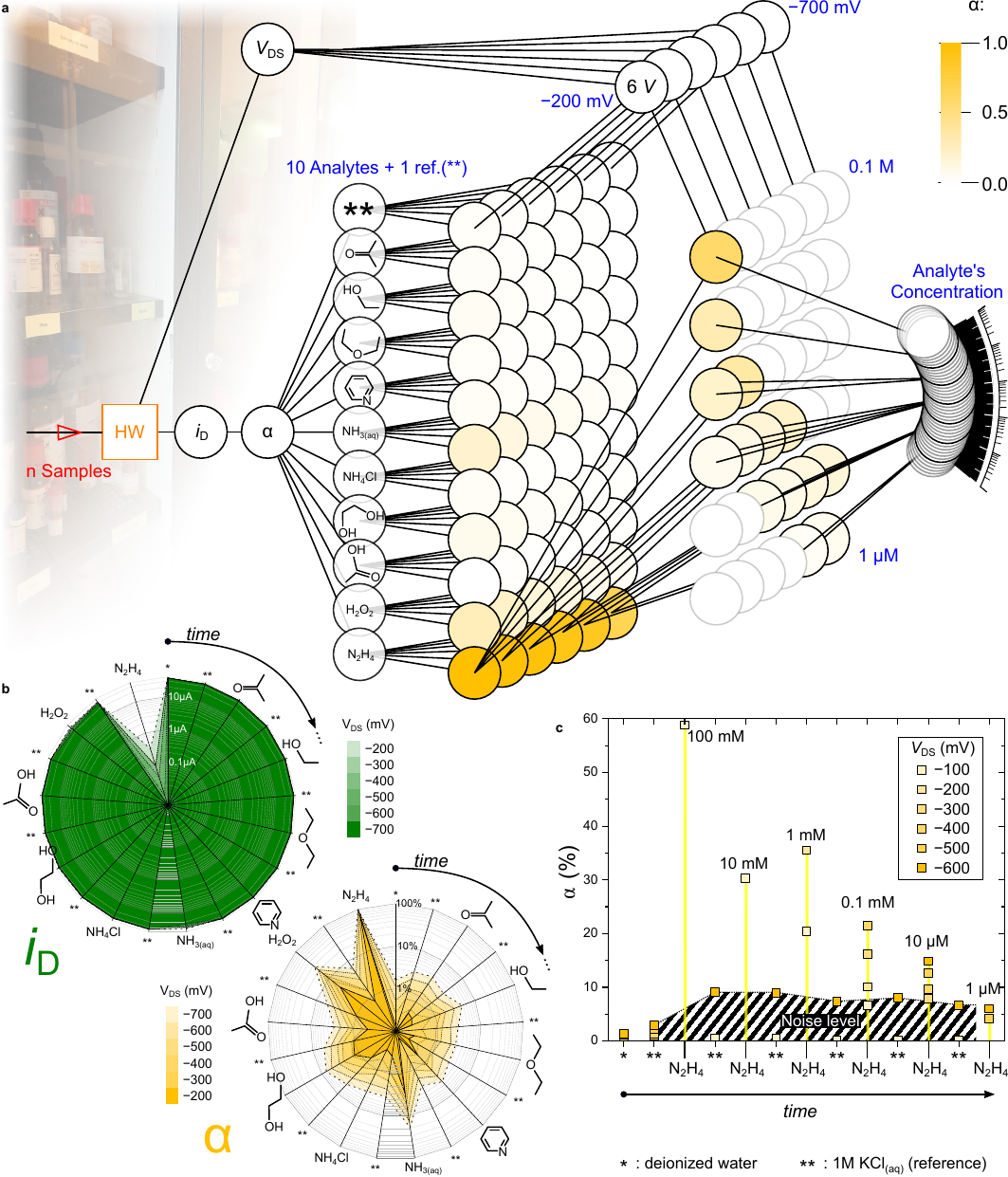}  \caption{\textbf{Metrology on Molecules with an Organic Electro-Chemical Transistor $\vert$ a,} Schematic of the classifier used in the study of Pecqueur \textit{et al.} 2017,\cite{Pecqueur2017} assessing the selectivity of the current modulation $\upalpha$ for an OECT, calculated from its drain current \textit{i}\textsubscript{D} values, upon the exposure of ten different analytes at a given concentration, and then a single analyte at different concentrations. Each analysis is simulatenously performed at multiple \textit{V}\textsubscript{DS} bias voltages to probe an analyte's concentration on the widest range.\cite{Pecqueur2017}. \textbf{b,} Kiviat diagrams for the drain current \textit{i}\textsubscript{D} values and the current modulation $\upalpha$ values, on a logarithmic scale, for the selectivity assessment of the response of an integrated Pt-gate OECT in different analyte solutions. On both charts, variables are sorted in chronological order with the test iterations starting from deionized water (*) followed by the reference electrolyte (**). In both charts, each analysis is simulatenously performed at multiple \textit{V}\textsubscript{DS} bias voltages, as presented in Fig.3.a from Pecqueur \textit{et al.} 2017.\cite{Pecqueur2017}. \textbf{c,} Sensitivity of the current modulation $\upalpha$ for the response of an integrated Pt-gate OECT exposed to electrolyte solutions containing hydrazine at different concentrations. Responses are sorted in chronological order, starting from deionized water (*) followed by the reference electrolyte (**), as presented in Fig.3.b from Pecqueur \textit{et al.} 2017.\cite{Pecqueur2017}.}
  \label{fig:fig1.1.1}
\end{figure}

A key aspect lies on how \textbf{choosing appropriate sets of molecules} to screen a detectors' \textbf{selectivity}:\\
Experimentally, one cannot interface the whole chemical space of all possibly existing molecules to a sensor, but only a restricted set of them. Chemicals (pure or diluted in a reference) can be chosen randomly in a laboratory environment, or specifically selected to demonstrate chemical specificity of a transducer. The main concern is on how an approach can be sufficiently objective to conclude on selectivity while exploring just a tiny fragment of such large chemical space.\cite{Fink2007,Reymond2012,Ruddigkeit2012,Reymond2015,Buehlmann2020}\\
In Pecqueur \textit{et al.} 2017,\cite{Pecqueur2017} "just" ten different chemicals have been tested, diluted at the same concentration in a \textbf{reference solution} (Fig.\ref{fig:fig1.1.1}). The choice for a reference is not universal and shall be paid attention to, so the transduction is enabled with no limitation other than the capability of specific targets to be detected. In the study, the reference is set to 1~M KCl\textsubscript{aq} to provide enough cations to deplete an OECT, in a solvent favoring the conducting ionomer swelling without threatening its integrity (ions and solvent could have been though for their relevance in a specific application, but it was not considered in the study).\cite{Pecqueur2017} Today, many studies have explained in detail important principles for an OECT transducer.\cite{Bernards2007,Kergoat2012,Inal2014,Rivnay2015,Nielsen2016,Friedlein2016,Kaphle2016,Inal2017,Friedlein2017,Friedlein2018,Rivnay2018,Zeglio2018,Flagg2019,Sophocleous2021,Paudel2021,Tropp2023} At the time of the experiments, platinum (Pt) was selected as metal to pattern OECTs with local gate electrodes (Pt being a standard in electrochemistry)\cite{Inzelt2015}. Usually, voltage and ion concentration set the limit of an OECT current. It was only after micro-fabricating and testing with the chosen reference that authors were aware of other works stating that Pt is a poor choice for a metal with a small-sized gate electrode to enable sufficient voltage drop across the electrolyte/channel interface.\cite{Cicoira2010,Tarabella2010,Yaghmazadeh2011} Therefore, the choice for ten different analytes was oriented towards identifying electrocatalytically redox-active compounds on Pt (like H\textsubscript{2}O\textsubscript{2}) to enable the transduction mechanism and compensate the poor selection of electrode metal. The set of tested molecules was then completed with counter-examples available on site. The study concludes that small Pt-gate OECTs are highly selective to hydrazine (N\textsubscript{2}H\textsubscript{4}), but other choices made on the analyte set could have greatly impacted the outcome for the selectivity assessment.\\[3pt]
The number of analytes to test was set considering the time required to test each of them, with "enough" chemical diversity to probe. In this study,\cite{Pecqueur2017} \textbf{small-sized} (<~80~g/mol) organic and inorganic analytes with \textbf{different chemical groups} (amine, imine, alcohol, ketone and carboxylic acids) \textbf{and properties} (aromaticity, polarity, ionicity and Br{\o}nsted acidity) were selected to be solubilized, at most at 100~mM, in the reference medium: a first selection parameter is therefore the solubility. To be systematic in the approach, the setup had to be adapted considering health and safety measures for all analytes, in particular N\textsubscript{2}H\textsubscript{4}.\cite{Matsumoto2016,Nguyen2021} To not discriminate any analyte, a same setup had to be used generically. However, complex setup adaptation may slower significantly the throughput for sequentially exposing analytes for selectivity/sensitivity assessment. No particular feature predicted the selectivity for N\textsubscript{2}H\textsubscript{4}, as the reactivity is assumed to be ruled by electrocatalysis on Pt, and this property does not scale with a known metric (such as Nernst potentials for redox activity or pK\textsubscript{a} constants for Br{\o}nsted acidity). The chosen positive examples were only based from the state-of-the-art literature knowledge: While NH\textsubscript{3} and H\textsubscript{2}O\textsubscript{2} showed some current modulation ($\upalpha$) at V\textsubscript{G}~=~{\textminus}200~mV, respectively 3.0 and 4.6\% at 0.1~M in 1~M KCl\textsubscript{aq}, N\textsubscript{2}H\textsubscript{4} shows 68.2\% \textbf{under exactly the same experimental conditions}.\\[3pt]
It is important for the mechanism to be \textbf{reversible} in all case for high-added value sensor aiming multiple uses. To verify this, it is important to expose the reference medium each time after exposing and verifying systematically the reversibility (Fig.\ref{fig:fig1.1.1}.b,c). Even if the chosen analytes do not degrade the transducer materials, it is worth noticing that partial poisoning may occur and limits sensitivity (Fig.\ref{fig:fig1.1.1}.c). It is therefore important to develop \textbf{well-defined rinsing protocols} to regenerate detectors to their pristine state. Such a practice conditions greatly a Kiviat diagram: a representation for selectivity tests (Fig.\ref{fig:fig1.1.1}.b). In such representation, the selectivity pattern of a measurement is highly depending on the exposed analyte sequencing when the axes are ordered chronologically. However, there is no strict rule on the axis ordering to respect in regard of the experiment, so a particular care shall be taken when visualizing such representation to see whether the axis order remains strictly the same when comparing different selectivity tests, to not bias the selectivity interpretation with such format. We can also observed in Fig.\ref{fig:fig1.1.1}.b that such a selectivity pattern depends highly on the \textbf{measurement} which is considered: whether we consider either the drain current (\textit{i}\textsubscript{D}) or the current modulation ($\upalpha$) as the quantity which carries the relevant information for selectivity assessment.

\subsection{Calculated Measurements: What, How Many, Why?}
\label{Ch2.1.2}

\begin{figure}
  \centering
  \includegraphics[width=1\columnwidth]{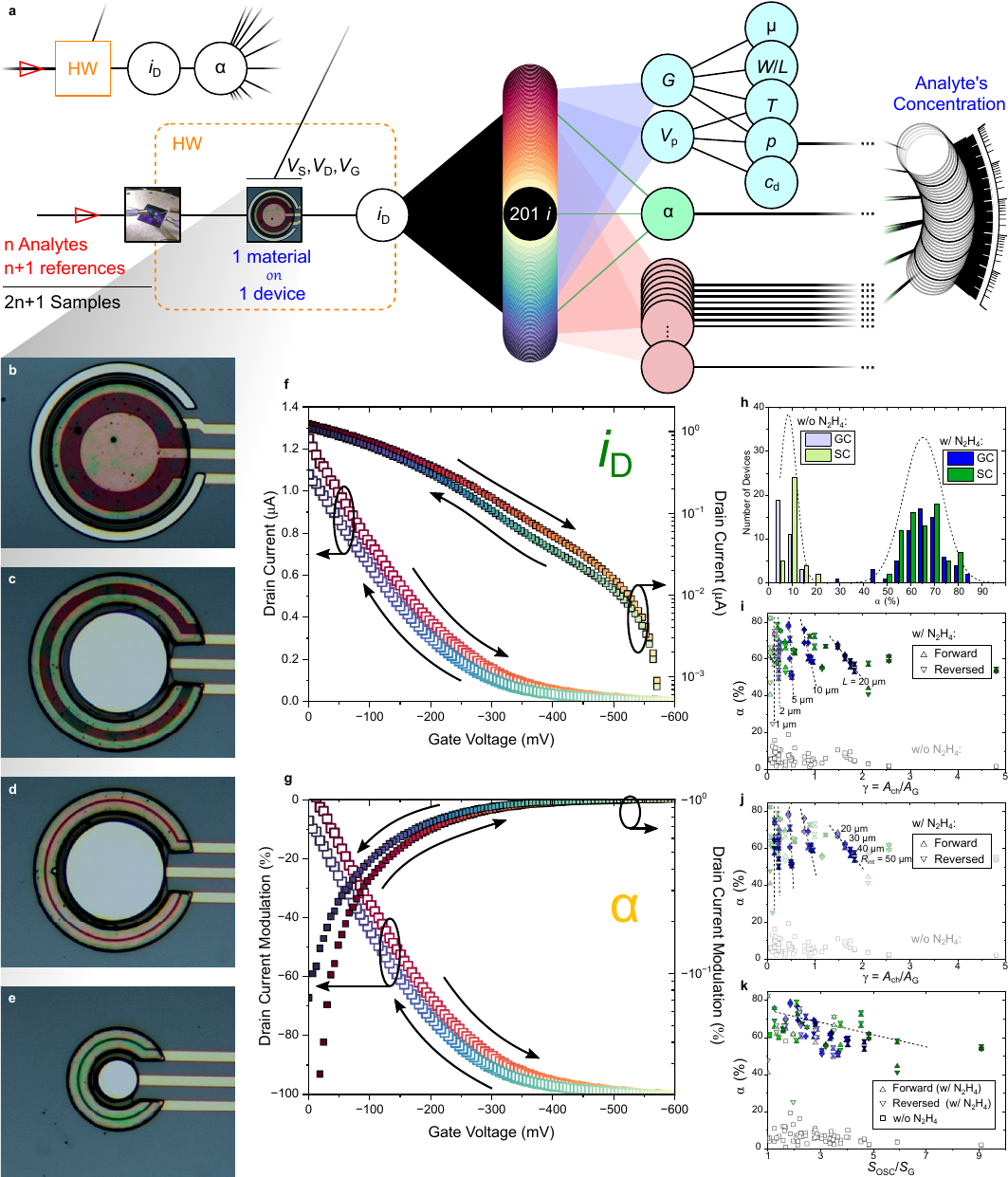}
  \caption{\textbf{Measurements from a Device in an Environment $\vert$ a,} Schematic of a generic classifier for one OECT device measured in Pecqueur \textit{et al.} 2017,\cite{Pecqueur2017} to classify a single analyte's concentration from the drain current \textit{i}\textsubscript{D} either from the doping density (p) using Bernards \& Malliaras' model (blue), the current modulation ($\upalpha$) directly calculated from specific \textit{i}\textsubscript{D} values (green), or eventually exploiting all measured \textit{i}\textsubscript{D} values (red) collected in a transfer characteristic. \textbf{b-e,} Different geometries of an integrated OECT where the gate is concentric with the CP channel, either polymer-centered (b) or gate-centered (\textbf{c}), each geometry featuring specific channel length (\textbf{d}) and specific central electrode diameter (\textbf{e}). \textbf{f-g,} As an information feature, absolute current value (\textit{i}\textsubscript{D} - \textbf{f}) and the relative current modulation ($\upalpha$ - \textbf{g}) plotted with the gate voltage applied during a reversed sweep, both in linear and logarithmic scales. \textbf{h-k,} Device geometry dependencies of a chosen measurement $\upalpha$ with the $\upgamma$ geometric factor, suggesting theoretical model limitations: dispersion of $\upalpha$ values (\textbf{h}), influence of device channel length \textit{L} (\textbf{i}) or the central electrode diameter \textit{R}\textsubscript{int} (\textbf{j}) on the dependency of $\upalpha$ with $\upgamma$ and identified correlation of $\upalpha$ with surface ratio (\textit{S}\textsubscript{OSC}/\textit{S}\textsubscript{G}) between the total area of CP and gate electrode (\textbf{k}), non-obvious in the physical model.}
  \label{fig:fig1.1.2}
\end{figure}

Selectivity patterns are extremely sensitive to the measurements considered to carry an information (Subs.\ref{Ch2.1.1}): a simple logarithmic conversion can \textbf{stretch differences} in a measurement between samples and \textbf{hide others} (Fig.\ref{fig:fig1.1.2}.f-g). In the scientific literature, this way to transform graphs can suggest a model for a technology to be acceptable or not, without having to explicit the information transformation on an axis' label of a graph, which may lead to misconceptions in the physical origin of an information. A high care for measuring \textbf{relevant physical values as information carrier} must be made.\\[3pt]
In the study of Pecqueur \textit{et al.} 2017 on evaluating the sensitivity of an OECT to N\textsubscript{2}H\textsubscript{4}, a parametric approach was exploited by using the current modulation $\upalpha$. First, as a transistor, an information carrier for an OECT can either be a raw current or a raw voltage value, depending on whether the three connectors are voltage- or current controlled. Here, OECTs are voltage controlled, and currents are recorded at different gate voltages (Fig.\ref{fig:fig1.1.2}). Depending on the model ruling a device behavior, measuring many current values under different voltage polarizations \textbf{may not bring as much information}. For instance, if a two-connector element is an ohmic resistor, each measurement at different voltage bias gives the same information. If two information values are requested, gathered within the charge carrier density (\textit{p}) and their mobility ($\upmu$), only different devices under different configurations could allow decoupling both information from the current value.\cite{Pecqueur2014} As non-linear devices, transistors are however more convenient for extracting many physical properties out of one device, by biasing gate (\textit{V}\textsubscript{G}) and drain (\textit{V}\textsubscript{D}) voltages \textbf{out of one single device}. However, false assumptions of physical models governing custom-made devices with may lead to false reporting of physical properties as information. An illustrative case is the "S"-shape obtained in OTFT/OFET's output characteristics, which is due to contact resistance limitations.\cite{Kim2013} As models from inorganic semiconductor physics have often been employed to fit OTFT/OFET data (which do not include a contact resistance term), the extracted physical parameters may be false, and may not gather the expected information if used as a sensor.\\[3pt]
\textbf{Different models} for OECTs have been reported in the literature in the last decade,\cite{Kaphle2016,Stoop2017,Faria2017,Friedlein2018,Athanasiou2019,Yamamoto2022,Cucchi2022,Weissbach2023,Skowrons2024,Bisquert2024,Bisquert2024a,Bisquert2025} with the one of Bernards \& Malliaras' reported in 2007 on the physical modeling of both steady and transient behaviors of an OECT drain current.\cite{Bernards2007} It describes the steady current dependency by \textbf{two electrical parameters}, themselves governed by \textbf{five physical properties} (Fig.\ref{fig:fig1.1.2}.a in blue).\cite{Bernards2007} It reports also on the relevance for the single $\upalpha$ parameter for sensing application, which involves only two current measurements, three to average the hysteresis which is often experimentally observed when reverse biasing an OECT for a transfer characteristics (Fig.\ref{fig:fig1.1.2}.a in green).\cite{Bernards2007} This stresses the point that a physical property in a device may express it sensitivities in fewer points, easier to access than fitting a larger set of datapoints on a physical model with an algorithm. This parametric method is however very sensitive to each device due to the multiple dependencies of the $\upalpha$ parameter with different experimental variables and intrinsic device properties. In Pecqueur \textit{et al.} 2017's study, this has been characterized by the investigation of different geometries (Fig.\ref{fig:fig1.1.2}.b-e), where the dependency of $\upalpha$ with an analyte was highly device-dependent (Fig.\ref{fig:fig1.1.2}.h-k).\cite{Pecqueur2017} A population of devices with different electrode geometries, areas and lengths will have a certain broadness in its $\upalpha$ value when exposing to an analyte sample under exactly the same conditions (Fig.\ref{fig:fig1.1.2}.h).\cite{Pecqueur2017} The measurement broadness shows linear dependencies with the channel length (\textit{L}) and with the inner-electrode diameter (\textit{R}\textsubscript{int}), despite the fact that $\upalpha$ is linearly correlated with \textit{p} as the sensitive physical property, and the $\upalpha$ parameter is a relative value, which cancels the measurement dependency to nominal physical properties of the devices, linearly dependent in the \textit{i}\textsubscript{D} expression. Still, the assumed model does not include the non-linear contribution of the CP over source and drain contacts (Fig.\ref{fig:fig1.1.2}.k) and strictly models the depletion of a long channel OECT.\cite{Bernards2007} The $\upalpha$ parameter shows some dependency with the geometry, according to a higher complexity model which shall include contact resistance effects.\cite{Kaphle2016} This study showed that if a parametric measurement is taken as an information descriptor in a sensing device, it is no ultimate figure of merit for benchmarking physical technology in a generic way, but specific to an architecture.\\[3pt]
\textbf{However, parametric measurements based on simpler transformations shall greatly increase the a sensing classifier throughput by increasing the dimensionality when including more voltage-dependent current values to enrich the information. while remaining easy to access for real-time measurement monitoring. This is detailed in the next section Sec.\ref{Ch2.2}.}

\section{Information Dimensionality: Widening the Keyhole}
\label{Ch2.2}

As metrology aims for selectivity out of single measurements (Sec.\ref{Ch2.1}), sensors can only quantify one piece of information in a chemical space. Either to differentiate multiple information, or to identify more complex ones, classifiers receptive field must be large. The sensing dimensionality can be increased by \textbf{two complementary methods}: either by gathering different sensing elements on an array, or considering measurements of single sensing elements as vectors and not numbers. In general, increasing the complexity of sensed information comes with complications: Optimal strategies have to be chosen in regard to their advantages and limitations.\\[3pt]
In the following, both methods specifically cover co-integrating different materials over conductimetric elements on a same substrate (Subs.\ref{Ch2.2.1}), and exploiting voltage, time and frequency dependencies in the multi-variate response of a single OECT transducer (Subs.\ref{Ch2.2.2}).\\

\textit{This section is mostly associated to works published in:} Pecqueur \textit{et al.} \textit{\href{https://doi.org/10.1016/j.orgel.2018.03.020}{Org. Electron.}} \textbf{57}, 232‒238 (2018), Pecqueur \textit{et al.} \textit{\href{https://doi.org/10.1002/aelm.201800166}{Adv. Electron. Mater.}} \textbf{4}(\textit{9}), 1800166 (2018), Boujnah \textit{et al.} \textit{\href{https://doi.org/10.1016/j.synthmet.2021.116890}{Synth. Met.}} \textbf{280}, 116890 (2021) \& Pecqueur \textit{et al.} \textit{\href{https://doi.org/10.3390/electronicmat4020007}{Electron. Mater.}} \textbf{4}(\textit{2}), 80‒94 (2023).\\

\subsection{More Pieces of Information with Many Sensing Elements}
\label{Ch2.2.1}

\begin{figure}
  \centering 
  \includegraphics[width=1\columnwidth]{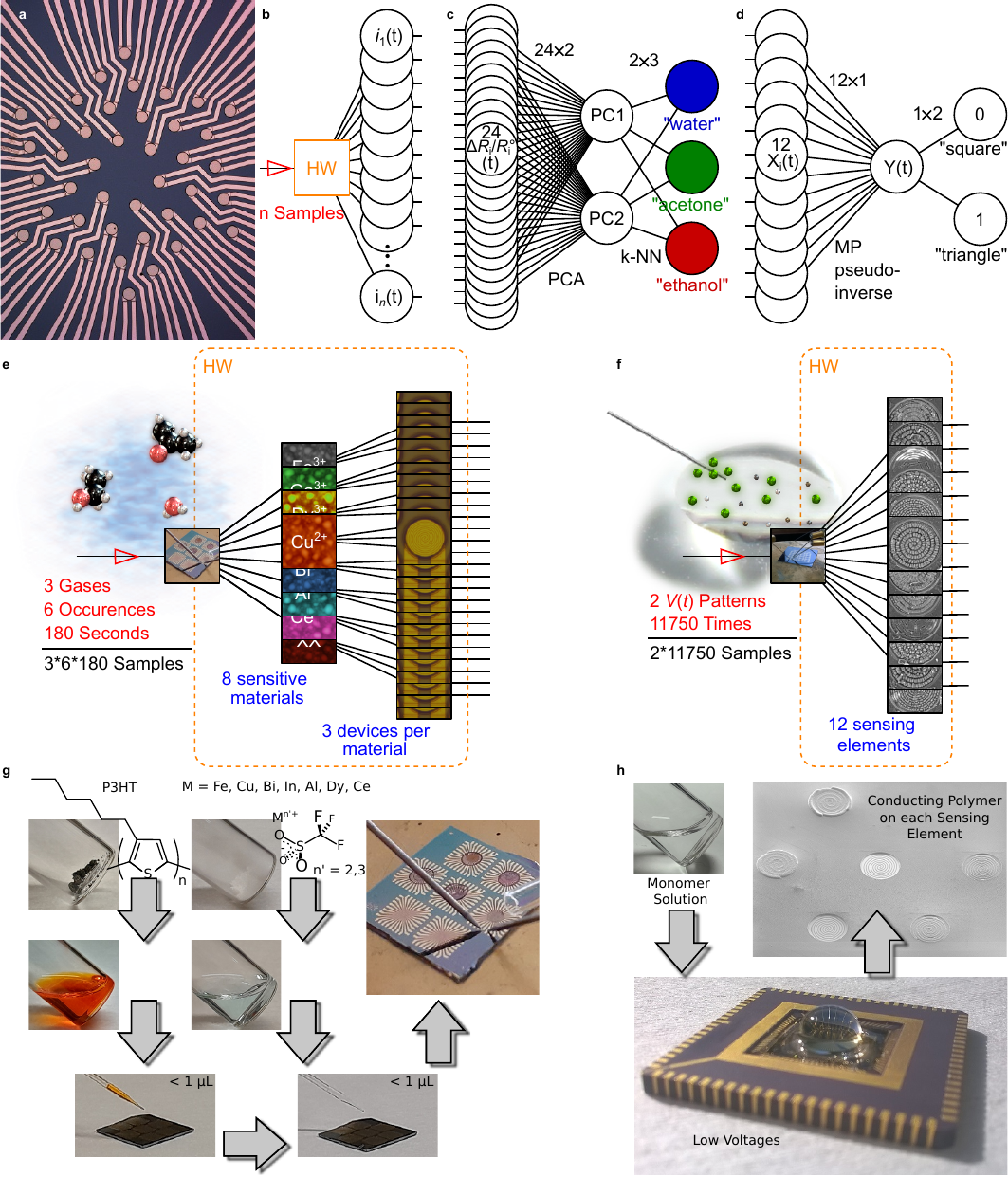}
  \caption{\textbf{More Information from More Physical Elements $\vert$ a,} Microscope picture of an array of 42 individually addressable conductimetric sensing elements (electrodes' geometry first disclosed in Pecqueur \textit{et al.} 2018)\cite{Pecqueur2018b}. \textbf{b,} Generic schematic of an array of amperometric sensing elements. \textbf{c,} Schematic of the classifier's output used in studies of Boujnah, Haj Ammar \textit{et al.} involving a multivariate data analysis technique to process multiple resistance modulations as information.\cite{Boujnah2021,HajAmmar2023,HajAmmar2024} \textbf{d,} Schematic of the classifier's output used in the study of Pecqueur \textit{et al.} 2018, involving a matrix pseudo-inversion to process multiple smoothed currents as information.\cite{Pecqueur2018b} \textbf{e,} Specific schematic of the classifier's input (hardware) used in studies of Boujnah, Haj Ammar \textit{et al.}.\cite{Boujnah2021,HajAmmar2023,HajAmmar2024} \textbf{f,} Array fabrication from successive drop-castings of a single commercial CP and multiple commercial metal triflates, as in studies of Boujnah, Haj Ammar \textit{et al.}.\cite{Boujnah2021,HajAmmar2023,HajAmmar2024} \textbf{h,} Specific schematic of the classifier's input (hardware) used in Pecqueur \textit{et al.} 2018.\cite{Pecqueur2018b} \textbf{g,} Array fabrication from successive electropolymerisations on individual sensing elements from a single solution containing an electroactive monomer, as in Pecqueur \textit{et al.} 2018.\cite{Pecqueur2018b}.}
  \label{fig:fig1.2.1}
\end{figure}

With this first method, one can use the chemical specificity of (macro)molecular conductors of electricity controlled by the affinity of electronic devices with their chemical environments. By gathering many materials, different environmental features can be sensed by an array of devices (Fig.\ref{fig:fig1.2.1}.a). CPs are great materials, as both organic detectors and electrical transducers. Their generic detection/transduction mechanism lies on how their affinity with molecules and ions affects a device conductance. CPs can either be dry- or wet-deposited.\cite{Mohammadi1986,Stussi1996,Malinauskas2001,Bone2023} Wet-deposition offers more flexibility for diversifying materials at lower energy and material costs. Some wet-deposition techniques suit better multi-material co-integration than others.\\[3pt]
Among \textbf{top-down wet-deposition} techniques, droplet deposition offers great advantages for small area co-integration (Fig.\ref{fig:fig1.2.1}.b,c,e,g). Conventional ink-jet printing is however very sensitive to the rheology of inks formulation and limits the number of formulations to co-integrate with a simple head.\cite{Gans2004,Glasser2019,Lukyanov2024} In Boujnah \textit{et al.} 2021,\cite{Boujnah2021} \textbf{sequential drop casting} was used to co-integrate different p-dopants on CP coatings. A significant advantage is that one can control the chemistry of the materials by simply compounding different precursors. The approach offers perspectives for development as it uses few quantities of sensitive materials. Different formulations can be made by associating materials at moderate cost as high chemical-engineering is not required for fine selectivity. In the study,\cite{Boujnah2021} a single CP is p-doped with different commercially available and very affordable (10~€/g, g-scale) Lewis acids (Fig.\ref{fig:fig1.2.1}.e,g). The process preserves the chemical integrity of previously deposited materials, as materials interface exclusively the device to be sensitized. Despite the lack of dopants' selectivity, the cross-reactivity of dopants with gases allows the classifier to recognize them, and this despite the large variability of the deposition process.\cite{Boujnah2021} However, an important limitation of top-down co-integration is the area of materials restricting their co-integration.\\[3pt]
Among \textbf{bottom-up wet-deposition} techniques, electroplating allows depositing an electrochemically active precursor on electrodes.\cite{Schlesinger2010} Electrographting concerns specifically small organic molecules stopping at the stage of monolayers.\cite{Belanger2011} Electropolymerization is a voltage-induced chain reaction enabling growths of $\uppi$-$\upsigma$-$\uppi$ conjugated macromolecules from thin- to thick-films:\cite{SchabBalcerzak2011} it is a technique of choice to integrate CPs with different chemistry at the scale of each device (Fig.\ref{fig:fig1.2.1}.b,d,f,h). In Pecqueur \textit{et al.} 2018,\cite{Pecqueur2018b} \textbf{electropolymerization} is used as a mask-less deposition technique to pattern a cross-linked poly(oligothiophene) on $\upmu$m-distant sensing elements. Used as OECTs, the electrical characteristics of all 12 devices showed to be very distinct to one another in terms of nominal current, current modulation at a given gate-voltage and time responses. Despite the fact that all were grown in the same conditions, on a same substrate and from the same monomer solution, large variability was observed. Surprisingly, these differences between sensing elements were all very relevant to recognize dynamical patterns of voltages induced by cation accumulation on CP coatings.\cite{Pecqueur2018b} 

This shows that the molecular identity of materials is not the only feature conditioning information out of CP sensing elements, and that different CP morphologies can be exploited for information classification.\\[3pt]
Both studies highlight the fact that diversifying the chemistry on the one hand,\cite{Boujnah2021} and the morphology on the other hand,\cite{Pecqueur2018b} on a population of CPs on a sensing array is a mean to increase the receptive field of classifiers. It however requires more time to characterize all sensing elements individually, and increases the complexity for physically interconnecting all elements on an array with a readout system.

\subsection{Squeezing Information out of a Single Sensing Element}
\label{Ch2.2.2}

\begin{figure}
  \centering
  \includegraphics[width=1\columnwidth]{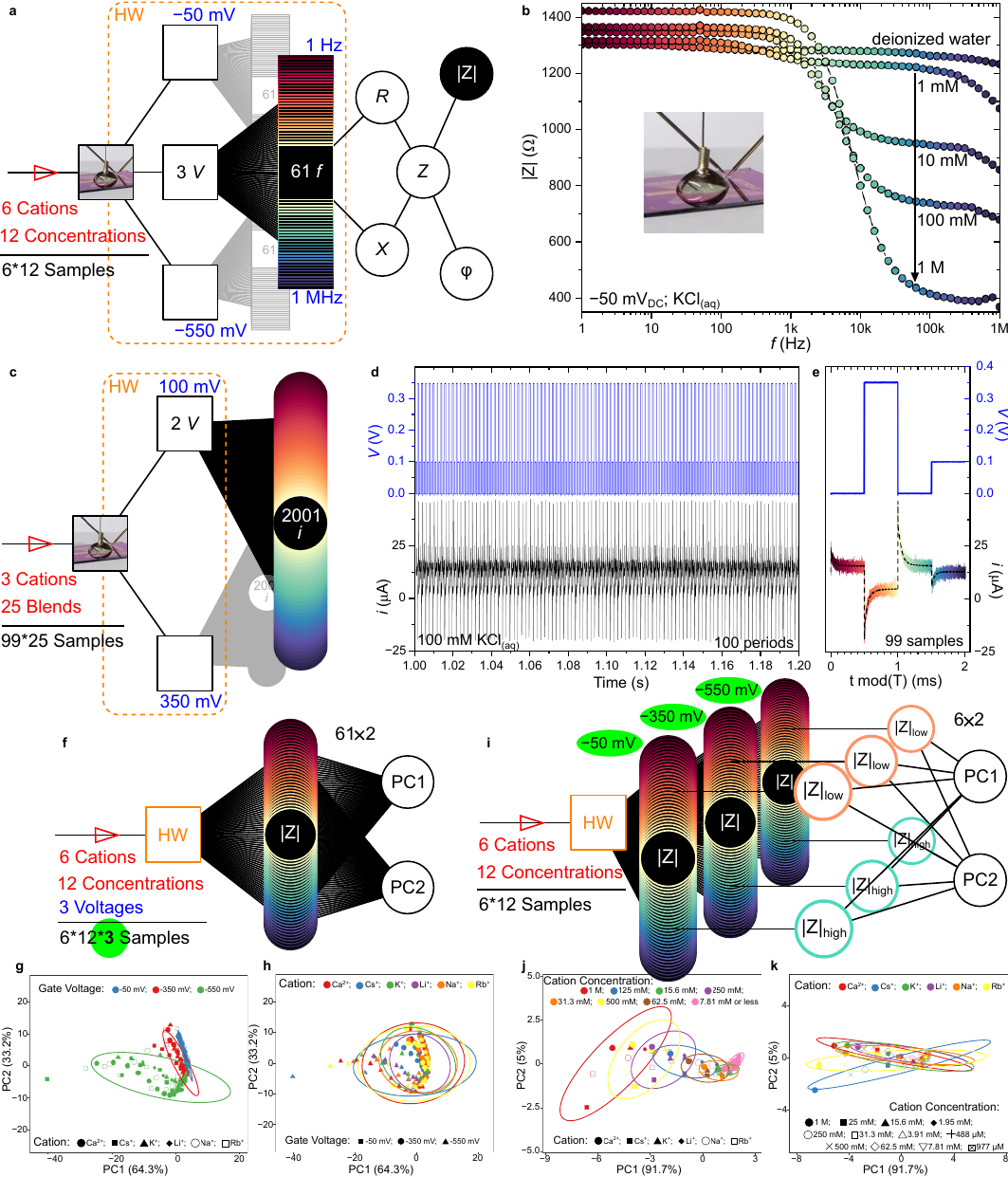}
  \caption{\textbf{More Information out of a Single Transistor Device $\vert$ a-b,} Classifier by Pecqueur \textit{et al.} on using an OECT in a diode-connected transistor configuration, with a resistive load at the source for high dimensional cation sensing:\cite{Pecqueur2018a} Classifier structure (\textbf{a}) and an illustrative set of impedance modulus data (\textbf{b}).\cite{Pecqueur2018a} \textbf{c-e,} Classifier by Pecqueur \textit{et al.} on using an OECT for majority-cation recognition in a blend:\cite{Pecqueur2023} Classifier structure (\textbf{c}) and an illustrative 0.2-second set of dynamical current data (\textbf{d}) and 2-millisecond stacking of the 99 last periods as a sub-dataset (\textbf{e})\cite{Pecqueur2023}. \textbf{f-h,} Classification by clustering of the dataset published in Pecqueur \textit{et al.} 2018,\cite{Pecqueur2018a} considering the different voltage-bias series as independent: Classifier structure (\textbf{f}), score projection grouped by voltage bias (\textbf{g}) or by cation nature (\textbf{h}). \textbf{i-k,} Classification by clutering of the dataset published in Pecqueur \textit{et al.} 2018,\cite{Pecqueur2018a} considering the different voltage-bias series as inter-dependent: Classifier structure (\textbf{i}), score projection grouped by cation concentration (\textbf{j}) or by cation nature (\textbf{k}).}
  \label{fig:fig1.2.2}
\end{figure}

With this second method, one can exploit the multi-sensitivity of materials to extract different piece of environmental information, like an analyzer. From a measurement on a single device, a classifier can program a sequence of different stresses to condition the readout mode of a single sensing element.\\[3pt]
To this aim, electrochemical impedance spectroscopy (EIS) has been studied on an OECT in an diode-connected transistor configuration by Pecqueur \textit{et al.} 2018.\cite{Pecqueur2018a} In this study, \textbf{a single device governed by two different physical mechanisms can express a two-dimensional response}. The multi-dimensionality of the response allows discriminating cations composing a solution, without involving ion-selective materials neither at the gate surface not in the semiconducting channel. Multi-dimensionality in the impedance response lies in the fact that, both the steady-state conductance of the OECT channel and the electrolyte conduction through capacitive coupling with the gate depend on the ionic content a gated solution.\cite{Lin2010,Coppede2014} However, laws ruling the polymer's electrochemical doping under direct current (DC) does not strictly have the same dependencies as the ion drift mobility in water under an alternating current (AC) signal. Therefore, an OECT can show different ion-dependent impedance shifts at low and at high frequencies, because different frequencies promotes transport through the device at specific area, which conduction is governed by different physical mechanisms (Fig.\ref{fig:fig1.2.2}.a-b). Furthermore, applying different DC components during impedance spectrocopy allows inducing different stresses on the dedoped channel without affecting the mobility of the electrolyte. Therefore, the ion-dependent low-frequency response of the device can be conditioned without impacting the ion-dependent high-frequency one. Probing the impedance at different frequencies and under different DC voltages generates an electrolyte-specific signature which depends on both the ion concentration and the nature of the cations.\cite{Pecqueur2018a}\\[3pt]
Despite OECTs' non-linearity, the rich ion-dependency of their impedance is expressed in the time-domain current output under dynamical voltage stress: In Pecqueur \textit{et al.} 2023,\cite{Pecqueur2023} an OECT stimulated with three-level voltage spikes contains \textbf{various information descriptors of the ion content in the multiple transience of it current characteritics} (Fig.\ref{fig:fig1.2.2}.c-e). At higher sampling rates than the periodicity of the two spike stimulations, few specific time delays are enough to generate a high-dimensional image of solutions, allowing to discriminate them by their content as an aqueous blend of three different cations.\cite{Pecqueur2023}\\[3pt]
In both studies however, the information dimensionality does not necessary scale with the method's throughput: \textbf{each datapoint in a time series, a spectrum or a current-voltage trace does not embed a unique information as measurements can be highly correlated}. When applying the multi-variate data analysis used in 2023's study on the dataset generated in 2017's (Fig.\ref{fig:fig1.2.2}.f-k), one observes that voltage has a significant impact on the overall data organization (Fig.\ref{fig:fig1.2.2}.g). But, considering each spectrum (61 dimensional vectors - Fig.\ref{fig:fig1.2.2}.f) independently does not promote recognizing cations just by frequency (Fig.\ref{fig:fig1.2.2}.h). However, only two impedance values taken at specific well-chosen frequencies for three different voltage values (six-dimensional vectors - Fig.\ref{fig:fig1.2.2}.i) allows discriminating analytes both by the electrolyte concentration (Fig.\ref{fig:fig1.2.2}.j) and by the nature of their cations (Fig.\ref{fig:fig1.2.2}.k).\\[3pt]
Overall, classifiers' receptive field can increase by integrating more sensing elements, or measuring more points with them. However \textbf{poor choices in materials to co-integrate, or naive settings for a stimulation protocol will mostly cost in implementing a too complicated screening technique or decrease the sampling rate to collect a complete dataset}.\\[3pt]
\textbf{Data must be pruned considering physical or phenomenological models. This is detailed in the next section Sec.\ref{Ch2.3}.}

\section{Models to rule the Information Structure or its Physical Carriers' Behavior}
\label{Ch2.3}

Improving a sensitive classifier aims at sensing better information, not just more pieces of it. Defining models which correlate environmental perturbations to measurements indicates on optimal materials and geometries to feature devices and how to stress them. A model is also required for calibration: setting constants in a transform to convert perturbations into quantitative data. In case the physical definition of sensed information is more trivial than its experimental conception, a \textbf{physical model} helps optimizing conditions prior investing on implementation. If information is not physically trivial and data are straightforward to generate, a \textbf{mathematical transform} can project the data organization and model its structure to orientate hardware optimization.\\[3pt]
In the following, both approaches of modeling specifically focus on the dynamical behavior of OECTs with a global gating immersed in an aqueous electrolyte: the multivariate impedance information is studied through EIS modeling at the scale of a single device (Subs.\ref{Ch2.3.1}), and the collective behavior of 12 different devices is exploited to discriminate frequency-modulated gate-voltage patterns using an algorithm (Subs.\ref{Ch2.3.2}).\\

\textit{This section is mostly associated to works published in:} Pecqueur \textit{et al.} \textit{\href{https://doi.org/10.1002/aelm.201800166}{Adv. Electron. Mater.}} \textbf{4}(\textit{9}), 1800166 (2018) \& Pecqueur \textit{et al.} \textit{\href{https://doi.org/10.1016/j.orgel.2019.05.001}{Org. Electron.}} \textbf{71}, 14‒23 (2019).\\

\subsection{From Physical Models \textit{ab initio}}
\label{Ch2.3.1}

\begin{figure}
  \centering
  \includegraphics[width=1\columnwidth]{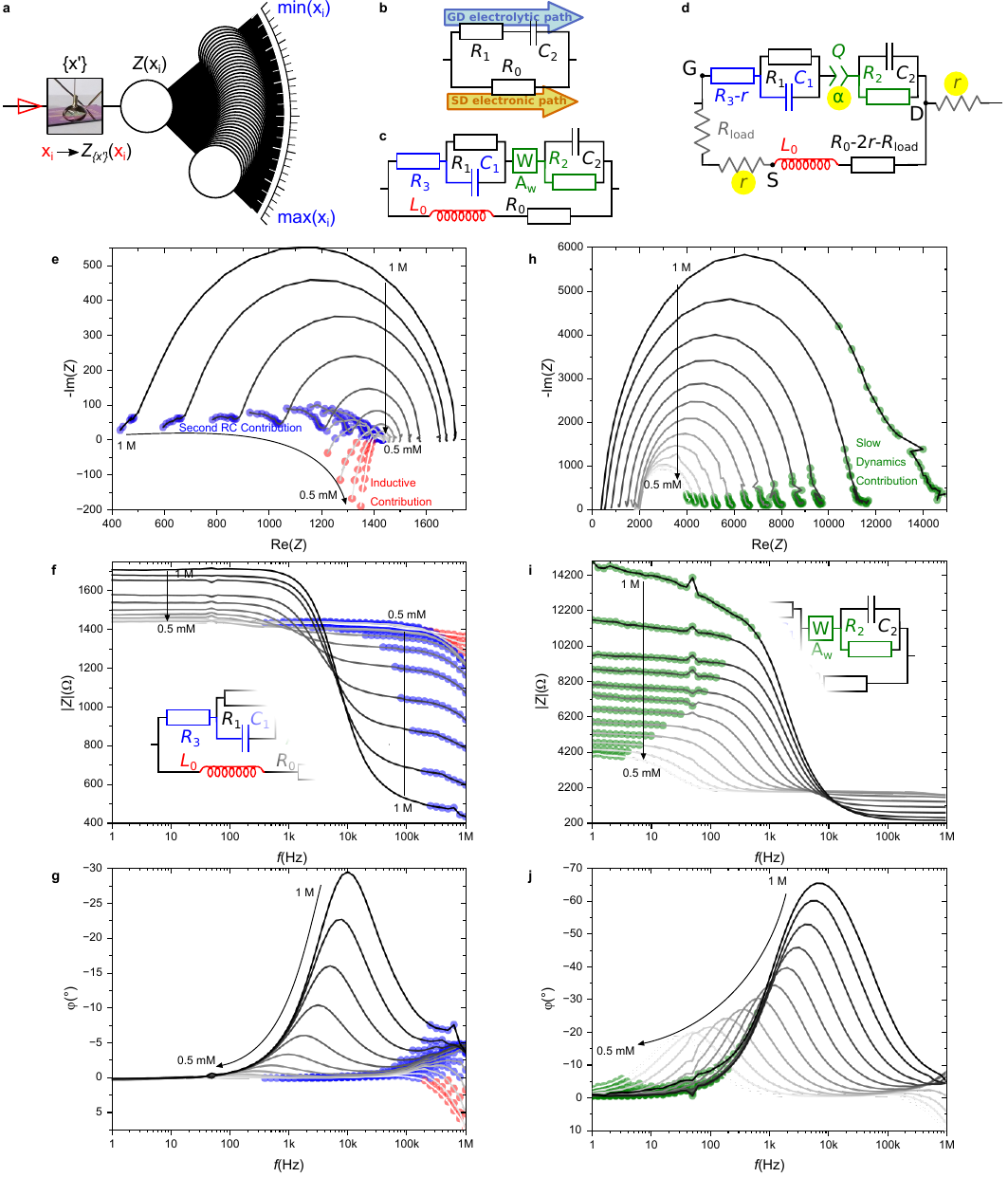}
  \caption{\textbf{An Equivalent Electrical Circuit as Model for Calibration $\vert$ a,} Schematics for a classifier using an OECT to evaluate ion concentration by mean of an impedance value. \textbf{b,} Three-parameter circuit proposed to model the major features of an OECT's impedance in a diode-connected transistor configuration.\cite{Pecqueur2018a} \textbf{c,} Eight-parameter circuit proposed to model the low- and high-frequency deviations experimentally evidenced by impedance spectroscopy.\cite{Pecqueur2019}. \textbf{d,} Further extensions of a parametric model including physically plausible elements, but converging to many solutions with a restricted dataset. \textbf{e-i,} Experimental impedance data highlighting concentration-dependent deviations of the three-parameter model, suggesting a revision of the initial model for using an OECT to recognize ion concentrations (from Pecqueur \textit{et al.})\cite{Pecqueur2018a,Pecqueur2019}. In each subfigures, the contribution of model's elements is highlighted with scatters of identical color as the element displayed in Fig.\ref{fig:fig1.3.1}.b over raw data plots. \textbf{e-g,} Impedance plots in Nyquist representation (\textbf{e}) and spectra in Bode diagrams in modulus (\textbf{f}) and phase (\textbf{g}) of an OECT biased at {\textminus}50~mV\textsubscript{DC} in KCl\textsubscript{aq} electrolytes of 12 different concentrations. \textbf{h-j,} Impedance plots in Nyquist representation (\textbf{h}) and spectra in Bode diagrams in modulus (\textbf{i}) and phase (\textbf{j}) of an OECT biaised at {\textminus}550~mV\textsubscript{DC} in CaCl\textsubscript{2,aq} electrolytes of 12 different concentrations.}
  \label{fig:fig1.3.1}
\end{figure}

This first approach aims at \textbf{understanding the correlation between perturbations and measurements} when the information to classify is physically trivial. If information is a physical variable, a deterministic physical equation provides such correlation, which can either be theoretically or empirically defined.\\[3pt]
In electrical/electronic engineering, \textbf{compact models} focus on the behavior of signals through permittive systems composed of electrical elements,\cite{Woltjer2007,Burghartz2013} while \textbf{finite-element models} focus on the organization of systems composed of physical elements. They both are generically physical as they feature variables with a unit. These models are therefore specific to classifiers and not to classification tasks. On sensing classifiers, compact models focus on the information carriers (the wave and how it propagates) and finite-element models on the physical classifier itself (how charges are dictated in a media).\\[3pt]
To specifically determine the signal propagation models ruling the ion-dependency observed in an OECT impedance,\cite{Pecqueur2018a} the dataset was fitted with an \textbf{electrochemical impedance model} in Pecqueur \textit{et al.} 2019 a year later (Fig.\ref{fig:fig1.3.1}).\cite{Pecqueur2019} It is important to mention that a quantitative model was proposed in the latter study only after the practical use of an OECT to recognize cation in the prior study:\cite{Pecqueur2018a,Pecqueur2019} This shows that \textbf{it is not required to have a quantitative model prior finding a classifier}. There are iterations between establishing classifier models and implementing them experimentally with benefits of the ones optimizing the others. As a matter of fact, the prior study on exploiting low- and high-frequency impedances was initiated from an assumption that, if a resistive load \textit{R}\textsubscript{load} was inserted on the source-drain path, the electrical conduction can be measured at low frequency and the the ionic one at high frequency (Fig.\ref{fig:fig1.3.1}.b). Physical models are perfectible until an experiment shows unexpected trends. While first models were attempted to fit the experimental data (Fig.\ref{fig:fig1.3.1}.b), it was observed that extra elements were required to be introduced in the equivalent circuit: such as an inductive element \textit{L} on the electronic path to model the positive reactance (Fig.\ref{fig:fig1.3.1}.c,e-g), or a Warburg element W on the ionic path to model the impedance transience at low frequency (Fig.\ref{fig:fig1.3.1}.c,h-j). From an empirically satisfying model, hypothesis were proposed for a physical mechanism.\\[3pt]
A very important point is that some \textit{ad hoc} elements shows to be responsible for specific ion dependencies in the impedance signal, such as the inductor, which shows a particular sensitivity to low ion concentrations \textit{x} by the value of the phase at high frequency (Fig.\ref{fig:fig1.3.1}.g). Without being certain of the physical origin of this inductor,\cite{Pecqueur2019} experimentally observing this trend suggested to focus specifically on exploiting $\upmu$s-scale transient phenomema and oriented the choice of information features to exploit in the further later study of cation blends classification with an OECT.\cite{Pecqueur2023} 

Although defining the physical model is not required, its allows for a qualitative correlation of the measurement \textit{Z} and the perturbation \textit{x}\textsubscript{i} (here,\cite{Pecqueur2019} the ion concentration for a specific cation i), providing that all other environmental variables (chemical nature of ions in a well defined reference) are fixed, and all parameters in the model are calibrated (\textit{R}\textsubscript{1}, \textit{R}\textsubscript{2}, \textit{C}\textsubscript{1}, \textit{C}\textsubscript{2}, \textit{L} and \textit{W}). A well structured function of \textit{x}\textsubscript{i}$\rightarrow$\textit{Z}(\textit{x}\textsubscript{i}) can be defined generically, as long as the model is exploited within specific boundaries defined by the calibrated parameters, set for specifically-fixed environmental variables. Upon technological increment, real environments are more complex and multiple of its properties are transient: fixed parameters must be periodically re-calibrated and models may require more parametrical flexibility (Fig.\ref{fig:fig1.3.1}.d) or total reconsideration of the architecture in order to adapt to potential new variations.\\[3pt]
At the genesis of a classifier aiming a well-defined recognition/quantification, \textbf{a physical approach is crucial to identify an initial architecture, and further refining its complexity helps understanding the physical origin of information carriers}. However, as an environment complexity increases, so increases the instability of a valid classifier model, \textbf{so classifier must adapt to the complexity of environments they are experienced to}.\\

\subsection{To Phenomenological Models \textit{in naturo}}
\label{Ch2.3.2}

\begin{figure}
  \centering
  \includegraphics[width=1\columnwidth]{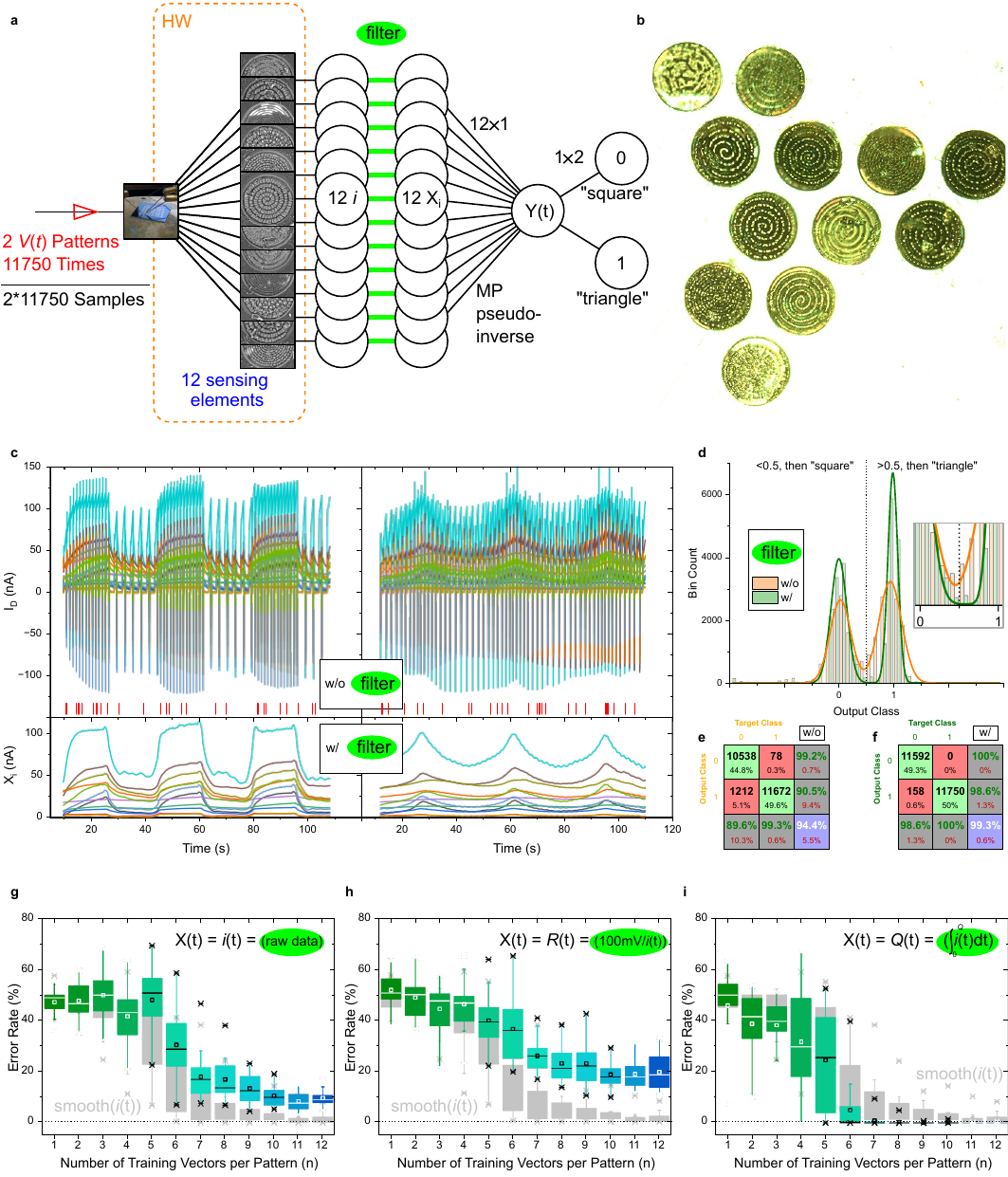}
  \caption{\textbf{Nondimensionalizing Models with Mathematical Transforms $\vert$ a,} Classifier by Pecqueur \textit{et al.} on using an array of 12 OECTs to discriminate two frequency-modulated voltage signals applied to a common electrolytic gate electrode.\cite{Pecqueur2018b} \textbf{b,} Highly contrasted microscope image of the 12-OECT array used in the same study.\cite{Pecqueur2018b} \textbf{c,} Dynamical characteristics of the 12 OECTs current responses under stimulation with two different gate voltage patterns: the raw current responses (I\textsubscript{D} at the top) is compared to the filtered information (X\textsubscript{i} at the bottom) inputted to the classifier (data from Pecqueur \textit{et al.} 2018)\cite{Pecqueur2018b}. \textbf{d,} Effect of the current smoothing filter, used to prepare the input information X\textsubscript{i} from the raw data \textit{i}\textsubscript{D}, on the statistical distribution of output vectors, after training the MP pseudo-inverse classifier with n~=~30 vectors/pattern. \textbf{e-f,} Confusion matrices for the binary recognition results displayed in Fig.\ref{fig:fig1.3.2}.d, without applying a smoothing filter (\textbf{g}) and with the application of the smoothing filter (\textbf{h}) as performed in Pecqueur \textit{et al.} 2018.\cite{Pecqueur2018b} \textbf{g-i,} Impact of the choice of the physical variable on the classification: Fig.4.a published in Pecqueur \textit{et al.} 2018 in the case of current raw data (\textbf{g}), resistance (\textbf{h}) and charge (\textbf{i}) was inputted to the classifier in the same conditions as in the study (in grey is displayed the statistics for the smooth current as performed in the study of Pecqueur \textit{et al.}).\cite{Pecqueur2018b}}
  \label{fig:fig1.3.2}
\end{figure}

This second approach aims at \textbf{optimizing a correlation between measurements and information classes} when classes are physically too complex to be understood as a function of physical measurements. First, a naive model structure is proposed, disregarding the relationship between measurements and classes. Since the relationship between measurements and perturbations is irrelevant to choose models, choice for environment samples to calibrate models can be of any physical complexity. If properly calibrated with enough application-relevant samples, such model is a practical solution for complex environment classification.\\[3pt]
This approach has been used by Pecqueur, Boujnah, Haj Ammar, Routier \textit{et al.} in different studies with an array of CP conductimetric elements for molecular recognition,\cite{Boujnah2021,Boujnah2022,HajAmmar2023,HajAmmar2024,Routier2024,Routier2024a}, 
with an array of OECTs for transient-voltage pattern recognition,\cite{Pecqueur2018b,Ghazal2024} or with a single OECT for cation-blends classification.\cite{Pecqueur2023} More complex models are used in these studies (such as artificial neural networks)\cite{Pecqueur2023,Ghazal2024,Routier2024}, but simple linear classifiers provide insight into understanding the relevance of different materials in an array,\cite{Pecqueur2018b,Boujnah2021,Boujnah2022,HajAmmar2023,HajAmmar2024,Routier2024,Routier2024a}  or feature in a trace signal,\cite{Pecqueur2023} when put together for pattern classification.\\[3pt]
In the first of these studies (Fig.\ref{fig:fig1.3.2}), the Moore-Penrose (MP) pseudo-inversion is used to calibrate a linear model connecting measurements of 12 OECTs to two classes of experienced gate-voltage activities.\cite{Pecqueur2018b} The MP pseudo-inversion is a matrix transform solution of a least-square linear regression in an n-dimensional space. In this study, measurements are 12 transient current values \textit{i}, low-pass filtered numerically (\textit{X}\textsubscript{i}) prior being classified (Fig.\ref{fig:fig1.3.2}.c). One may be attempted to define a physical translation to the effect of the smoothing on fed data, but \textbf{algorithms transform data, not physical quantities}: there is no need for dimensional consistency like in a physical equation. However, classification is very sensitive to data preparation, especially linear classifiers, which may require non-linear transformations. In the study, filtering data by smoothing gives better results than feeding raw data directly (Fig.\ref{fig:fig1.3.2}.c-g). If resistance or charge was used instead of current values, performances would have been very different with the same raw data and the same model (Fig.\ref{fig:fig1.3.2}.g-i).\\[3pt]
This illustrates that \textbf{naive models can support the qualitative understanding of how relevant physical measurements can be for figures of merit to discriminate given classes}, and how to improve a sensing array's classification. Here specifically,\cite{Pecqueur2018b} the very distinct morphologies of materials on the array (Fig.\ref{fig:fig1.3.2}.b) promotes statistical distributions (Fig.\ref{fig:fig1.3.2}.c), impacting classification in different way (Fig.\ref{fig:fig1.3.2}.d-f). Identified distributions are nominal currents (conductance), current modulations (tranconductance), characteristic time for charge/discharge (volume), charge/discharge asymmetry (density). The fact that the linear classifier exploits better \textit{q}(t) than \textit{R}(t) or \textit{i}(t) suggests a relevant dispersion of property promoting pattern classification can origin from the ones impacting \textit{q} directly (such as characteristic time constants) than \textit{R} or \textit{i}. This is coherent with the identity of patterns being different by dynamics: different frequency modulations imprint specifically different charging profiles at a given time than resistances or currents.\\[3pt]
\textbf{If well calibrated, structure of mathematical models can promote an understanding of how measurements are relevant in data to classifying complex environments. This detailed in the next section Sec.\ref{Ch2.4}.}

\section{Calibration Methods to use a Given Model}
\label{Ch2.4}

For both physical and mathematical models (Sec.\ref{Ch2.3}), calibration (or training, tooling, conditioning) is required to set models' parameters to output quantitative vector coordinates from measured perturbations. For classifications involving high-dimensional measurements (Sec.\ref{Ch2.2}), calibrations may be complex and require well-defined algorithms: either for physical models such as equivalent electrical circuits requiring a fitting algorithm, or artificial neural networks requiring a learning rule. Depending on its operation, an algorithm can perform well on a given classifier model, or not. Among them, \textbf{some require supervision} of a user to define true classes on a set of exemplary environment samples (Sec.\ref{Ch2.1}), and \textbf{some do not require supervision}.\\[3pt]
In the following, both supervised and unsupervised calibrations are compared to detail on how the earlier one is better suited when "one knows what to look for" (Subs.\ref{Ch2.4.1}) while the latter one when "one knows how to look at" (Subs.\ref{Ch2.4.2}).

\textit{This section is mostly associated to peer-reviewed results published in} Pecqueur \textit{et al.} \textit{\href{https://doi.org/10.1002/aelm.201800166}{Adv. Electron. Mater.}} \textbf{4}(\textit{9}), 1800166 (2018), Boujnah \textit{et al.} \textit{\href{https://doi.org/10.1016/j.synthmet.2021.116890}{Synth. Met.}} \textbf{280}, 116890 (2021), Pecqueur al. \textit{\href{https://doi.org/10.3390/electronicmat4020007}{Electron. Mater.}} \textbf{4}(\textit{2}), 80‒94 (2023), Haj Ammar \textit{et al.} \textit{\href{https://doi.org/10.3390/eng4040141}{Eng}} \textbf{4}(\textit{4}), 2483‒2496 (2023) \& Haj Ammar \textit{et al.} \textit{\href{https://doi.org/10.3390/electronics13030497}{Electronics}} \textbf{13}(\textit{3}), 497 (2024) \textit{among other information from the state-of-the-art literature.}\\

\subsection{Supervised Training: to Find Specific Information Classes}
\label{Ch2.4.1}

\begin{figure}
  \centering
  \includegraphics[width=1\columnwidth]{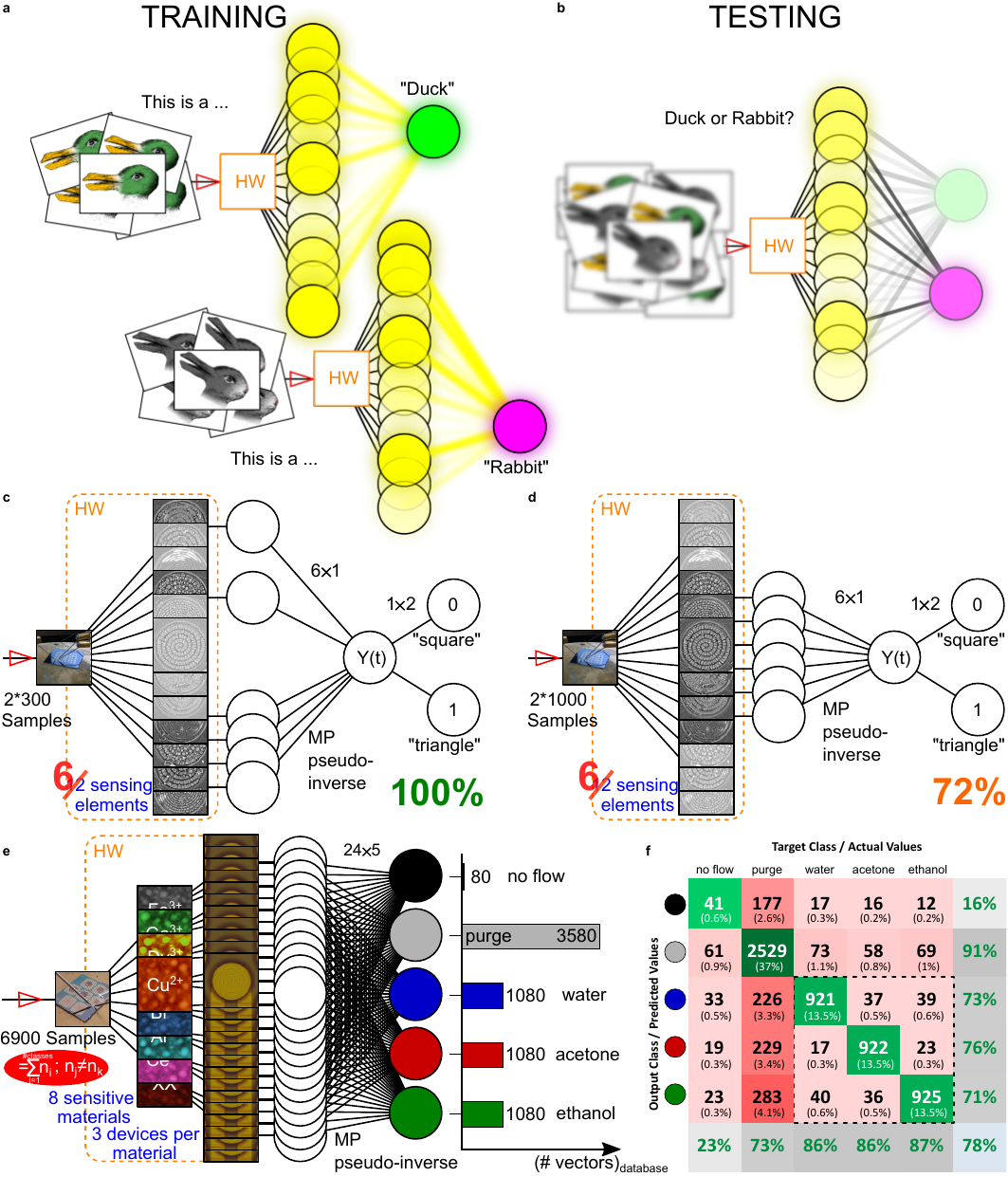}
  \caption{\textbf{Supervised Calibration $\vert$ a,} Supervised training a linear classifier for a two-class recognition (weights optimization to associate a known output class to a presented training vector). \textbf{b,} Testing unknown vectors on the calibrated system which shall project most on its corresponding output class. \textbf{c,} Half-pruned sensing array classifier able to recognize most of the data in Pecqueur \textit{et al.} 2018 (case 1 from Fig.S4 in the aforementioned reference)\cite{Pecqueur2018b}. \textbf{d,} Half-pruned sensing array classifier able to recognize least of the data in Pecqueur \textit{et al.} 2018 (case 5 from Fig.S4 in the aforementioned reference)\cite{Pecqueur2018b}. \textbf{e,} Supervised classifier used in Haj Ammar \textit{et al.} 2024 and disproportion of some classes in the whole database and the training subset.\cite{HajAmmar2024}. \textbf{f,} Confusion matrices for classifying Boujnah \textit{et al.} 2021's data with Haj Ammar \textit{et al.} 2024's classifier showing a correlation of between the number of samples displayed in Fig.\ref{fig:fig1.4.1}.e and the prediction rates for each class (right column), showing that classifiers recognition rates can be conditioned by the structure of a training dataset (here, more a class is represented in a training dataset, higher probability it has to be predicted, independent from the qualities of a sensing array nor the model to recognize this class among others.\cite{Boujnah2021,HajAmmar2024}.}
  \label{fig:fig1.4.1}
\end{figure}

Supervised learning relates to classifiers' calibration from \textbf{known examples} to optimize parametric models prior using them on unknown samples (Fig.\ref{fig:fig1.4.1}.a-b). "Known examples" refers strictly to the use of semantic labels along with samples and not to a user's actual awareness of shared physical properties between samples of a given class. Users must \textbf{have access to a "large-enough" dataset} to train a classifier prior using it, and know the actual class of each sample composing the training subset. This presents two weaknesses: The first is on \textbf{samples accessibility in large numbers}, unlike classification cases on fragile samples (like cell cultures) or hostile environments (with limited access and duration exposures). Training subsets shall size models complexity (neural networks are highly parametric and require large training datasets). The second weakness is on \textbf{the necessity to define classes in the first place}, which is complicated if classes are atypical (such as polluted environments) or unsuspected (such as detecting novel environmental threats). Supervised learning is however \textbf{very performing once properly trained}.\\[3pt]
A first example is on identifying two different gate-voltage frequency-modulation out of an OECT array's response.\cite{Pecqueur2018b} Once a dataset was generated, a subset was used to calibrate a model to differentiate "square" and "triangle" measured vectors by a MP pseudo-inversion. Class names are completely independent from the actual environment spectral signature (Fig.\ref{fig:fig1.4.1}.c-d), but the \textbf{calibration requires to know classes with a certain number of samples}. If a third pattern would be introduced in the dataset, the calibration would need to be re-initiated. The study extensively focused on the classifier's performance with the training: First, the error rate depends on both the size of a sensing array and the number of vectors available to train the classifier. \textbf{The more users know from their database, the better performances are}. However, in case sensing elements are too few, the algorithm never succeeds calibrating: typically, a binary classification requires at least two sensing elements to not overfit.\cite{Pecqueur2018b} Moreover, the study showed that \textbf{error rates differ with different sets of sensing elements of a same size}: with six sensing elements, a material set can either recognize the full dataset with only 300 known examples per class, or just 72\% of it with 1000 known examples per class. Despite a relationship between the size of a training subset and the error rate, \textbf{sensing arrays shall be composed of complemental elements for specific classifications}: randomly picking elements is extremely risky for the recognition quality. Also, the algorithm does not inform on how the quality of each element conditions the error rate, neither suggests improvement strategies. \textbf{This calibration framework can help to identify good sensing materials or signal features, but it will not explain correlations}.\\[3pt]
A second implementation of the MP pseudo-inversion was performed on a dataset of conductimetric sensing elements with doped CPs to recognize solvent vapors.\cite{HajAmmar2024} This study showed that a \textbf{non equi-representation of the classes in a dataset is not a limiting factor for the classification}. 

This study took into account that "purge" classes are over-represented in the training subset (as random pick from the full dataset), and "no flow" are under-represented in the training subset. This would be adequate for classifications where samples of one class are less available than others. However, since this information is not explicit with the error rates, false estimation of classifiers performance can occur and may bias its significance interpretability (Fig.\ref{fig:fig1.4.1}.e,f).\\[3pt]
\textbf{In hardware, supervised calibration can be implemented if examples are known. However, classes must be few but well-defined, and all of them sufficiently represented according to the chosen model's complexity.}

\subsection{Non-Supervised Training: to Map Environmental Complexity}
\label{Ch2.4.2}

\begin{figure}
  \centering
  \includegraphics[width=1\columnwidth]{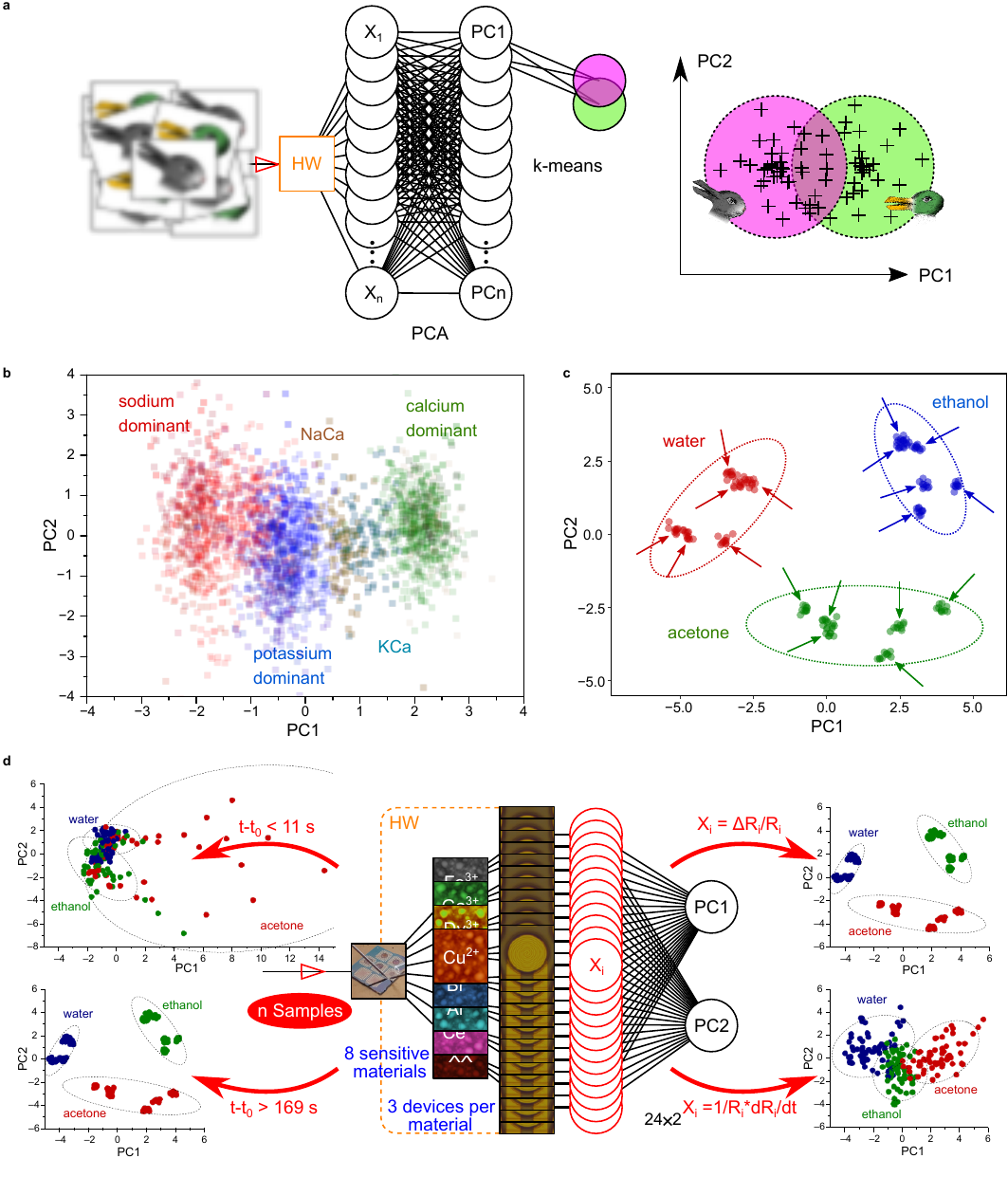}
  \caption{\textbf{Non-Supervised Calibration $\vert$ a,} Classification by PCA and k-means clustering on the first two principal components as a linear classifier to discriminate k~=~2 classes. \textbf{b,} PCA score on the first two principal components, discriminating analytes by main component composition, and where five different classes can be identified (from Pecqueur \textit{et al.} 2023)\cite{Pecqueur2023}. \textbf{c,} PCA score on the first two principal components, where k-means clustering identifies three clusters of data characteristic of the composition, while subclusters of different iterations can be identified in each category (from Boujnah \textit{et al.} 2021)\cite{Pecqueur2023}. Arrows indicate six subclusters within each of the k~=~3 classes. \textbf{d,} Impact on the PCA score of both the data sampling which conditions the quality of the raw information and the information feature preprocessed to input to the PCA classifier (from Haj Ammar \textit{et al.} 2023)\cite{HajAmmar2023}.}
  \label{fig:fig1.4.2}
\end{figure}

Non-supervised learning relates to classifiers' calibration \textbf{without requiring information from an end-user} (Fig.\ref{fig:fig1.4.2}.a). The calibration is done by an algorithm which exploits only data organization without correlating them to known examples. Moreover, it does not require defining a specific number of classes nor to define them. It is therefore particularly suitable for autonomous inference with no \textit{a priori} knowledge of an environment. It is however very sensitive to data preparation and requires an understanding of the physical significance of pertubations in regard to classes to recognize. A non-supervised calibration which is extensively studied for non-metrological sensing is principal component analysis (PCA) with by k-means clustering.\cite{Ding2004,Clayman2020,Sadeghi2024}\\[3pt]
Two initially independent studies on using PCA for multivariate data analysis on CP sensing elements were performed with conductimetric arrays for volatile solvent recognition,\cite{Boujnah2021,Boujnah2022,HajAmmar2023,HajAmmar2024,Routier2024,Routier2024a} and for dynamical analysis of OECTs' dynamical response for ionic composition identification.\cite{Pecqueur2023} The nature of both studies shows that PCA can be used with simple measurements for multi-material studies as well as multi-feature signal extraction on a single CP material. In the case of ion-blends classification with an OECT (performed before the other one, but published much later)\cite{Kozic2020} the ability for PCA to not require inputting the number of classes allows \textbf{using this methodology for semi-quantitative analysis}. Here,\cite{Pecqueur2023} PCA suggests a number of class to define data by the distribution of a specific projection of them. Different compositions of Na\textsuperscript{+}\textsubscript{aq}, K\textsuperscript{+}\textsubscript{aq} and Ca\textsuperscript{2+}\textsubscript{aq} were prepared and specific values of the current decays under gate voltage spike stimulation were used as measurements. The analysis defines clusters of datapoints in (PC1;PC2) that depend on the nature of the electrolyte with which the OECTs were characterized (Fig.\ref{fig:fig1.4.2}.b). The analysis did not identify each of the 25 blends by 25 individual clusters. However, it showed five domains which were representative from five composition classes, while only three salts were used. Indeed, solutions showing for high concentrations of (Na\textsuperscript{+}+Ca\textsuperscript{2+}) and (K\textsuperscript{+}+Ca\textsuperscript{2+}) were gathered in a specific domain in (PC1;PC2). This was an unexpected result which demonstrated that \textbf{unsupervised calibration is adequate to identify unsuspected classes}. This was further demonstrated in the later case of multi-material analysis,\cite{Boujnah2021} when \textbf{PCA hierarchized three clusters} by environment nature, each structured into six subclusters characteristic from the experiments repetitions (Fig.\ref{fig:fig1.4.2}.c). In addition to the data separation by PCA, the k-means method defines the "suitable" number of clusters to label, and then defines a confidence ellipsoid with no supervision of the end-user.\\[3pt]
Further PCA studies on the same raw dataset have shown that clustering quality is extremely sensitive to the data collection and preparation.\cite{HajAmmar2023,HajAmmar2024} The initial study supposed the information relies on the currents' steady response as relative value of resistance modulation due to material's chemical doping modulation,\cite{Ferchichi2020,Boujnah2021} Haj Ammar \textit{et al.} studied the full transience to see if relevant information can be extracted in the transient.\cite{HajAmmar2023} The study concluded that PCA can exploit transient currents if data are transformed as resistance drift and not as resistance modulation to feed the PCA (Fig.\ref{fig:fig1.4.2}.d). This shows that \textbf{clear guidelines shall be provided for an end-user to collect relevant data} if a given measurement is chosen to feed unsupervised calibration. A last study on the same dataset focused on a parametric function to condition the measurement prior feeding to the PCA.\cite{HajAmmar2024} The study showed that an end-user can condition data separability by the control of an adimensional parameter conditioning the measurements filtering. As an optimum for such parameter is extremely sensing-element and case dependent, it would be convenient that classifiers supervise such extra parametrization autonomously as part of their calibration.\\[3pt]
\textbf{In hardware, non-supervised calibration can be implemented even if classes are non-defined. However, users shall be well-guided for data collection, and preparation for proper classifications.}

\chapter{Evolutionary Semiconductors}
\label{Ch3}

\textit{This chapter gathers aspects of my research involving voltage-pattern classifications with electrically-induced modifications of conducting-polymer structures using electropolymerization.\\
The discussion focuses on two levels: on modifying individual sensing elements in an array of electrodes (Sec.\ref{Ch3.1}) and on their interconnecting electrodes with a readout system (Sec.\ref{Ch3.2}).}\\

\section{Neurogenesis of Conducting Polymer Elements}
\label{Ch3.1}

The previous chapter describes how \textbf{sensing classifiers have to adapt} during their whole operational lifetime: they need specific sensitivity from their material according to the uniqueness of environment sets they will experience (Sec.\ref{Ch2.1}), the information features they need to extract may change according to the effect of new analytes to include for efficient discrimination (Sec.\ref{Ch2.2}), that models may increase in complexity level with more inputted  information (Sec.\ref{Ch2.3}) and that calibration risks to take longer with measuring new perturbations for the classifier to train recognizing new classes (Sec.\ref{Ch2.4}).\\
Unlike living organisms which move, heel and grow, \textbf{adaptation} in a conventional electronic classifier is only at the software level. If updating software to promote models adaptation and telecommunicating promotes calibrating with shared databases in a generic manner, software have no effect on the physics of detectors. In a living classifier, a physical turnover of cells is ensured when they die, and regenerate new functional cells, sometimes specifically to the needs of the system. In consumer electronics, hardware resources are oversized from their manufacture, if contact breaks or gate dielectric shorts, devices are no more functional, discarded and poorly recyclable when new versions of themselves are on sale.\\[3pt]
In the aim of presenting neurogenesis with CPs in sensing classifier to promote physical adaptation of electronic hardware in their environment, three different aspects are presented: what are the two main motivations for CP neurogenesis (Subs.\ref{Ch3.1.1}), how can it be physically implemented using the electrochemical reactivity of CP precursors (Subs.\ref{Ch3.1.2}) and up to what extend CP neurogenesis can modifying sensing elements' figure-of-merits (Subs.\ref{Ch3.1.3}).

\textit{This section is mostly associated to peer-reviewed results published in} Pecqueur \textit{et al.} \textit{\href{https://doi.org/10.1002/aelm.201800166}{Adv. Electron. Mater.}} \textbf{4}(\textit{9}), 1800166 (2018), Boujnah \textit{et al.} \textit{\href{https://doi.org/10.1016/j.synthmet.2021.116890}{Synth. Met.}} \textbf{280}, 116890 (2021), Ghazal \textit{et al.} \textit{\href{https://doi.org/10.1002/aelm.202100891}{Adv. Electron. Mater.}} \textbf{8}(\textit{3}), 2100891 (2022) \& Ghazal \textit{et al.} \textit{\href{https://doi.org/10.1002/aelm.202100891}{Biosens. Bioelectron.}} \textbf{237}, 115538 (2023) \textit{among other information from the state-of-the-art literature \& early results of M. T. Nauto, S. Lenfant and S. Pecqueur.}\\

\subsection{Why Changing an Array of Sensing Elements \textit{in operando}}
\label{Ch3.1.1}

When information classes are physically non-trivial in non metrological classifiers, their dimensionality in a properly-defined vector base cannot be easily estimated. Therefore, two problems arise: First, theorizing on a minimum number of sensing elements to efficiently recognize classes is unfeasible \textit{a priori}. Second, estimating which complementary features each sensing element shall be specific to is hard to picture. An example is the human odor receptive field, which is a particularly high dimensional information space.\cite{MadanyMamlouk2004,Khan2007,Castro2013,Meister2015,Endo2022} In such space, defining a vector base of primary scents is particularly hard, at the opposite of colors with a specific wavelength for color vision. As several sets of molecules in different concentrations may induce similar odoriferous patterns, sample populations belonging to a same class of smells project in such space as a distribution within a domain. If one aims at identifying a specific number of specific smells, a required number of necessary receptors for discrimination depends first on the number of smells to discriminate and on the distance and size of class domains in such space. \textit{a priori}, it would rather be complex to model the distribution of information classes by the constitutional complexity of all possible sets of molecules able to induce a specific smell. \textbf{A practical way to estimate it is through experiments}.\\[3pt]
In two independent studies involving CPs on similar conductimetric arrays (Fig.\ref{fig:fig3.1.1}),\cite{Pecqueur2018b,Boujnah2021} experiments were conducted to classify two or three physically non-trivial information patterns. Through the experiments, both studies evidenced two important trends: training conditions optimal numbers of sensing elements to optimize classification (Fig.\ref{fig:fig3.1.1}.a,c-d,f),\cite{Pecqueur2018b} and a given number of sensing elements conditions choices for materials to compose an array (Fig.\ref{fig:fig3.1.1}.b,e,g).\cite{Boujnah2021}\\[3pt]
Previously in Subs.\ref{Ch2.4.1}, differences in error rates were observed in Pecqueur \textit{et al.} 2018's study,\cite{Pecqueur2018b} after training specific sets of sensing elements with a same number (n) of training vectors per patterns. To reduce the time required to calibrate a classifier, the number of training vectors to provide should be the smallest to converge to the smallest error rate. In this study, an evolution is observed at different training stages, on the optimal error rate per sensing element as the number of training vectors per pattern increases (Fig.\ref{fig:fig3.1.1}.c-d). A statistical minimum error of 20\% is observed with four sensing elements if only three samples per pattern are provided to calibrate the binary classification with the MP pseudo-inversion (Fig.\ref{fig:fig3.1.1}.c). Too few sensing elements do not give a sufficiently rich picture to discriminate triangular modulations from squared modulations of frequency signals, however, too many sensing elements appear to be more a hindrance for the classification, approaching the 50\% performance of a random binary classifier (Fig.\ref{fig:fig3.1.1}.c). If a hundred time more examples are added to the training database, the classifier becomes much better performing, but more importantly, the lowest statistical error rate is reached with a higher number of eleven sensing elements (Fig.\ref{fig:fig3.1.1}.d). This study shows that for a practical classification application, \textbf{the best number of sensing elements to feature in an array does not depend only on the task, but also on the use} (Fig.\ref{fig:fig3.1.1}.f).\\[3pt]
In another study, Boujnah \textit{et al.} used such sensing elements coated with different doped CPs to recognize vapors of three different solvents.\cite{Boujnah2021} As above, the relationship between classes to discriminate and materials' sensitivities is not obvious enough to identify a model preferring one material over another one: here, Lewis acid dopants are not particularly selective to molecules but have different reactivities in regard to their electron-accepting chemistry.\cite{Pecqueur2014,Pecqueur2016} Similar to above, the classifier's performance has been studied depending on the number of sensing materials composing a pruned array to identify how many and what materials are important for recognition: Systematically, more materials lowers the error rate (Fig.\ref{fig:fig3.1.1}.e).\cite{Boujnah2021} There are materials which are statistically better performing than others disregarding how many materials compose the array. (for instance in Fig.\ref{fig:fig3.1.1}.e, the Fe(OTf)\textsubscript{3}-doped polymers are the worst choice when composing an array with the other seven materials).\cite{Boujnah2021} However, the ranking for eight materials differs with their number in an array (for instance in Fig.\ref{fig:fig3.1.1}.e, an array with undoped polymers is statistically the best choice when composed of two elements or less, but collective contributions of more elements compensate it, to prefer Ce(OTf)\textsubscript{3}-doped polymers on larger arrays).\cite{Boujnah2021} This shows that \textbf{for a classifier in which the sensing dimensionality increases during training, "good" materials composing an array have to adapt to the complementarities of new sensing elements composing an array} (Fig.\ref{fig:fig3.1.1}.g).\\[3pt]
\textbf{To improve classification by structural adaptation, one must establish a protocol where sensing elements can evolve: creating during an application (Subs.\ref{Ch3.1.2}) and modifying when more elements generate (Subs.\ref{Ch3.1.3}).}

\begin{figure}
  \centering
  \includegraphics[width=1\columnwidth]{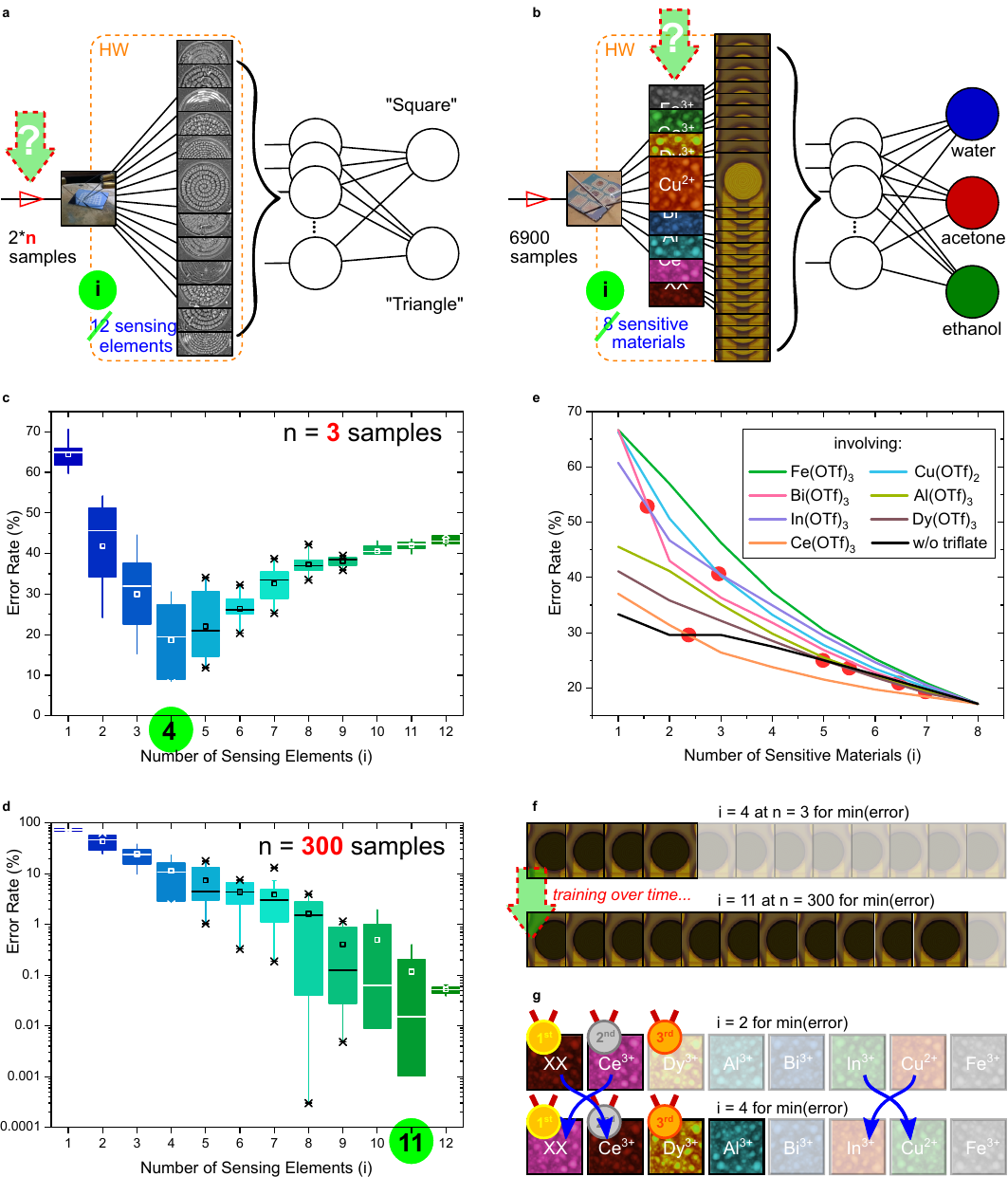}
  \caption{\textbf{Necessity for Sensing Arrays to Evolve while Classifying $\vert$ a-b,} Studying the impact of pruning sensing arrays of classifiers from Pecqueur \textit{et al.} (\textbf{a})\cite{Pecqueur2018b} and Boujnah \textit{et al.} (\textbf{b})\cite{Boujnah2021}, depending on the number of training vectors (classifier structure \textbf{a}) and the type of materials of sensing elements (classifier structure \textbf{b}). \textbf{c-d,} Evolution of the optimal number of sensing elements with a classification, from four sensing elements when 2$\times$3 vectors are in the training dataset, to 11 sensing elements when 2$\times$300 vectors are in the training dataset (from Pecqueur \textit{et al.})\cite{Pecqueur2018b}. \textbf{e,} Impacts of different materials on recognition performances. The red dots indicate ranking inversions between materials as the size of a material set increases (from Boujnah \textit{et al.})\cite{Boujnah2021}. \textbf{f-g,} Consequences for an optimal classifier's sensing array to evolve during a classification: by increasing in size as the number of training dataset increases (\textbf{f}) and that during such evolution, materials may need to adapt as their contributions change with the introduction of new materials on sensing elements (\textbf{g}).}
  \label{fig:fig2.1.1}
\end{figure}

\subsection{Creating New Elements}
\label{Ch3.1.2}

\begin{figure}
  \centering
  \includegraphics[width=1\columnwidth]{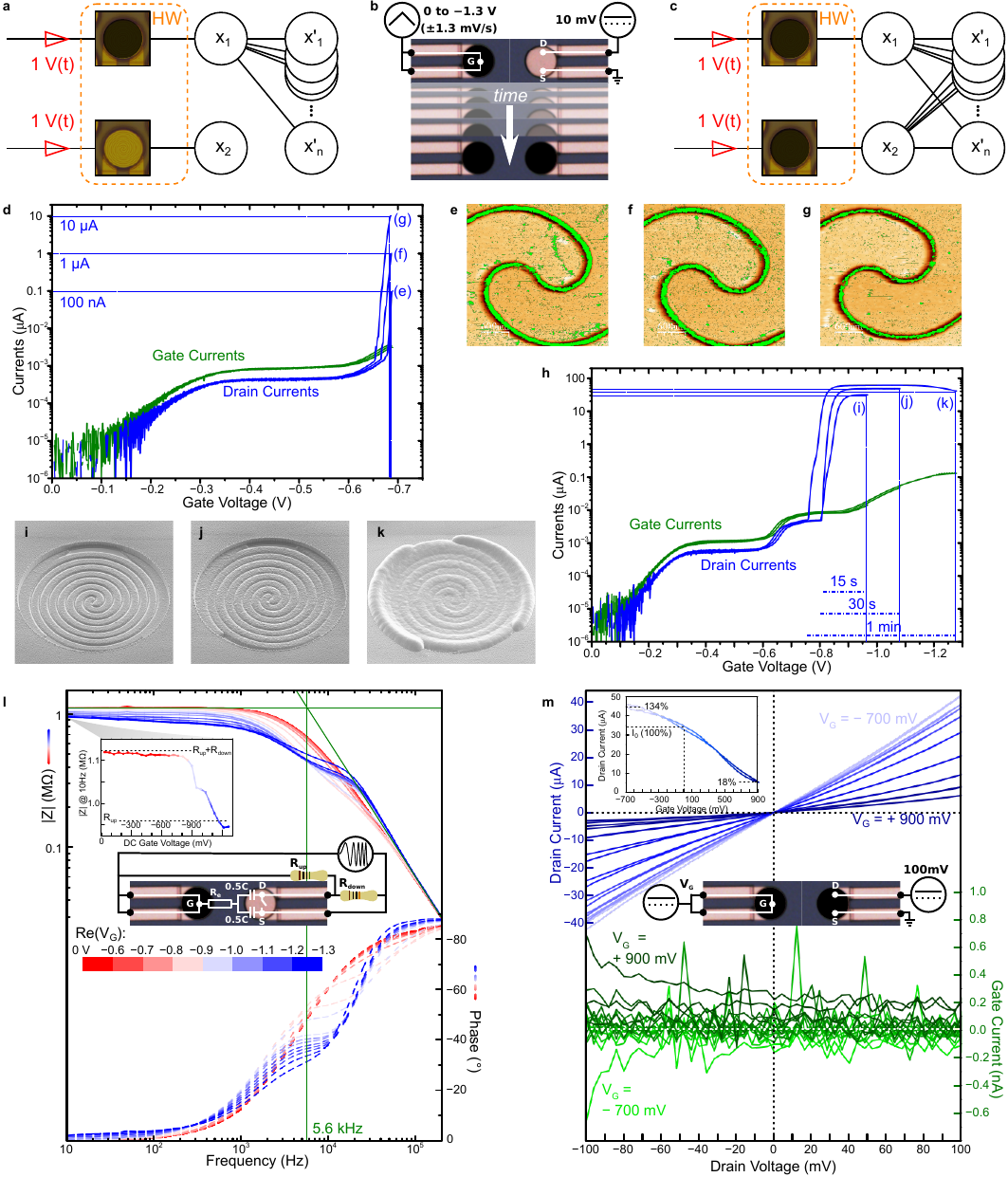}
  \caption{\textbf{Electropolymerization as Neurogenerative Process to create Sensing Elements $\vert$ a-c,} Schematics for an evolving classifier, using neurogenesis as physical mechanism to increase the number of sensing elements on an array. \textbf{a,} A juvenile two-dimensional classifier. \textbf{b,} Voltage-control to grow a CP on an uncoated element thanks to a coated one in a common electroactive solution, with \textit{in operando} amperometric readout of the neurogenerated element's state. \textbf{c,} Fig.\ref{fig:fig2.1.2}.a's classifier at a mature stage. \textbf{d,} "Pseudo-transfer" characteristics of three uncoated elements, grown as depicted in Fig.\ref{fig:fig2.1.2}.b and with a stop condition set at different current levels for the neurogenerated sensing elements. \textbf{e-g,} Conductance map (poorly-conductive sites in green) stacked on topographic Atomic Force Microscopy (AFM) images for the three elements characterized in Fig.\ref{fig:fig2.1.2}.d (100~nA in \textbf{e}, 1~$\upmu$A in \textbf{f} and 10~$\upmu$A in \textbf{g}). \textbf{h,} "Pseudo-transfer" characteristics of three uncoated elements, grown as depicted in Fig.\ref{fig:fig2.1.2}.b, with a stop condition set at different durations after the grow starts. \textbf{i-k,} SEM images for the three elements characterized in Fig.\ref{fig:fig2.1.2}.h (15~s in \textbf{i}, 30~s in \textbf{j} and 1~min in \textbf{k}). \textbf{l,} \textit{in operando} impedimetric readout during neurogenesis on a sensing element with voltage-ramped impedance spectroscopy. \textbf{m,} Output characteristics of a neurogenerated element using another element as a gate electrode in 100~mM NaPSS\textsubscript{aq}.}
  \label{fig:fig2.1.2}
\end{figure}

Promoting the formation of new electrical transducers with electricity is both a way to mix energy supply with information generation, but also technology manufacturing and sensing hardware utilization, for which the use of an electrical device usefully lowers its own fabrication footprint.\\[3pt]
To do this, \textbf{electropolymerization} allows growing polymers on an electrode, out of monomer-containing solutions.\cite{Yamazaki1965,Funt1966} It requires electrodes to be already patterned on a sensing array, and as a wet process, electrodes must be exposed in a phase able to support charged species. This phase requires \textbf{at most only four components}: (1) monomers yielding to CPs (further details in the later Subs.\ref{Ch4.1.1}), (2) electrolytes, if monomers are no ions themselves, (3) solvents if electrolytes are no fluids themselves and (4) counter redox-active agents if neither electrolytes nor solvents nor the outer environment are electro-active themselves.\\[3pt]
Supplied voltages to electrogenerate CPs is low and at the same level as to read them (despite both are not related the one another). On a similar system as the readout circuit, CP neurogenesis can be done sequentially while training a supervised classifier or concomitantly in an unsupervised classifier if environments to classify are electropolymerizable. Despite researches on new monomers are investigated on electrolytically-referenced voltage setups, there is \textbf{no need for potentio-/galvanostat equipments}. To ensure the electropolymerization rate control, the counter reaction on the other electrode must be not limiting.\\[3pt]
For each electropolymerization on sensing elements, a third electrode (which can be another sensing element already coated with CPs as pictured in Fig.\ref{fig:fig2.1.2}.a-c) serves as an auxiliary electrode to promote the redox counter-reaction in the electrolyte solution phase. In a three-electrode setup with each S and D electrodes of a sensing element to be coated and a G electrode from an already coated sensing element, a DC-voltage polarization routine similar to a transistor transfer characteristic allows monitoring the conductance of the sensing element while coating a CP (Fig.\ref{fig:fig2.1.2}.d). Practically, during electropolymerization, the sensing element's conductance changes from the noise level (100~fA in Fig.\ref{fig:fig2.1.2}.d) to 10\textsuperscript{4}-10\textsuperscript{5} times more current for the sensing element reaching the resistance of the contact lines (about 200~$\Upomega$), with very few nanometers of a CP shorting the gap (Fig.\ref{fig:fig2.1.2}.d-g). When such an evolving material shorts two electrodes, they \textbf{instantly} form a pair in a conductimetric element.\cite{Genies1988} It is therefore required for the electrodes to be close enough so few material is required for the process and several coatings can be made on different sensing elements on the same array out of the same volume of electrolyte. Upon growths, coatings are uniform over the electrodes: at the stage when both electrodes are shorting, the polymer forms multiple conductive paths all along the width \textit{W} of the electrode (Fig.\ref{fig:fig2.1.2}.d-g). Furthermore material can be deposited at the scale of a few seconds (Fig.\ref{fig:fig2.1.2}.h) to increase the overall CP thickness and fill a 2~$\upmu$m deep cavity over the S \& D electrodes and in the \textit{L} = 200~nm to 2~$\upmu$m gap they form.\\[3pt]
The energy spent in the process is quasi-proportional to the quantity of material to deposit. This electrochemically synthetic process is usually investigated on mm\textsuperscript{2} to cm\textsuperscript{2} sized electrodes, but scaling down sensing arrays and narrowing the gap \textit{L} between S \& D electrodes scales down resources consumption. It also \textbf{consumes only what it needs}, so the electrolyte environments can promote further more coatings on other sensing elements from the same array, or for other arrays on other substrates.\\[3pt]
Reading sensing elements' completion state can also be performed \textit{in operando} by voltage-ramped impedance spectroscopy , similarly as in Ghazal \textit{et al.} 2021:\cite{Ghazal2021} Impedance spectroscopy is performed repeatedly while increasing slowly the DC component of the signal applied between a coated sensing element and a sensing element to grow on (Fig.\ref{fig:fig2.1.2}.i). If two well-chosen resistors are placed in the setup as depicted in Fig.\ref{fig:fig2.1.2}.i, the low frequency impedance shows the transition between open and short circuits. The higher frequency impedance indicates the shortage as well, by a sudden change in capacitance (Fig.\ref{fig:fig2.1.2}.i).\\[3pt]
After a CP shorts pair of electrodes in an electrolyte, they form an OECT with the counter-electrode in the case of EDOT leading to PEDOT in a PSS containing electrolyte. The electrogenerated OECT can be operated both as an accumulation and a depletion mode p-type transistor (Fig.\ref{fig:fig2.1.2}.m). The electrogenerated device can therefore be used in an \textbf{ion sensitive classifier} (as in the previous Sec.\ref{Ch2.2}) or as a \textbf{molecule sensitive classifier} (as in the previous Sec.\ref{Ch2.1}).

\subsection{Modifying Existing Elements}
\label{Ch3.1.3}

\begin{figure}
  \centering
  \includegraphics[width=1\columnwidth]{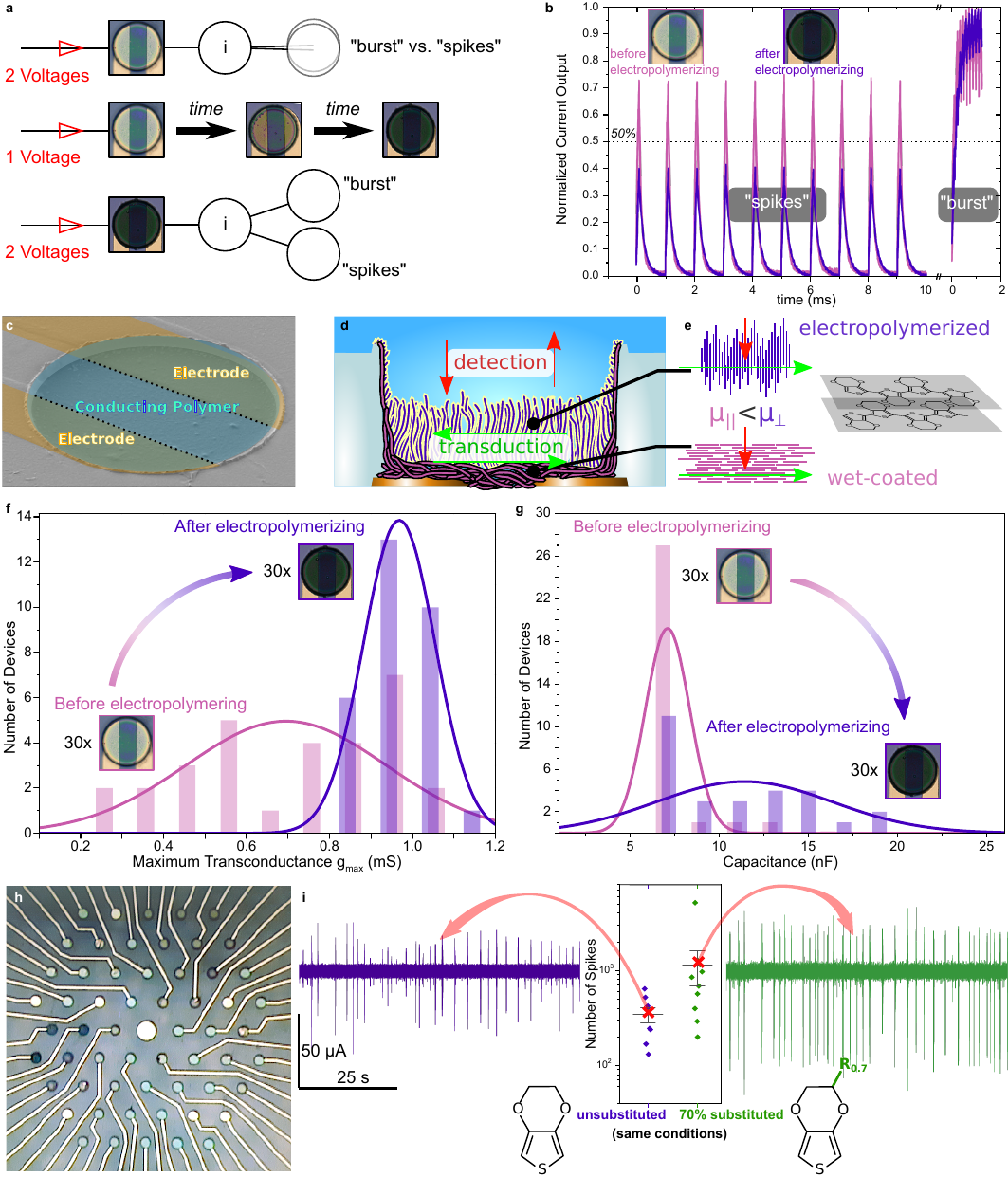}
  \caption{\textbf{Electropolymerization to Tune Sensing Elements' Properties $\vert$ a,} Schematics for an evolving classifier, using neurogenesis to recognize voltage pulse rate on two levels. \textbf{b,} Modification of the voltage-spike filtering of an OECT with electropolymerization (from Ghazal \textit{et al.})\cite{Ghazal2021}. \textbf{c,} Scanning Electron Microscope (SEM) picture of an OECT featuring spin-coated PEDOT:PSS in a 28~$\upmu$m diameter cavity (from Ghazal \textit{et al.})\cite{Ghazal2021}. \textbf{d,} Schematic cross-section of an OECT featuring an electropolymerized coating (blue fibers) over a wet-coated CP (pink fibers). \textbf{e,} Alignment/misalignment of both organizations with the detected ion transport (red arrow) and the transducing electron transport (green arrow) with different macromolecular packings (as interpreted by Ghazal \textit{et al.})\cite{Ghazal2021}. \textbf{f,} Reduction of the transconductance dispersion for a population of OECTs with electropolymerization (from Ghazal \textit{et al.})\cite{Ghazal2021}. \textbf{g,} Impact on the capacitance dispersion with the same population of OECTs (from Ghazal \textit{et al.})\cite{Ghazal2021}. \textbf{h,} Microscope picture of five populations of electrodes, electro(-co-)polymerized with different molar ratios of EDOT and a glycolated derivative.\cite{Ghazal2023} \textbf{i,} Statistical rates of sensed spikes with two populations of electrodes, electropolymerized with EDOT and electro(-co-)polymerized with EDOT and a glycolated derivative (from Ghazal \textit{et al.})\cite{Ghazal2023}.}
  \label{fig:fig2.1.3}
\end{figure}

The previous subsection (Subs.\ref{Ch3.1.2}) refers to the ability of electropolymerization to enable conductimetric sensitivity of uncoated electrode pairs in a cavity. Fast changes in device conductance are achieved with narrow inter-electrodes gaps: From noise level to shorts. It is therefore important to verify if materials' properties are \textbf{dispersed}, \textbf{controllable}, and particularly if sensing elements' \textbf{sensitivities} can be tuned.\\[3pt]
When aiming a particular conducting-polymer chemical structure, the first thing to consider is that many so-called "PEDOT" are reported in the literature: Not specifically on modified structures of poly(ethylenedioxythiophene) derivatives with different chemical functions,\cite{Fenoy2021} but the material-processing significantly affects the physical properties by the way monomeric units organize as a supra-macro-molecular material\cite{Gueye2016,Rivnay2016,Genovese2017,Gueye2020}. Moreover, electropolymerized coatings are quite different from the ones from spin-coated formulations: Light absorption coefficients of electropolymerized PEDOT are high due to more disorder compared to spin coated PEDOT, the latter being very transparent and extensively studied to replace transparent oxide-based electrodes in electronics,\cite{Dauzon2020} and can specifically achieve high transparency on low impedance electrodes.\cite{Susloparova2021} When Ghazal \textit{et al.} electrodeposit PEDOT from EDOT in NaPSS\textsubscript{aq} on top of devices precoated with a Clevios\textsuperscript{\textregistered} PEDOT:PSS formulation, sensing elements darken quickly during electropolymerization (Fig.\ref{fig:fig2.1.3}.a).\cite{Ghazal2021}\\[3pt]
As both spin-coated and electropolymerized PEDOT can reach outstanding electrical conductivities, both typically up to 10\textsuperscript{2} to 10\textsuperscript{3}~S/cm,\cite{Zhang2015,Castagnola2014} it is without surprise that electropolymerizing PEDOT on top of sensing elements, which are already patterned with spin-coated PEDOT, has a significant impact on sensing elements' electrical properties: In Ghazal \textit{et al.}, gradual changes in the effective capacitance is observed after electropolymerizing strictly on the device area for a few seconds.\cite{Ghazal2021} This shows that if electropolymerized in NaPSS\textsubscript{aq}, materials participate to the ion accumulation in their bulk as OMIEC like spin-coated PEDOT:PSS does.\cite{Rivnay2015} Additionally, OECT transconductance increases gradually with electropolymerization.\cite{Ghazal2021} Thanks to both modifications, device sensitivity to dynamical voltage patterns changes significantly.\cite{Ghazal2021} Therefore, sensing elements modification by PEDOT electrodeposition could be a way to program their ability to discriminate voltage activities by rate of events (Fig.\ref{fig:fig2.1.3}.a-b).\\[3pt]
The contrast between both transconductance and capacitance originates from different limitations in the ion/electron coupling between the material and its environment: capacitance reflects the intimacy of the electrical conductor with the solution and how maximized is the Helmholtz layers formed between both, whereas transconductance translates how much an electric field through the solution conditions this Helmholtz layer modulating the material's electrical conductance. Supervising their tuning requires understanding how both independently affect information classification. This is crucial with electropolymerization, as large property dispersions are expected,\cite{Ghazal2019} like in the previous Subs.\ref{Ch1.3.2}.\cite{Pecqueur2018b} While reducing the transconductance dispersion of a 30-OECT array using electropolymerization, the capacitance dispersion of these same devices increases (Fig.\ref{fig:fig2.1.3}.f-g).\cite{Ghazal2021} If at the scale of an array, several dispersions condition a classifier's performance,\cite{Pecqueur2018b} which can be affected with electropolymerization, \textbf{it is important to understand how modifying electropolymerizations' conditions helps at controlling each dispersions}.\\[3pt]
In another study, Ghazal \textit{et al.} aimed at using an electropolymerized sensing arrays for \textit{in vitro} neural activity classification (Fig.\ref{fig:fig2.1.3}.h-i).\cite{Ghazal2023,Ghazal2023a} Classifying relevant information patterns is quite complex for such application, and sometimes, translating what physical properties to change on sensing elements is more complex than experimenting to improve classifiers. In Ghazal \textit{et al.} 2023,\cite{Ghazal2023} PEDOT's partial glycolation was made by copolymerizing EDOT with a glycolated-EDOT derivative. Compounding PEDOT with glycols have positive effects on sensing arrays' cell cultures.\cite{Jimison2012} As glycolation improves materials' hydrophilicity, the choice for this strategy was initially motivated to increase sensing element bulk capacitance by increasing its affinity with water. After the custom-made monomer synthesis,\cite{Perepichka2002} several tests on different codeposited polymers did not show significant improvements of sensing elements surface capacitance. However, the material did improve neural activity detection, but by diminishing PEDOT's Young modulus on sensing elements, so choice was good but for mechanical reasons. \textbf{If sensing elements modification by electropolymerization shall be supervised, it should not exclusively be by the only mean of an \textit{a priori} understanding of physical mechanism ruling detection/transduction, particularly when the environment has complex boundaries, and electropolymerization modifies a CP physics in different ways}.

\section{Morphogenesis of Conducting Polymer Interconnections}
\label{Ch3.2}

Adjusting materials' sensitivity in a sensing array is of equal importance as the programability of a classifier software to control the strength of individual weights between sensing inputs and output classes. Aiming to materialize this strength by a physical link between electrodes connected to sensing elements in an array and others electrodes connected to a readout circuit, electropolymerization is studied as an electrochemical mechanism to implement a functionality common to many naturally intelligent systems, opposed to our conventional electronic paradigm which does not exploit metamorphism as a computational resource, and where only electrons move, not matter.

\textit{This section is mostly associated to peer-reviewed results published in} Janzakova \textit{et al.} \textit{\href{https://doi.org/10.1038/s41467-021-27274-9}{Nat. Commun.}} \textbf{12}, 6898 (2021), Janzakova \textit{et al.} \textit{\href{https://doi.org/10.1002/advs.202102973}{Adv. Sci.}} \textbf{8}(\textit{24}), 202102973 (2021), Kumar \textit{et al.} \textit{\href{https://doi.org/10.1038/s41598-022-10082-6}{Sci. Rep.}} \textbf{12}, 6395 (2022), Scholaert \textit{et al.} \textit{\href{https://doi.org/10.1088/2634-4386/ac9b85}{Neuromorphic Comput. Eng.}} \textbf{2}(\textit{4}), 044010 (2022) \& Baron \textit{et al.} \textit{\href{https://doi.org/10.1149/2754-2734/ad9bcb}{ECS Adv.}}, \textbf{3}(\textit{4}), 044001 (2024) \textit{among other information from the state-of-the-art literature \& early results of C. Scholaert, E. H. Balaguera and A. Baron.}\\

\subsection{Growing Interconnections}
\label{Ch3.2.1}

\begin{figure}
  \centering
  \includegraphics[width=1\columnwidth]{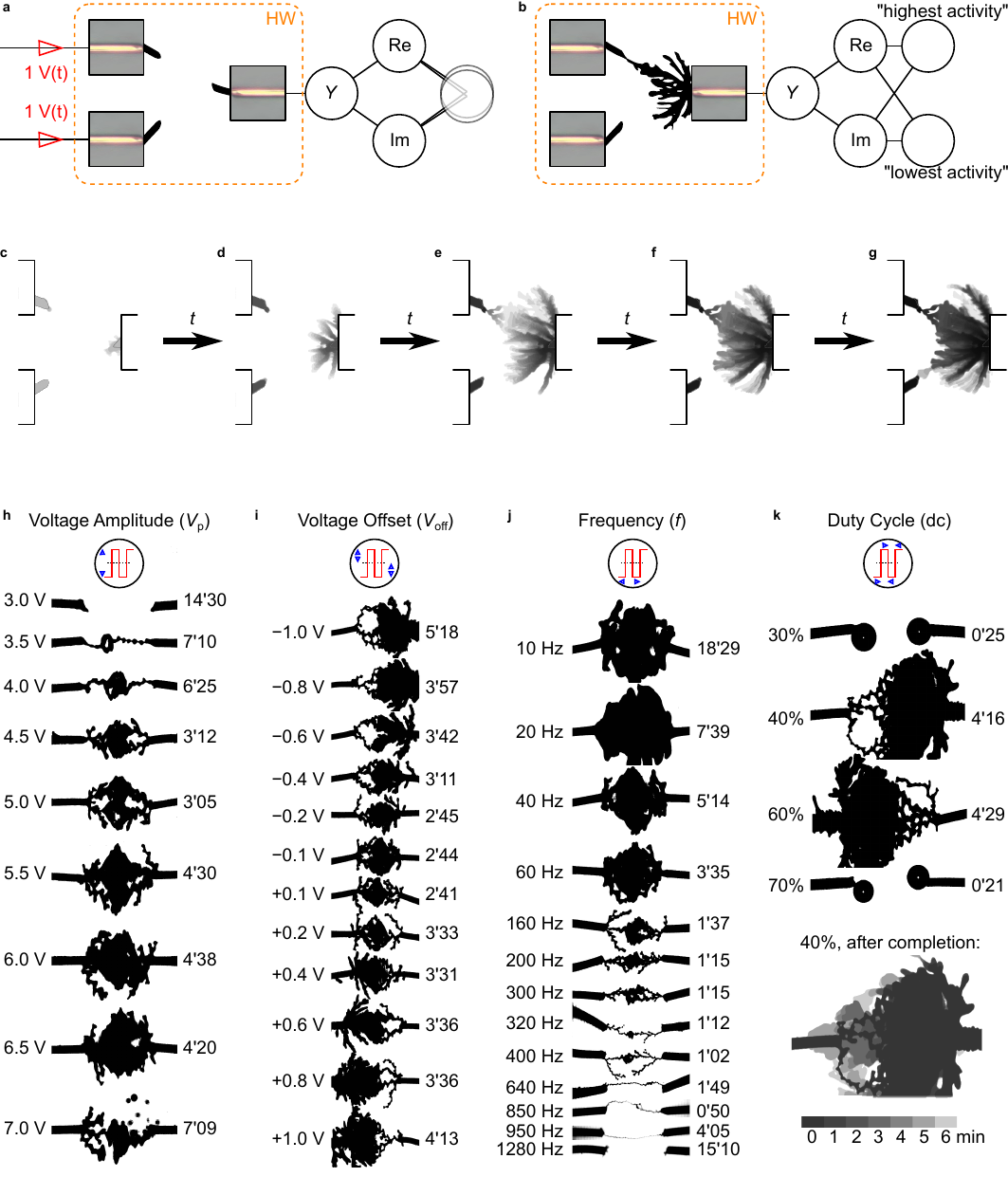}
  \caption{\textbf{Dynamical Voltage Control for the Electrochemical Growth of Conducting Polymer Dendrites $\vert$ a-b,} Schematics for an evolving classifier, using CPDs morphogenesis as physical mechanism to program the interconnectivity between two input nodes and an output class based on difference of voltage activity. \textbf{c-g,} Morphology of simultaneous growths from two input wires experiencing not the same voltage activity, at multiple time during their growth (from Janzakova \textit{et al.} 2021)\cite{Janzakova2021,Janzakova2021b}. \textbf{h-k,} Pulse voltage parameter dependency of a dendritic PEDOT morphology, under different voltage amplitudes \textit{V}\textsubscript{p} (\textbf{h}), voltage offsets \textit{V}\textsubscript{off} (\textbf{i}), signal frequencies \textit{f} (\textbf{j}) and signal duty-cycles dc (\textbf{k}). Where not mentioned, the standard parameters for all growths are \textit{V}\textsubscript{p}~=~5~V, \textit{V}\textsubscript{off}~=~0~V, \textit{f}~=~80~Hz and dc~=~50\%. For all cases, the varied experimental parameter and the growth time is display respectively on the left and the right hand side of each picture (from Janzakova \textit{et al.} 2021)\cite{Janzakova2021,Janzakova2021b}.}
  \label{fig:fig2.2.1}
\end{figure}

Initially inspired by the works of Inagi and workers,\cite{Koizumi2016,Ohira2017,Watanabe2018,Koizumi2018,Inagi2019} and pioneered by Dang, Akai-Kasaya \textit{et al.},\cite{Dang2014} using CPs was foreseen in 2018 as a way to mimic the brain to implement topological plasticity in neuromorphic electronic hardware \cite{Pecqueur2018c}. Confirming observation of Eickenscheidt \textit{et al.},\cite{Eickenscheidt2019} Conducting Polymer Dendrites (CPDs) show very \textbf{rich morpholologies} upon their growth in a solution when directly polarizing electrodes with a periodic voltage wave (Fig.\ref{fig:fig2.2.1}). Janzakova \textit{et al.} implemented these growth in a \textbf{water-based electrolyte} containing EDOT, NaPSS and benzoquinone (BQ).\cite{Janzakova2021b} Electropolymerization under an AC signal in water showed the same \textbf{directivity from PEDOT growths} as in the earlier work of Akai-Kasaya \textit{et al.} performed with acetonitrile as cosolvent.\cite{AkaiKasaya2020} On free-standing wires, correlations between morphological features of the growths were established with the parametrizing of the pulse voltage waveform, as Watanabe \textit{et al.} evidenced that such growths can show some affinity with a solid substrate.\cite{Watanabe2018} Compared to DC electropolymerization, growth rates are higher.\cite{Janzakova2021b} Formation of suspended particles was visible on the camera,\cite{Janzakova2021} which cannot be evidenced in DC electropolymerization. Such \textbf{particles may have a significant contribution in the formation of CPDs},\cite{Kumar2022} in addition to the applied dynamical voltage pattern.\\[3pt]
When applying a waveform of voltage pulses, \textbf{voltage amplitude, offset, frequency and duty cycle} are four elementary features which have distinct contributions in the growths.\cite{Janzakova2021b} Despite an apparently high \textbf{stochasticity in the growth}, typical trends were observed over the multiple experiments (Fig.\ref{fig:fig2.2.1}.h-k). Growths are \textbf{voltage-activated above a threshold} far above the solvent electrochemical window, without observing any bubbling: typically, growth were observed when a difference of potential higher than 3~V was applied on both gold electrodes. Beyond 6.5~V of applied voltage amplitude, water bubbling starts to occur on the electrodes. \textbf{A DC component in the signal promotes growth asymmetry} (which can be induced by a voltage offset, or by a duty-cycle different than 50\%). The morphologies present different features in number and size of branches. \textbf{The lower the voltage, the higher CPDs' dendricity}. The waveform dynamics has a strong impact on the size of the branches, as \textbf{increasing the frequency up to about 1~kHz leads to thinner and wire-like growths} (filament). The \textbf{growth rate is also highly signal dependent}: the higher the frequency, the faster the growth, as well as the higher the voltage amplitude, the faster the growth. Overall, \textbf{growths are highly tunable from isotropic to directional} depending on the applied signal, but whatever the applied signal, \textbf{growths always lead to shorting both electrodes at a certain point}, typically between one to ten minutes for a distance of 240~$\upmu$m.\cite{Janzakova2021,Janzakova2021b}\\[3pt]
By tuning the four waveform parameters, a practical stimulation of a three-wire junction showed that \textbf{morphogenesis can be conditionned to connect an input electrode experiencing more activity than another one} (Fig.\ref{fig:fig2.2.1}.c-g).\cite{Janzakova2021b} The input pulse activity is low enough to not promote electropolymerization by itself. As a results, the applicative potentials of CPDs morphogenesis for \textit{in materio} voltage classification is \textbf{not restricted by the redox activity of a chosen monomer, nor the size of the sensed voltage} to input in hardware.

\subsection{Electrical Interconnectivity after a Physical Contact}
\label{Ch3.2.2}

\begin{figure}
  \centering
  \includegraphics[width=1\columnwidth]{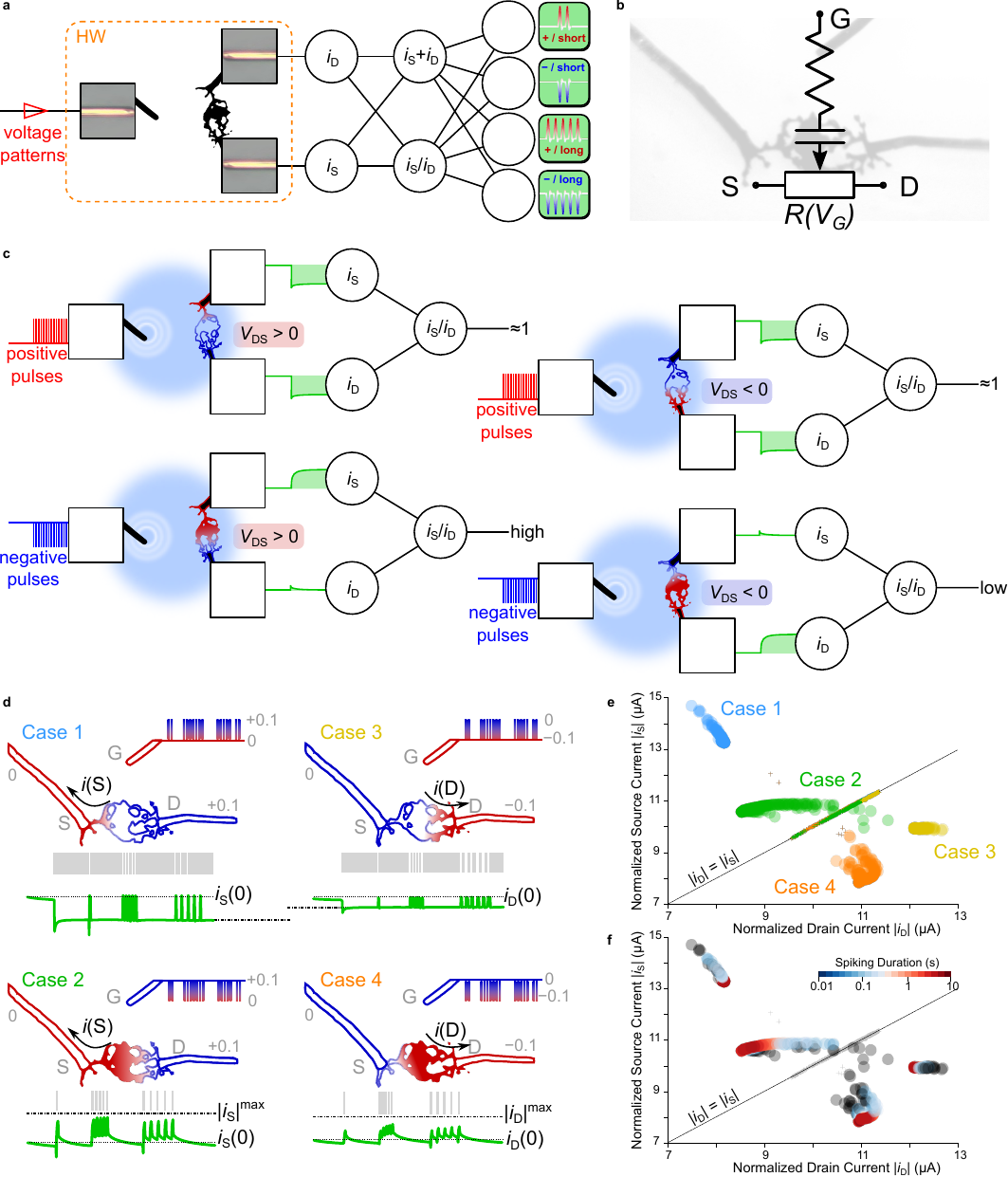}
  \caption{\textbf{Multi-Sensitivity in Conducting Polymer Dendrites $\vert$ a,} Schematic for a hardware classifier which uses CPDs' heterogeneity to classify a voltage past-experience on two terminals by the difference of ion-electron coupling on a single CPD structure. \textbf{b,} Generic equivalent circuit of a compact electrical model for a conventional OECT where a local-contact gating point is limiting in the case of a CPD. \textbf{c,} Four different polarization cases of gate spike projections on contacted CPDs biases with a voltage, outputting different source and drain signals which highly depend on the ion charge accumulation profile (blue and red colors) on a CPD (from Scholaert \textit{et al.})\cite{Scholaert2022}. \textbf{d,} Source and drain current traces recorded under the four different cases (from Scholaert \textit{et al.})\cite{Scholaert2022}. \textbf{e,} Projection of the normalized source and normalized drain  current values (experimental) showing the pulse polarization dependency (data from Scholaert \textit{et al.})\cite{Scholaert2022}. \textbf{f,} Same data showing the dependency with the duration of spike bursts (data from Scholaert \textit{et al.})\cite{Scholaert2022}.}
  \label{fig:fig2.2.2}
\end{figure}

Once two CPDs grow until contacting, they merge into a single object. Moving the wires which hold them does not break it right away. CPDs are quite \textbf{brittle}, do not elongate nor shrink and bend until a certain curvature before breaking. They are difficult to study out of their electrolyte solution, since attempts to pull them out from liquids leads to a collapse of the branches. The bridging they form is \textbf{electrically conductive} and their material can be considered more as a metal than an intrinsic semiconductor. The fact that they grow preferentially on already-grown materials than on chemically unmodified gold area is an experimental clue of this.\cite{Janzakova2021b} The current drawn from two \textbf{merged CPDs is ohmic} from the first millivolt.\cite{Janzakova2021b} Their DC conduction relates somehow to their morphology, as thinner CPDs are less conductive than thicker ones under the same conditions. CPD's material seems far \textbf{less conductive than wet-deposited PEDOT:PSS} from commercial formulations: A 240~$\upmu$m long PEDOT CPD filament shows about 10 to 100~k$\Upomega$ resistance when grown in 1~mM NaPSS in water, leading to a material's conductivity about 0.1 to 10~S$\cdot$cm\textsuperscript{-1}.\cite{Janzakova2021b}\\[3pt]
Free-standing in an electrolyte solution, \textbf{CPDs behave as OECTs} when using a third wire as gate electrode if coated with a dendrite itself or if made of silver (a clean gold wire seems to have no effect).\cite{Janzakova2021a,Scholaert2022} As material's conductivity is quite poor compared commercial PEDOT:PSS formulations, CPDs are only partially p-doped: \textbf{CPDs behave as both accumulation and depletion p-type OECTs}.\cite{Janzakova2021a,Scholaert2022} Upon gate voltage modulation, CPDs do not seem to change in morphology: neither when electrochemically doped nor dedoped, unlike in PEDOT:PSS actuators.\cite{Hara2014,Nguyen2018,Hu2019a,Rohtlaid2021}. The fact they behave as OECTs was particularly appealing in regard to earlier works made on using OECTs for neuromorphic computing.\cite{Gkoupidenis2015,Gkoupidenis2015a,Gkoupidenis2016,Gkoupidenis2017} An essential focus on why using CPDs as OECTs for unconventional information processing was on observing whether it was possible to \textbf{retrieve the past voltage information experienced during their growth, characteristic of their specific morphology.}\\[3pt]
When two CPDs merge, it is particularly difficult to access to such information with just DC currents and with two wires only: Janzakova \textit{et al.} showed that shorted-CPDs \textbf{conductance can further be reinforced as they weld slightly better overtime},\cite{Janzakova2021b} but they behave more or less as resistors. When using them as an OECT with a third wire, it is possible to \textbf{modulate the value of this resistance}, which CPDs' morphological state can be retrieved in the transfer-characteristic's hysteresis:\cite{Janzakova2021a} \textbf{Filamentary dendrites tend to show less hysteresis than fractal structures}.\cite{Janzakova2021a} Reaching the steady-state, the signal is also morphology-dependent, as a similar relationship with thin-film OECTs is evidenced:\cite{Tarabella2010,Cicoira2010} \textbf{larger gates tend to modulate more the doping level of smaller channels}.\cite{Janzakova2021a,Scholaert2022} Since no swelling nor growth is observed when characterized under low voltage, \textbf{stimulating contacted CPDs with transient voltages can be a way to read CPDs without writing further on them}, and accessing to morphology-dependent information from both dynamical and steady-state currents.\\[3pt]
In a voltage-controlled transient regime, \textbf{transient ionic currents can be much higher than the electrical one} for wire-like CPDs polarized at 100~mV of voltage bias. Therefore, exploiting both properties in CPD-OECTs is preferred for massive and highly fractal CPDs over small CPD filaments.\cite{Janzakova2021a,Scholaert2022} Under a periodic wave of voltage spikes, source and drain output currents have a \textbf{specific signature} from the applied voltage input promoted by CPDs' morphological asymmetry.\cite{Scholaert2022} Unlike short-channel integrated OECTs,\cite{Pecqueur2018b} polymer's structural heterogeneity affects far more independently currents drawn either from the source or the drain. Furthermore, growths originate from each contact point and not from the shorted branches defining the OECT channel. Therefore, \textbf{charge transience associated to doping modulation will influence far more the current outputted from the contact it origins than from the other contact}. \\[3pt] 
This was studied by Scholaert \textit{et al.},\cite{Scholaert2022} upon stimulating CPDs under $\pm$100~mV source-drain biases with $\pm$100~mV gate voltage spikes (Fig.\ref{fig:fig2.2.2}).\cite{Scholaert2022} \textbf{Cations and anions do not imprint equally on CPDs} disregarding their polarity. Since CPDs are not symmetrical, the ion charge/discharge time constants of both wires are not identical. Coupled with the polarity-dependent electrochemical doping promoting high conductivity of one side but low one on the other side, polarity inversion on CPDs does not lead to the same information as the opposite projection of an opposite spiking event. This property is not intrinsic to CPDs (presumably, it could be obtained from concentric thin-film OECTs)\cite{Pecqueur2017}, but it demonstrates how \textbf{non-ideal devices enrich recurrent systems} for classifications. The study also shows that implementing short-term plasticity on CPDs (as OECTs)\cite{Gkoupidenis2015} allows \textbf{identifying events with dynamically-dependent conditions}, so short-lasting bursts of spikes are discriminated from long lasting ones in addition to their polarity.\cite{Scholaert2022}

\subsection{Electrochemical Interconnectivity while Growing}
\label{Ch3.2.3}

\begin{figure}
  \centering
  \includegraphics[width=1\columnwidth]{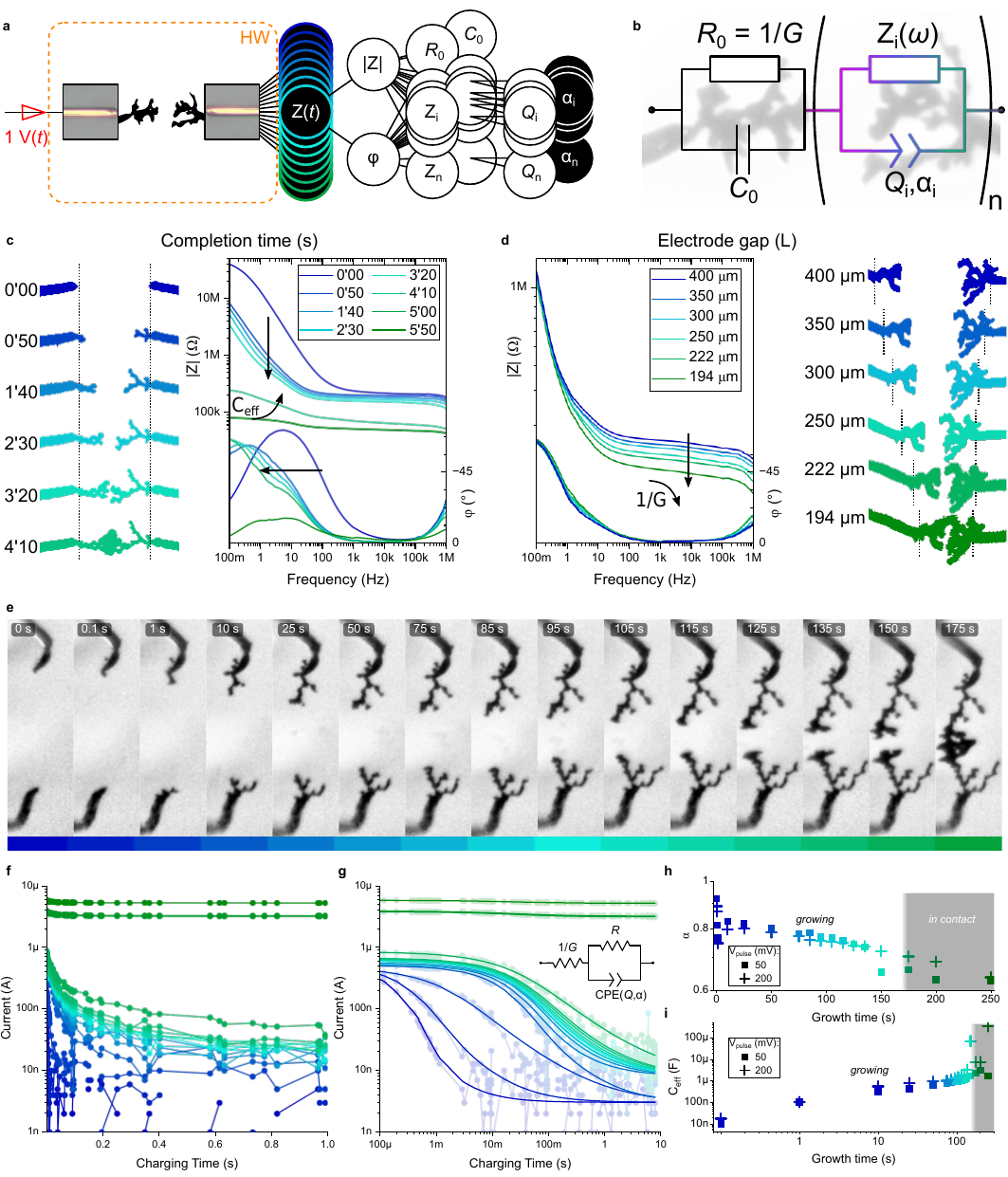}
  \caption{\textbf{Conducting Polymer Dendrites' Non-Ideality $\vert$ a,} Schematic for a hardware classifier which uses CPD morphogenesis to classify a voltage past-experience on two terminals by the nature of its electrochemical relaxations. \textbf{b,} Generic equivalent circuit proposed to model CPDs' multiple relaxations as serial CPE-containing Voigt circuits (from Baron \textit{et al.})\cite{Baron2024b}. \textbf{c,} Dependency of a CPDs' electrochemical impedance spectrum with its growth duration (data from Baron \textit{et al.})\cite{Baron2024a}. \textbf{d,} Dependency of a CPDs' electrochemical impedance spectrum with the gap between gold wires (data from Baron \textit{et al.})\cite{Baron2024a}. \textbf{e,} Pictures of a CPD growth at multiple stages (data from Baron \textit{et al.})\cite{Baron2025}. \textbf{f,} Current transience after a voltage impulse for a CPD at the different stages displayed in Fig.\ref{fig:fig2.2.3}.e (data from Baron \textit{et al.})\cite{Baron2025}. \textbf{g,} Same experimental data (scatters) in double log\textsubscript{10} scale graph, fitted with the model displayed in Fig.\ref{fig:fig2.2.3}.b (lines) in the time domain (data from Baron \textit{et al.})\cite{Baron2025}. \textbf{h-i,} Dependency of the dispersion coefficient (\textbf{k}) and the effective capacitance (\textbf{l}) as time-domain fitted parameters from the model displayed in Fig.\ref{fig:fig2.2.3}.b and the experimental data for the CPDs displayed in Fig.\ref{fig:fig2.2.3}.c (data from Baron \textit{et al.}).\cite{Baron2025}}
  \label{fig:fig2.2.3}
\end{figure}

Considering that most of the morphology-dependent information of a CPD-OECT outputs from the ionic currents under a gate electrode dynamics, observing comparable information on CPDs as pseudo-capacitors before they physically contact was envisioned in Baron \textit{et al.}'s studies (Fig.\ref{fig:fig2.2.3}.a).\cite{Baron2024,Baron2024a,Baron2024b,Baron2025}\\[3pt]
CPDs were studied by EIS to determine the nature of their electrochemical relaxation(s).\cite{Baron2024,Baron2024a,Baron2024b} An apparent \textbf{non-ideal relaxation}, in regards to conventional Debye capacitors, was already identified by Janzakova \textit{et al.},\cite{Janzakova2021b} supposed to be morphology dependent (three morphologies were compared). EIS on a CPD-OECT set in a diode-connected transistor configuration was also done by Scholaert \textit{et al.},\cite{Scholaert2022} but relaxations' non-ideality was not observed because of the use of a silver electrode. Therefore, these relaxations were studied on the much simpler two-electrode setup with free-standing gold wires, each of them featuring CPDs at different growth stages.\cite{Baron2024,Baron2024a,Baron2024b,Baron2025}
By repeatedly interrupting a CPD growth on a pair of gold wires to record the electrochemical impedance from one wire to the other, it was possible to observe the morphological dependency of the voltage signal projection from one CPD to another in a common electrolyte.\cite{Baron2024} \textbf{The EIS spectra show gradual changes with the growth}, from pristine gold wires to their completed phase when they contact (Fig.\ref{fig:fig2.2.3}.c). \textbf{The variation of the distance between CPDs is not the reason of these changes} (Fig.\ref{fig:fig2.2.3}.d), so the changes are more likely to be attributed to the electrodeposited mass or volume rather than the electrochemical gap reduction.\cite{Baron2024a} The study also shows \textbf{at least two relaxations in the electrochemical system}, requiring deeper understanding of the underlying mechanisms.\\[3pt]
Later in a work involving Distributed Relaxation Time (DRT) analysis,\cite{Baron2024b} Baron \textit{et al.} evidenced that more relaxations are hidden in the raw impedance data, implying the two major relaxations visible in the phase diagram \textbf{could not be simply assigned to both CPDs independently} (Fig.\ref{fig:fig2.2.3}.b). The study showed that three relaxations vary in a same way all along the analysis: During a \textbf{nucleation phase} (0 to 1~s approximately), relaxations' time constant shift down, then they increase further during the \textbf{growth phase} (1 to 100~s approximately) until both dendrites merge and behave more or less as a resistor (\textbf{contacting phase}). The study confirms that a CPD's \textbf{impedance is stable} for at least half an hour, and the slight phase-shift presumably supposed from evaporation does not explain the whole impedance change upon growth. Studying different CPD morphologies to assign the physical origins of their relaxations was also attempted (Fig.\ref{fig:fig2.2.3}.f-i).\cite{Baron2024b} Moreover, it showed \textbf{the thinner the CPD, the less conductive the junction}. Additionally, \textbf{asymmetrizing the junction allows the third relaxation being more obvious in the phase diagram} without using the DRT, confirming abilities for CPDs to engrave in their morphology various properties conditioning dynamical-signals filtering.\cite{Baron2024b}\\[3pt]
\textbf{These signal filters are complex by the number but also by the nature of relaxation}. The apparent Constant Phase Elements (CPEs) suggests fractional capacitances, as the result of Havriliak-Negami relaxations with diffusion-limited electrochemical processes. Supposing this fractional character can be conditioned by CPDs' tunable morphology implies that \textbf{charge/discharge non-ideality can greatly be controlled by CPDs' past experience engraved in the morphology}. This was evidenced in the time domain under voltage-impulse stimulations on CPDs grown at multiple stages and fitting the charge-discharge current decays with Mittag-Leffler functions accounting for different fractional capacitor values at growth stages (Fig.\ref{fig:fig2.2.3}.g-i).\cite{Baron2025} The study confirmed that \textbf{non-ideality is not due to voltage non-linearity, as long as CPDs are operated below 200~mV and as long as CPDs remain fairly symmetrical}. In these conditions, fitting the apparent single-relaxation on impulse-voltage current decays clearly confirms the non-ideality in regard to an exponential decay (Fig.\ref{fig:fig2.2.3}.f). With the growth, the effective-capacitance \textit{C}\textsubscript{eff}~=~\textit{R}\textsuperscript{(1/$\upalpha$-1)}\textit{Q}\textsuperscript{1/$\upalpha$} increases and the dispersion coefficient $\upalpha$ decreases as CPDs expand from the wires (\textit{R} and \textit{Q} being respectively the serial electrolytic resistance and the CPDs' anomalous capacitance in the electrolyte).\cite{Baron2025} This was exploited to modify the volatility of an electrical synapse via dendritic morphogenesis as voltage-conditioned programming mechanism to adapt the filtering ability of a junction to pass voltage spikes through or not.\cite{Baron2025} Although results do not currently exploit the programming of the dispersion coefficient $\upalpha$ independently from the quasi-capacitance \textit{Q}, \textbf{this first implementation of an evolving CPE-limited electrochemical synapse using CPD morphogenesis is an essential ingredient to physically build Fractional-Leaky Fire-Integrate Neurons,\cite{Teka2014,Teka2017} with adaptive fading-memory time-windows}.

\chapter{Preliminary Classifiers}
\label{Ch4}

\textit{This chapter introduces three on-going developments of environmental classifiers where technological capabilities of electropolymerization are challenged to understand how organic semiconductors can enhance chemical information projection in sensing hardware.}\\

\section{On Small Organic Electronic 'Tongues' featuring Large Receptive Fields}
\label{Ch4.1}

Co-integrating many sensitive materials at a small scale is technically difficult \cite{Shulaker2017}. It requires finding a generic technique which preserves each material chemical sensitivity all along a deposition process without cross-contamination. Increasing sensing arrays' receptive field by material co-integration on a given area implies \textbf{downscaling all materials while preserving their functionality}. However, most top-down deposition techniques have technical limitations in feature size, volume, number of co-integrated materials, not to mention material, energy costs. Downscaled to particle sizes, materials' stability can be threatened by different forces than the ones controlling large-sized thin-films. This is of a particular concern when multi-material co-integration is sufficiently large to significantly increase a sensing array's added-value. However, not much information is available in the state-of-the-art literature on proper co-integration techniques and environment exposure conditions to preserve materials' stability over sufficiently long time for full calibration or training for classification.\\[3pt]
In this section is detailed how \textbf{electro(-co-)polymerization} can be used to generically design chemically different macromolecular conductors of electricity (Subs.\ref{Ch4.1.1}), what are the practical limitations to consider for co-integrating them (Subs.\ref{Ch4.1.2}) and what specificities for a \textbf{conductimetric tongue} prototype shall be considered to minimize intrusiveness of an environment upon exposure and not threaten the integrity of an electro(-co-)polymerized sensing micro-array sensing array (Subs.\ref{Ch4.1.3}).\\

\textit{This section is mainly associated to information from the cited state-of-the-art literature \& early results of M. T. Nauto, P. Moustiez, A. Baron, D. Gu\'{e}rin and S. Pecqueur.}\\

\newpage

\subsection{Electro(-co-)polymerization for Multi-Material co-Integration}
\label{Ch4.1.1}

\begin{figure}
  \centering
  \includegraphics[width=1\columnwidth]{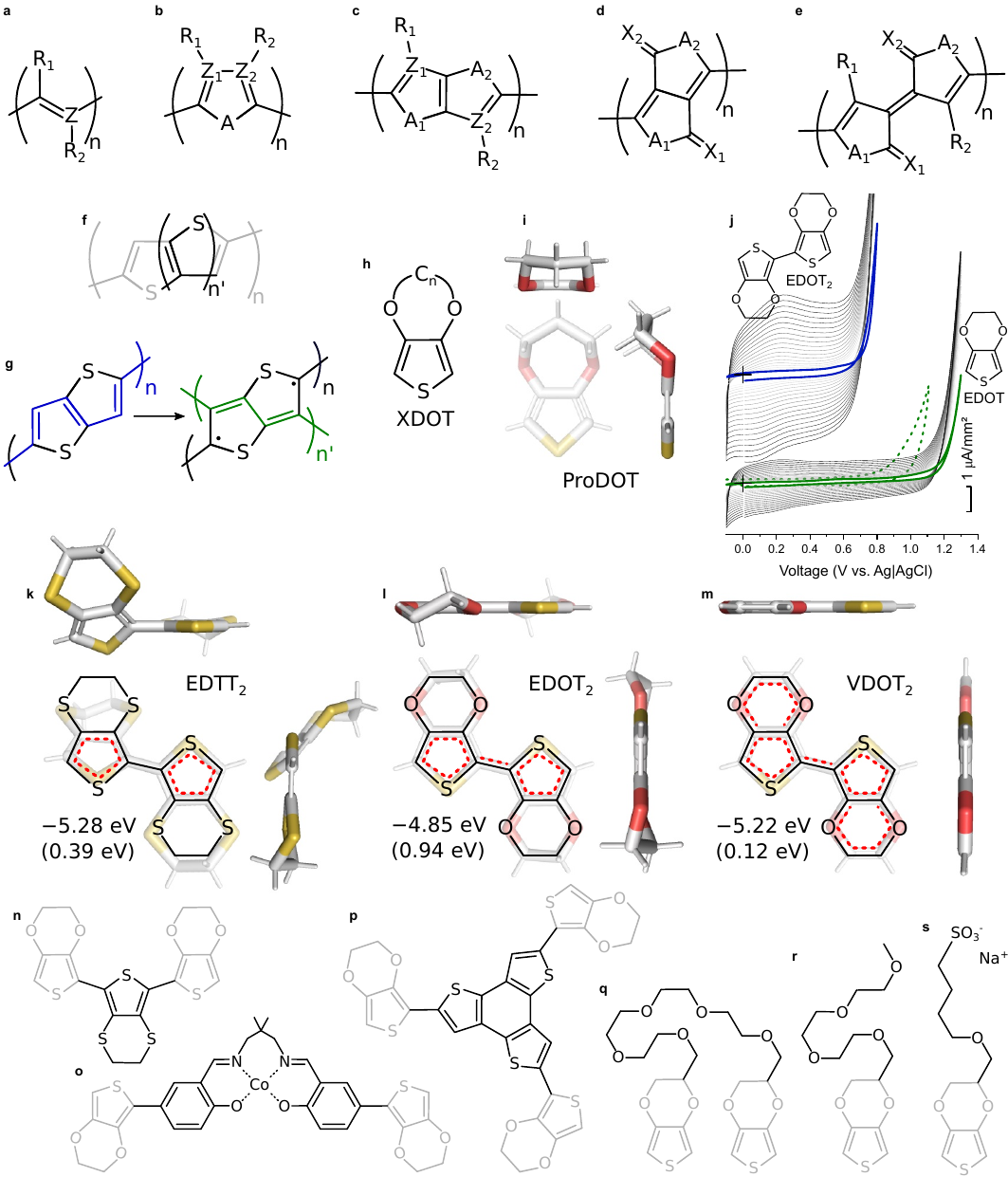}
  \caption{\textbf{Versatility of Electropolymerization $\vert$ a-e,} Various types of linear polymeric structures which can be electrogenerated by electropolymerization form the diprotonated monomer of the different monomeric units. \textbf{f,} Increasing the number of aromatic cycles of a monomer as a possibility to modify optoelectronic properties. \textbf{g,} Higher voltage activation of a secondary electropolymerization mode can change polymers' electronic property out of a same electroactive monomer. \textbf{h,} Generic structure of 3,4-alkylenedioxythiophene monomers\cite{Heywang1992,Dietrich1994,Kumar1998}. \textbf{i,} Crystal structure of ProDOT\cite{Kumar1998,Kumar1999}. \textbf{j,} Cyclic voltammetry at 100~mV/s of 10~mM EDOT (green) and 10~mM of its dimer (blue) showing different oxidation potentials in propylene carbonate (the dashed curve shows the first cycle of EDOT in water). \textbf{k-m,} Crystal structure of EDTT\textsubscript{2} \cite{Turbiez2005a}, EDOT\textsubscript{2} \cite{Raimundo2001a} and VDOT\textsubscript{2} \cite{Leriche2006a} dimers and their calculated first ionization energy (in brackets is the difference with the calculated first ionization energy of their corresponding monomers)\cite{Turbiez2005,Xue2019,Leriche2006}. \textbf{n-s,} Various EDOT end-capped monomers which feature different functionnalities: to produce an isoelectronic and neutral polymer of PEDOT (\textbf{n})\cite{Spencer2005}, to sentitize PEDOT derivatives to NO in gas phase (\textbf{o})\cite{Holliday2006}, to increase PEDOT stability (\textbf{p,q})\cite{Ringk2016,Perepichka2002}, to increase PEDOT's cation affinity in water (\textbf{q,r})\cite{Perepichka2002} or to homogenize PEDOT's p-doping (\textbf{s})\cite{Stephan1998,Cutler2005,Yano2019}.}
  \label{fig:fig3.1.1}
\end{figure}

Electropolymerization was used by organic electronic communities as the main mean to study CPs in the 80's.\cite{Horwitz1988,Goldsby1989,Hoferkamp1989,Roncali1990,Roncali1990a} Later advances on chemical designs to promote solubility strongly promoted wet deposition as mean to control material's homogeneity over large-area.\cite{Calvert1983,Patil1987,Kumar1998a} \textbf{If electropolymerization is less investigated now}, it is still highly valued to gather different chemistries on individual devices without cross-contaminating polymers during the deposition process. As mask-less patterning technique, it uses voltage activation on electrodes of any sizes and typically below 1.4~V to grow high added-value materials by oxidation (anodic polymerization) or reduction (cathodic polymerization), locally and \textbf{with the least material spoilage}.\\[3pt]
It is no more representative of a dominant topic in electrochemistry researches, more focused on specific challenges now such as energy storage/conversion (fuel cells, supercapacitors and batteries).\cite{Badwal2014,Mathis2019,Little2020} The community is however still very involved in polymers to research on non-fluorinated ion exchange membranes.\cite{Esmaeili2019,Lv2024} Electrogenerated CPs are still investigated in other fields, such as for surface dewetting,\cite{Dione2023,Diouf2023,Fradin2020,Mortier2017,Diouf2017,RamosChagas2016,Godeau2016,Mortier2016,Mortier2015}  or catalysis to immobilize organometallic complexes.\cite{Hong2013,Hong2012}\\[3pt]
Different monomeric units can compose a CP: from linear (such as polyacetylenes, derived from Fig.\ref{fig:fig3.1.1}.a), to cyclic (mostly 5-membered ring aromatics such as polypyrroles or polythiophenes, derived from Fig.\ref{fig:fig3.1.1}.b) and specifically polycyclic (such as thieno[3,2-b]thiophenes or thiazolo[4,5-d]thiazole, derived from Fig.\ref{fig:fig3.1.1}.c, diketopyrro[3,4-c]pyrroles derived from Fig.\ref{fig:fig3.1.1}.d and isoindigos derived from Fig.\ref{fig:fig3.1.1}.e). Thiophene units can also be more complex by repeating rings on the monomer (example in Fig.\ref{fig:fig3.1.1}.f) and a single monomer could lead to different polymers according to the applied voltage (scheme in Fig.\ref{fig:fig3.1.1}.g and specific case by Shi \textit{et al.}\cite{Shi2016}). Different units promote different properties independently on monomers and polymers (such as electrochemical potentials and solvent affinity).\\[3pt]
Among many published structures, \textbf{EDOT is truly exceptional} (structure in Fig.\ref{fig:fig3.1.1}.h for n~=~2): as it is stable in moist air, soluble in water, but leads to a CP which is not water-soluble. It is quite unique as most neutral oligothiophenes are insoluble in water, and doped polymers are ionic and have high affinity in polar solvents. EDOT has four free-lone pairs on two oxygens to form hydrogen bonds with water, on a light structure of 142~g/mol. Unlike ProDOT (structure in Fig.\ref{fig:fig3.1.1}.i), it is liquid at room temperature.\cite{Kumar1998} Upon oligomerisation, non-covalent bonds form between adjacent anti-conformal thiophenes (EDOT\textsubscript{2} crystal structure in Fig.\ref{fig:fig3.1.1}.l): the dimer is already insoluble in water. The ethylenedioxy bridge induces a very specific distorsion on the thiophenic structure: at the opposite to the vinylenedioxy VDOT derivative (VDOT\textsubscript{2} crystal structure in Fig.\ref{fig:fig3.1.1}.m),\cite{Leriche2006} the sp\textsuperscript{3} carbons of the bridge force both oxygens slightly out of the aromatic plane of the thiophenic units. Despite this distorsion, vicinal thiophenes are coplanar unlike its EDTT isoelectronic analogue, where sulfurs hinder the anticonformal configuration (EDTT\textsubscript{2} crystal structure in Fig.\ref{fig:fig3.1.1}.k).\cite{Turbiez2005} As a consequence, the oxidation point decreases substantially by oligomerizing EDOT (voltammetry in Fig.\ref{fig:fig3.1.1}.j) compared to VDOT or EDTT (theoretical ionization energies in Fig.\ref{fig:fig3.1.1}.k-m): the EDOT\textsubscript{3} trimer is even barely stable in its oligomeric form.\cite{Roncali2005} Other XDOT monomers have comparable properties,\cite{Groenendaal2003} despite the fact that longer bridges are out of the thiophenic plane and diminishes greatly the interchain interactions in the polymer.\cite{Dietrich1994,Kumar1998} It affects greatly the ion permeability and stability in a solvent.\cite{Dietrich1994,Kumar1998} Despite PEDOT was reported thirty years ago,\cite{Jonas1991,Heywang1992} \textbf{there is still nothing like it}!\cite{Elschner2010}\\[3pt]
To add further functionalities on PEDOT derivatives, one can either electrocopolymerize EDOT (or its dimer) with another monomer to form random copolymers (special care on both monomers' oxidation matching shall be considered).\cite{Guo2019,Hu2019} One can also form alternated copolymers by \textbf{end-capping a monomer with EDOTyl groups}.\cite{Sotzing1997} Regioregular copolymers are expected when monomers are symmetrical. Various EDOTyl end-capped monomers are reported for various properties: very few examples are displayed in Fig.\ref{fig:fig3.1.1}.n-s. End-capping monomers with EDOTyl (or ProDOTyl) usually decreases significantly the oxidation potential in the range of EDOT\textsubscript{2} (except if the central moiety is a really strong electron acceptor), so both strategies could be compatible but are rarely used concomitantly in the literature.\\[3pt]
\textbf{Rapid production of chemically tunable CPs on micro-sensors from monomer solutions and exclusively by voltage activation as lab-on-a-chip represents a powerful strategy to try-and-test many combinations of materials on arrays with very low amount of resources}.

\subsection{Miniaturizing Conductimetric Transducers with Specific Conducting Polymers}
\label{Ch4.1.2}

\begin{figure}
  \centering
  \includegraphics[width=1\columnwidth]{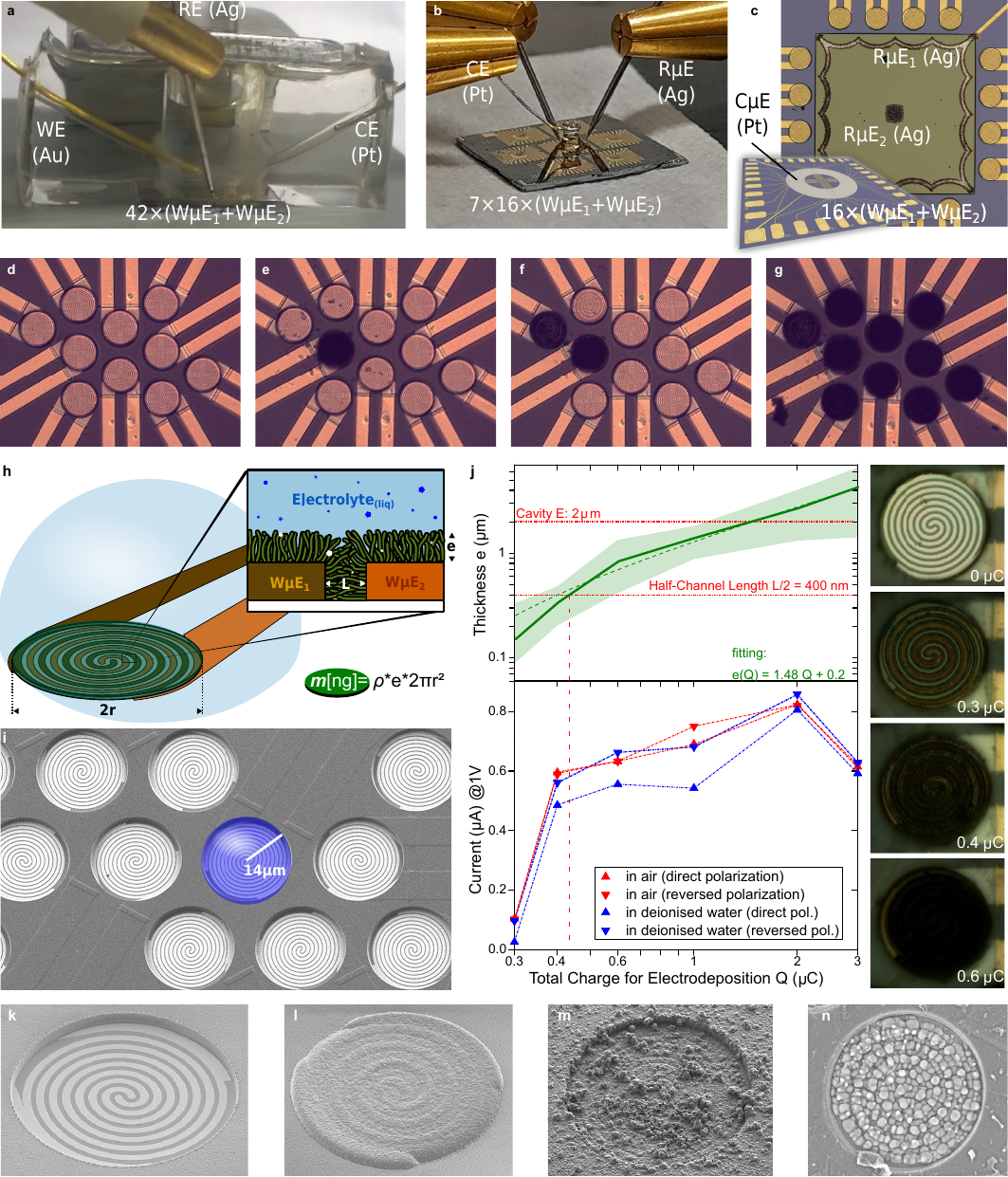}
  \caption{\textbf{Electropolymerizing on a Chip $\vert$ a,} Closed cell with an array of 42 pairs of W$\upmu$E, featuring an Ag RE in a glass frit, a Pt wire as CE and a macroscopic Au wire as WE for about 100~$\upmu$L of electroactive solution (substrate size: 1.1$\times$1.1~cm\textsuperscript{2}. \textbf{b,} Open drop electropolymerization on an array of pairs of micro-electrodes for one $\upmu$L of electroactive solution, featuring an electroplated Ag RE on the substrate (R$\upmu$E). \textbf{c,} SEM image of a 16$\times$(W$\upmu$E\textsubscript{1}+W$\upmu$E\textsubscript{2}) array, featuring two electroplated Ag R$\upmu$E equidistant to each W$\upmu$E and off the field with the larger-sized in-plane CE (C$\upmu$E). \textbf{d-g,} Iterative electropolymerizations on different conductimetric W$\upmu$E\textsubscript{1}+W$\upmu$E\textsubscript{2} elements integrated on a same chip. \textbf{h,} Structure of a 2\textit{r}~=~28~$\upmu$m conductimetric element featuring a pair of interlaced spiral-shaped electrodes as W$\upmu$E. \textbf{i,} SEM image of an array of 2\textit{r}~=~28~$\upmu$m elements, stacked under a theoretical hemispherical drop containing enough monomers at 12~mM to coat a single element. \textbf{j,} Coulommetry on conductimetric elements and relationship to polymer thickness and conductance. \textbf{k-n,} SEM images of an element before electropolymerization (\textbf{k}), after polymer coating from EDOT in 100~mM NaPSS\textsubscript{aq} (\textbf{l}), from EDOT\textsubscript{2} in propylene carbonate with 100~mM EmimOTf (\textbf{m}) and a cross-linked oligothiophene in acetonitrile with 100~mM NBu\textsubscript{4}PF\textsubscript{6} (from Pecqueur \textit{et al.})\cite{Pecqueur2018b} (\textbf{n}).}
  \label{fig:fig3.1.2}
\end{figure}

As bottom-up deposition, electropolymerization can theoretically be applied to patterning conducting-polymers selectively on \textbf{electrodes of any size}, and on \textbf{arrays of any pitch}. Practically, reducing electrodes' surface in an electrochemical setup, as well as narrowing distances between them hinder greatly repeatability. It is however of a prior importance to understand how to control these boundaries to downsize sensing micro-arrays and use only very few high added-value monomers along multiple processes.\\[3pt]
Electropolymerization is performed either in a cell (Fig.\ref{fig:fig3.1.2}.a) or in an open drop setup (Fig.\ref{fig:fig3.1.2}.b) to confine an electrolyte solution carrying electroactive mono/oligomer(s). Using \textbf{a closed-cell setup} allows using volatile solvents. However, tedious cell cleaning is required between each monomer deposition to not cross-contaminate elements' coating (Fig.\ref{fig:fig3.1.2}.d-g). The \textbf{open-drop setup} suits better multi-material co-integration, but is limited to non-volatile solvents which dewet enough on the substrate carrying the working-micro-electrodes (W$\upmu$E) array. A two-wire power generator or a three-wire potentio/galvanostat with a reference-electrode (RE) can be used to electropolymerize on each W$\upmu$E. The \textbf{two-electrode configuration} requires adding a redox agent (such as benzoquinone)\cite{Ghazal2021,Ghazal2023} in the electrolyte to pin the solution's potential on the counter-electrode's (CE) applied voltage. It is simpler to implement, but contaminates the electrodeposited polymer with the agent. Using a \textbf{potentio/galvanostat} is far more adapted for a qualitative coating and for metrology \textit{in operando}. However, it requires to integrate a non-soluble redox agent on a RE specifically, to not contaminate the solution composition and to control its position near each W$\upmu$E while being substantially away from the larger-sized CE (Fig.\ref{fig:fig3.1.2}.c). The polarization can either be \textbf{static or dynamic}. Static electrodepositions are either \textbf{galvanostatic} (current-controlled) to ensure the growth rate steadiness, or \textbf{potentiostatic} (voltage-controlled) to select only the most electro-reactive species for reaction on W$\upmu$E.\\[3pt]
Limits for CP electropolymerization  on individual elements can be illustrated with a few numbers (Fig.\ref{fig:fig3.1.2}.h-j): On typical 28~$\upmu$m diameter (2\textit{r}) elements featuring a pair of interlaced spiral-shaped electrodes, optimal conduction is reached for a polymer thickness above half the channel length (\textit{L}). So, when \textit{L}~=~400~nm,\cite{Pecqueur2018b,Boujnah2021} \textbf{only 1/4~ng is required per element} (assuming 2~g/cm\textsuperscript{3} for CPs). Such amount can be contained in a hemispherical droplet of the same diameter as the micro-sensing element itself (same molar weight at EDOT\textsubscript{3} at 12~mM): \textbf{only about 50~pL of electroactive solution}. In practice, electropolymerized PEDOT on such structure leads to less than 100~$\Upomega$ resistance (from 100~mM NaPSS\textsubscript{aq}). The large spirals yield \textit{W}/\textit{L} over 10\textsuperscript{3} (\textit{W}~=~412~$\upmu$m) despite the element's small size. In case \textit{W}/\textit{L}~=~1 with the same \textit{L}, the resistance would be still be about 100 k$\Upomega$, low enough to measured it with an ohmmeter. As such element could fit a 400~nm diameter area, 4900 of them could be prepared with the same amount of polymer as one single \textit{W}/\textit{L}~=~1000, 2\textit{r}~=~28~$\upmu$m element. If they have a minimal spacing of 400~nm between them in a hexagonal lattice, they would cover an area of only 1700~$\upmu$m\textsuperscript{2} (a circle of less than 47~$\upmu$m in diameter).\\[3pt]
On conductimetric elements, material's conductance \textbf{can be monitored \textit{in operando}} with either a bipotentiostat or two grounded voltage generators simultaneously electropolymerizing on both W$\upmu$E.\cite{Kankare1992} Coulommetry or impedancemetry can be used in case of a passive one-electrode sensor to monitor the growth. But as the measured currents scale with deposition rates, downsizing a sensing element rapidly limits co-integration. Instead, \textbf{measured currents scale with the electrical conductance of a material with conductimetric elements}. So, a fine control of few materials can be achieved with an appropriate choice of copolymerized monomers to tune coatings' conductivity appropriately in an array.\\[3pt]
In practice, challenges to control the quality of electropolymerized elements at low scale relies strongly on the \textbf{affinity of the electropolymerized coating with solvents/electrodes} and the handling process for cleaning (Fig.\ref{fig:fig3.1.2}.k-n). Coating PEDOT from EDOT and NaPSS in water is usually quite simple (Fig.\ref{fig:fig3.1.2}.k-l). However because most oligothiophenes are not water-soluble, organic solvent must be used for electro(-co-)polymerization. Because \textbf{polymers can swell, be easily suspended and be washed away at such scale} (Fig.\ref{fig:fig3.1.2}.m), specific formulations shall be developped, considering solvents' Hansen's solubility parameters (in addition to volatility, toxicity, surface tension), low surfactant abilities of well-chosen electrolytes and polythiophene cross-linking with appropriate comonomers. It should however be mentioned that \textbf{all these factors have severe impacts on the film morphology}, hence the electro-sensitivity (Fig.\ref{fig:fig3.1.2}.n).\\[3pt]
\textbf{Very few results on these specific aspects are currently available in the literature}.

\subsection{Limitations for Quality Assessment with an Organic Electronic 'Tongue' Classifier}
\label{Ch4.1.3}

\begin{figure}
  \centering
  \includegraphics[width=1\columnwidth]{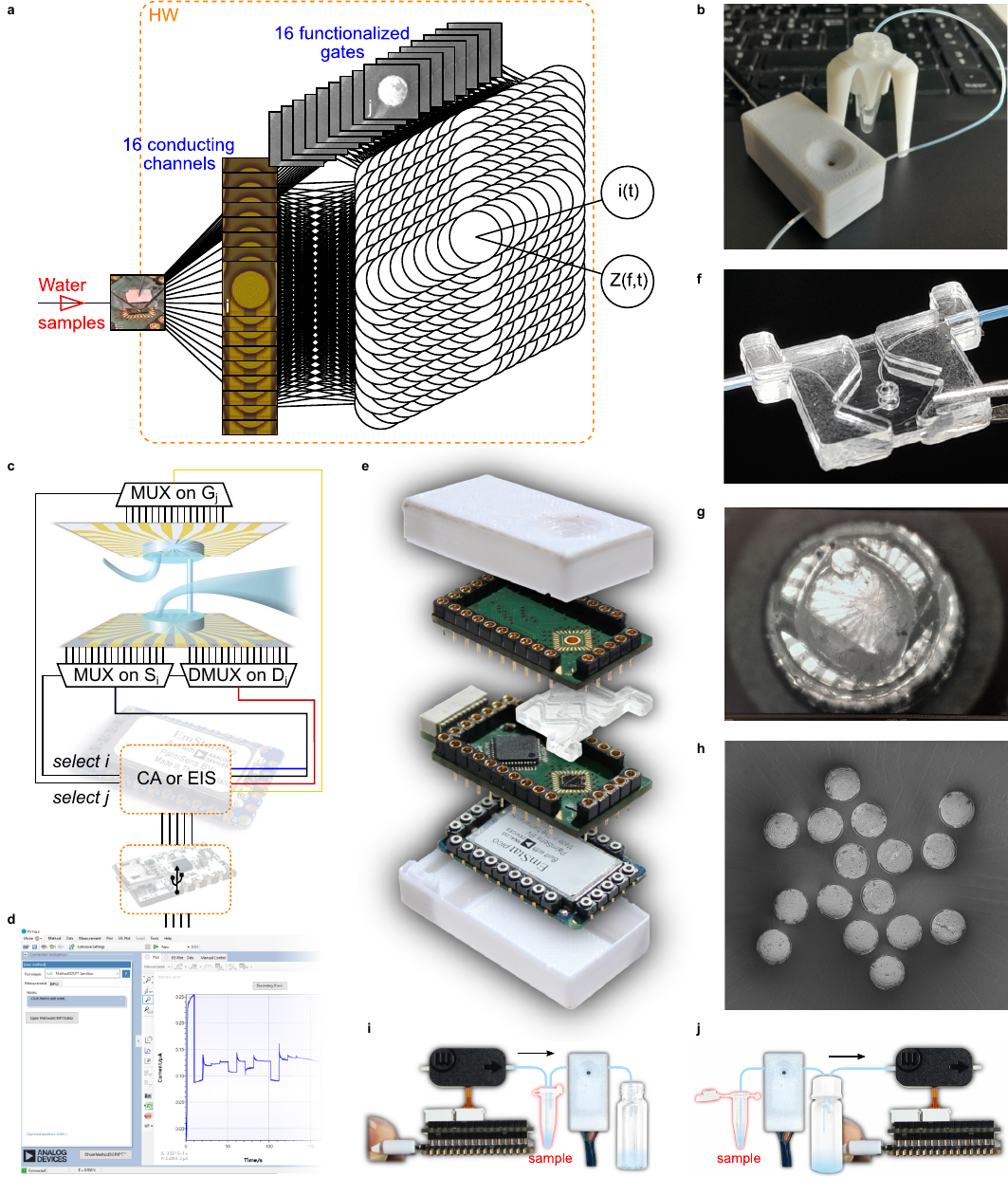}
  \caption{\textbf{16-Gates$\times$16-channels OECT analyzer $\vert$ a,} Classifier's input layer. \textbf{b,} Photograph of the functional prototype. \textbf{c,} Schematic for multiplexing on both OECT gate and channel substrates with an \textit{EmStat Pico} board. \textbf{d,} Screenshot of a sample amperometric recording multiplexed over 16 channels with \textit{PSTrace}. \textbf{e,} Exploded view of an axonometric projection of the prototype. \textbf{f,} Photograph of the spacer to expose a solution on both OECT gate and channel arrays without damaging wire bonding with the PCB shields nor wetting the electronics. \textbf{g,} Microscope image of the upper window showing the PDMS cavity. \textbf{h,} SEM image of a 16-channel array. \textbf{i,j,} "Push-in" (\textbf{i}) and "Pull-out" (\textbf{j}) configurations of a sample to the prototype with a commercial micro-pump.}
  \label{fig:fig3.1.3}
\end{figure}

Gathering all sensitive materials \textbf{on the smallest surface possible} is the key to generate chemical information which is \textbf{not altered by spatial anisotropy}, such as in colloidal solutions (with insoluble aggregates, immiscible droplets or gas bubbles) but also monophasic domain with any potential gradients (concentration, temperature, pressure). Downsizing an input array is necessary but not sufficient to establish reliable fingerprints of an environment's chemistry. \textbf{To ensure the least heterogeneity in time}, samples must be exposed in a systematic manner to a time-invariant transducer. The characteristic time to read different information from individual sensing element yields the overall duration to train a classifier. Therefore, even for static sample classification, optimizing the information accessibility in a classifier is crucial.\\[3pt]
In a conventional electronic classifier with serial communication (Fig.\ref{fig:fig3.1.3}), the input port of a readout circuit can be \textbf{demultiplexed and multiplexed} on and from the individually connected sensing elements overtime. For a conductimetric input layer, an amperometric or impedimetric Analog-Front-End (AFE) can serve as transducer (Fig.\ref{fig:fig3.1.3}.a-d). As each connection between the sensing nodes and the transducer may absorb or dissipate energy, it is required for all lines to be as short as possible in an heterogeneously co-integrated electronic hardware, and the whole system being packaged adequately (Fig.\ref{fig:fig3.1.3}.b). \textbf{The readout system shall be as close as possible to the sensing array in an integrated system} (Fig.\ref{fig:fig3.1.3}.e). To this end, the PalmSens\textsuperscript{\textregistered} \textit{EmStat Pico} is quite adequate, as it features a 1.8$\times$3-cm\textsuperscript{2} programmable dual-channel galvano/potentio/impedimetic readout under potentio/galvanostatic setup,\cite{emStatPico} using AnalogDevice\textsuperscript{\textregistered} \textit{ADuCM355} 5$\times$6-mm\textsuperscript{2} 16-bit AFE microcontroler.\cite{ADuCM355} The transducer is calibrated to measure absolute DC currents from 0.1 to 3000~$\upmu$A (5.5~pA resolution) and impedance from 10 to 10\textsuperscript{8}~$\Omega$ between 0.016 and 200~kHz (99\%-accuracy above 80~pF). The board features general purpose inputs/outputs (GPIOs) to address 5V-de/multiplexers to control the readout of different input nodes overtime. \textbf{Coupled with a de/multiplexing stage, these different specificities are compliant for reading amperometric sensing micro-arrays}. In case of an OECT array, two AnalogDevice\textsuperscript{\textregistered} \textit{ADG726} can address independently the source-drain channels from the gate electrodes (Fig.\ref{fig:fig3.1.3}.c), to generate 16$\times$16-dimensional environment fingerprint of either \textit{I}\textsubscript{SD}\textsuperscript{(i)}(\textit{V}\textsubscript{G}\textsuperscript{(j)}) or |\textit{Z}|\textsubscript{SD}\textsuperscript{(i)}(\textit{V}\textsubscript{G}\textsuperscript{(j)}), out of only 2$\times$16 materials: 16 different CP electropolymerized on the interlaced spiral elements (Fig.\ref{fig:fig3.1.3}.h) and 16 co-integrated materials on the gate electrodes (which could be electropolymerized as well).\\[3pt]
In such case where environments require long exposures to a classifier for a high-dimensional readout, samples require to be isolated from perturbations by means of a chamber and an actuator:\\[3pt]
$\bullet $ The chamber shall be sized to the sensing micro-array(s) and designed in a inert material which is not permeable to any component of the environment to be analyzed. In case of a polydimethylsiloxane (PDMS) chamber casted in a customized mold (Fig.\ref{fig:fig3.1.3}.e-g), the chamber \textbf{must be treated prior use}. In the case of fluorinated pollutants detection, chlordecone dispersed in water has a particularly high affinity with surfaces, leading to false interpretation of the class readout when contaminating a chamber. \textbf{Reducing the size of the chamber to lower the amount of potential contaminants} at a given concentration is therefore of importance: in the case of the chamber displayed in Fig.\ref{fig:fig3.1.3}.e-g (separated in two parts such as displayed in Fig.\ref{fig:fig3.1.3}.c), the volume is about 0.1 $\upmu$L, which is negligible compared to the inner volume for the tubings.\\[3pt]
$\bullet $ The actuator must dispense environments to be analyzed. For electropolymerized sensing elements, the integrity of coatings is really sensitive to shearing of the solution brought in the chamber (as dimensions are downsized, fluids' velocity is high). \textbf{Very few information about CP material stability under dynamical conditions is available in the literature}. To better control this mechanical stress, fluids' dynamics shall be controlled with actuators. If integrated to the readout system to supervise the exposure after each analysis, piezoelectric actuators are very adequate in regards to the flows and volumes,\cite{Li2022,Liu2022,Wang2019,Zhang2013} among other Micro Electro-Mechanical Systems (MEMSes) which can be used to design integrated micro-pumps.\cite{Laser2004,Wang2018,Luo2023} Commercial piezoelectric micro-pumps are available for both gas and liquids. Two different exposure modes are possible: "push-in" and "pull-out" (as depicted in Fig.\ref{fig:fig3.1.3}.i-j). While "push-in" may damage the electrical circuitry in case of leaks, "pull-out" may introduce air bubbles in the chamber, influencing greatly the readout.\\[3pt]
\textbf{Many of these aspects associated to development are not reported in the scientific literature, despite their importance to evaluate the feasibility for evaluating materials sets for chemical classification}.\\[3pt]
\textit{Stable setups need to be further developed on the long term for reliable information generation}.

\section{On Metamorphic Organic Electronic 'Noses' with Adaptive Receptive Fields}
\label{Ch4.2}

Electropolymerization is powerful to co-integrate many materials (Sec.\ref{Ch4.1}). But integrating useless materials limits the classification. Increasing sensing array's dimensionality increases superlinearly calibration's complexity: In the best case when all materials project orthogonally information, finding eigenvectors for the simplest linear classifiers requires one matrix inversion, which itself has between $O(n^{2})$ and $O(n^{3})$ computational complexity.\cite{Pan1987,Petkovic2009} So, calibrating n times a m-dimensional classifier is systematically simpler than calibrating one single n$\times$m-dimensional "super-classifier". Therefore, each classifier shall focus on what they are trained for to be quickly calibrated and be designed accordingly to their future experience. This implies that \textbf{sensitive arrays should be customized during operation, out of the factory}, with the least number of sensing elements for specialized usage. Such customization shall depend either on the number and nature of classes an end-user wishes to classify and on how many and what kind of environments. This is opposed to the conventional electronic manufacturing practices where resources (information generators, processors and memory) are oversized \textit{a priori}, mass-produced and standardized.\\[3pt]
In this section is detailed how CP cross-sensitivities can be exploited (Subs.\ref{Ch4.2.1}) with electropolymerization on a circuit board (Subs.\ref{Ch4.2.2}) to supervise "sparser but better" manufacturing while sensing (Subs.\ref{Ch4.2.3}).\\

\textit{This section is mainly associated to information from the cited state-of-the-art literature \& early results of L. Routier, S. K. Koudjou, A. K. Taka, M. Delarue and S. Pecqueur.}\\

\subsection{Benchmarking Multiple Materials' Complementarity on a same Sensing Arrays}
\label{Ch4.2.1}

\begin{figure}
  \centering
  \includegraphics[width=1\columnwidth]{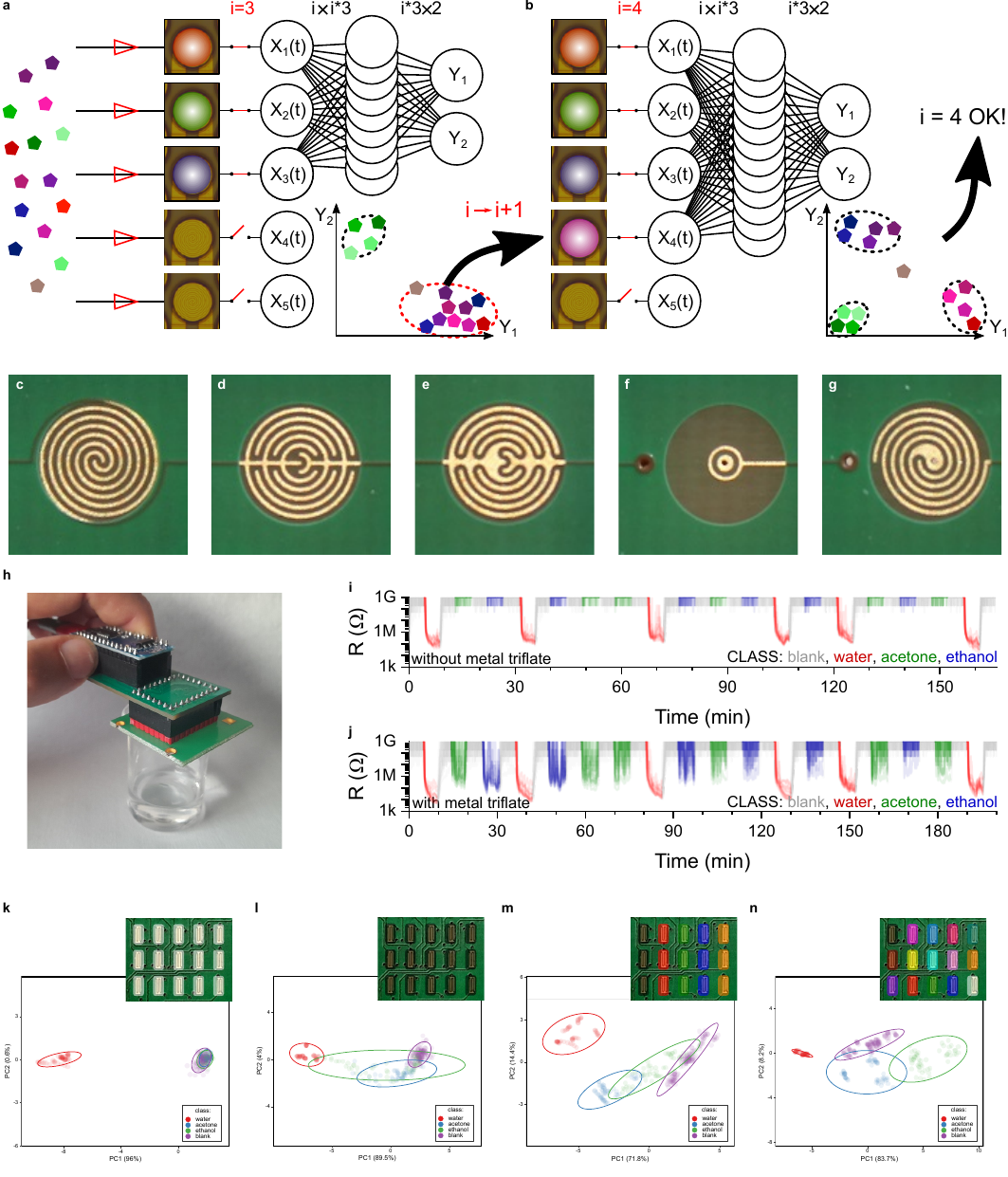}
  \caption{\textbf{Materials' Complementarity for Classification $\vert$ a-b,} Scheme of a binary classifier with i~=~3 input nodes (\textbf{a}) evolving to i~=~4 for better classification (\textbf{b}), at a higher computational expense. \textbf{c-g,} Five different designs of input nodes on a FR4-PCB: With the highest channel width over length (\textit{W}/\textit{L}) (\textbf{c}), with a minimized \textit{W} (\textbf{d}), with a uniform \textit{L} (\textbf{e}), with a minimized size as a via (\textbf{f}) and with the largest inductive reactance (\textbf{g}). \textbf{h,} Photograph of static sampling volatile analytes to an \textit{ATMega168p}-controlled \textit{AD5933} analyzer featuring a multiplexing/demultiplexing stage over 15 individual input nodes on a FR4 substrate. \textbf{i,j,} Dynamical resistance acquisitions with the 15-dimensional prototype: coating of the input nodes with hygroscopic polythiophene without (\textbf{i}) and with doping (\textbf{j}) with a metal triflate salt. \textbf{k-n,} PCA performed after exposing the same sequence of samples in the same experimental conditions and database size and sampling conditions for 15 input nodes of a polythiophene without triflates (\textbf{k}), with the same triflate over the 15 nodes (\textbf{l}), five different triflates (\textbf{m}) and 14 different triflate plus one node without triflate (\textbf{n}), demonstrating different contributions of the dopants on a same CP.}
  \label{fig:fig3.2.1}
\end{figure}

Complexity for defining what is 'useless' as a material comes with the classification itself, as the lack of collective contribution in a set of materials to discriminate multiple environments. Therefore, it is no intrinsic property of materials, but an interdependent property between them to sense differences in their fingerprints. Materials can have identical abilities to detect a pattern but different fields of perception to sense other ones.
In classifications by variance using PCA, certifying the suitability of doped polymer sets comes only after generating large datasets, and choosing to fabricate them \textit{a fortiori}.\cite{Boujnah2021} Knowledge of an appropriate set of only two materials out of eight to recognize acetone from ethanol from water requires testing all eight materials exposed to all three chemicals, repeatedly for a large enough number of time.\cite{Boujnah2021} If target classes are well-defined and require specifically selective materials, a manufacturing strategy can be envisioned. Instead, if classifications are not determined \textit{a priori}, severe compromise in their complementarity must be found \textbf{to cover the broadest perception field} with no redundant materials (Fig.\ref{fig:fig3.2.1}.a-b). \textbf{To investigate on large sets of materials and environments, high-throughput methods must be identified}.\\[3pt]
$\bullet $ The first concern is \textbf{choosing the nature of sensitive materials to screen}. To compose \textit{a priori} a superset of materials to test, stable materials shall be prioritized. Materials' sensitivity can be tuned by their chemistry or their physics, in an uncontrolled way or by design. In the case of CPs, chemistry can be designed at the level of the dopants and the monomeric units. While varieties of monomer structures are available (Fig.\ref{fig:fig3.1.1}), dopants also impact strongly the materials' response and can be tuned in different ways.\cite{Pecqueur2014} The selection of materials composing a superset shall be based on an \textit{a priori} understanding of a transduction principle,\cite{Ferchichi2020} where testing promotes furthermore understanding on its interaction with specific classes.\cite{Boujnah2021} The total number of materials of a superset is defined by the complexity of the classification: They can be more than the number of classes to recognize, but few enough in regard to the time necessary to train the classifier in recognizing each class exposed several times than the number of classes to recognize. A drift on a physical property of a single material can greatly affect the information features input for classification.\\[3pt]
$\bullet $ The second concern is the \textbf{availability and cost of materials, and in particular, the substrate of electrodes} to evaluate complementarity of sensitive materials. As a solution, commercial Printed Circuit Boards (PCBs) of ENIG-coated (Electroless nickel immersion gold) copper contacts are effective, on both on epoxy/fiber-glass composite (FR-4) and polyimide substrates. If ENIG is proposed as an anti-corrosion treatment, it may also serve as non-redox active working electrodes for electropolymerization. 

However, it shall be considered that current commercial standards are set to 5~mil (127~$\upmu$m) as space feature to pattern electrode channels (Fig.\ref{fig:fig3.2.1}.c-g): Highly conductive CP coatings must be prioritize in regards to the resulting resistance which can be measured.\\[3pt]
$\bullet $ The third concern is on the \textbf{systematicity of the multi-material evaluation. Large dataset per materials must be generated}. Measuring them one after the other may generate systematic variabilities which are not to be attributed to the classes to recognize. In order to not over-interpret the different materials' complementarity, materials shall be screened simultaneously via multiplexing on a sensing array exposed to a single sample (Fig.\ref{fig:fig3.2.1}.h-n). This also requires to prioritize multi-material co-integration methods which preserve the integrity of all materials deposited in an array.\\[3pt]
$\bullet $ The fourth concern is on the generic aspect of multi-environment screening, requiring to anticipate measurements of more complex samples. One can differentiate applications where environments are sampled in packages (products quality assessment) from transient and open environments (natural condition air/water quality assessment). Materials complementarity must be evaluated in conditions which are \textbf{as close as possible to the environmental conditions a sample will be presented to an end-user}. Concerns about how the sample is exposed (dynamically or statically) in a stable manner (non-invasivness and autonomy) must be thought ahead prior multi-material selection (Fig.\ref{fig:fig3.2.1}.h-n).

\subsection{\textit{in Operando} Material Co-integration on a Single Board / with a Single Chip}
\label{Ch4.2.2}

\begin{figure}
  \centering
  \includegraphics[width=1\columnwidth]{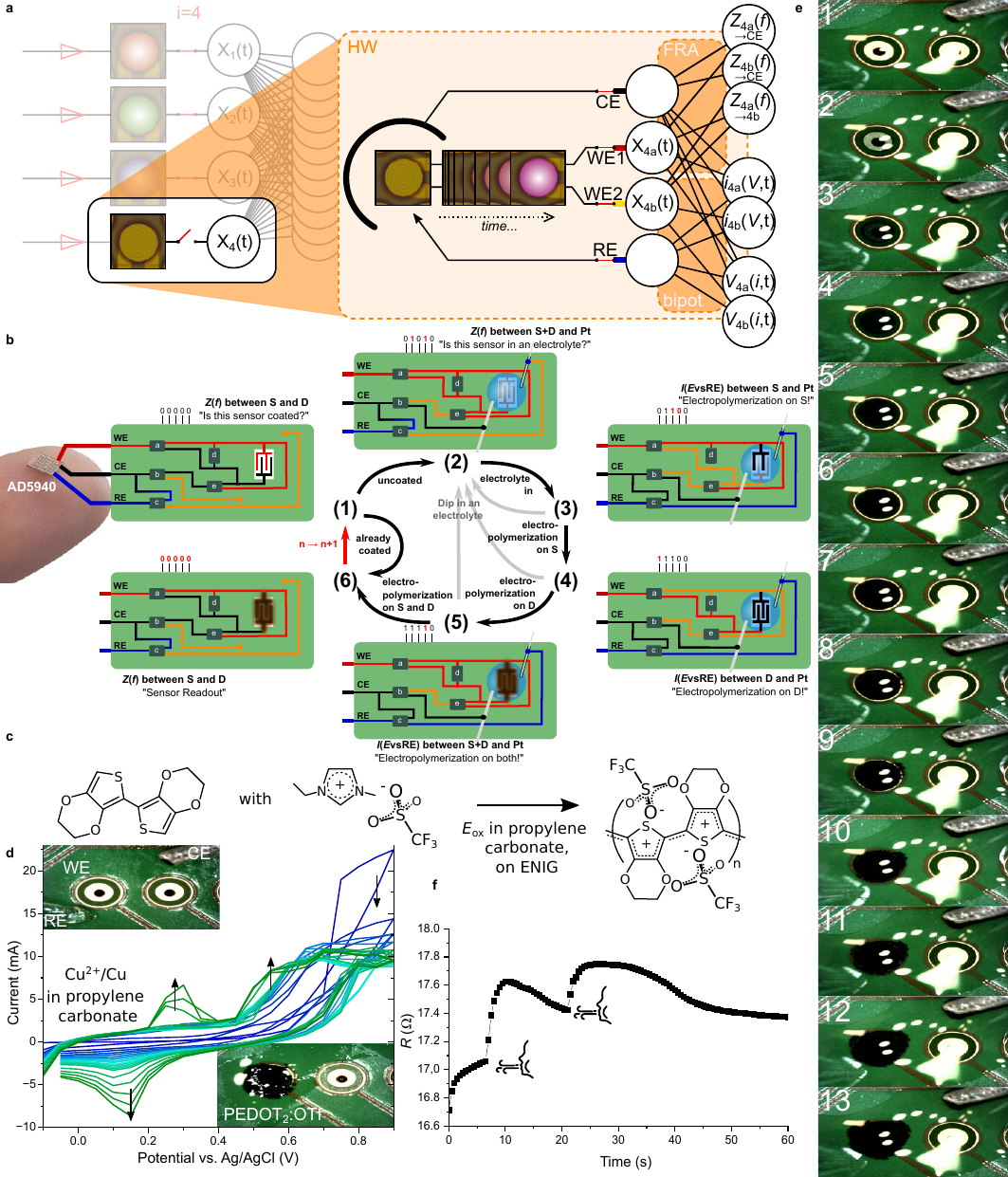}
  \caption{\textbf{Electrochemical Manufacturing on a Circuit Board $\vert$ a,} Zoom-in on a single adaptive input node featuring four electrodes projecting information on a large dimension via a bipotentio/bigalvanostat and an electrochemical impedance spectrometer. \textbf{b,} 6-step procedure for supervising a sensing-element coating, starting from verifying the conductive state of an element (step 1), ensuring the interfacing with an electrolyte solution (step 2), electroplating on each contact individually and simultaneously (steps 3 to 5) and assessing the sensing element's response under different conditions (step 6). \textbf{c,} A reaction to electroplate a polythiophene from a small drop on a commercial circuit board without corroding the contacts. \textbf{d,} Cyclic voltammetry of EDOT\textsubscript{2} in a 100~mM EmimOTf electrolyte with propylene carbonate as solvent, using an ENIG-coated two-electrode pad as working electrode, a Pt wire as counter electrode and an Ag/AgCl wire as reference electrode (as inset, photographs of the dried system before and after the cyclic voltammetry. \textbf{e,} Photographs of the system after multiple cycles from {\textminus}0.1~V to 0.9~V (50~mV/s). \textbf{f,} Resistance response of the dried two-electrode pad coated with electrodeposited PEDOT\textsubscript{2}:OTf, biased with 10~mV, and resistance transient observed when mouth blowing over the board during the measurement.}
  \label{fig:fig3.2.2}
\end{figure}

As material selection depends on user-defined classifications, their integration during a device use is examined (Fig.\ref{fig:fig3.2.2}.a). Electropolymerizing on electrical systems is an elegant way to imprint multiple sensitivities in a circuit. Two scales shall be considered: \textbf{on a board and on a chip co-integration}. For material-screening, on board co-integration is crucial to minimize development duration and costs.\\[3pt]
Electroplating directly on commercially available PCBs must consider contacts' electroactivity. At the opposite of polyanilines, polypyrroles and polythiophenes are anodically deposited: non-noble metals like copper or nickel can be \textbf{corroded}. A good quality ENIG finish on the contacts is therefore required for electropolymerization. Monomers must be adequately chosen to lower the applied potential. EDOT oxidizes at about 1.3~V in aprotic solvents and about 0.8~V in water with a potentiostat (Fig.\ref{fig:fig3.1.1}.i). Its dimer is better to oxidize at lower voltage, able to not corrode even iron.\cite{Akoudad1998}.\\[3pt]
Such low-potentials for material manufacturing are compliant with low voltage electronics without involving step-up voltage regulators, ensuring a \textbf{high control of the growth kinetics}. Aspects related to energy-consumption can be dealt simply by down-scaling sensing elements (Subs.\ref{Ch4.1.2}). For a board co-integration level, \textbf{remarkable system size reduction} can be achieved thanks to the current advances in integrated circuits (ICs) developments dedicated to electrochemical sensing. For instance, researches with miniaturized electrical transducers involve AnalogDevice\textsuperscript{\textregistered}'s \textit{AD5940/1} as a potentio/galvanostat,\cite{Yu2022,Kim2022,Bill2023,Wu2024} as an impedance spectrometer,\cite{Asiain2022,Tasneem2022} or as both.\cite{Min2023,Tang2023}. Other IC can be exploited for electrochemistry on a chip when impedance spectrometry is not required, like Texas Instrument\textsuperscript{\textregistered}'s \textit{LMP91000}.\cite{LMP91000,Hoilett2020,Lee2022} AnalogDevice\textsuperscript{\textregistered} \textit{ADuCM355} features also an \textit{ARM Cortex-M3} microcontroller with the \textit{AD5940/1}'s core and an AFE for bipotentiostat control, all on a 5$\times$6 mm\textsuperscript{2} System on Chip (SoC).\cite{ADuCM355,Manoharan2023} Used with a sensing array, prototype miniaturization is strongly limited by the footprint of the analog multiplexers/demultiplexers, for which currently available shelf electronics are limited to dual 16-channels on quite a larger size. Unless comprising active-matrix addressing of the sensing elements,\cite{Oh2019,Liu2020,Bao2022} restricting studies to \textbf{i~=~16} sensing elements is required for multi-material investigations on small classifiers.\\[3pt]
All these specifications are compliant with digitally-automated supervision for multi-material co-integration on a board. For each specifically-addressed uncoated element, several steps can be sequenced to modify sensing elements with CPs using either a impedance spectrometer, a potentio/galvanostat or a bipotentio/bigalvanostat (Fig.\ref{fig:fig3.2.2}.b): A spectrometer could control coatings' initiation or abortion by inspecting impedance between contacts, or control the interfacing with an electrolyte by the impedance through a counter-electrode. A potentiostat can ensure electrodepositing the material on each contact individually or both of them simultaneously. A biopotentiostat could even monitor sensing element's conductance \textit{in operando} and \textbf{aboard deposition at a stage when the channel is just about to short or reach a specific resistance}.

This feature can greatly control the sensing qualities of conductimetric elements.\\[3pt]
A major challenge for electropolymerizing on a board relies on the monomers availability, supplied to a large sensing-element channel by an electrolyte solution drop. This requires using oligothiophenes which have a particularly \textbf{high solubility} (Fig.\ref{fig:fig3.2.2}.c). With a drop of 425~mM of EDOT\textsubscript{2} in propylene carbonate, it is possible to short a via-sized sensing element and visualize the impact of the coating under cyclic voltammetry, but only after tenth of minutes (Fig.\ref{fig:fig3.2.2}.d-e). The fact that such coating leads to 10-20~$\Upomega$ conductance despite the size channel and appears to be sensitive to an aerial environment (Fig.\ref{fig:fig3.2.2}.d-e) is an encouraging perspective to pursue electropolymerization on a board.

\subsection{Closed-Loop \& Mass-Flows upon Supervising Electrolyte Solution \& Analyte Exposures}
\label{Ch4.2.3}

\begin{figure}
  \centering
  \includegraphics[width=1\columnwidth]{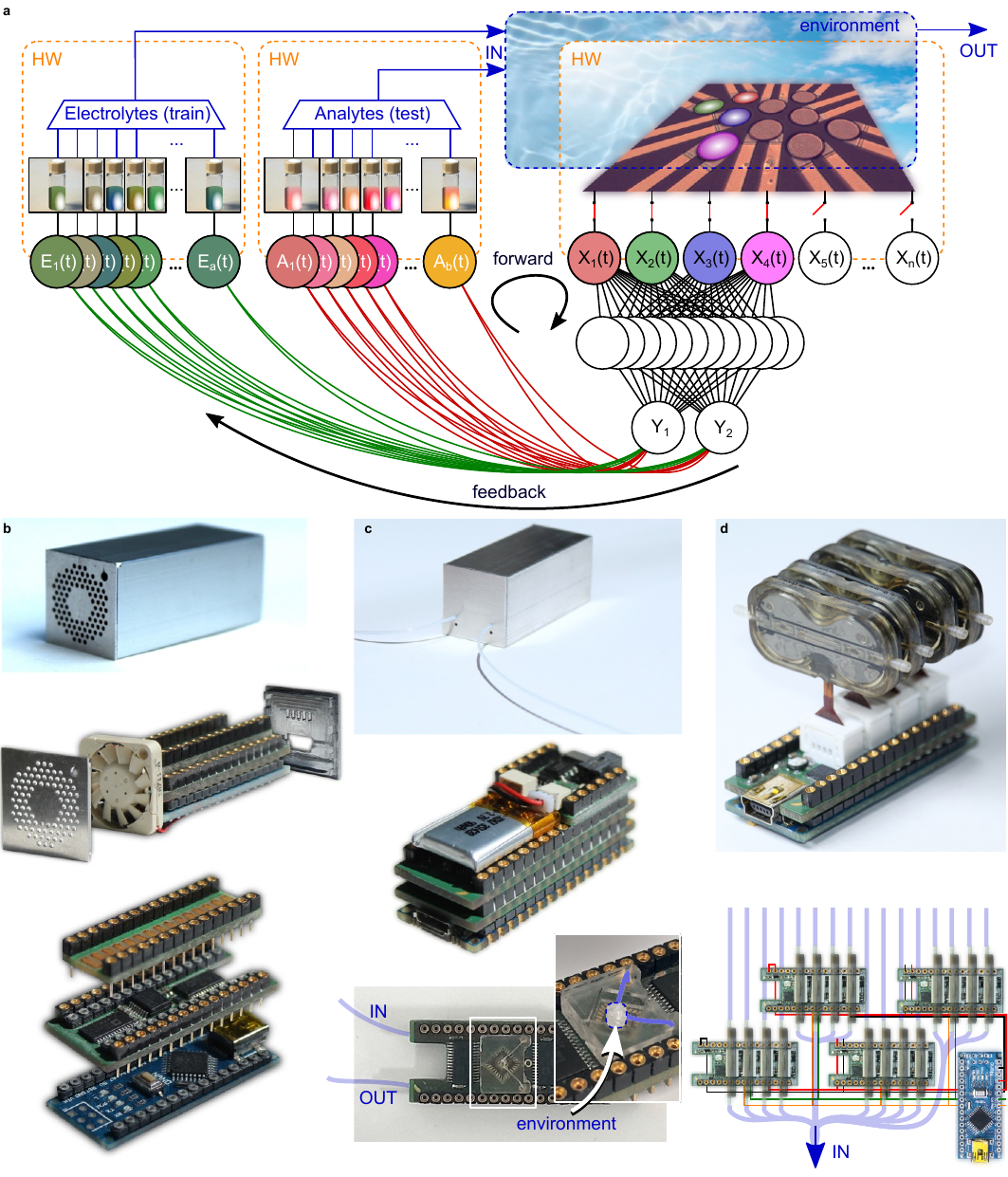}
  \caption{\textbf{Material Supply Included in a Classifier $\vert$ a,} Scheme of a classifier supervising its exposure to different electrolyte and analyte environments based on its recognition with a specific receptive field. \textbf{b,} Photograph and exploded view of axonometric projections of an electronic nose prototype featuring an on board fan. This electronic nose prototype is not adapted for an ARF. \textbf{c,} Photograph of an electronic tongue prototype featuring common inlet and outlet for electrolyte and analyte exposures over a sensing array in a PDMS molded cavity. This electronic tongue prototype can feature an ARF if communicating with electronics supervising exposures. \textbf{d,} Photograph of a prototype for four co-integrated Bartels' \textit{MP6} piezoelectric micro-pumps adapted from Bartels' \textit{mp-Highdriver4} driver, and schematics for controlling four of them with one single microcontroler to pilot 16 micro-pumps. This prototype is no standalone sensing system, but can complete a sensing classifier to feature an ARF.}
  \label{fig:fig3.2.3}
\end{figure}

In an electropolymerization process on sensing elements with different monomers, managing numerous electroactive solutions shall be accounted. Considering also various samples for assessing recognition, controlling the environment transience for adaptive receptive field (ARF) classifiers should be clearly defined, \textbf{considering the electrolyte and analyte exposing system as full part of the classifier} (Fig.\ref{fig:fig3.2.3}.a). Explainable classifiers can quantify the relevance of different materials co-integrated together, \textit{a posteriori} from iteratively exposing uncoated elements to \textbf{different electrolyte solutions} and assessing their response with \textbf{different analytes}. Depending on the recognition performance, \textbf{selected electrolyte solutions} could be preferred to reinforce the coverage of an array perception field towards specific features, and \textbf{specific analytes} could be tested more (or less) frequently according to the difficulty (or ease) to discriminate specific classes in this perceptive field. In case environment exposures are fully supervised by the classifier with digital control of temporal multiplexing for both electrolyte and analyte sample collections (Fig.\ref{fig:fig3.2.3}.a), the material supply to the classifier shall be accounted for the functionality of the system, on the same level as computational resources and energetic expenses. Major concerns lie on \textbf{how can environments be exposed and of what kind can they be}.\\[3pt]
The way environments are exposed to a sensing classifier depends on their nature as \textbf{fluids}:\\[3pt]
$\bullet $ \textbf{ARF with gases} is more complex to implement than for liquids: electropolymerization does not occur in gas phase and electrolytes need to be supported in a condensed phase medium in standard conditions (no plasma). Therefore, iterating cycles of training and testing would impose several cycles of dewetting on a sensing layer, threatening electropolymerized materials' stability. \textit{Adaptive receptive field on electronic noses are more a long-term exploration}. Gas-phase samples are however less invasive than liquids. Also, pumping is not necessarily required in the case of gases, and miniaturized fans can control sample flows without dedicated AFE, only by pulse width modulation (PWM) with a single digital port from the microcontroler (Fig.\ref{fig:fig3.2.3}.b). This setup is particularly suitable for air quality assessment, where the rotation speed of the fan controlled via PWM modulation could further increase the sensing dimensionality of the classifier, and measure modulations of resistance/conductance of CPs under different dynamical regimes.\cite{HajAmmar2023}\\[3pt]
$\bullet $ \textbf{ARF with liquids} would not require drying CPs repeatedly and challenge less the stability of the electropolymerized materials. Exposing simultaneously electrolytes and analytes could also be performed with a common inlet (Fig.\ref{fig:fig3.2.3}.c), which can be beneficial in case electrolytes must be prepared with tested analytes, such as such for molecularly imprinted polymers.\cite{Sharma2012,Cieplak2016,Suriyanarayanan2017,Georgescu2018,MoreiraGoncalves2021,ELSharif2022,Seguro2022,Dykstra2022,Zidaric2023}
In such configuration, a single cartridge of electrolyte would be required in conjunction with multiple analytes to diversify the sensing nature of each element. Otherwise, the flow through multiple cartridges of electrolytes would be independently multiplexed with the one through multiple cartridges of analytes. As the system scales up with the number of electrolyte and analyte environments to expose, it is required to anticipate optimizing a minimum volume of fluids to expose, particularly for each sample to be exposed multiple times.\cite{Boujnah2021} \textbf{Using sensing element gathered in an array on the smallest area possible is therefore a prerequisite}. The pumping system must be scaled with the sensitive layer to supply: different microfluidics prototyping technologies are commercially available for multiplexing fluids, such as Bartels' \textit{MP6} piezoelectric micro-pumps, compliant with both air and water, and for which 16 of them can be multiplexed with four LED drivers by a single microcontroler (Fig.\ref{fig:fig3.2.3}.d).\\[3pt]
\textit{Long-term investigations on classifiers' ARF will require a high level of interdisciplinarity and focus on specific challenges on environmental recognition.}

\section{On Metamorphic Organic Electronic Networks Self-Calibrating Sensing Arrays}
\label{Ch4.3}

Gathering the right sensing elements by electropolymerizing useful materials is crucial for classification (Sec.\ref{Ch4.2}). However, too large numbers of sensitive materials may be required for fine recognition tasks. Instead of addressing specifically chosen CP, element by element in an array, a different approach could be to co-integrate electrically-loose sensing materials and \textbf{let systems connecting only the most relevant elements to the readout circuit}.\\[3pt]
In the following, the implementation of CPD morphogenesis is studied to this aim: In particular, under what framework could it interface conventional electrical systems (Subs.\ref{Ch4.3.1}), how parallelizing many growths at small scale (Subs.\ref{Ch4.3.2}) and how designing proper electrolyte media to do it (Subs.\ref{Ch4.3.3}).\\

\textit{This section is mainly associated to information from the cited state-of-the-art literature \& early results of A. Baron, C. Scholaert, D. Gu{\'e}rin, E. Kowalewska, L. Brulin and S. Pecqueur.}\\

\subsection{Framework for Integrating Evolutionary Chips in Electrical Systems}
\label{Ch4.3.1}

\begin{figure}
  \centering
  \includegraphics[width=1\columnwidth]{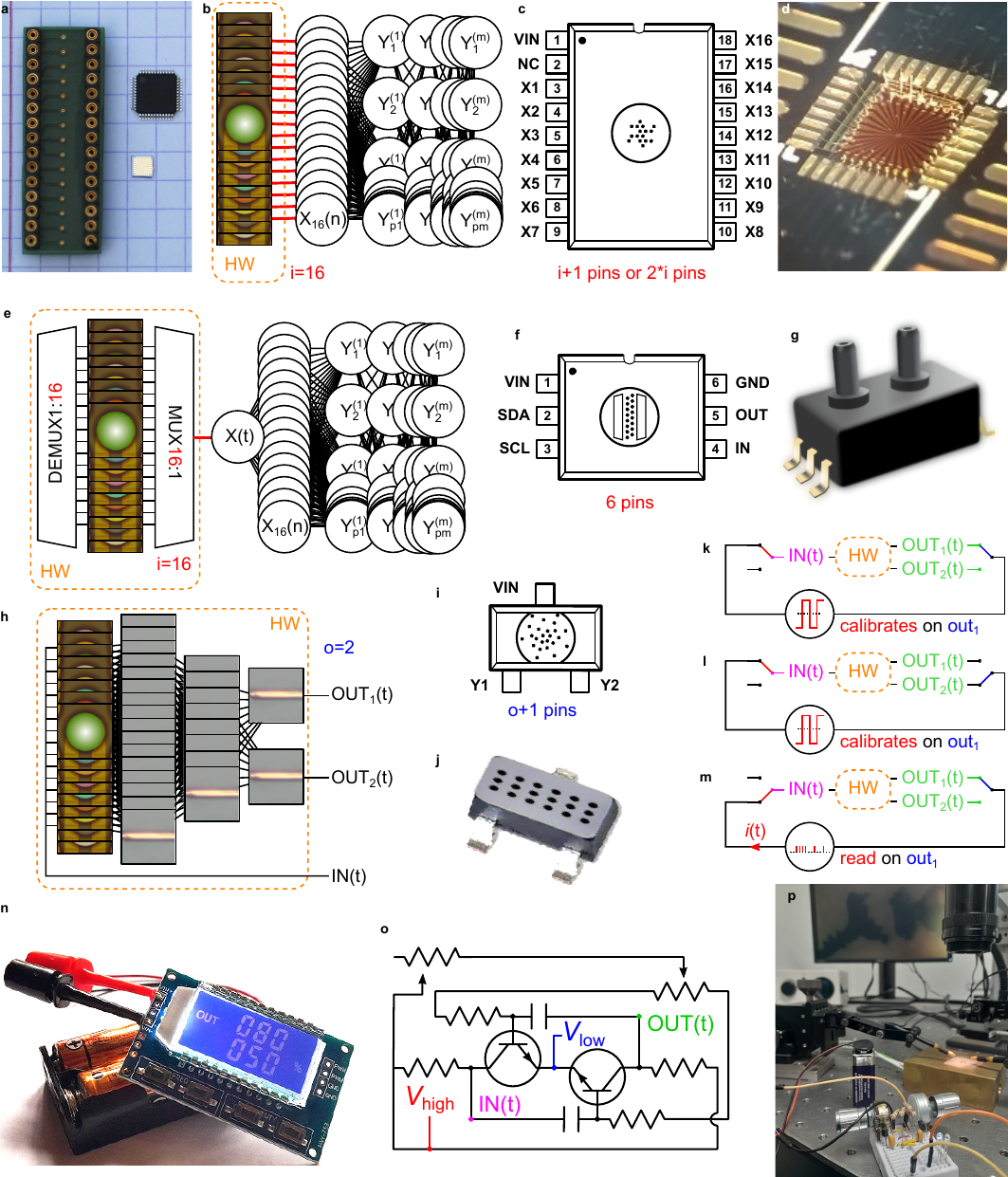}
  \caption{\textbf{Systems to Program Metamorphic Classifiers $\vert$ a,} Photograph of 15 sensing elements on a PCB, a 48-lead package of a dual 1:16 analog multiplexer, and a Si die of 16 sensing elements. \textbf{b,} Classifier structure of an analog multi-sensing hardware parallelized to a software classifier. \textbf{c,} DIP-18 package for the classifier depicted in Fig.\ref{fig:fig3.3.1}.a. \textbf{d,} Photograph of a 16 sensing element die wire-bonded on a PCB. \textbf{e,} Digital entanglement on physical classifiers to parallelize information. \textbf{f,} DIP-6 package for the classifier depicted in Fig.\ref{fig:fig3.3.1}.e. \textbf{g,} Surface Mounted Devce (SMD) package (based on the XFDM-050KPDSR pressure sensor) to embed the classifier depicted in Fig.\ref{fig:fig3.3.1}.e. \textbf{h,} Evolving classifier featuring sensing and information processing resources merged on the same layered hardware. \textbf{i,} SOT23 package for the classifier depicted in Fig.\ref{fig:fig3.3.1}.h. \textbf{g,} SMD package (based on the MICS-5524 sensor and a classical SOT23 package) to embed the classifier depicted in Fig.\ref{fig:fig3.3.1}.h. \textbf{k-m,} Electrical setup to calibrate/read Fig.\ref{fig:fig3.3.1}.k's classifier. \textbf{n,} Example of low-cost PWM to enable CPD morphogenesis. \textbf{o,} Adaptation of an astable multivibrator as low-complexity waveform generator to enable CPD morphogenesis. \textbf{p,} CPD morphogenesis with an astable multivibrator supplied with a 9~V battery. }
  \label{fig:fig3.3.1}
\end{figure}

In conventional electronics, analog sensors are standalone without computational nor memory resources. Therefore, arrays of them are big and require many connections (Fig.\ref{fig:fig3.3.1}.a-d). Readout systems must be specifically designed according to them, but their integration is always limited by the capability to address the pins of all sensing elements to address them individually. If manufactured on CMOS, addressing could be made on the same substrate, like for charge-coupled devices (Fig.\ref{fig:fig3.3.1}.e-g). The footprint is much smaller, however without computational nor memory resources, data must be sent serial to a software classifier. This limits information processing on hardware. They are also expensive to manufacture and incompatible with development at a material screening phase to integrate several hundreds of different sensitive materials.\cite{Niimura2006}\\[3pt]
To reduce sensing components' footprint while not limiting parallelism with a serial bus, \textbf{the full classifier must be packaged with the sensing array}. To reduce development time, both parts shall be made with compatible materials \& processes. Electropolymerization is therefore ideal to co-integrate multiple sensitive materials at small scale and program their interconnectivity with a readout system via CPD morphogenesis (Fig.\ref{fig:fig3.3.1}.h-j). Diminishing systems' complexity by not involving external CMOS devices, diminishing the number of connections between components and shorting contact lines to lithograph shall greatly diminish costs. It also diminishes development time to feature a full classifier on each array, instead of testing material sets, interchangeably plugged with a readout circuits (as previously in Fig.\ref{fig:fig3.2.3}.b-c). If all material resources required for morphogenesis are in the package with the sensing array, the chip requires only low voltage to specifically program output pins and read them (Fig.\ref{fig:fig3.3.1}.k-m). A transient voltage is required to map dendritic paths of CPDs between resistive sensing elements and output pins to monitor admittance differences between output pins with the input after engraving a network of CPDs. In this approach, information related to calibration is stored locally in hardware without saving individual training vectors accessible in a normative database on a local register nor on an external server. Classes strictly depend on what environments end-users define manually by the environments to discriminate, without setting semantic labels as training vector metadata for supervised learning. \textbf{It is a strategy which ensures a total independency for end-users from external resources and guaranties their personal data privacy.}\\[3pt]
\textbf{Analog systems to interface such device could be of an extreme simplicity to program CPD growths and read admittance with periodical voltage signals}: PWM can be generated with digital microcontrolers or specifically dedicated network analyzers,\cite{ADuCM355,AD5933} low-cost boards directly supplied with AA batteries (Fig.\ref{fig:fig3.3.1}.n), simpler analog circuits (such as NE555) or even much simpler circuits like astable multivibrators (Fig.\ref{fig:fig3.3.1}.o). Although it may affect the growth, signal non-ideality in regard to perfect voltage impulses is no limitation for morphogenesis (Fig.\ref{fig:fig3.3.1}.p): such approach could really be adapted to \textbf{embed information processing resources near sensors (Fig.\ref{Ch4.3.2}) with low energy/material costs in minimalist electric system designs}.

\subsection{Co-Integrating Dendrites on \& in an Array}
\label{Ch4.3.2}

\begin{figure}
  \centering
  \includegraphics[width=1\columnwidth]{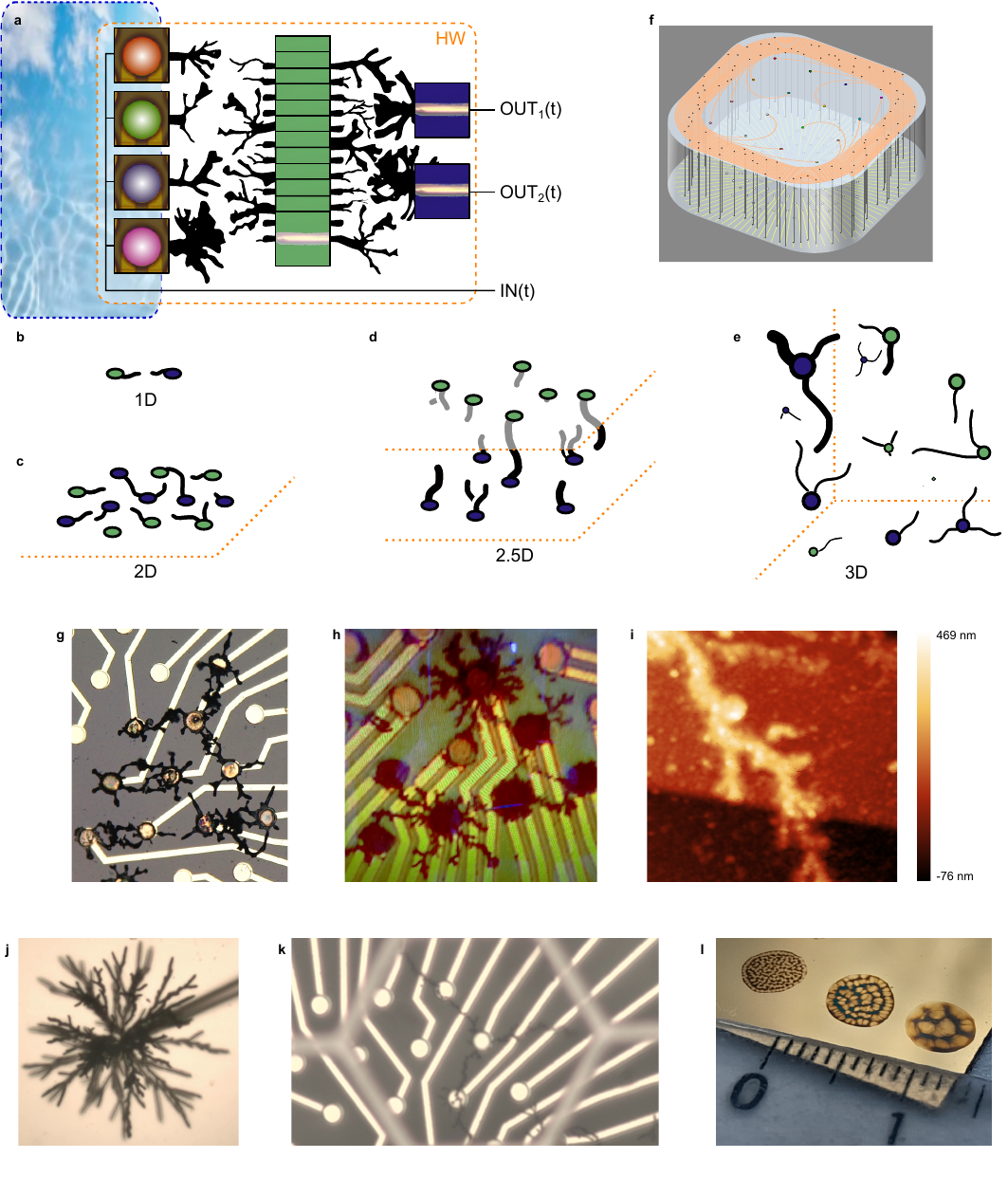}
  \caption{\textbf{Structuring networks of dendrites $\vert$ a,} Physical classifier using CPDs' admittance between nodes to program a sensing array interconnectivity with a readout circuit. \textbf{b,} Schematics of the free-standing gold wire setup (no co-integration) \textbf{c-e,} Perspective schematics of three topological categories of CPDs to co-integrate them in the bulk: coplanar (\textbf{c}), layered (\textbf{d}) and dispersed (\textbf{e}). \textbf{f,} Example of spatially-deconvolutive input layer for a 2.5D topology, featuring 16 different materials at the top (exposed to an outer environment) projecting information towards a readout layer on 96 positions, optimized to lower spatial anisotropy. \textbf{g,} Optical microscope picture of a network of oriented CPDs on a micro-electrode array (diameter: 30~$\upmu$m - from Scholaert \textit{et al.})\cite{Scholaert2024}. \textbf{h,} Optical microscope picture of a network of 2D isotropic growths on an array of conductimetric elements (diameter: 28~$\upmu$m). \textbf{i,} AFM image of a dendritic topology on Parylene C (area: 15$\times$15~$\upmu$m\textsuperscript{2}). \textbf{j,} Optical microscope picture of a 3D isotropic growth on a gold wire. \textbf{k,} Optical microscope picture of oriented CP filaments on a micro-electrode array (diameter: 30~$\upmu$m), visualized through a honey-comb structured counter electrode on glass. \textbf{l,} Photograph of disordered CP two dimensional patterns on a gold surface.}
  \label{fig:fig3.3.2}
\end{figure}

The ability to use CPD morphogenesis as calibration mechanism \textit{in materio} on sensing arrays depends on the control of their growth by voltage activation from individually polarizable contacts. A global electrochemical interconnection ensures the signal propagation at all points of a network when supplied by pulse voltages from a sensitive layer (exposed to an outer environment) to a readout layer at the other end of the network (deep down in the circuit), without having to individually address the inner floating points of the network (Fig.\ref{fig:fig3.3.2}.a). In such a topology, \textbf{CPD morphogenesis embeds non-volatile memory in a liquid-state machine} (in a sense of a spike-voltage fed reservoir computer)\cite{Maass2004,Lukosevicius2009}.\\[3pt]
To physically implement such classifiers, four configurations to grow CPDs are considered (Fig.\ref{fig:fig3.3.2}.b-e):\\
$\bullet $ In a \textbf{1D configuration} (Fig.\ref{fig:fig3.3.2}.b), one input connects one output at the fashion of formerly presented results on two \textbf{free-standing gold wires} to avoid any interaction with a surface (Sec.\ref{Ch3.2}). If different networking can be expected in such a configuration,\cite{Janzakova2023} this setup is unsuitable for interconnecting an array of sensing elements with a readout circuit.\\
$\bullet $ In a \textbf{3D configuration} (Fig.\ref{fig:fig3.3.2}.e), a \textbf{dispersion} of electrically conductive particles in an electroactive electrolyte allows working in a phase where growth is not influenced by any interaction with a surface, such as on free-standing wires. Despite this structure has great advantages in term of physical implementation by its simplicity, the lack of control of the spatial dispersion of different floating points in a 3D volume requires a deeper understanding of CPD morphogenesis in colloids (which is not reported yet in the literature with CPs).\\ 
$\bullet $ In a \textbf{2D configuration} (Fig.\ref{fig:fig3.3.2}.c), \textbf{coplanar} electrodes populate a two-dimensional substrate as both input and output nodes. Different studies used such setup to integrate various CP topologies on a lithographically patterned array of electrodes.\cite{Petrauskas2021,Cucchi2021,Cucchi2021a} It is a rather convenient implementation when sensitive materials are located on another substrate, to be hard-wired to the input nodes of a patterned electrode substrate, with a drop of electroactive solution deposited over the array of nodes. The electrogenerated topologies are highly two dimensional due to their affinity with the substrate.\cite{Watanabe2018} They can however show various topologies: highly oriented between polarized input and output nodes (Fig.\ref{fig:fig3.3.2}.g)\cite{Scholaert2024} or isotropic around anodic nodes (Fig.\ref{fig:fig3.3.2}.h). Studying CPDs on a substrate however offers the advantage to characterize them more easily, where AFM allows observing these objects typically below a micrometer in diameter without collapsing when transferred in air (Fig.\ref{fig:fig3.3.2}.i). The main disadvantage of confining the input and output nodes in a two dimensional plane is the limited parallelism.\\
$\bullet $ In a \textbf{2.5D configuration} (Fig.\ref{fig:fig3.3.2}.d), two-dimensional \textbf{layers} of electrodes are electrochemically interconnected in a \textbf{stack}, analogically to layered software architectures of feed-forward neural networks. A stack of electrode arrays offers both the advantage of three-dimensional interconnectivity layer-to-layer and control of nodes' spatial dispersion on each layer. An input layer of conductimetric sensing elements at the top layer exposed to an environment (Fig.\ref{fig:fig3.3.2}.a) can project its signal deeper down to the next layer. A first prerequisite for such input layer is to spatially deconvolute the different sensing nodes for the morphogenesis to not be conditioned by the materials arrangement on the input layer surface (Fig.\ref{fig:fig3.3.2}.f). The ability for different interconnections to grow shall depend on the electric signal and the physical property of electrodes separated by an electrolyte solution on two vicinal layers: dimensions, distances, roughness, material. To interconnect layers, out-of-plane growths must be made (Fig.\ref{fig:fig3.3.2}.j), with the CPD growths to be sensitive to the topology of the facing layer to interconnect (Fig.\ref{fig:fig3.3.2}.k).\\[3pt]
Stacks of several layers under the sensing array offers a great perspective to embed computing and memory resources on the same area as the sensing elements at the fashion of \textbf{System-In-Package (SiP) technologies,\cite{Wang2023} but in an extremely lower-cost fabrication process} (in material, energy and manufacturing equipment investment), by wetting arrays of vias with an electrolyte medium. However, among all configurations, the \textbf{2.5D layered topology imposes severe limitations on the electrolyte medium}. First, CPD morphogenesis through an interlayer contains a much lower quantity of free monomers to electropolymerize than a spherical drop. Growth may be prematurely stopped at their nucleation phase in 2D (Fig.\ref{fig:fig3.3.2}.l), so monomers' saturation concentration must be high. Second, the electrolyte medium must be processable and structurable: liquids are not convenient to handle, evaporates overtime and hardly structurable by lithography.\\ \textbf{Deeper investigations in formulating electroactive electrolytes for CPD morphogenesis must be performed (Subs.\ref{Ch4.3.3})}.

\subsection{Growing Dendrites in Various Media}
\label{Ch4.3.3}

\begin{figure}
  \centering
  \includegraphics[width=1\columnwidth]{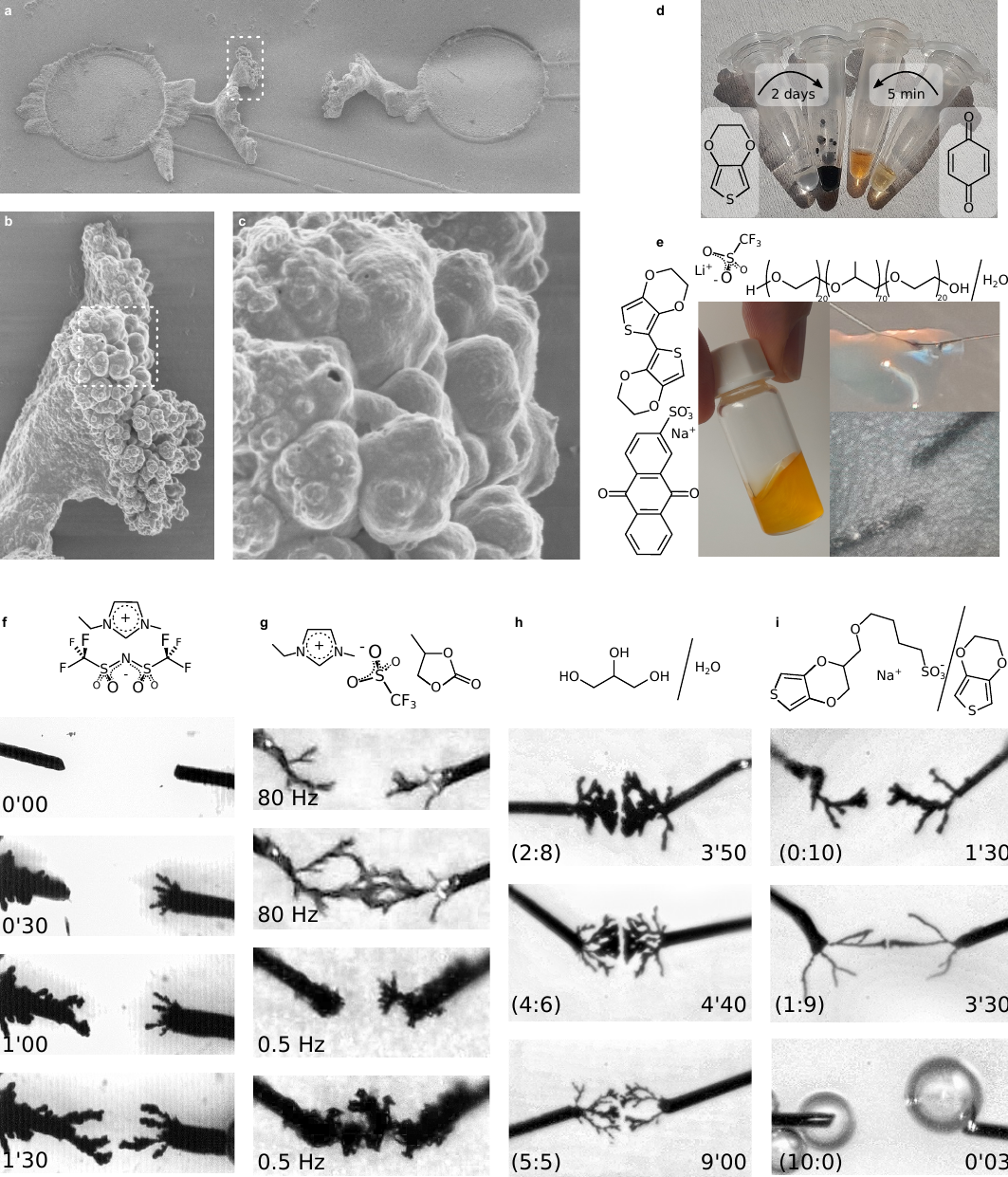}
  \caption{\textbf{Conducting Polymer Dendrites in Various Media $\vert$ a-c,} SEM images of PEDOT:PSS-based CPDs grown between two lithographically patterned gold electrodes (electrode active area: 30~$\upmu$m in diameter, tracks are passivated by a Parylene C layer). \textbf{d,} Reactivity of EDOT and BQ in ambient. From left to right, a fresh aliquot of EDOT in EmimOTf with no solvent (transparent), the same composition aged two days under sunlight (dark), an aliquot of BQ in water aged five minutes under sunlight (orange) and the same composition but fresh (yellow). \textbf{e,} Photographs and microscope image of an attempt to grow CPDs in a hydrogel. Electropolymerization is observed at a macro-scale but no CPD structure is evidenced at a micro-scale (chemical structure of the different components used for the attempt). \textbf{f} Growth of a PEDOT:NTf\textsubscript{2} CPD in an IL (EmimNTf\textsubscript{2}) as both a solvent and an electrolyte. \textbf{g} Growth of a PEDOT:OTf CPD in PC as an organic solvent and EmimOTf as an electrolyte. \textbf{h} Influence of a ratio of glycerol with water as co-solvents in the morphology of PEDOT:PSS CPDs. \textbf{i} Influence of a ratio of EDOTS with EDOT as co-monomers in the morphology of a CPD.}
  \label{fig:fig3.3.3}
\end{figure}

Dendrites are very versatile by themselves during their evolution, even without having to change their chemistry (Sec.\ref{Ch3.2}). They form very specific structures at different stages of their growth, depending on their completion and conditioning the applied electric field influencing electropolymerization. Dendrites have also fractal structures at their apex, with micrometer sized buds featuring specifically the front of each branches and not their side surface (Fig.\ref{fig:fig3.3.3}.a-c). These properties are highly material-dependent and require a profound understanding of the correlation between CPD structures and growth-media chemistry.\\[3pt]
To integrate them in a three dimensional hardware, many adaptations must be thought to structure the electro-active growth-medium in regard to the physical requirements for structuring it for 2.5D and 3D topological systems (as depicted in Fig.\ref{fig:fig3.3.2}.b-c), while preserving their stability over time. Even in the case of 1D (Sec.\ref{Ch3.2}) and 2D growths (Sub.\ref{Ch4.3.2}), the stability of the EDOT monomer and the BQ precursor must be considered: EDOT may undergo cationic polymerization overtime in the presence of Lewis acids or non-coordinating anions in ambient, while BQ mother solutions must be stored in a freezer prior usage over a few hours without observing noticeable changes of the solutions (Fig.\ref{fig:fig3.3.3}.d).\\[3pt]
A growing interest for processing and operating PEDOT:PSS in hydrogels is observed in the recent scientific literature.\cite{Lu2019,Zhang2020,Bhat2021,Yang2023,Zhou2023,Zhao2024a,Zhao2024} Hydrogel-based electrolytes would be an interesting field of research to invetigate on CPD integration for information processing \textit{in materio} on a sensing array. If results in this direction are only preliminary up to now (Fig.\ref{fig:fig3.3.3}.e), they motivate on a more profound understanding of the underlying physical mechanisms ruling CPD growths.\\[3pt]
Different chemical variations of an electroactive  electrolyte solution have been performed so far:\\
$\bullet $ First, \textbf{CPDs can grow in highly ionic media}. In case no neutral molecule is used a solvent, but an electrolyte as an ionic liquid (IL), PEDOT CPDs can grow between two AC-polarized electrodes.\cite{Chen2023} Some ILs may oxidize monomers directly to their polymeric form (such as pure EmimOTf). In the case of EmimNTf\textsubscript{2}, the solution remains translucent overtime and forms CPDs between two suspended gold wires (Fig.\ref{fig:fig3.3.3}.f). To use it as a supporting medium is however not convenient, as it tends to wet on many surfaces and conventional ILs are quite fluid. It however shows that CPDs can grow in highly concentrated electrolyte where the Helmholtz layer at the CPD/medium interfaces is completely screened by the free ion content of the medium.\\
$\bullet $ Second, \textbf{CPDs can grow in organic solvents}. Many works have studied dynamical growths of PEDOT in acetonitrile.\cite{Koizumi2016,Ohira2017,Eickenscheidt2019,AkaiKasaya2020,Cucchi2021} Acetonitrile is however too volatile to be conveniently used for system integration. Instead, PC has a remarkably low vapor pressure and a high boiling point.\cite{Pokorny2017} It is also a polar solvent extensively used for lithium-ion batteries and has very low toxicity.\cite{Xing2018,1987} CPD growths in PC show thinner structures than in water with the same voltage conditions (Fig.\ref{fig:fig3.3.3}.g). Along with its larger electrochemical window than water, no bubbling is observed in PC at low frequencies (typically at 0.5~Hz - Fig.\ref{fig:fig3.3.3}.g).\\
$\bullet $ Third, \textbf{CPDs can grow differently under the same electrochemical conditions, involving the same voltage patterns and ionic/electroactive compounds at the same concentration, but replacing only a part of the solvent by another "chemically close" substitute} (Fig.\ref{fig:fig3.3.3}.h). When glycerol is used as a neutral, non acidic but protic hydroxylated solvent to replace a fraction of water, CPDs grow slower and thinner. Glycerol is less polar, it lowers the medium's permittivity, its surface tension, increases its viscosity and increases PEDOT:PSS' conductance. The main physical property for slower and thinner growths is still unidentified. However, it confirms that introducing inert additives does not lead to the same morphological properties nor the growth kinetics, but does not disable morphogenesis neither.\\
$\bullet $ Fourth, \textbf{modifying the structure of an electroactive monomer precursor can change the morphology of a CPD}. In particular, Koizumi \textit{et al.} showed that increasing the length of an alkyl chain on the ethylenedioxy-bridge of EDOT produces dendrites which appear less fractal.\cite{Koizumi2016} A similar effect is observed by substituting a fraction of EDOT with EDOTS: a sulfonated derivative of EDOT used to produce a self-doped PEDOT derivatives.\cite{FrancoGonzalez2017,Gu2024} However, replacing totally EDOT by EDOTS leads to no CPD on free-standing gold wires (Fig.\ref{fig:fig3.3.3}.i). This suggests that only small structural modifications on a few of the electroactive monomers can lead to a total redesign of the CPDs networking on a system.\\[3pt]
\textbf{All these different variations controlled by an electroactive medium chemistry should be studied in the scientific literature to understand the control of CPD morphogenesis with specifically-designed media.}\\[3pt]
\textit{This mechanism shall be globally understood to exploit it for information processing on the long-term.}

\bibliographystyle{natsty-doilk-on-jour}  
\bibliography{refT2}  

\newpage

\begin{appendices}
\chapter{Publication List}
\label{Ch6}
\nopagebreak
\section{Doctoral Thesis (1)}
\label{Ch6.1}

\begin{longtable}{ l p{14cm} }
T1. & \underline{\textbf{S.~Pecqueur}}, Lewis Acid/Base Theory Applied on Evaluation of New Dopants for Organic Light-Emitting Diodes $\vert$ \textit{Friedrich-Alexander Erlangen-Nuremberg University} - July 2, 2014 \url{https://opus4.kobv.de/opus4-fau/frontdoor/index/index/docId/5005}. \\
 &  \\[1pt]
\end{longtable}

\section{Contributions to Peer-Reviewed Articles in International Journals (29)}
\label{Ch6.2}
\rightline{(*: corresponding authors)}

\begin{longtable}{ l p{13.5cm} }
\hypertarget{A29}{A29}. & A.~Baron, E.~H.~Balaguera \& \textbf{S.~Pecqueur}*, Correlation between Electrochemical Relaxations and Morphologies of Conducting Polymer Dendrites $\vert$ \textit{ECS Adv.} \textbf{3}(4), 044001 (2024) \url{https://doi.org/10.1149/2754-2734/ad9bcb} \\
 &  \\[1pt]
\hypertarget{A28}{A28}. & L.~Routier, A.~Westrelin, A.~Cerveaux, G.~Louis, T.~Horlac'h, P.~Foulon, K.~Lmimouni, \textbf{S.~Pecqueur} \& B.~Hafsi*, Single-Point Calibration Process Based Integrated Electrical Impedance Analyzer For Multi-Selective Gas Detection $\vert$ \textit{Discov. Appl. Sci.} \textbf{6}, 403 (2024) \url{https://doi.org/10.1007/s42452-024-06102-x} \\
 &  \\[1pt]
\hypertarget{A27}{A27}. & L.~Routier, A.~Westrelin, A.~Cerveaux, P.~Foulon, G.~Louis, T.~Horlac'h, K.~Lmimouni, \textbf{S.~Pecqueur} \& B. Hafsi*, Portable Multiplexed System-Based AD5933 Impedance Analyzer: Toward Multiselective Gas Recognition $\vert$ \textit{IEEE Sens. Lett.} \textbf{8}(7), 5502304 (2024) \url{https://doi.org/10.1109/LSENS.2024.3415789}  \\
 &  \\[1pt]
\hypertarget{A26}{A26}. & W.~Haj~Ammar, A.~Boujnah, A.~Baron, A.~Boubaker, A.~Kalboussi, K.~Lmimouni \& \textbf{S.~Pecqueur}*, A Temporal Filter to Extract Doped Conducting Polymer Information Features from an Electronic Nose $\vert$ \textit{Electronics} \textbf{13}(3), 497 (2024) \url{https://doi.org/10.3390/electronics13030497}  \\
 &  \\[1pt]
\hypertarget{A25}{A25}. & M.~Ghazal, A.~Kumar*, N.~Garg, \textbf{S.~Pecqueur} \& F.~Alibart*, Neuromorphic Signal Classification using Organic Electrochemical Transistor Array and Spiking Neural Simulations $\vert$ \textit{IEEE Sens. J.} \textbf{24}(6), (2024) \url{https://doi.org/10.1109/jsen.2024.3353307}  \\
 &  \\[1pt]
\hypertarget{A24}{A24}. & K.~Janzakova, I.~Balafrej, A.~Kumar, N.~Garg, C.~Scholaert, J.~Rouat, D.~Drouin, Y.~Coffinier, \textbf{S.~Pecqueur} \& F.~Alibart*, Structural plasticity for neuromorphic networks with electropolymerized dendritic PEDOT connections $\vert$ \textit{Nat. Commun.} \textbf{14}, 8143 (2023) \url{https://doi.org/10.1038/s41467-023-43887-8}  \\
 &  \\[1pt]
\hypertarget{A23}{A23}. & W.~Haj~Ammar, A.~Boujnah, A.~Boubaker, A.~Kalboussi, K.~Lmimouni \& \textbf{S.~Pecqueur}*, Steady vs. Dynamic Contributions of Different Doped Conducting Polymers in the Principal Components of an Electronic Nose's Response $\vert$ \textit{Eng} \textbf{4}(4), 2483-2496 (2023) \url{https://doi.org/10.3390/eng4040141}  \\
 &  \\[1pt]
\hypertarget{A22}{A22}. & M.~Ghazal, A.~Susloparova, C.~Lefebvre, M.~Daher~Mansour, N.~Ghodhbane, A~ Melot, C.~Scholaert, D.~Gu\'{e}rin, S.~Janel, N.~Barois, M.~Colin, L.~Bu\'{e}e, P.~Yger, S.~Halliez, Y.~Coffinier*, \textbf{S.~Pecqueur}* \& F.~Alibart, Electropolymerization Processing of Side-Chain Engineered EDOT for High Performance Microelectrode Arrays $\vert$ \textit{Biosens. Bioelectron.} \textbf{237}, 115538 (2023) \url{https://doi.org/10.1016/j.bios.2023.115538}  \\
 &  \\[1pt]
\hypertarget{A21}{A21}. & \textbf{S.~Pecqueur}*, D.~Vuillaume, \v{Z}.~Crljen, I.~Lon\v{c}ari\'{c} \& V.~Zlati\'{c}*, A Neural Network to Decipher Organic Electrochemical Transistors' Multivariate Responses for Cation Recognition $\vert$ \textit{Electronic Materials} \textbf{4}(2), 80‒94 (2023) \url{https://doi.org/10.3390/electronicmat4020007}  \\
 &  \\[1pt]
\hypertarget{A20}{A20}. & M.~Ghazal, C.~Scholaert, C.~Dumortier, C.~Lefebvre, N.~Barois, S.~Janel, M.~Ça\u{g}atay Tarhan, M.~Colin, L.~Bu\'{e}e, S.~Halliez, \textbf{S.~Pecqueur}, Y.~Coffinier, F.~Alibart* \& P.~Yger*, Precision of neuronal localization in 2D cell cultures by using high-performance electropolymerized microelectrode arrays correlated with optical imaging $\vert$ \textit{Biomed. Phys. Eng. Express} \textbf{9}, 035013 (2023) \url{https://doi.org/10.1088/2057-1976/acb93e}. \\
 &  \\[1pt]
\hypertarget{A19}{A19}. & A.~Boujnah*, A.~Boubaker, \textbf{S.~Pecqueur}, K.~Lmimouni \& A.~Kalboussi, An electronic nose using conductometric gas sensors based on P3HT doped with triflates for gas detection using computational techniques (PCA, LDA, and kNN) $\vert$ \textit{J. Mater. Sci.: Mater. Electron.} \textbf{33}, 27132‒27146 (2022) \url{https://doi.org/10.1007/s10854-0209376-2}. \\
 &  \\[1pt]
\hypertarget{A18}{A18}. & C.~Scholaert, K.~Janzakova, Y.~Coffinier, F.~Alibart \& \textbf{S.~Pecqueur}*, Plasticity of Conducting Polymer Dendrites to Bursts of Voltage Spikes in Phosphate Buffered Saline $\vert$ \textit{Neuromorphic Comput. Eng.} \textbf{2}(4), 044010 (2022) \url{https://doi.org/10.1088/2634-4386/ac9b85}. \\
 &  \\[1pt]
\hypertarget{A17}{A17}. & A.~Kumar*, K.~Janzakova, Y.~Coffinier, \textbf{S.~Pecqueur} \& F.~Alibart, Theoretical modeling of dendrite growth from conductive wire electropolymerization $\vert$ \textit{Sci. Rep.} \textbf{12}(6395), 1‒11 (2022) \url{https://doi.org/10.1038/s41598-022-10082-6}. \\
 &  \\[1pt]
\hypertarget{A16}{A16}. & K.~Janzakova, A.~Kumar, M.~Ghazal, A.~Susloparova, Y.~Coffinier, F.~Alibart \& \textbf{S.~Pecqueur}*, Analog Programing of Conducting-Polymer Dendritic Interconnections and Control of their Morphology $\vert$ \textit{Nat. Commun.} \textbf{12}(6898), 1‒11 (2021) \url{https://doi.org/10.1038/s41467-021-27274-9}. \\
 &  \\[1pt]
\hypertarget{A15}{A15}.  &  M.~Ghazal, M.~Daher~Mansour, C.~Scholaert, T.~Dargent, Y.~Coffinier, \textbf{S.~Pecqueur}* \& F.~Alibart*, Bio-inspired adaptive sensing through electropolymerization of organic electrochemical transistors $\vert$ \textit{Adv. Electron. Mater.} \textbf{8}(3), 2100891 (2022) \url{https://doi.org/10.1002/aelm.202100891}. \\
 &  \\[1pt]
\hypertarget{A14}{A14}.  &  K.~Janzakova, M.~Ghazal, A.~Kumar, Y.~Coffinier, \textbf{S.~Pecqueur}* \& F.~Alibart*, Dendritic organic electrochemical transistors grown by electropolymerization for 3D neuromorphic engineering $\vert$ \textit{Adv. Sci.} \textbf{8}(24), 2102973 (2021) \url{https://doi.org/10.1002/advs.202102973}. \\
 &  \\[1pt]
\hypertarget{A13}{A13}.  &  K.~Ferchichi*, \textbf{S.~Pecqueur}, D.~Gu\'{e}rin, R.~Bourguiga \& K.~Lmimouni, High rectification ratio in polymer diode rectifier through interface engineering with Self-Assembled Monolayer $\vert$ \textit{Electron. Mater.} \textbf{2}(4), 445‒453 (2021) \url{https://doi.org/10.3390/electronicmat2040030}. \\
 &  \\[1pt]
\hypertarget{A12}{A12}.  &  A.~Boujnah, A.~Boubaker, A.~Kalboussi, K.~Lmimouni \& \textbf{S.~Pecqueur}*, Mildly-Doped Polythiophene with Triflates for Molecular Recognition $\vert$ \textit{Synth. Met.} \textbf{280}, 116890 (2021) \url{https://doi.org/10.1016/j.synthmet.2021.116890}. \\
 &  \\[1pt]
\hypertarget{A11}{A11}.  &  K.~Ferchichi*, \textbf{S.~Pecqueur}, D.~Gu\'{e}rin, R.~Bourguiga \& K.~Lmimouni, Organic doped diode rectifier based on Parylene-electronic beam lithography process for Radio frequency applications $\vert$ \textit{Org. Electron.} \textbf{97}, 106266 (2021) \url{https://doi.org/10.1016/j.orgel.2021.106266}. \\
 &  \\[1pt]
\hypertarget{A10}{A10}.  &  A.~Susloparova, S.~Halliez, S.~Begard, M.~Colin, L.~Bu\'{e}e, \textbf{S.~Pecqueur}, F.~Alibart, V.~Thomy, S.~Arscott, E.~Pallecchi \& Y.~Coffinier*, Low impedance and highly transparent microelectrode arrays (MEA) for in vitro neuron electrical activity probing $\vert$ \textit{Sens. Actuator B-Chem.} \textbf{327}, 128895 (2021) \url{https://doi.org/10.1016/j.snb.2020.128895}. \\
 &  \\[1pt]
\hypertarget{A9}{A9}.  &  K.~Ferchichi, R.~Bourguiga, K.~Lmimouni \& \textbf{S.~Pecqueur}*, Concentration-control in all-solution processed semiconducting polymer doping and high conductivity performances $\vert$ \textit{Synth. Met.} \textbf{262}, 116352 (2020) \url{https://doi.org/10.1016/j.synthmet.2020.116352}. \\
 &  \\[1pt]
\hypertarget{A8}{A8}.  &  D.~Przyczyna, \textbf{S.~Pecqueur}, D.~Vuillaume \& K.~Szaci{\l}owski*, Reservoir computing for sensing - an experimental approach $\vert$ \textit{Int. J. Unconv. Comput.} \textbf{16}(3‒4), 267‒284 (2019) \url{https://www.oldcitypublishing.com/journals/ijuc-home/ijuc-issue-contents/ijuc-volume-14-number-3-4-2019/ijuc-14-3-4-p-267-284/}. \\
 &  \\[1pt]
\hypertarget{A7}{A7}.  &  V.~Athanasiou*, \textbf{S.~Pecqueur}, D.~Vuillaume \& Z.~Konkoli*, On a generic theory of the organic electrochemical transistor dynamics $\vert$ \textit{Org. Electron.} \textbf{72}, 39‒49 (2019) \url{https://doi.org/10.1016/j.orgel.2019.05.040}. \\
 &  \\[1pt]
\hypertarget{A6}{A6}.  &  \textbf{S.~Pecqueur}*, I.~Lon\v{c}ari\'{c}, V.~Zlati\'{c}, D.~Vuillaume \& \v{Z}.~Crljen*, The Non-Ideal Organic Electrochemical Transistors Impedance $\vert$ \textit{Org. Electron.} \textbf{71}, 14‒23 (2019) \url{https://doi.org/10.1016/j.orgel.2019.05.001}. \\
 &  \\[1pt]
\hypertarget{A5}{A5}.  &  \textbf{S.~Pecqueur}, D.~Vuillaume \& F.~Alibart*, Perspective: Organic electronic materials and devices for neuromorphic engineering $\vert$ \textit{J. Appl. Phys.} \textbf{124}(15), 151902 (2018) \url{https://doi.org/10.1063/1.5042419}. \\
 &  \\[1pt]
\hypertarget{A4}{A4}.  &  \textbf{S.~Pecqueur}*, M.~Mastropasqua~Talamo, D.~Gu\'{e}rin, P.~Blanchard, J.~Roncali, D.~Vuillaume \& F.~Alibart*, Neuromorphic time-dependent pattern classification with organic electrochemical transistor arrays $\vert$ \textit{Adv. Electron. Mater.} \textbf{4}(9), 1800166 (2018) \url{https://doi.org/10.1002/aelm.201800166}. \\
 &  \\[1pt]
\hypertarget{A3}{A3}.  &  \textbf{S.~Pecqueur}*, D.~Gu\'{e}rin, D.~Vuillaume \& F.~Alibart*, Cation Discrimination in Organic Electrochemical Transistors by Dual Frequency Sensing $\vert$ \textit{Org. Electron.} \textbf{57}, 232‒238 (2018) \url{https://doi.org/10.1016/j.orgel.2018.03.020}. \\
 &  \\[1pt]
\hypertarget{A2}{A2}.  &  \textbf{S.~Pecqueur}*, S.~Lenfant, D.~Gu\'{e}rin, F.~Alibart \& D.~Vuillaume, Concentric-electrode organic electrochemical transistors: case study for selective hydrazine sensing $\vert$ \textit{Sensors} \textbf{17}(3), 570 (2017) \url{https://doi.org/10.3390/s17030570}. \\
 &  \\[1pt]
\hypertarget{A1/A1'}{A1/A1'}.  &  \textbf{S.~Pecqueur}, A.~Maltenberger, M.~A.~Petrukhina, M.~Halik, A.~Jaeger, D.~Pentlehner \& G.~Schmid*, Wide Band-Gap Bismuth-based p-Dopants for Opto-Electronic Applications $\vert$ \textit{Angew. Chem. Int. Ed.} \textbf{55}(35), 10493‒10497 (2016) \url{https://doi.org/10.1002/anie.201601926} Bismut-haltige p-Dotanden mit gro{\ss}er Bandl\"{u}cke für optoelektronische Anwendungen $\vert$ \textit{Angew. Chem.} \textbf{128}(35), 10649‒10653 (2016) \url{https://doi.org/10.1002/ange.201601926}. \\
\end{longtable}

\section{Contributions to International Conference Oral Presentations (26)}
\label{Ch6.3}
\rightline{(presenting authors \underline{underlined})}

\begin{longtable}{ l p{14cm} }
\hypertarget{O26}{O26}. & L.~Routier, A.~Westrelin, A.~Cerveaux, P.~Foulon, G.~Louis, T.~Horlac'h, K.~Lmimouni, \textbf{S.~Pecqueur} \& \underline{B.~Hafsi}, Portable Multiplexed System Based AD5933 Impedance analyzer: Towards Multi-Selective Gas Recognition $\vert$ \textit{2024 IEEE 23\textsuperscript{rd} Int'l Conf. on Sensors (IEEE-Sensors 2024)}, Kobe/Japan - Oct. 21, 2024. \\
 &  \\[1pt]
\hypertarget{O25}{O25}. & \underline{\textbf{S.~Pecqueur}}, A.~Baron, C.~Scholaert, M.~Toledo-Nauto, P.~Moustiez, L.~Routier, D.~Gu{\'e}rin, K.~Lmimouni, Y.~Coffinier, B.~Hafsi \& F.~Alibart, Transience and Disorder of Organic Semiconductors in Future-Emerging Sensing $\vert$ \textit{4\textsuperscript{th} Workshop on Neuromorphic Organic Devices (NOD2024)}, invited talk, Paris/France - Oct. 9, 2024. \\
 &  \\[1pt]
\hypertarget{O24}{O24}. & A.~Baron, E.~H.~Balaguera \& \underline{\textbf{S.~Pecqueur}}, A Compact Electrochemical Model for a
Conducting Polymer Dendrite Impedance $\vert$ \textit{17\textsuperscript{th} Int'l Workshop on Impedance Spectroscopy (IWIS 2024)}, Chemnitz/Germany - Sept. 26, 2024, \url{https://doi.org/10.1109/IWIS63047.2024.10910272}. \\
 &  \\[1pt]
\hypertarget{O23}{O23}. & \underline{\textbf{S.~Pecqueur}}, Conducting-Polymer Neuro/Morphogenesis as in materio Computing Mechanism: Evolving Electronics at Almost No Cost $\vert$ \textit{2024 IEEE 24\textsuperscript{th} Int'l Conf. on Nanotechnology (IEEE-NANO 2024)}, invited talk SIS20/ID355, Gij\'{o}n/Spain - July 11, 2024. \\
 &  \\[1pt]
\hypertarget{O22}{O22}. & \underline{C. Scholaert}, K. Janzakova, Y. Coffinier, \textbf{S. Pecqueur} \& F. Alibart, From a Single Dendrite to Dendritic Networks: Towards Multi-Terminal OECTs $\vert$ \textit{2024 MRS Spring Meeting}, talk SB10.09.03,  Seattle/US - Apr. 25, 2024. \\
 &  \\[1pt]
\hypertarget{O21}{O21}. & \underline{F.~Alibart}, K.~Janzakova, C.~Scholaert, I.~Balafrej, A.~Kumar, D.~Drouin, J.~Rouat, \textbf{S.~Pecqueur} \& Y.~Coffinier, Structural Plasticity with PEDOT-Based Dendritic Electropolymerization for Neuromorphic Engineering $\vert$ \textit{2024 MRS Spring Meeting}, invited talk SB10.03.04, Seattle/US - Apr. 23, 2024. \\
 &  \\[1pt]
\hypertarget{O20}{O20}. & \underline{A.~Baron} \& \textbf{S.~Pecqueur}, Electrochemical Morphogenesis of Conducting Polymers for Evolvable Electronics $\vert$ \textit{8\textsuperscript{th} Baltic Electrochemistry Conference "Finding New Inspiration 2" (BEChem 2024)}, Tartu/Estonia - April 16, 2024. \\
 &  \\[1pt]
\hypertarget{O19}{O19}. & \underline{C.~Scholaert}, K.~Janzakova, Y.~Coffinier, \textbf{S.~Pecqueur} \& F.~Alibart, From a single dendrite to dendritic networks: towards multi-terminal OECTs $\vert$ \textit{3\textsuperscript{rd} Workshop on Neuromorphic Organic Devices (NOD2023)}, Bad Schandau/Germany - Sept. 18, 2023. \\
 &  \\[1pt]
\hypertarget{O18}{O18}. & \underline{K.~Janzakova}, M.~Ghazal, A.~Kumar, Y.~Coffinier, \textbf{S.~Pecqueur}, \& F. Alibart, Development of 3D organic polymer dendrites as neuromorphic device $\vert$ \textit{15\textsuperscript{th} International Symposium on Flexible Organic Electronics (ISFOE22)}, Thessaloniki/Greece - July 7, 2022. \\
 &  \\[1pt]
\hypertarget{O17}{O17}. & \underline{M.~Ghazal}, C.~Scholaert, M.~Daher~Mansour, S.~Janel, N.~Barois, S.~Halliez, T.~Dargent, Y.~Coffinier, \textbf{S.~Pecqueur} \& F.~Alibart, Neurites Whispering at Adaptive Sensors-High Spike-Signal-to-Noise Ratio Recorded with Electropolymerized Microelectrode Arrays $\vert$ \textit{2022 Virtual MRS Spring Meeting \& Exhibit}, talk SB06.15.04 - May 24, 2022. \\
 &  \\[1pt]
\hypertarget{O16}{O16}. & \underline{F.~Alibart}, M.~Ghazal, K.~Janzakova, A.~Kumar, C.~Scholaert, Y.~Coffinier \& \textbf{S.~Pecqueur}, From Bio-Sensing to Neuromorphic Engineering with Electropolymerized PEDOT:PSS Iono-Electronic Materials $\vert$ \textit{2022 Virtual MRS Spring Meeting \& Exhibit}, invited talk EQ11.15.02 - May 23, 2022. \\
 &  \\[1pt]
\hypertarget{O15}{O15}. & \underline{F.~Alibart}, M.~Ghazal, K.~Janzakova, A.~Kumar, A.~Susloparova, S.~Halliez, M.~Colin, L.~Bu\'{e}e, D.~Gu\'{e}rin, T.~Dargent, Y.~Coffinier \& \textbf{S.~Pecqueur}, Merging Bio-Sensing and Neuromorphic Computing with Organic Electro Chemical Transistors $\vert$ \textit{Spring's European Material Research Society Conf. 2021 (eMRS 2021 Spring)}, invited talk R.VIII.1 - June 3, 2021. \\
 &  \\[1pt]
\hypertarget{O14}{O14}. & \underline{A.~Kumar}, K.~Janzakova, Y.~Coffinier, D.~Gu\'{e}rin, F.~Alibart \& \textbf{S.~Pecqueur}, Governing Conducting-Polymer Micro-Objects' Fractality for Unconventional Computing $\vert$ \textit{2020 Virtual MRS Fall Meeting \& Exhibit}, talk F.SM05.01.05 - Nov. 27, 2020. \\
 &  \\[1pt]
\hypertarget{O13}{O13}. & \underline{F.~Alibart}, M.~Ghazal, K.~Janzakova, A.~Kumar, A.~Susloparova, S.~Halliez, M.~Colin, L.~Bu\'{e}e, D.~Gu\'{e}rin, T.~Dargent, Y.~Coffinier \& \textbf{S.~Pecqueur}, Merging Bio-Sensing and Neuromorphic Computing with Organic Electro Chemical Transistors $\vert$ \textit{2020 Virtual MRS Fall Meeting \& Exhibit}, invited talk F.SM05.04.02 - Nov. 27, 2020. \\
 &  \\[1pt]
\hypertarget{O12}{O12}. & \underline{K.~Janzakova}, M.~Ghazal, A.~Kumar, Y.~Coffinier, D.~Gu\'{e}rin, \textbf{S.~Pecqueur} \& F.~Alibart, Organic Electrochemical Transistors Based on Electropolymerized Dendritic Structures $\vert$ \textit{2020 Virtual MRS Fall Meeting \& Exhibit}, talk F.FL01.06/SM05.05.04 - Nov. 27, 2020. \\
 &  \\[1pt]
\hypertarget{O11}{O11}. & \underline{A.~Susloparova}, M.~Ghazal, D.~Gu\'{e}rin, Y.~Coffinier, T.~Dargent, F.~Alibart \& \textbf{S.~Pecqueur}, Low-Impedance Electropolymerized Coatings on Microelectrodes for Higher Neuro-Transduction $\vert$ \textit{2020 Virtual MRS Fall Meeting \& Exhibit}, talk F.FL01.04.03 - Nov. 27, 2020. \\
 &  \\[1pt]
\hypertarget{O10}{O10}. & \underline{M.~Ghazal}, A.~Susloparova, S.~Halliez, M.~Colin, L.~Bu\'{e}e, Y.~Coffinier, \textbf{S.~Pecqueur}, T.~Dargent \& F.~Alibart, Post-Fabrication Optimization Technique of Organic Electrochemical Transistor (OECT) for Electrophysiology by Electropolymerization $\vert$ \textit{2020 Virtual MRS Fall Meeting \& Exhibit}, talk F.FL01.05.03 - Nov. 27, 2020. \\
 &  \\[1pt]
\hypertarget{O9}{O9}. & K.~Ferchichi, \textbf{S.~Pecqueur}, D.~Gu\'{e}rin, R.~Bourguiga \& \underline{K.~Lmimouni}, Organic rectifier diode with very low turn-on voltage for RF Energy harvesting in Smart Textiles Applications $\vert$ \textit{T\'{e}l\'{e}com 2019 \& 11\textsuperscript{\`emes} JFMMA}, Saïda/Marocco - June 12, 2019. \\
 &  \\[1pt]
\hypertarget{O8}{O8}. & \underline{\textbf{S.~Pecqueur}}, D.~Gu\'{e}rin, D.~Vuillaume \& F.~Alibart, Material's Variability enabling Neuromorphic Pattern Recognition in Organic Electrochemical Transistor Networks $\vert$ \textit{9\textsuperscript{th} Int'l Conf. on Molecular Electronics 2018 (elecMol 2018)}, Paris/France - Dec. 17, 2018. \\
 &  \\[1pt]
\hypertarget{O7}{O7}. & \underline{\textbf{S.~Pecqueur}}, D.~Gu\'{e}rin, D.~Vuillaume \& F.~Alibart, Dynamical Neuromorphic Computing with Electropolymerized Organic Electrochemical Transistors $\vert$ \textit{Fall's European Material Research Society Conf. 2018 (eMRS 2018 Fall)}, invited talk M.12.1, Warsaw/Poland - Sep. 17, 2018. \\
 &  \\[1pt]
\hypertarget{O6}{O6}. & K.~Ferchichi, \textbf{S.~Pecqueur}, D.~Gu\'{e}rin, R.~Bourguiga \& \underline{K.~Lmimouni}, Organic rectifier diode with very low turn-on voltage for RF Energy harvesting in Smart Textiles Applications $\vert$ \textit{2018 SPIE Optics + Photonics 2018}, San Diego/USA - Aug. 19, 2018, \url{https://doi.org/10.1117/12.2321016}.  \\
 &  \\[1pt]
\hypertarget{O5}{O5}. & \underline{\textbf{S.~Pecqueur}}, S.~Lenfant, D.~Gu\'{e}rin, F.~Alibart \& D.~Vuillaume, Dual Sensing in a Single Organic Electrochemical Transistor (OECT) $\vert$ \textit{Organic Bioelectronics in Italy 2017 (OrBItaly 2017)}, Cagliari/Italy - Oct. 25, 2017. \\
 &  \\[1pt]
\hypertarget{O4}{O4}. & \underline{\textbf{S.~Pecqueur}}, S.~Lenfant, D.~Gu\'{e}rin, F.~Alibart \& D.~Vuillaume, Concentric-electrode organic electrochemical transistors: case study for selective hydrazine sensing $\vert$ \textit{6\textsuperscript{th} Int'l Conf. on Materials and Applications for Sensors and Transducers (IC-MAST 2016)}, Athens/Greece - Sep. 27, 2016, \url{https://doi.org/10.1088/1742-6596/939/1/012017}. \\
 &  \\[1pt]
\hypertarget{O3}{O3}. & \underline{\textbf{S.~Pecqueur}}, A.~I.~Borrachero~Conejo, S.~Bonetti, S.~Toffanin, G.~Generali, V.~Benfenati \& M.~Muccini, Selective Self-Assembled Monolayer to Passivate Organic Cell Stimulating and Sensing Transistor (OCSTs) $\vert$ \textit{2015 IEEE 15\textsuperscript{th} Int'l Conf. on Nanotechnology (IEEE-NANO 2015)}, Roma/Italy - July 27, 2015, \url{https://doi.org/10.1109/NANO.2015.7388937}. \\
 &  \\[1pt]
\hypertarget{O2}{O2}. & \underline{A.~I.~Borrachero~Conejo}, S.~Bonetti, S.~Karges, A.~Pistone, S.~D.~Quiroga, M.~Natali, I.~Grishin, \textbf{S.~Pecqueur}, M.~Caprini, G.~Generali, M.~Muccini, S.~Toffanin \& V.~Benfenati, An Organic Transistor Architecture for Stimulation of Calcium Signaling in Primary Rat Cortical Astrocytes $\vert$ \textit{2015 IEEE 15\textsuperscript{th} Int'l Conf. on Nanotechnology (IEEE-NANO 2015)}, Roma/Italy - July 27, 2015, \url{http://doi.org/10.1109/NANO.2015.7388936}. \\
 &  \\[1pt]
\hypertarget{O1}{O1}. & \underline{G.~Schmid}, \textbf{S.~Pecqueur} \& M.~Halik, Differentiation between Redox Chemistry and Lewis Acid/Base Model in Organic Semiconductor Doping $\vert$ \textit{6\textsuperscript{th} Int'l Symposium on Technologies for Polymer Electronics (TPE14)}, Illmenau/Germany - Dec. 20, 2014. \\
\end{longtable}

\section{Contributions to International Conference Poster Presentations (12)}
\label{Ch6.4}
\rightline{(presenting authors \underline{underlined})}

\begin{longtable}{ l p{14cm} }
\hypertarget{P12}{P12}. & E.~Vercoutere, S.~Kenne, C.~Morchain, \textbf{S.~Pecqueur} \& \underline{B.~Hafsi}, Vapour Recognition Based on Deep-Convolutional Neural Network: Portable Impedance Analyzer $\vert$ \textit{2024 IEEE 23\textsuperscript{rd} Int'l Conf. on Sensors (IEEE-Sensors 2024)}, Kobe/Japan - Oct. 21, 2024, \url{https://doi.org/10.1109/SENSORS60989.2024.10785096}.\\
 &  \\[1pt]
\hypertarget{P11}{P11}. & \underline{M.~Toledo~Nauto}, B.~Le~Cacher~de~Bonneville, H.~Kanso, M.-E.~Gourdel, C.~Reverdy, C.~Gasse, P.-L.~Saadi, J.-C.~Rain, \textbf{S.~Pecqueur} \& Y.~Coffinier, Biofunctionalized OECT for the Detection of Chlordecone $\vert$ \textit{5\textsuperscript{th} Workshop of IEEE Sensors France Chapter}, Grenoble/France - June 13, 2024. \\ 
 &  \\[1pt]
\hypertarget{P10}{P10}. & \underline{Z.~Oumekloul}, \textbf{S.~Pecqueur}, A.~de~Maistre, P.~Pernod, K.~Lmimouni, A.~Talbi \& B.~Hafsi, Design and Fabrication of Surface Acoustic Wave Gas Sensor Based on Electrochemical Process $\vert$ \textit{Symposium on Design, Test, Integration \& Packaging of MEMS/MOEMS (DTIP2022)}, Pont-\`{a}-Mousson/France - July 11, 2022. \\
 &  \\[1pt]
\hypertarget{P9}{P9}. & \underline{C.~Scholaert}, K.~Janzakova, M.~Ghazal, M.~Daher~Mansour, C.~Lefebvre, S.~Halliez, Y.~Coffinier, \textbf{S.~Pecqueur} \& F.~Alibart, Dendritic-like PEDOT:PSS electrodes for 2D in-vitro electrophysiology $\vert$ \textit{2022 MEA Meeting}, T\"{u}bingen/Germany - July 7, 2022. \\
 &  \\[1pt]
\hypertarget{P8}{P8}. & \underline{M.~Ghazal}, C.~Scholaert, C.~Lefebvre, N.~Barois, S.~Janel, M.~Ça\u{g}atay Tarhan, M.~Colin, L.~Bu\'{e}e, S.~Halliez, \textbf{S.~Pecqueur}, Y.~Coffinier, F.~Alibart \& P.~Yger, Accurate neurons localization in 2D cell cultures by using high performance electropolymerized microelectrode arrays correlated with optical imaging $\vert$ \textit{2022 MEA Meeting}, T\"{u}bingen/Germany - July 7, 2022. \\
 &  \\[1pt]
\hypertarget{P7}{P7}. & A.~Boujnah, A.~Boubaker, A.~Kalboussi, \underline{K.~Lmimouni} \& \textbf{S.~Pecqueur}, An Electronic Nose with one Single Conducting Polymer? How Mild-Doping Tunes P3HT's Chemo-Sensitivity for Molecular Recognition $\vert$ \textit{10\textsuperscript{th} Int'l Conf. on Molecular Electronics 2021 (elecMol 2021)}, PO93 - T8, Lyon/France - Nov. 29, 2021. \\
 &  \\[1pt]
\hypertarget{P6}{P6}. & K.~Ferchichi, \textbf{S.~Pecqueur}, D.~Gu\'{e}rin, R.~Bourguiga \&  \underline{K.~Lmimouni}, Organic doped diode rectifier based on Parylene-electronic beam lithogrpahy process for Radio frequency applications $\vert$ \textit{10\textsuperscript{th} Int'l Conf. on Molecular Electronics 2021 (elecMol 2021)}, PO33 - T3, Lyon/France - Nov. 29, 2021. \\
 &  \\[1pt]
\hypertarget{P5}{P5}. &  \underline{M.~Ghazal}, M.~Daher~Mansour, S.~Halliez, Y.~Coffinier, T.~Dargent, \textbf{S.~Pecqueur} \& F.~Alibart, Post-fabrication optimization technique of organic electrochemical transistor (OECT) by electropolymerization for electrophysiology $\vert$ \textit{Technologies for Neuroengineering - Nature Conference}, Virtual, May 26, 2021. \\
 &  \\[1pt]
\hypertarget{P4}{P4}. &  \underline{M.~Ghazal}, T.~Dargent, \textbf{S.~Pecqueur} \& F.~Alibart, Addressing Organic Electrochemical Transistors for Neurosensing and Neuromorphic Sensing $\vert$ \textit{2019 IEEE 18\textsuperscript{th} IEEE Conference on Sensors (IEEE-Sensors 2019)}, Montreal/Canada - Oct. 27, 2019, \url{https://doi.org/10.1109/SENSORS43011.2019.8956648}. \\
 &  \\[1pt]
\hypertarget{P3}{P3}. & \textbf{S.~Pecqueur}, I.~Lon\v{c}ari\'{c}, V.~Zlati\'{c}, D.~Vuillaume \&  \underline{\v{Z}.~Crljen}, Optimized Model for Non-ideal Organic Electrochemical Transistors Impedance $\vert$ \textit{Int'l Conf. Nano Materials \& Devices (Nano-M\&D2019)}, Paestum/Italy - June 4, 2019. \\
 &  \\[1pt]
\hypertarget{P2}{P2}. &  \underline{S.~Karges}, S.~Bonetti, A.~I.~Borrachero~Conejo, A.~Pistone, S.~D.~Quiroga, M.~Natali, I.~Grishin, \textbf{S.~Pecqueur}, F.~Mercuri, M.~Caprini, G.~Generali, M.~Muccini, S.~Toffanin \& V.~Benfenati, An Organic Device for Stimulation and Optical Read-out of Calcium Signaling in Primary Rat Cortical Astrocytes $\vert$ \textit{XII European Meeting on Glial Cells in Health and Disease (Glia2015)}, San Sebastian/Spain - July 15, 2015. \\
 &  \\[1pt]
\hypertarget{P1}{P1}. &  \underline{\textbf{S.~Pecqueur}}, M.~Halik \& G\"{u}nter~Schmid, Differentiation between Redox Chemistry and Lewis Acid/Base Model in Organic Semiconductor Doping $\vert$ \textit{Plastic Electronics 2013 (PE2013)}, Dresden/Germany - Oct. 8, 2013. \\
\end{longtable}

\section{International Patent Applications (8)}
\label{Ch6.5}

\begin{longtable}{ l p{14cm} }
\hypertarget{B8}{B8}. & F.~Kessler, A.~Maltenberger, \textbf{S.~Pecqueur}, D.~Pentlehner, S.~Regensburger \& G.~Schmid, Organic electronic component having a charge generation layer and use of a zinc complex as a p-dopant in charge generation layers $\vert$ Pat. Appl. WO/2017/055283(A1) - Apr. 6, 2017. \\
 &  \\[1pt]
\hypertarget{B7}{B7}. & F.~Kessler, \textbf{S.~Pecqueur} \& G.~Schmid, Organic heterocyclic alkali metal salts as n-dopants in organic electronics $\vert$ Pat. Appl. WO/2016/192902(A1) - Dec. 8, 2016. \\
 &  \\[1pt]
\hypertarget{B6}{B6}. & F.~Kessler, \textbf{S.~Pecqueur} \& G.~Schmid, Proazaphosphatranes as n-dopants in organic electronics $\vert$ Pat. Appl. WO/2016/116205(A1) - July 28, 2016. \\
 &  \\[1pt]
\hypertarget{B5}{B5}. & F.~Kessler, A.~Maltenberger, \textbf{S.~Pecqueur}, S. Regensburger \& G. Schmid, Organic electronic component, use of a zinc complex as a p-dopant for organic electronic matrix materials $\vert$ Pat. Appl. WO/2016/050705(A1) - Apr. 7, 2016. \\
 &  \\[1pt]
\hypertarget{B4}{B4}. & A.~Maltenberger, \textbf{S.~Pecqueur}, S.~Regensburger \& G.~Schmid, Method for producing an organic electronic component, and organic electronic component $\vert$ Pat. Appl. WO/2016/050330(A1) - Apr. 7, 2016. \\
 &  \\[1pt]
\hypertarget{B3}{B3}. & A.~Maltenberger, \textbf{S.~Pecqueur} \& G.~Schmid, P-doping cross-linking of organic hole transporters $\vert$ Pat. Appl. WO/2015/185440(A1) - Dec. 10, 2015. \\
 &  \\[1pt]
\hypertarget{B2}{B2}. & A.~Maltenberger, \textbf{S.~Pecqueur}, G.~Schmid \& J.-H.~Wemken, Metal complexes as p-type dopants for organic electronic matrix materials $\vert$ Pat. Appl. WO/2013/182383(A1) - Dec. 12, 2013. \\
 &  \\[1pt]
\hypertarget{B1}{B1}. & A.~Kanitz, \textbf{S.~Pecqueur}, G.~Schmid \& J.-H.~Wemken, Organic electronic components having organic superdonors having at least two coupled carbene groups and use thereof as an n-type dopants $\vert$ Pat. Appl. WO/2013/153025(A1) - Oct. 17, 2013. \\
\end{longtable}

\chapter{curriculum vit\ae}
\label{Ch7}

\section{eg{\={o}} et \={\i}nsum}
\label{Ch7.1}

\centering
\Large Dr.-Ing. S\'{e}bastien Pecqueur \\\normalsize
\textit{french-born on the 10\textsuperscript{th} of February 1987 in Beuvry, France.}\\[-5pt]
\justify Since October 2019, I am tenure employed as Charg\'{e} de Recherche de Classe Normale by the French National Centre for Scientific Research (\href{https://www.cnrs.fr}{CNRS}) at the formerly-named National Institute for System Engineering Sciences (\href{https://www.insis.cnrs.fr}{INSIS}), now CNRS-Ing{\'e}nieurie. My research project was defended at the CNRS' National Comity for Micro/nanotechnologies \& Micro/nano-photonic/electronic/electromagnetic/energy/electric systems (\href{https://www.cnrs.fr/comitenational/sections/section.php?sec=08}{CoNRS 8}), to take place at the Institute of Electronics, Microelectronics and Nanotechnology in Lille (\href{https://www.iemn.fr/}{IEMN~-~UMR8520}). Since then, I am team-member of the Nanostructures, nanoComponents \& Molecules (\href{https://www.iemn.fr/la-recherche/les-groupes/groupe-ncm}{NCM}) group at IEMN, on a research topic related to organic electronics, electrochemistry \& sensing.\\[-10pt]

\centering
\textit{IEMN. av. Poincar\'{e}, cit\'{e} scientifique, CS 60069, 59652 Villeneuve d’Ascq Cedex, France (office 145) \\ +33 (0)3.20.29.79.08 $\vert$ sebastien.pecqueur[at]iemn.fr}\\

\section{alma mater studia}
\label{Ch7.2}

\centering
\textit{From 2011 till 2014,}\\
\Large Friedrich-Alexander Erlangen-N\"{u}rnberg Universit\"{a}t $\vert$ Erlangen, Germany \\\normalsize
\textit{Doctorate in Material Science Engineering, under the supervision of Prof. Dr. rer. nat. Marcus Halik, graduated "mit Auszeichnung Bestanden"  - highest honors.}\\[-5pt]
\justify Title: "Lewis Acid-Base Theory Applied on Evaluation of New Dopants for Organic Light-Emitting Diodes".\\[-10pt]

\centering ---\\[2pt]
\textit{From 2007 till 2010,}\\
\Large \'{E}cole Nationale Sup{\'e}rieure de Chimie Physique de Bordeaux $\vert$ Pessac, France \& Universit\'{e} de Bordeaux I $\vert$ Talence, France\\\normalsize
\textit{Engineering degree in Physical Chemistry and Master's degree in Chemistry}\\[-5pt]
\justify Major in Micro- \& Nanotechnologies from the ENSCPB Engineering school.\\ Major in Synthesis \& Property of Inorganic Materials from the Unversity of Bordeaux I.\\[-10pt]

\section{per\={\i}ti\ae}
\label{Ch7.3}

\centering
\textit{From 2016 till 2019,}\\
\Large Institut d'\'{E}lectronique de Micro\'{e}lectronique et de Nanotechnologie (IEMN) \\ Villeneuve d'Ascq, France \\\normalsize
\textit{Postdoctoral Research Fellow.}\\[-5pt]
\justify Nine months of researches on organic electrochemical transistors for neurosensing in the ERC-CoG \href{https://cordis.europa.eu/project/id/773228}{773228} IONOS project (PI: Dr. Alibart).\\
Three years of researches on organic electrochemical transistors for neuromorphic sensing in the EU/FET-open \href{https://cordis.europa.eu/project/id/664786}{664786} RECORD-IT project, (PI: Dr. Vuillaume).\\[-10pt]

\centering ---\\[2pt]
\textit{From 2014 till 2015,}\\
\Large ETC srl $\vert$ Bologna, Italy \\\normalsize
\textit{Postdoctoral Research Fellow.}\\[-5pt]
\justify 13 months of researches on organic cell-stimulating \& sensing transistors for neurosensing  in the EU/MSCA \href{https://cordis.europa.eu/project/id/316832}{316832} OLIMPIA project (PI: M. Muccini) and developments of other organic electronic stack devices, extended stay by three months.\\[-10pt]

\centering ---\\[2pt]
\textit{From 2011 till 2014,}\\
\Large Siemens AG - Corporate Technology $\vert$ Erlangen, Germany \\\normalsize
\textit{Doctoral Student.}\\[-5pt]
\justify 39 months of researches and developments on charge-transfer doping in materials for organic light-emitting diodes (PI: Dr. rer. nat. Schmid).\\[-10pt]

\centering ---\\[2pt]
\textit{In 2010,}\\
\Large Cambridge Display Technology Ltd $\vert$ Godmanchester, The United Kingdom \\\normalsize
\textit{Master Student Intern.}\\[-5pt]
\justify Six months of researches and developments on materials \& processes for organic electronic devices (PI: Dr. Wilson).\\[-10pt]

\centering ---\\[2pt]
\textit{In 2009,}\\
\Large Philips NV $\vert$ Eindhoven, The Netherlands \\\normalsize
\textit{Engineering Student Intern.}\\[-5pt]
\justify Five months of researches and developments on materials \& processes for infrared lamps manufacturing (PI: Dr. Van Sprang).\\[-10pt]

\chapter{Research Advising and Supervision}
\label{Ch8}

\section{Postdoctoral Researchers \& Research Engineers (3)}
\label{Ch8.1}

\centering
\Large Dr. Paul Moustiez \\ \normalsize
\textit{under my supervision as a Postdoctoral Research Fellow since 6.2024.}\\[-5pt]
\justify Paul has a PhD in Engineering from the University of Lille, France. I am supervising Paul's experiments on micro-fabricating conductimetric elements and electropolymerizing on them at small scale for chemical recognition (whose data contributed to the Fig.\ref{fig:fig3.1.2} of the monograph).\textsuperscript{\hyperlink{O25}{O25}} His stay is financed by the French National Research Agency for the \hyperlink{Sensation}{SENSATION} project.\\[-5pt]

\centering --- \\
\Large Dr. Zakariae Oumekloul \\ \normalsize
\textit{co-supervised with Dr. Hafsi as a Research Engineer from 9.2021 till 8.2022.}\\[-5pt]
\justify Zak has a PhD in Engineering from the University Moulay Ismail - Mekn{\`e}s, Marocco. I supervised Zak's experiments on electropolymerizing surface-acoustic wave micro-systems for gas sensing.\textsuperscript{\hyperlink{P10}{P10}} Later, Zak obtained a tenured research position as a Ma{\^i}tre-de-Conf{\'e}rence at the Universit{\'e} Polytechnique Hauts-de-France, France. His stay was financed by the IEMN laboratory.\\[-5pt]

\centering --- \\
\Large Dr. rer. nat. Anna Susloparova \\ \normalsize
\textit{co-supervised with Dr. Coffinier \& Dr. Alibart as a Research Engineer from 10.2019 till 8.2020.}\\[-5pt]
\justify Anna has a PhD in Physics from Justus-Liebig-Universit{\"a}t Gie{\ss}en, Germany. I supervised Anna's experiments on electro-co-polymerization for electrochemical impedance tuning on micro-electrode arrays (whose data contributed to the Fig.\ref{fig:fig2.1.3} of the monograph).\textsuperscript{\hyperlink{A16}{A16},\hyperlink{A22}{A22},\hyperlink{O10}{O10}-\hyperlink{O11}{O11},\hyperlink{O13}{O13},\hyperlink{O15}{O15}} Her stay was financed by the European Research Council for the \hyperlink{Ionos}{IONOS} project.\\[-5pt]

\newpage

\section{Doctoral Researchers (9)}
\label{Ch8.2}

\centering
\Large Antoine Baron \\ \normalsize
\textit{under my direction \& supervision as a Ph.D. candidate at the University of Lille since 1.2023.}\\[-5pt]
\justify I am directing Antoine's thesis on investigating conducting polymer morphogenesis for information storage \textit{in materio} for non-metrological sensing hardware (whose data contributed to the Fig.\ref{fig:fig2.2.3}, Fig.\ref{fig:fig3.3.2} and Fig.\ref{fig:fig3.3.3} of the monograph).\textsuperscript{\hyperlink{A26}{A26},\hyperlink{A29}{A29},\hyperlink{O20}{O20},\hyperlink{O24}{O24}-\hyperlink{O25}{O25}} His stay is financed by the French National Research Agency for the \hyperlink{Sensation}{SENSATION} project.\\[-5pt]

\centering --- \\
\Large Louis Routier \\ \normalsize
\textit{co-supervised with Dr. Hafsi and directed by Prof. Lmimouni as a Ph.D. candidate at the University of Lille since 10.2021.}\\[-5pt]
\justify I supervised Louis' experiments on machine-assisted fabrication and characterization of electropolymerized sensing arrays on circuit boards (whose data contributed to the Fig.\ref{fig:fig3.2.2} of the monograph).\textsuperscript{\hyperlink{A27}{A27}-\hyperlink{A28}{A28},\hyperlink{O25}{O25}} His stay was financed with a fellowship from the University of Lille.\\[-5pt]

\centering --- \\
\Large Ewelina Kowalewska \\ \normalsize
\textit{directed by Prof. Szaci{\l}owski as a Ph.D. candidate at AGH Krakow in Poland since 10.2021.}\\[-5pt]
\justify I supervised Ewelina's experiments during a month-long visit in 10.2023 at IEMN on the investigation of conducting dendritic morphogenesis \& electrochemical characterizations in organic solvents (whose data contributed to the Fig.\ref{fig:fig3.3.3} of the monograph). Her visit was financed by a stipend from the EU Erasmus+ program.\\[-5pt]

\centering --- \\
\Large Wiem Haj Ammar \\ \normalsize
\textit{directed by Dr. Boubaker as a Ph.D. candidate at the University of Monastir in Tunisia since 10.2021.}\\[-5pt]
\justify I supervised Wiem's work during two visits at IEMN (eight months overall), on conducting polymers' feature extraction for the sensitivity assessment in volatile molecule recognition (whose data contributed to the Fig.\ref{fig:fig1.4.1} and Fig.\ref{fig:fig2.1.1} of the monograph).\textsuperscript{\hyperlink{A23}{A23},\hyperlink{A26}{A26}} Her two visits were financed by a stipend from the University of Monastir and by the French National Research Agency for the \hyperlink{Sensation}{SENSATION} project.\\[-5pt]

\centering --- \\
\Large Dr. Corentin Scholaert \\ \normalsize
\textit{co-supervised with Dr. Coffinier and directed by Dr. Alibart as a Ph.D. candidate at the University of Lille from 10.2021 till 12.2024.}\\[-5pt]
\justify I supervised Corentin's experiments on signal classification of transient voltage activity in water with conducting polymer dendrites (whose data contributed to the Fig.\ref{fig:fig2.2.2}, Fig.\ref{fig:fig3.3.2} and Fig.\ref{fig:fig3.3.3} of the monograph).\textsuperscript{\hyperlink{A15}{A15},\hyperlink{A18}{A18},\hyperlink{A20}{A20},\hyperlink{A22}{A22},\hyperlink{A24}{A24},\hyperlink{O16}{O16}-\hyperlink{O17}{O17},\hyperlink{O19}{O19},\hyperlink{O21}{O21}-\hyperlink{O22}{O22},\hyperlink{O25}{O25},\hyperlink{P8}{P8}-\hyperlink{P9}{P9}} Later, Corentin pursued his researches as a Research Engineer in a related topic at IEMN, France. His stay was financed by the European Research Council for the \hyperlink{Ionos}{IONOS} project and a stipend from the Hauts-de-France region.\\[-5pt]

\newpage \centering
\Large Dr. Kamila Janzakova \\ \normalsize
\textit{co-supervised with Dr. Alibart and directed by Dr. Coffinier as a Ph.D. candidate at the University of Lille from 10.2019 till 3.2023.}\\[-5pt]
\justify 
I supervised Kamila's experiments on conducting polymer dendrite morphogenesis in water under pulse waves (whose data contributed to the Fig.\ref{fig:fig2.2.1} of the monograph).\textsuperscript{\hyperlink{A14}{A14},\hyperlink{A16}{A16}-\hyperlink{A18}{A18},\hyperlink{A24}{A24},\hyperlink{O12}{O12}-\hyperlink{O16}{O16},\hyperlink{O18}{O18}-\hyperlink{O19}{O19},\hyperlink{O21}{O21}-\hyperlink{O22}{O22}} Later, Kamila pursued her researches as a Post-doctoral Research Fellow at the LAAS laboratory in Toulouse, France. Her stay was financed by the European Research Council for the \hyperlink{Ionos}{IONOS} project.\\[-5pt]

\centering --- \\
\Large Dr. Aicha Boujnah \\ \normalsize
\textit{directed by Prof. Kalboussi as a Ph.D. candidate at the University of Monastir in Tunisia from 2018 till 2023.}\\[-5pt]
\justify I supervised Aicha's work during three visits at IEMN (eight months overall), on investigating doped polymers and assessing their sensitivity for solvent vapor recognition (whose data contributed to the Fig.\ref{fig:fig1.4.2} and Fig.\ref{fig:fig2.1.1} of the monograph).\textsuperscript{\hyperlink{A12}{A12},\hyperlink{A19}{A19},\hyperlink{A23}{A23},\hyperlink{A26}{A26},\hyperlink{P7}{P7}} Later, Aicha had a Research Engineer position at CEA Leti, France. Her three visits were financed by a stipend from the University of Monastir.\\[-5pt]

\centering --- \\
\Large Dr. Mahdi Ghazal \\ \normalsize
\textit{co-supervised with Dr. Alibart and directed by Dr. Coffinier as a Ph.D. candidate at the University of Lille from 9.2019 till 12.2022.}\\[-5pt]
\justify I supervised Mahdi's experiments on electro-co-polymerization for electrochemical impedance tuning on organic electrochemical transistors and micro-electrode arrays (whose data contributed to the Fig.\ref{fig:fig2.1.3} of the monograph).\textsuperscript{\hyperlink{A14}{A14}-\hyperlink{A16}{A16},\hyperlink{A20}{A20},\hyperlink{A22}{A22},\hyperlink{A25}{A25},\hyperlink{O10}{O10}-\hyperlink{O13}{O13},\hyperlink{O15}{O15}-\hyperlink{O18}{O18},\hyperlink{P5}{P5},\hyperlink{P8}{P8}-\hyperlink{P9}{P9}} Later, Mahdi had an R{\&}D Application Engineer position at MaxWell Biosystems AG, Switzerland. His stay was financed by the European Research Council for the \hyperlink{Ionos}{IONOS} project.\\[-5pt]

\centering --- \\
\Large Dr. Khaoula Ferchichi \\ \normalsize
\textit{directed by Prof. Lmimouni and Prof. Bourguiga as a Ph.D. candidate at the University of Lille and the University of Carthage, Tunisia, from 12.2015 till 3.2019.}\\[-5pt]
\justify I supervised Khaoula's experiments on Lewis acid doping in conducting polymers to control diodes rectification ratio.\textsuperscript{\hyperlink{A9}{A9},\hyperlink{A11}{A11},\hyperlink{A13}{A13},\hyperlink{O6}{O6},\hyperlink{O9}{O9},\hyperlink{P6}{P6}} After, Khaoula obtained a tenured research position as a Ma{\^i}tre-de-Conf{\'e}rence at the Universit{\'e} du Littoral C{\^o}te d'Opale, France. Her stay was financed by a fellowship from the University of Carthage.\\[-5pt]

\newpage

\section{Graduated Masters/Engineers (4)}
\label{Ch8.3}

\centering
\Large Milan Toledo Nauto \\ \normalsize
\textit{co-supervised with Dr. Coffinier as an Engineer from 9.2022 till 11.2024.}\\[-5pt]
\justify I supervised Milan's experiments on conducting-polymer co-integration and electrical characterization on sensing micro-arrays for water quality recognition (whose data contributed to the Fig.\ref{fig:fig3.1.3} of the monograph).\textsuperscript{\hyperlink{O25}{O25},\hyperlink{P11}{P11}} Later, Milan pursued a PhD degree in a related topic at the Universit{\'e} Paris Cit{\'e} in Paris, France. Her stay was financed by the French National Research Agency for the \hyperlink{Sens-PestOLS}{SENS-PESTOLS} project.\\[-5pt]

\centering --- \\
\Large Rajarshi Sinha \\ \normalsize
\textit{co-supervised with Dr. Coffinier as an Engineer from 9.2021 till 5.2022.}\\[-5pt]
\justify I supervised Raj's experiments on micro-fabricating sensing arrays for water quality recognition. Later, Raj had a process integration engineer position at NXP Semiconductors, the Netherlands. His stay was financed by the French National Research Agency for the \hyperlink{Sens-PestOLS}{SENS-PESTOLS} project.\\[-5pt]

\centering --- \\
\Large Najami Ghodhbane \\ \normalsize
\textit{co-supervised with Dr. Coffinier and Dr. Alibart as an Engineer from 9.2022 till 10.2022.}\\[-5pt]
\justify I supervised Najami's experiments on electro-co-polymerization for electrochemical impedance tuning on micro-electrode arrays (whose data contributed to the Fig.\ref{fig:fig2.1.3} of the monograph).\textsuperscript{\hyperlink{A22}{A22}} Her stay was financed by the European Research Council for the \hyperlink{Ionos}{IONOS} project.\\[-5pt]

\centering --- \\
\Large Dr. Julien Couturier \\ \normalsize
\textit{co-supervised with Dr. Vuillaume as an Engineer from 5.2017 till 11.2018.}\\[-5pt]
\justify I supervised Julien's experiments on fabricating and characterizing organic field-effect transistors. Later, Julien obtained a PhD in environmental sciences from the University Aix-Marseille, France. His stay was financed by the European Commission for the \hyperlink{Record-It}{RECORD-IT} project.\\[-5pt]

\newpage

\section{Master/Engineering Students (9)}
\label{Ch8.4}

\centering
\Large Marie Delarue, Aymar Kingoum Taka \& Sandra Kenne Koudjou \\ \normalsize
\textit{co-supervised with Dr. Hafsi as ICAM Eng. Interns from 2.2023 till 6.2024.}\\[-5pt]
\justify I successively supervised Marie, Aymar and Sandra's experiments during three five-month long internships on characterizing doped conducting polymers' resistance sensitivity to volatile solvents using a multiplexed AD5933/ATmega168p micro-controlled prototype (whose data contributed to the Fig.\ref{fig:fig3.2.1} of the monograph).\textsuperscript{\hyperlink{P12}{P12}} Marie, Aymar and Sandra's results are the fruit of a longer collaboration with the ICAM Engineering school involving indirectly many more interns.\\[-5pt]

\centering --- \\
\Large Mona Ghazal \\ \normalsize
\textit{co-supervised with Dr. Alibart as an Intern from 2.2022 till 7.2022.}\\[-5pt]
\justify I supervised Mona's work on simulating a circuit model of an organic electrochemical transistor featuring electrogenic cells at its surface. Later, Mona pursued a PhD degree in physics at the {\'E}cole Normale Sup{\'e}rieure Paris-Saclay, France. Her stay was financed by the European Research Council for the \hyperlink{Ionos}{IONOS} project.\\[-5pt]

\centering --- \\
\Large Luc Brulin \\ \normalsize
\textit{co-supervised with Dr. Alibart as an Intern from 2.2022 till 7.2022.}\\[-5pt]
\justify I supervised Luc's experiments on signal time-dependency during conducting-polymer morphogenesis and on two-dimensional morphogenetic generations (whose data contributed to the Fig.\ref{fig:fig3.3.2} of the monograph). His stay was financed by the European Research Council for the \hyperlink{Ionos}{IONOS} project.\\[-5pt]

\centering --- \\
\Large Ambre Alleon \& Amira Bousbaa \\ \normalsize
\textit{co-supervised with Dr. Dargent as Interns during 5.2022.}\\[-5pt]
\justify I supervised Ambre \& Amira's experiments on structuring conducting polymer micro-spheres by flow focusing. Later, both Ambre and Amira pursued a PhD degree on different topics in biotechnologies.\\[-5pt]

\centering --- \\
\Large Dr. Corentin Scholaert \\ \normalsize
\textit{co-supervised with Dr. Alibart as an Intern from 2.2021 till 7.2021.}\\[-5pt]
\justify I supervised Corentin's experiments on implementing dendritic morphogenesis to pattern the channel of microfabricated organic electrochemical transistors. Later, Corentin pursued a PhD degree on the same project. His stay was financed by the European Research Council for the \hyperlink{Ionos}{IONOS} project.\\[-5pt]

\centering --- \\
\Large Dr. Mahdi Ghazal \\ \normalsize
\textit{co-supervised with Dr. Alibart as an Intern from 2.2019 till 7.2019.}\\[-5pt]
\justify I supervised Mahdi's experiments on characterizing electropolymerized organic electrochemical transistors and building a readout circuit to record electrophysiological activities.\textsuperscript{\hyperlink{P4}{P4}} Later, Mahdi pursued a PhD degree on the same project. His stay was financed by the European Research Council for the \hyperlink{Ionos}{IONOS} project.\\[-5pt]

\chapter{Scientific Responsibilities, Collaborations and Impacts}
\label{Ch9}

\section{Funded Projects (4)}
\label{Ch9.1}

\centering
\Large The \hypertarget{Biosurv}{BIOSURV} project \textit{funded by} the CNRS/PEPS \\ \normalsize
\textit{\textbf{Biosurveillance de milieux naturels: comparaison de diff{\'e}rentes techniques de d{\'e}tection de stress chez les plantes}, started from 5.2024, Coord.: L. Nicolas.}\\[-5pt]
\justify I participate on studying the possibility to record plant's chemical stresses (Arabidopsis thaliana and Glyceria maxima) with a conducting-polymer electronic-nose prototype in a controlled environment (phytotron).\\[-5pt]

\centering ---\\
\Large The \hypertarget{Sensation}{SENSATION} project \textit{funded by} the ANR/JCJC \\ \normalsize
\textit{\textbf{SENSing hardware burnt to feel physically non-trivial InformATION}, \\started from 10.2022, Coord.: \textbf{S. Pecqueur}, 275~203 €.} \href{https://anr.fr/Projet-ANR-22-CE24-0001}{ANR-22-CE24-0001}\\[-5pt]
\justify I coordinate a national project on studying conducting-polymer morphogenesis as an electrochemical mechanism to program sensing hardware recognizing environmental patterns.\\[-5pt]

\centering ---\\
\Large The \hypertarget{Sens-PestOLS}{SENS-PESTOLS} project \textit{funded by} the ANR/PRCE \\ \normalsize
\textit{\textbf{SENSing PESTicides On Labelled Surfaces}, started from 5.2021, Coord.: J.-C. Rain.} \href{https://anr.fr/Projet-ANR-20-CE04-0013}{ANR-20-CE04-0013}\\[-5pt]
\justify I lead the "Sensor" work-package and participate as an organic electronic technologist on developing a sensing platform exploiting multiplexed organic electrochemical transistors for pollution classification.\\[-5pt]

\centering ---\\
\Large The \hypertarget{MicroMacro}{MICROMACRO} project \textit{funded by} the CNRS/MITI \\ \normalsize
\textit{\textbf{"Microcapteurs de Macrocages" pour des outils non-intrusifs de monitoring {\'e}l{\'e}mentaire dynamique à l'{\'e}chelle des micro-syst{\`e}mes vivants}, started from 3.2021 till 12.2022, Coord.: \textbf{S. Pecqueur}, 30~000 €.}\\[-5pt]
\justify I coordinated a project on synthetizing an electropolymerizable oligothiophene graphted with cryptophanes to condition its affinity with cations for heavy metal identification and recognition.\\[-5pt]

\textbf{In addition, I have also had the direct support from ...} \\[2pt]
- In 2024, the University of Lille who supported the expenses for a two-week visit of Dr. Enrique H. Balaguera as an invited lecturer, which strengthened our collaboration.\textsuperscript{\hyperlink{A29}{A29},\hyperlink{O24}{O24}}
 \\[2pt]
  - In 2022, The LST Master program of the University in Lille who welcomed topics on conducting polymer micro-structuring for Ambre Alleon \& Amira Bousbaa.\\[2pt]
- In 2021, I had a six-people shared support of 60~000 € from IEMN's laboratory to initiate a joint-activity on electropolymerizing on sensing transducers, which allowed co-supervising Dr Oumekloul.\\[2pt]
- Since 2020, I have the support of the ICAM engineering school in Lille who welcomes each semester the involvement of two to three students working with Dr. Hafsi and myself on the conducting polymer electronic nose topic, among who, Marie Delarue, Aymar Kingdoum Taka \& Sandra Kenne Koudjou.\\[2pt]
- In 2019, I had a three-people shared support of 4~500 € from IEMN's biosystems department to initiate a joint-activity on conducting polymer micro-structuring.\\[2pt]
- The CNRS/INSIS/IEMN since 10.2019 for an initial 25~000 € support along with my tenure appointment.\\

\textbf{Indirect supports and involvements in projects before tenure appointment includes ...}\\[2pt]
- In 2023, the EU Erasmus+ program granted a stipend to Ewelina Kowalewska for a one-month visit at IEMN.\\[2pt]
- In 2021, the University of Lille granted a full-fellowship to Louis Routier for enrolling a PhD program under the direction of Prof. Lmimouni at IEMN.\\[2pt]
- In 2021, the Hauts-de-France region granted a stipend to Dr. Scholaert for enrolling a PhD program under the direction of Dr. Alibart at IEMN.\\[2pt]
- From 2019 till 2024, the ERC-CoG \href{https://cordis.europa.eu/project/id/773228}{773228} \hypertarget{Ionos}{IONOS} project (grantee: Dr. Alibart), allowed me researching for nine months as a postdoctoral research fellow under the supervision of Dr. Alibart, and then co-supervising Dr. Ghazal, Dr. Janzakova, Dr. Susloparova, Dr. Scholaert, Luc Brulin, Najami Ghodhbane \& Mona Ghazal.\\[2pt]
- Between 2019 and 2022, the University of Monastir in Tunisia granted four stipends to Dr. Boujnah and Wiem Haj Ammar for two- to three-month visits at IEMN.\\[2pt]
- Between 2016 and 2019, the University of Carthage in Tunisia granted a fellowship to Dr. Ferchichi for enrolling a PhD program under the direction of Prof. Lmimouni at IEMN.\\[2pt]
- Between 2016 and 2019, the EU FET-open \href{https://cordis.europa.eu/project/id/664786}{664786} \hypertarget{Record-It}{RECORD-IT} project (PI: Dr. Vuillaume), allowed me researching during three years as a postdoctoral research fellow under the supervision of Dr. Vuillaume with Dr. Alibart, and then co-supervising Dr. Couturier.\\[2pt]
- From 2016, the CNRS \& the Renatech Network for supporting IEMN's cleanroom, with many fabrication/characterization tools, supervised by engineers \& technicians of invaluable expertise \& experience.\\[2pt]
- Prior 2016, the EU MSCA \href{https://cordis.europa.eu/project/id/316832}{316832} \hypertarget{Olimpia}{OLIMPIA} project (PI: M. Muccini) and the different corporate environments I worked in: ETC, Siemens, CDT and Philips where I had the spark for research!

\section{Communities \& Conflict of Interests}
\label{Ch9.2}

- Peer Reviewer for Scientific Journals: the American Chemical Society (\href{https://connect.acspubs.org/physical-chemistry-authors}{J. Phys. Chem.}), the Royal Chemical Society (\href{https://www.rsc.org/journals-books-databases/about-journals/journal-of-materials-chemistry-c}{J. Mater. Chem. C}), the Institute of Physics (\href{https://publishingsupport.iopscience.iop.org/journals/neuromorphic-computing-engineering/about-neuromorphic-computing-engineering/}{Neuromorphic Comput. Eng.}), the Institute of Electrical and Electronics Engineers (\href{https://ieee-cas.org/publication/ieee-transactions-agrifood-electronics}{Trans. Agrifood Electron.}), the Springer Nature Group (\href{https://www.nature.com/ncomms/aims}{Nat. Commun.}), John Wiley \& Sons, Inc. (\href{https://onlinelibrary.wiley.com/page/journal/15213773/homepage/productinformation.html}{Angew. Chem. Int. Ed.} / \href{https://onlinelibrary.wiley.com/page/journal/2199160x/homepage/productinformation.html}{Adv. Electron. Mater.} / \href{https://onlinelibrary.wiley.com/page/journal/18626319/homepage/productinformation.html}{Phys. Status Solidi a} / \href{https://scijournals.onlinelibrary.wiley.com/hub/journal/10970126/homepage/productinformation}{Polym. Int.} / \href{https://onlinelibrary.wiley.com/page/journal/26884062/homepage/productinformation.html}{Small Struct.}), Elsevier B.V. (\href{https://www.sciencedirect.com/journal/organic-electronics}{Org. Electron.} / \href{https://www.sciencedirect.com/journal/analytical-biochemistry}{Anal. Biochem.} / \href{https://www.sciencedirect.com/journal/trends-in-food-science-and-technology}{Trends Food Sci. Technol.}) \& the Multidisciplinary Digital Publishing Institute (\href{https://www.mdpi.com/journal/materials/about}{Materials} / \href{https://www.mdpi.com/journal/sensors/about}{Sensors}).\\[2pt]
- External Reviewer for the French National Research Funding Agency (\href{https://anr.fr/}{ANR}).\\[2pt]
- Copilot of the "IoT makes Sense" joint-activity at IEMN.\\[2pt]
- Contributed to CNRS' GdR OERA's white paper.

\newpage

\begin{figure}
  \centering
  \includegraphics[width=1\columnwidth]{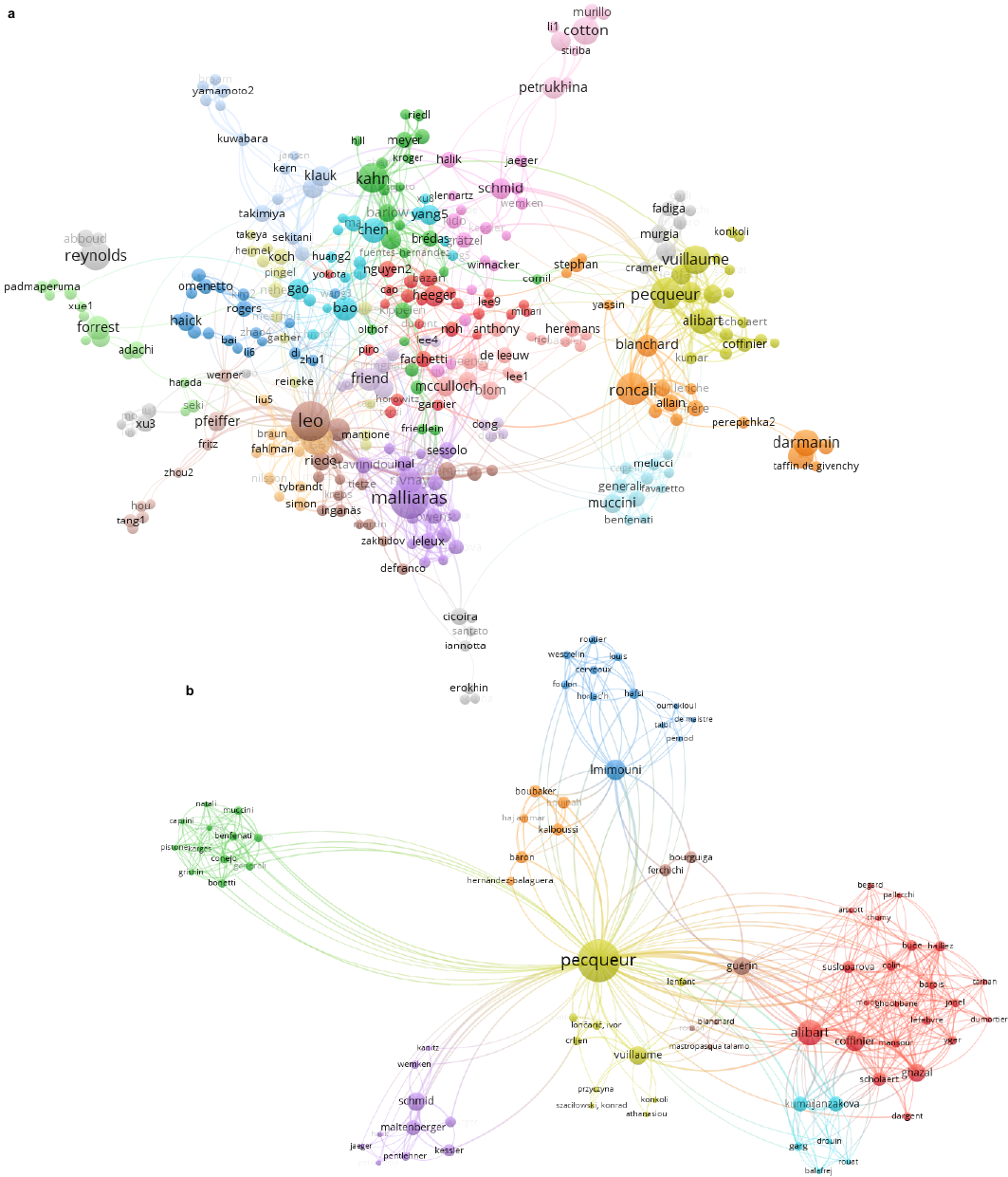}
  \caption{\textbf{Communities \& Collaborators $\vert$ a,} Color map of authors constituting my database of publications (+5k items) I compiled since my doctorate. \textbf{b,} Color map of the subset restricted to the publications I contributed to. Both maps are generated with VOSviewer 1.6.20 with input data cleaned by lastname grouping except firstname initials not matching. The database was not manipulated.}
  \label{fig:figD.3.1}
\end{figure}

\end{appendices}
\end{document}